\documentclass[12pt]{article} 
\usepackage[paper=letterpaper,margin=.8in]{geometry}

\usepackage{amsmath}
\usepackage{setspace}
\usepackage{amsthm}
\usepackage{url}
\theoremstyle{plain}
\usepackage{amssymb}
\usepackage{tensor}
\usepackage{graphicx}
\usepackage{subcaption}
\usepackage{color}
\usepackage{tikz}
\usepackage[numbers,sort&compress]{natbib}
\usepackage[utf8]{inputenc}
\usepackage{braket}
\usepackage{upgreek}
\usepackage[normalem]{ulem}
\usepackage[colorlinks = true, linkcolor = blue, citecolor = red, urlcolor  = blue, anchorcolor = blue]{hyperref}
\newtheorem*{theorem*}{Theorem}
\newtheorem*{definition*}{Definition}

\newcommand{\csch}{\text{csch }}
\newcommand{\cale}{\left(\mathcal{E}_R\right)}
\newcommand{\cali}{\mathcal{I}}
\newcommand{\calm}{\left(\mathcal{M}_R\right)}
\newcommand{\calk}{\mathcal{K}}
\newcommand{\psifhh}{\Psi_{f_{HH}}}
\newcommand{\hhket}{\ket{\text{HH}}}
\newcommand{\hhbra}{\bra{\text{HH}}}
\newcommand{\calc}{\mathcal{C}}
\newcommand{\calo}{\mathcal{O}}
\newcommand{\cals}{\mathcal{S}}
\newcommand{\calctilde}{\tilde{\mathcal{C}}}
\newcommand{\calh}{\mathcal{H}}

\newcommand{\cala}{\mathcal{A}}
\newcommand{\calv}{\mathcal{V}}
\newcommand{\calhobs}{\mathcal{H}_{\text{obs}}}
\newcommand{\calhobsR}{\mathcal{H}_{\text{obs,R}}}
\newcommand{\rhopl}{\rho_{\text{Pl}}}
\newcommand{\iphir}{i \overrightarrow{\partial} \phi_R}

\DeclareMathOperator{\tr}{Tr}

\numberwithin{equation}{section}

\begin{document}
	
 \begin{flushright}
MIT-CTP/5805
\end{flushright}
	
	\thispagestyle{empty}

	\vspace*{2.5cm}
\begin{center}

{\bf {\LARGE Chaos and the Emergence of the \\ Cosmological Horizon}}\\

\begin{center}

\vspace{1cm}

{\bf David K. Kolchmeyer and Hong Liu}\\
 \bigskip \rm

\bigskip 

Center for Theoretical Physics,\\Massachusetts Institute of Technology, Cambridge, MA 02139, USA

\rm
  \end{center}

\vspace{2.5cm}
{\bf Abstract}
\end{center}
\begin{quotation}
\noindent

We construct algebras of diff-invariant observables in a global de Sitter universe with two observers and a free scalar QFT in two dimensions. We work in the strict $G_N \rightarrow 0$ limit, but allow the observers to have an order one mass in cosmic units. The observers are fully quantized. In the limit when the observers have infinite mass and are localized along geodesics at the North and South poles, it was shown in previous work \cite{CLPW} that their algebras are mutually commuting type II$_1$ factors. Away from this limit, we show that the algebras fail to commute and that they are type I non-factors. Physically, this is because the observers' trajectories are uncertain and state-dependent, and they may come into causal contact. We compute out-of-time-ordered correlators along an observer's worldline, and observe a Lyapunov exponent given by $\frac{4 \pi}{\beta_{\text{dS}}}$, as a result of observer recoil and de Sitter expansion. This should be contrasted with results from AdS gravity, and exceeds the chaos bound associated with the de Sitter temperature by a factor of two. We also discuss how the cosmological horizon emerges in the large mass limit and comment on implications for de Sitter holography.

\end{quotation}

	\pagebreak

	{
		\hypersetup{linkcolor=black}
		\tableofcontents
	}


	\pagebreak
	
\section{Introduction}

In recent years, AdS holography has been re-examined from the perspective of emergent von Neumann (vN) algebras \cite{Leutheusser:2021frk,Leutheusser:2021qhd}. The prototypical example is a pair of holographic CFTs in the thermofield double state. In the large $N$ limit, the algebra of single-trace operators in one CFT becomes type III$_1$ above the Hawking-Page temperature $T_{HP}$ and type I below $T_{HP}$. Physically, this corresponds to the presence or absence of a horizon, or to a connected or disconnected bulk. Among other achievements, the algebraic approach to holography has led to a better understanding of generalized entropy \cite{Chandrasekaran:2022eqq, Witten:2021unn, Faulkner:2024gst, Jensen:2023yxy,Kudler-Flam:2023qfl,Akers:2024bel} and quantum chaos \cite{Ouseph:2023juq, Furuya:2023fei,Gesteau:2023rrx}, a more general diagnostic for the connectivity of bulk spacetimes \cite{Engelhardt:2023xer}, a criteron for the existence of stringy horizons \cite{Gesteau:2024rpt}, and improved descriptions of the error-correcting properties of spacetime \cite{Kang:2018xqy, Faulkner:2020hzi, Gesteau:2021jzp,Faulkner:2022ada}.

The notion of subregion-subregion duality in AdS/CFT \cite{Bousso:2012sj,Czech:2012bh,Bousso:2012mh,Hubeny:2012wa,Almheiri:2014lwa,Headrick:2014cta,Soni:2024oim} has been generalized to subregion-subalgebra duality in \cite{Leutheusser:2022bgi}. The boundary dual of a bulk subregion may not have a simple geometric description as the algebra of a boundary subregion \cite{Engelhardt:2023xer}. An example is a causal diamond in global AdS that does not touch the asymptotic boundary. A better understanding of its emergent type III$_1$ algebra\footnote{To be more precise, we are asking whether one can construct a subsystem code with complementary recovery \cite{Harlow:2016vwg} where the two complementary bulk subalgebras are a time-band algebra and its commutant, an interior causal diamond. For related results, see \cite{Bahiru:2022oas,Bahiru:2023zlc}.} would shed light on the emergence of spacetime in closed universes, where there are no boundaries.

Recent works have studied physically interesting vN algebras in closed universes with positive cosmological constant \cite{CLPW, Witten, Chen:2024rpx, Kudler-Flam:2024psh, Kaplan:2024xyk, Fewster:2024pur, Gomez:2023jbg}. Because there are no spatial boundaries to dress local operators to, the spacetime must have nontrivial features. These features can be a part of the classical background, such as a rolling inflaton \cite{Kudler-Flam:2024psh, Chen:2024rpx}, or they can be quantum \cite{Kaplan:2024xyk}. A simple setup involves global de Sitter space with two antipodal observers carrying clocks, such that local QFT operators can be dressed to any point on either observer's worldline.\footnote{de Sitter space with a single observer is sensible as well, as long as the configuration of the matter particles is such that Gauss's law is obeyed. When two observers are present, Gauss's law permits the matter QFT to be in the vacuum.} This was first studied in \cite{CLPW}. We refer to the observers as right and left observers. The causal development of an observer's worldline defines their static patch. A key finding is that the algebra of right static patch observables dressed to the right observer $\cala_R$ is a type II$_1$ factor whose commutant is the analogously defined algebra for the left observer $\cala_L$. This implies that the entire spacetime algebraically factorizes into two static patches, and that the entropy of a reduced density matrix on one of the static patches is well-defined. Moreover, the authors of \cite{CLPW} showed that this entropy is the generalized entropy of the cosmological horizon and that the Hartle-Hawking state has maximum entropy. This is consistent with the earlier ideas that the static patch is described by a finite-dimensional Hilbert space whose size is given by the Gibbons-Hawking entropy, and that the empty static patch has a maximally mixed density matrix \cite{Banks:2023uit,Bousso:2000md, Bousso:2000nf, Banks:2000fe, Banks:2001yp, Parikh:2004wh,Banks:2005bm, Banks:2006rx, Banks:2018ypk, Susskind:2021omt, Susskind:2021dfc,Dong:2018cuv,Lin:2022nss}.

In the analogous AdS context, $\cala_L$ and $\cala_R$ describe the left and right exterior wedges of a two-sided black hole, which is known to be nonperturbatively dual to two copies of a holographic CFT \cite{Maldacena:2001kr}. The algebraic factorization of $\cala_L$ and $\cala_R$ is inherited from the tensor product structure of the nonperturbative dual. From a bottom-up perspective, more evidence in favor of a factorized dual comes from the fact that in AdS, there are no traversable wormholes connecting decoupled asymptotic boundaries \cite{Gao:2016bin}. This ensures that operators on the left boundary always commute with operators on the right boundary. Furthermore, in the toy model of JT gravity with matter, where metric fluctuations can be quantized exactly, the boundary algebras continue to factorize \cite{Penington:2023dql,Kolchmeyer:2023gwa}.

Compared to AdS, the factorizing properties of $\cala_L$ and $\cala_R$ in the de Sitter setup of \cite{CLPW} seem much more mysterious, if not coincidental. This is because any small perturbation in the vicinity of an observer will deflect their trajectory. Due to the expansion of de Sitter space, the observer will eventually come into causal contact with the other observer. This is analogous to the shockwave de Sitter geometry studied in \cite{Aalsma:2020aib}, where the two static patches come into causal contact. The authors of \cite{CLPW} avoided this scenario by working in the limit where $G_N \rightarrow 0$ and the mass of the observers is infinite, such that they cannot be moved away from their respective podes. The limit $G_N \rightarrow 0$ is taken first so that the observers do not backreact on the spacetime. In this paper, 
we refer to the infinite-mass limit as the semiclassical limit.

There exist proposals for de Sitter holography that place a holographic screen along an observer's worldline \cite{Anninos:2011af,Narovlansky:2023lfz}. If the algebra of quantum fields along an observer's worldline emerges from a nonperturbative holographic dual, it is important to study this algebra in as many backgrounds and parameter ranges as possible. Previously, \cite{Witten} studied a family of backgrounds with different cosmological constants. In the present paper, we fix the background to be global de Sitter with a given cosmological constant, and we continue to work in the $G_N \rightarrow 0$ limit, but we allow the observers to have an order one mass in cosmic units. Thus, the two-observer setup of \cite{CLPW} is the semiclassical limit of our setup. Another scenario is when the mass of the observers scales as $\frac{1}{G_N}$ in the $G_N \rightarrow 0$ limit such that they backreact on the spacetime. The classical phase space that describes the relevant geometries was quantized in \cite{Verlinde:2024znh}. In future work, we would like to study this scenario from an algebraic perspective.

We now describe our setup and main results in more detail.

{\bf \vspace{.3 cm} \noindent Dynamical Observers \vspace{.3 cm}}

Because the two observers in our setup have order one masses (and hence nonzero Compton wavelengths), their states should be described using wavefunctions in a Hilbert space rather than classical geodesics. Following \cite{CLPW}, we model an observer as a relativistic particle whose mass is promoted to a continuous quantum number, restricted to be above some cutoff. The observer can be thought of as a clock, or a detector, which has different rest masses in different quantum states. For concreteness and computational simplicity, we take the matter QFT to be a free massive scalar field, and we work in two dimensions. The three-dimensional isometry group of dS$_2$ is gauged, reflecting the gravitational constraints. The physical Hilbert space is the gauge-invariant subspace of the tensor product of two observer Hilbert spaces and the QFT Hilbert space.\footnote{The exact definition of the physical Hilbert space is more subtle, and is the subject of Section \ref{sec:observers}.} We demonstrate how any local QFT operator, including higher-spin operators, can be dressed to an observer. These dressed operators are localized on the observer's worldline. The algebra $\cala_R$ is defined to be the algebra generated by all such local operators dressed to the right observer together with the right observer's Hamiltonian $H_R$ (namely, the operator that measures their mass). The algebra $\cala_L$ is defined analogously.

A key difference between our setup and \cite{CLPW} is that $\cala_R$ and $\cala_L$ do not commute with each other. This is because our observers are not confined to the North and South poles, so they may come into causal contact. We will argue that the commutant of $\cala_R$ is generated by the left observer's Hamiltonian. Intuitively, this is because the observers may explore the entire spacetime, and in quantum field theory, the algebra associated to an entire spacetime (or to an entire Cauchy slice) has trivial commutant. Because $\cala_R$ is by definition a vN algebra, $\cala_R$ is equal to its double-commutant, or the commutant of the commutative algebra generated by the left observer's Hamiltonian. We then find that $\cala_R$ is a direct integral of type I$_\infty$ factors, and the center is generated by the left observer's Hamiltonian. The gravitational constraints are so strong that the left observer's Hamiltonian is in the right observer's algebra. This is also true in pure JT gravity \cite{Penington:2023dql,Kolchmeyer:2023gwa}. In JT gravity with matter, the two boundary algebras factorize. In contrast, including matter in our de Sitter setup does not factorize the observers' algebras. From this perspective, the two observers may not be viewed as a partition of a quantum system into two independent subsystems.

Each type I$_\infty$ factor that appears in the direct integral refers to the algebra of all operators that act on the subspace of the physical Hilbert space in which the left observer has a definite energy. The unique trace on each factor simply counts the states in this subspace. Note that $\cala_R$ does not have a unique trace because it is a direct integral of factors. In contrast, \cite{CLPW} worked in the semiclassical limit. They found that the observer's algebra is a type II$_1$ factor, which has a unique trace. This trace does not explicitly count states in a Hilbert space, and its corresponding entropy is a generalized entropy.

To reiterate a key point, our results on the algebraic structure of $\cala_R$ crucially depend on the statement that the commutant of $\cala_R$ (which we denote by $\cala_R^\prime$) is generated by the left observer's Hamiltonian. We can only show this explicitly in the special case that the matter QFT is a massless chiral scalar, for reasons explained in Section \ref{sec:algebra}. We briefly sketch our argument here. Let $\ket{\Psi}$ be a state in the physical Hilbert space in which the matter fields are in the Bunch-Davies vacuum (we call the space of such states the vacuum sector). Let $\calo \in \cala_R^\prime$. Define $\ket{\Phi} := \calo \ket{\Psi}$, and assume for the sake of a proof by contradiction that $\ket{\Phi}$ has nonzero overlap with states outside the vacuum sector. We construct an operator $\calc_R \in \cala_R$ that obeys $\calc_R \ket{\Psi} = 0$ and $\calc_R \ket{\Phi} \neq 0$. It follows that
\begin{equation}
    0 = \calo \calc_R \ket{\Psi} =  \calc_R \calo \ket{\Psi} =  \calc_R \ket{\Phi} \neq 0,
\end{equation}
which is a contradiction. Thus, operators in $\cala_R^\prime$ map the vacuum sector to the vacuum sector. We also show that the vacuum sector contains a dense subspace of states that are cyclic with respect to $\cala_R$. Thus, the action of an operator in $\cala_R^\prime$ on any state in the Hilbert space is completely fixed by how it acts on the vacuum sector. Basis states in the vacuum sector are labeled by a single energy, $\ket{E}$. Using similar techniques, we show that an operator in $\cala_R^\prime$ cannot change the energy. Thus, the most general action of an operator $\calo \in \cala_R^\prime$ on $\ket{E}$ is $\calo \ket{E} = f_\calo(E) \ket{E}$ for some function $f_\calo(E)$. It follows that the action of $\calo$ on the Hilbert space is completely fixed to be $\calo = f_\calo(H_L)$, which is a function of the left observer's Hamiltonian.

We believe that this argument can be generalized to the case where the matter QFT is arbitrary. The vacuum sector is special due to the gravitational constraints. For instance, the classical trajectories of two observers in the vacuum sector are completely fixed. The observers must sit at each other's antipodes. In states with matter excitations, the observers can be in more general configurations. Thus, it is not surprising that operators in $\cala_R$ obey special identities only when acting on the vacuum sector.

{\bf \vspace{.3 cm} \noindent Emergence of cosmological horizon \vspace{.3 cm}}

The type II$_1$ algebra found in \cite{CLPW} (and its associated cosmological horizon) emerges from our type I algebra in the semiclassical limit. This is analogous to, but has important differences from, how the type III$_1$ algebra of the right wedge and the horizon of a two-sided black hole in AdS emerges from the type I algebra of a boundary CFT in the large $N$ limit for temperatures $T > T_{HP}$. 

Unlike the two type I algebras associated to a pair of finite-$N$ holographic CFTs, here $\cala_R$ and $\cala_L$ do not commute and hence do not factorize. So even the factorizing structure of the type II$_1$ algebras is emergent. Thus, it is unlikely that quantum gravity in global de Sitter space is nonperturbatively described by a pair of factorizing type I algebras.

In our setup, the role of $N$ is played by the observer's mass. To take the semiclassical limit, we project onto the subspace of the physical Hilbert space where the masses of the observers are restricted to be in a microcanonical window of order one size centered around $\Lambda$, which is a large parameter that is precisely analogous to $N$. The Hamiltonians $H_R$ and $H_L$ are ill-defined in the semiclassical limit $\Lambda \rightarrow \infty$, but the operators $H_R - \Lambda$ and $H_L - \Lambda$ are well-defined. In both our work and in \cite{CLPW}, the observer's vN algebra is generated by operators that are localized on their worldline. In the semiclassical limit, the operators on the worldline are only able to generate operators that are localized within the observer's cosmological horizon, resulting in a type II$_1$ algebra. This algebra has a nontrivial commutant, which includes operators in the other static patch that are dressed to the other observer. In our work, the observer's trajectory is dynamical and subject to quantum uncertainty. Because the observer's algebra away from the semiclassical limit has an almost-trivial commutant, the notion of horizon does not make sense in this regime. To recover the type II$_1$ algebra in the semiclassical limit, one should consider operators in $\cala_R$ whose expression in terms of the generators only depends on $H_R$ and $\Lambda$ through the combination $H_R - \Lambda$. In the semiclassical limit, these operators become operators in the type II$_1$ algebra.\footnote{In Section \ref{sec:hhstate}, we demonstrate that correlators of simple operators in the Hartle-Hawking state reduce to the expected semiclassical correlators in the semiclassical limit.} This is analogous to how simple, $N$-independent combinations of single trace operators in a holographic CFT define the causal wedge at large $N$.\footnote{In our setup, there also exist operators in $\cala_R$ which, in the semiclassical limit, become operators outside the observer's cosmological horizon. The expression for these operators in terms of the generators necessarily has a more general dependence on $\Lambda$. Roughly speaking, the expression should involve local operators on the observer's worldline interspersed with time evolution of order the scrambling time generated by $H_R - \Lambda$. The scrambling time is logarithmic in $\Lambda$. The local operators backreact on the observer, and the deflection of their trajectory becomes significant after a scrambling time. In AdS/CFT, there are analogous algorithms that take advantage of gravitational backreaction to reconstruct operators beyond the causal wedge \cite{Levine:2020upy,Engelhardt:2021mue}.} Thus, the type II$_1$ algebras associated to the two static patches in \cite{CLPW} emerge in the semiclassical limit from the type I algebras $\cala_R$ and $\cala_L$. That is, the cosmological horizons emerge from a description that has no notion of a horizon.

To explicitly see how a semiclassical type II$_1$ algebra emerges from a type I algebra, it is important to understand the definition of the Hartle-Hawking state away from the semiclassical limit. Note that this state has been conjectured to make sense in any background \cite{Witten}. In \cite{CLPW}, the Hartle-Hawking state may be defined algebraically as the state whose expectation value furnishes the unique trace (up to normalization) on the type II$_1$ factors associated to the static patches. Because $\cala_L$ and $\cala_R$ are not factors in our work, this definition cannot be used. Let $\calo_L$ be a local QFT operator dressed to the left observer, and let $\calo_R$ be the same local QFT operator dressed to the right observer. We define the Hartle-Hawking state to be the unique state $\ket{\Psi_{\text{HH}}}$ that obeys\footnote{We have checked that this is true for various operators in our free scalar field theory, and we expect it to be true for general operators in a general QFT.}
\begin{equation}
	\calo_L \ket{\Psi_{\text{HH}}} =  \Theta \left(\calo_R\right)^\dagger \Theta \ket{\Psi_{\text{HH}}}
	\label{eq:hhpropintro}
\end{equation}
and does not depend on the local QFT operator under consideration. Here, $\Theta$ is the spacetime CPT operator. If the algebras $\cala_L$ and $\cala_R$ were factors obeying $\cala_L^\prime = \cala_R$, then \eqref{eq:hhpropintro} would imply that the Hartle-Hawking state is tracial and has maximum entropy, in agreement with \cite{CLPW}. We instead find that the Hartle-Hawking state is not a trace on $\cala_R$ or $\cala_L$, in contrast to \cite{CLPW, Witten, Chen:2024rpx}. The Hartle-Hawking state is only tracial in the semiclassical limit. Furthermore, the Hartle-Hawking state defines a probability distribution for measurements of an observer's mass. In \cite{CLPW}, the distribution corresponds to the canonical ensemble at the de Sitter temperature, $p(m) \sim e^{- 2 \pi m}$. In our setup, we derive a more general expression. Define $s > 0$ as follows,
\begin{equation}
    s := \sqrt{m^2 - \frac{1}{4}}.
\end{equation}
Our more general distribution has support on $m > \frac{1}{2}$. It is
\begin{equation}
    p(s) \, ds = \frac{s \tanh \pi s}{4 \pi^2 \cosh^2 (\pi s)} \, ds,
\end{equation}
which decays as $e^{- 2 \pi m}$ for large masses, in agreement with semiclassical expectations.  

{\bf \vspace{.3 cm} \noindent Chaos induced from dressing \vspace{.3 cm}}

To probe the putative holographic dual, we compute correlators of a dressed scalar field in the Hartle-Hawking state. The results agree with \cite{CLPW} in the semiclassical limit of large observer mass, as expected. Using these correlators, we demonstrate the chaotic behavior of the putative holographic dual. In particular, we study the out-of-time-ordered correlator (OTOC) of four operators along an observer's worldline in the semiclassical limit. The time separation between the two pairs of operators is of order the scrambling time, which is logarithmic in the observer's mass in cosmic units. We demonstrate that the OTOC decays on this timescale, and we find that the Lyapunov exponent, which is the exponent in the leading correction to the OTOC, is $\frac{4 \pi}{\beta_{\text{dS}}}$. Note that in QFT on de Sitter space, vacuum correlation functions of operators restricted to the North Pole obey the KMS condition for inverse temperature $\beta_{\text{dS}} = 2 \pi$. If one naively assumes that these correlators emerge from a large $N$ chaotic quantum mechanical theory in a thermal state with $\beta = \beta_{\text{dS}}$, one would expect that the Lyapunov exponent in our OTOC should be no greater than $\frac{2 \pi}{\beta_{\text{dS}}}$ due to the chaos bound of \cite{Maldacena:2015waa}. Because the Lyapunov exponent we find exceeds the chaos bound by a factor of two, one of the assumptions in \cite{Maldacena:2015waa} must not apply to the putative holographic dual.\footnote{A natural proposal is that the putative dual is actually in the infinite-temperature state \cite{Lin:2022nss,Rahman:2024vyg}, so there is no bound.} The factor of two can be explained from simple geometric considerations, and is expected to be present in higher dimensions.\footnote{This exponent has been observed in a related but different context in three dimensions \cite{Aalsma:2020aib}. See equation (3.29). There, the decay of the OTOC is due to a dynamical metric rather than a dynamical observer.} In the semiclassical limit, the putative dual must exhibit thermal correlators at inverse temperature $\beta_{\text{dS}}$ and a Lyapunov exponent of $\frac{4\pi}{\beta_{\text{dS}}}$. We also describe another difference between the OTOCs we study and OTOCs in AdS involving cases when the OTOC may grow initially before decaying.

In summary, this work presents a bottom-up study of the algebraic and chaotic features of de Sitter space with fully dynamical observers.\footnote{Other bottom-up studies of chaos and scrambling in de Sitter include \cite{Milekhin:2024vbb,Aalsma:2020aib,Susskind:2021esx}.} We find that the factorizing type II$_1$ algebras of \cite{CLPW} (and their associated horizons) emerge from two non-factorizing type I algebras. We also find features of the OTOC that are puzzling from the perspective of AdS holography. We hope that our results can eventually be reconciled with top-down proposals involving the double-scaled SYK model \cite{Narovlansky:2023lfz}.

{\bf \vspace{1 cm} \noindent Plan of the paper \vspace{.3 cm}}

We summarize the contents of the remainder of this paper. In Section \ref{sec:CLPW}, we review the aspects of \cite{CLPW} that are necessary for making comparisons with our work. In Section \ref{sec:freescalar}, we review the quantization of a massive free scalar in dS$_2$. In Section \ref{sec:observers}, we discuss our model of an observer and construct a gauge-invariant Hilbert space that describes two observers and a matter QFT. In Section \ref{sec:gaugeinvariant}, we describe how local QFT operators can be dressed to an observer's worldline. In Section \ref{sec:hhstate}, we introduce the Hartle-Hawking state and compute correlation functions, including various OTOCs. In particular, we show that the OTOC may be interpreted as an eikonal scattering amplitude (in analogy to \cite{Shenker:2014cwa}), and we explain the origin of the $\frac{4 \pi}{\beta_{\text{dS}}}$ Lyapunov exponent. In Section \ref{sec:algebra} we define the algebras $\cala_R$ and $\cala_L$, argue for their algebraic properties, and explain the difference between the traces that appear in and away from the semiclassical limit. In Section \ref{sec:candidatetypeii}, which is separate from the main discussion in this paper, we present a modified setup in which the Hartle-Hawking state furnishes a trace on the observer's algebra, which appears to be type II$_\infty$. The semiclassical limit of this algebra is real and commutative, and we speculate that it may be related to large $N$ factorization in the putative holographic dual or gauged time-reversal symmetry in the bulk. The appendices contain various technical results, such as a calculation that directly links a 6j symbol of $\mathfrak{so}(2,1)$ to the eikonal scattering amplitude that appears in the OTOC.

\section{The setup of CLPW}

\label{sec:CLPW}

\begin{figure}[h!]
	\centering
	\begin{tikzpicture}
		
		\draw[line width=1.5pt] (0,0) rectangle (5,5);
		
		\draw[line width=1.5pt] (0,0) -- (5,5);
		\draw[line width=1.5pt] (5,0) -- (0,5);
		
		\fill (0,3.75) circle (3pt);
		
		\fill (5,2.5) circle (3pt);
		
		\draw[->, thick] (-0.2,2.5) -- (-0.2,3.75);
		
		\node[left] at (-0.4,3.125) {$p$};
		
		\node[right] at (5,1.25) {$N$};
		
		\node[left] at (0,1.25) {$S$};
		
	\end{tikzpicture}
	\caption{There are observers located at the North and South poles of de Sitter space (labeled $N$ and $S$), and each observer carries a clock. We sometimes refer to the North (resp. South) observer as the right (resp. left) observer. The dots represent the points along the observers' worldlines where the clocks read zero. The boost isometry that preserves the wedges moves the dots along the worldlines, and we work in a gauge where the right dot is fixed at the location shown. The location of the left dot is given by $p$, measured in units of proper time. In the text, we use a time coordinate $t$ that measures proper time along the North and South pole geodesics and always increases to the future, or toward the top of the diagram.}
	\label{fig:penrose}
\end{figure}
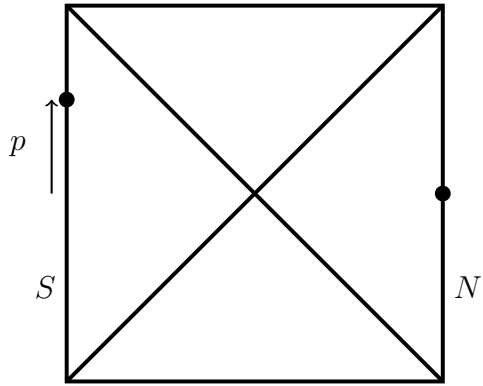

\vspace{10pt}

In this section, we review the aspects of \cite{CLPW} that are most relevant for this paper. More details may be found in \cite{CLPW,Witten}. We consider a classical background given by global de Sitter space with two observers moving along geodesics at the North and South poles. Each observer carries a clock. Classically, the configuration of an observer's clock is specified by the point on their worldline where it reads zero,\footnote{We assume throughout this paper that the clocks are forward-going clocks, such that the reading on the clock increases to the future of the Penrose diagram.
} as shown in Figure \ref{fig:penrose}. We choose to quantize the clocks. Furthermore, there is a QFT propagating on this background.  The classical background is preserved by a boost isometry, as well as rotations when $d > 2$.\footnote{We will comment on the discrete isometries shortly.} For simplicity, we will restrict our attention to $d = 2$. The boost isometry is a diffeomorphism that must be gauged, and we choose a gauge that freezes the point where the right clock reads zero. The clock configurations are thus specified by a single time, $p$. Hence, the Hilbert space is $L^2(\mathbb{R}) \otimes \calh_{\text{QFT}}$. Let $q$ be canonically conjugate to $p$, such that $[q,p] = i$. The left Hamiltonian $H_L$, which is defined to be the generator of the left observer's time evolution,\footnote{To be precise, acting with $e^{-i H_L t}$ advances the left observer's clock by $t$, which decreases $p$ by $t$.} is equal to $q$. Due to our gauge choice, the right Hamiltonian $H_R$ acts by a combined transformation on $p$ and the QFT. We have that\footnote{In most of this paper, we work away from the semiclassical limit and let $H_R$ and $H_L$ refer to the observers' masses. In this section, $H_R$ and $H_L$ refer to the differences between the observers' masses and an infinite additive constant, $\Lambda$. This is the more natural definition of Hamiltonian in the semiclassical limit. In the semiclassical limit, the observers' masses are strictly infinite, and it only makes sense to discuss fluctuations of the mass.}
\begin{equation}
H_L = q, \quad H_R = -h + q,
\label{eq:2.1}
\end{equation}
where $h$ generates the static patch time translation isometry on $\calh_{\text{QFT}}$. Let $\varphi_R(t)$ (resp. $\varphi_L(t)$) be a local scalar QFT operator on the right (resp. left) geodesic at time $t$, where $t$ measures proper time and increases to the future (that is, towards the top of the Penrose diagram).\footnote{This time variable coincides with the $t$ coordinate in \eqref{eq:metric}.} Our conventions for $h$ are such that
\begin{equation}
\varphi_R(t) = e^{i h t} \varphi_R(0) e^{- i h t}, \quad \varphi_L(t) = e^{-i h t} \varphi_L(0) e^{i h t}.
\label{eq:varphiR}
\end{equation}
If we write $h = h_R - h_L$,
 then \eqref{eq:2.1} implies that
 \begin{equation}
     H_R + h_R =  H_L + h_L, 
 \end{equation}
 which says that the left static patch energy (which is the sum of observer and matter energies) must agree with the right static patch energy. This is Gauss's law. We define $\phi_R(\tau)$ to be the local QFT operator $\varphi_R(t)$ evaluated at the spacetime point where the right observer's clock reads $\tau$. We define $\phi_L(\tau)$ analogously. We have that
\begin{equation}
\label{eq:2.3}
    \phi_R(\tau) := \varphi_R(\tau), \quad \phi_L(\tau) := \varphi_L(\tau+p).
\end{equation}
For convenience, we define 
\begin{equation}
\phi_R := \phi_R(0), \quad \quad \phi_L := \phi_L(0).
\end{equation} 
From these definitions, it follows that
\begin{equation}
    \phi_R(\tau) = e^{-i H_R \tau} \phi_R e^{i H_R \tau}, \quad \phi_L(\tau) = e^{- i H_L \tau} \phi_L e^{ i H_L \tau}.    \label{eq:timeevolvewithhamiltonian}
\end{equation}
One may verify that $H_R$ and $\phi_R$ commute with $H_L$ and $\phi_L$. The right observer's algebra is generated by $H_R$, $\phi_R$, and all other local QFT operators in a small tube around the observer's worldline that are dressed to the observer \cite{Strohmaier:2023opz,Witten}. The algebra includes all QFT operators that are dressed to the observer and contained within the observer's static patch.\footnote{One way to ``dress'' a local operator to an observer is to evaluate the operator at the antipode of the observer. This is excluded from the observer's algebra because it cannot be generated by local QFT operators in a small tube around the observer's worldline.} The left observer's algebra is defined analogously. These algebras are crossed-product algebras, as shown in \cite{CLPW,Witten}. The crossed-product algebra is a Type II$_\infty$ factor that becomes Type II$_1$ after applying a projection onto the subspace where $H_L > 0$ and $H_R > 0$. In order to facilitate a comparison with other calculations in this paper, we choose to not apply this projection. Later in this paper, we will treat the observers as relativistic quantum particles. These particles move along classical geodesics only in the large mass limit. We interpret the variable $q$ as an energy fluctuation measured with respect to a large reference mass $\Lambda$ that goes to infinity. Hence, $q \in (-\infty, \infty)$.

There is a distinguished state, $\hhket$, that furnishes a trace on the crossed-product algebra. Up to normalization, it is defined by\footnote{Because $q \in (-\infty,\infty)$ in this discussion, the function $e^{- \pi q}$ specifies a non-normalizable state. It is better to refer to \eqref{eq:hh} as the Hartle-Hawking weight. The Hartle-Hawking weight becomes the Hartle-Hawking state after restricting to a subspace where $q$ is bounded from below.} \cite{Witten}
\begin{equation}
\hhket := e^{- \pi q} \otimes \ket{0},
\label{eq:hh}
\end{equation}
where $\ket{0}$ is the Bunch-Davies vacuum of the QFT. It follows that\footnote{For readers familiar with Tomita-Takesaki theory \cite{Witten:2018zxz}, \eqref{eq:2.5} can be recast as follows. Let $a \in \cala$, the type III$_1$ algebra of the left static patch. Using $S a \ket{0} = a^\dagger \ket{0}$ as well as $S =  J \Delta^{1/2}$ and $J^2 = 1$, it follows that $ \Delta^{1/2} a \ket{0} = J a^\dagger J \ket{0}$. Note that $J a^\dagger J \in \cala^\prime$, which is the right static patch algebra. In our notations, $e^{h \pi} = \Delta^{1/2}$.}
\begin{equation}
\label{eq:2.5}
\phi_L \hhket = e^{ h \pi} \varphi_L(0) e^{ - h \pi } e^{- \pi q} \otimes \ket{0} =  \varphi_L( i \pi )  e^{- \pi q} \otimes \ket{0} =  \varphi_R( 0 )  e^{- \pi q} \otimes \ket{0} = \phi_R \hhket.
\end{equation}
Thus, we have that
\begin{equation}
H_L \hhket = H_R \hhket, \quad \phi_L \hhket = \phi_R \hhket.
\label{eq:lr}
\end{equation}
We define the trace as follows:
\begin{equation}
\tr( \cdots ) := \hhbra \cdots \hhket.
\end{equation}
Because the left operators commute with the right operators, one may use \eqref{eq:lr} to demonstrate the cyclic property of the trace on expressions built from $H_R$ and $\phi_R$. The two-point function is given by

\begin{equation}
	\label{eq:2.8}
\tr( e^{-i H_R \tau_2} \phi_R e^{-i H_R \tau_1} \phi_R ) = \int_{-\infty}^\infty dq \, e^{-2 \pi q - i q (\tau_2 + \tau_1)} \braket{0| \varphi_R(0) \varphi_R(\tau_1)  |0}.
\end{equation}
Higher-point functions may be computed analogously. Although \eqref{eq:2.8} is not convergent, in Section \ref{sec:two-pointfunction}, we will interpret \eqref{eq:2.8} as a special case of a more general convergent expression, \eqref{eq:twodomains}, where the observer's mass is naturally bounded from below. This two-point function may be viewed as a physically-motivated regulator for \eqref{eq:2.8}. Another way to regulate is to project the algebra onto the $q > 0$ subspace. With this regulator, the two-point function is given in (3.32) of \cite{Witten}. To facilitate a direct comparison of finite quantities, we consider the slightly modified expression
\begin{equation}
\tr( e^{-i H_R \tau_2} \phi_R e^{-i H_R \tau_1} \phi_R \Theta_{-\frac{\epsilon}{2},\frac{\epsilon}{2}}(H_R) ) = \int_{-\frac{\epsilon}{2}}^{\frac{\epsilon}{2}} dq \, e^{-2 \pi q - i q (\tau_2 + \tau_1)} \braket{0| \varphi_R(0) \varphi_R(\tau_1)  |0},
\label{eq:comparefirst}
\end{equation}
where the function $\Theta_{a,b}(x)$ is equal to $1$ for $x \in (a,b)$ and zero otherwise. We will directly compare this with \eqref{eq:6.17} in Section \ref{sec:two-pointfunction}. Our method of regulating the $q$ integral in \eqref{eq:comparefirst} is equivalent to computing the correlator in a microcanonical Hartle-Hawking state instead of the Hartle-Hawking state (this is analogous to the microcanonical TFD in \cite{Chandrasekaran:2022eqq}).

Also, note that \eqref{eq:2.8} is not obviously invariant under $\tau_2 \leftrightarrow \tau_1$. To understand why \eqref{eq:2.8} arises from a trace, one may replace $e^{- i H_R \tau}$ with a smeared function $ \int_{-\infty}^\infty du \, e^{-(u-\tau)^2}e^{- i H_R u}$ and then apply the discussion that leads to equation (3.45) in \cite{Witten:2021unn}.

\begin{figure}[h!]
    \centering
    \includegraphics[width=0.3\linewidth]{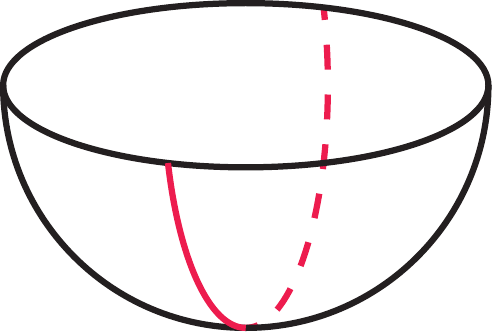}    
    \caption{A geodesic through the Euclidean hemisphere that connects the North and South poles of de Sitter. This geodesic represents a semiclassical Euclidean path integral calculation of the Hartle-Hawking state.}
    \label{fig:hh}
\end{figure}

The Hartle-Hawking state is usually understood to be the state prepared by the Euclidean path integral with a single boundary. For a QFT on global de Sitter, this state is the Bunch-Davies vacuum, prepared using the Euclidean path integral on a hemisphere. This was extended to include observers in \cite{Witten}. In the Lorentzian spacetime, the observers sit at the North and South poles. Their geodesics may be naturally continued into Euclidean signature such that they meet, as shown in Figure \ref{fig:hh}. The semiclassical path integral for a particle of mass $m$ to propagate from the North to the South pole is dominated by a geodesic saddle. Because the length of the geodesic in Euclidean signature is $\pi$, the result is $e^{- \pi m}$. Because the mass of a clock is conjugate to its time, this explains how a semiclassical Euclidean path integral accounts for the $e^{- \pi q}$ factor in \eqref{eq:hh}.

\subsection{Gauging the discrete isometries}

\label{sec:gaugediscrete}

Because gravity is a theory of gauged diffeomorphisms, any isometries of the background must be gauged \cite{Harlow:2023hjb}. Assuming that the two observers are distinguishable (such that a rotation by $\pi$, which would exchange the observers, is not an isometry), the isometries of the background are generated by boosts, time-reversal, and parity. The gauge-fixing choice made in the previous subsection (and shown in Figure \ref{fig:penrose}) is only sufficient to gauge-fix the boosts. Parity leaves the configurations of the observers intact and only acts nontrivially on the QFT. Note that parity preserves each static patch.\footnote{For dimensions $d > 2$, each point away from the poles on the Penrose diagram in Figure \ref{fig:penrose} refers to an $S_{d-2}$. In our setting, $d = 2$, each point away from the poles refers to two points. The action of parity simply interchanges these points.} Thus, if $\cala$ is the Type III QFT algebra of one of the static patches, then the gauge-invariant algebra that goes into the crossed-product construction is the subalgebra of $\cala$ that is invariant under the parity automorphism, which is also Type III.

Unlike the other isometries discussed above, time-reversal is implemented by an antiunitary operator on the Hilbert space. The Hilbert space after gauge-fixing only the boosts is $L^2(\mathbb{R}) \otimes \calh_{QFT}$. The time reversal operator $T$ commutes with $q$ but anticommutes with $p$. A general state in the Hilbert space may be written as $\ket{\psi(p)} \in \calh_{QFT}$, or a function of the time shift $p$ that is valued in the QFT Hilbert space. The inner product of two such states is then
\begin{equation}
	\label{eq:ip}
	\int_{-\infty}^\infty dp \, \braket{\psi_1(p)|\psi_2(p)}.
\end{equation}
The states invariant under time reversal obey
\begin{equation}
	\label{eq:trsymmetry}
	\ket{\psi(-p)} = T \ket{\psi(p)},
\end{equation}
and for such states, the inner product \eqref{eq:ip} will always be real. Thus, as pointed out in \cite{Harlow:2023hjb}, the gauge-invariant Hilbert space is real. We may use \eqref{eq:trsymmetry} to rewrite \eqref{eq:ip} as an integral over $p > 0$, which amounts to working in a gauge where the timeshift $p$ is positive. The operators $\phi_R$, $\phi_L$, $H_R$, $H_L$, and the state $\ket{\text{HH}}$ are all invariant under $T$. However, $e^{-i H_R \tau}$ is not invariant.

Let $\cala_R$ be the (complex) von Neumann algebra acting on $L^2(\mathbb{R}) \otimes \calh_{QFT}$ that arises from imposing only the boost gravitational constraint. The fully gauge-invariant algebra associated to the right observer is the $*$-subalgebra of $\cala_R$ that is invariant under all the discrete isometries. This subalgebra is a real von Neumann algebra that acts on a real Hilbert space. A real von Neumann algebra is a $*$-subalgebra of the algebra of all bounded operators on a real Hilbert space that contains the identity and is weakly closed. The double-commutation theorem implies that a real von Neumann algebra is equal to its double-commutant.\footnote{See Theorem 4.3.8 of \cite{Li2003}, as well as \cite{Li1995}.}

When gauging a symmetry, one may consider a new background that includes a nontrivial profile for the gauge field. Background gauge fields for discrete gauge symmetries have been discussed in \cite{Harlow:2023hjb}. In this paper, we focus on a single background, namely the one considered by \cite{CLPW} and depicted in Figure \ref{fig:penrose}. It would be interesting to study algebras in these other related backgrounds in future work. 

In the remainder of this paper, we generalize the setup of \cite{CLPW} such that the observers are relativistic particles that are no longer confined to classical geodesics. We will identify an analogue of the Hartle-Hawking state whose definition is similar to \eqref{eq:hh}, except that $e^{- \pi q}$ is replaced by a more general function that decays like $e^{- \pi q}$ at large $q$. We will define dressed operators analogous to $\phi_R, \phi_L, H_R, H_L$ such that \eqref{eq:lr} holds. However, in our more general setup, we will find that $\hhket$ does not furnish a trace, and that $\phi_L$ and $\phi_R$ do not commute.

In this paper, we mostly focus on the case where only the subgroup $SO^+(2,1)$ of the Lorentz group is gauged. We will thus construct a complex von Neumann algebra associated to an observer. We will also comment on how the states and operators we consider transform under the discrete isometries. Passing from the complex von Neumann algebra to the real von Neumann algebra that is invariant under the discrete isometries is straightforward because the discrete isometries form a finite-dimensional group, which is easier to gauge than the noncompact Lie group $SO^+(2,1)$.

\section{Free scalar in dS$_2$}
\label{sec:freescalar}

In this section, we review the quantization of a free scalar field in dS$_2$. The action is
\begin{equation}
S = - \frac{1}{2} \int d^2 x \sqrt{-g} \, \left(g^{ab} \partial_a \varphi \partial_b \varphi + m^2 \varphi^2\right).
\end{equation}
We work in global coordinates, such that the metric is
\begin{equation}
\label{eq:metric}
ds^2 = -dt^2 + d\theta^2 \, \cosh^2 t, \quad t \in \mathbb{R}, \quad \theta \sim \theta + 2 \pi.
\end{equation}
The field $\varphi$ obeys the Klein-Gordon (KG) equation,
\begin{equation}
\left(-\nabla^2 + m^2\right) \varphi = 0.
\end{equation}
The isometry group of dS$_2$ is the three-dimensional Lorentz group $O(2,1)$. The representation that a single-particle state belongs to depends on the mass. For $m \ge \frac{1}{2}$, the states transform in a principal series representation. For $\frac{1}{2} > m > 0$, states transform in the complementary series. For $m = 0$, single-particle states transform in the discrete series. In this section we restrict our attention to $m \ge \frac{1}{2}$. We will also consider the case $m = 0$ in section \ref{sec:chiralscalar}.

Single-particle states are given by solutions to the KG equation, and their inner product is defined using the KG inner product. Given two solutions $f_1$ and $f_2$, the KG inner product is
\begin{equation}
\braket{f_1|f_2}_{KG}  := i \cosh t \int_0^{2 \pi} d\theta \left[ (f_1)^* \partial_t f_2 - f_2 \partial_t (f_1)^*\right], \label{eq:KGinner}
\end{equation}
where the choice of $t$ is arbitrary. The ``positive-frequency'' solutions that actually belong to the Hilbert space of single-particle states are such that this inner product is positive definite. We define a CPT operator $\Theta$ which maps $f$ to $\Theta f$, defined by
\begin{equation}
(\Theta f)(t,\theta) := (f(-t,-\theta))^*.
\end{equation}
It follows that $\Theta$ is antiunitary:
\begin{equation}
\braket{\Theta f_1 | \Theta f_2}_{KG} = \braket{f_2 | f_1}_{KG}.
\end{equation}
The antiunitary time-reversal operator $T$ is defined by
\begin{equation}
	(T f)(t,\theta) = (f(-t,\theta))^*,
	\label{eq:3.7}
\end{equation}
while the unitary parity operator $P$ is defined by
\begin{equation}
	(P f)(t,\theta) = f(t,-\theta).
	\label{eq:3.8}
\end{equation}
Charge conjugation is trivial in our setup, so $\Theta = PT$.

We denote the positive-frequency solutions that span the Hilbert space of a single particle of mass $m$ by $\psi_n^s(t,\theta)$, where $n \in \mathbb{Z}$ labels the momentum around the spatial circle, and  $s \ge 0$ is related to the mass by
\begin{equation}
m^2 = \frac{1}{4} + s^2.
\label{eq:masss}
\end{equation}
See \eqref{eq:modefunction} for an explicit definition. These solutions obey the following:
\begin{align}
	\label{eq:prop1}
\psi_n^s(t,\theta) &= e^{i n \theta} \psi_n^s(t,0),
\\
\psi_n^s(t,0) &= \psi_{-n}^s(t,0),
\label{eq:prop3}
\\
\psi_n^s(t,\theta) &= (\Theta \psi_n^s)(t,\theta),
\label{eq:prop312}
\\
	\label{eq:prop4}
\braket{\psi_{n_1}^s|\psi_{n_2}^s}_{KG} &= \delta_{n_1,n_2}.
\end{align}
Thus, a parity transformation maps a right-moving excitation to a left-moving excitation, $n \rightarrow - n$. The same is true for time reversal.
There are physically inequivalent ways to choose the $\psi^s_n$ solutions such that the above properties are obeyed. These different choices correspond to different ``$\alpha$-vacua.'' See \cite{Spradlin:2001pw,Mottola:1984ar,Allen:1985ux} for further discussion. We will select the Bunch-Davies vacuum, which is prepared by the Euclidean path integral on the half-sphere.

The scalar field operator may be expanded as follows:
\begin{equation}
	\varphi(t,\theta) = \sum_{n \in \mathbb{Z}} \psi^s_n(t,\theta) a_n + (\psi^s_n(t,\theta))^* a_n^\dagger,
	\label{eq:scalarfield}
\end{equation}
where the raising and lowering operators obey
\begin{equation}
	[a_n,a^\dagger_m] = \delta_{nm}.
\end{equation}
Using \eqref{eq:prop1} to \eqref{eq:prop4}, it follows that $\varphi$ and $\Pi := \cosh t \, \partial_t \varphi$ obey the equal-time canonical commutation relation
\begin{equation}
[ \varphi(\theta) , \Pi(\theta^\prime) ] = i \delta(\theta - \theta^\prime).
\end{equation}
The Hilbert space of this QFT, $\calh_m$,  is given by the Fock space of oscillators acting on the vacuum $\ket{0}$, which is normalized by $\braket{0|0} = 1$. As mentioned previously, the set of single-particle states $a^\dagger_n \ket{0}$ transforms as a principal series representation of the Lorentz group. A general $N$-particle state,
\begin{equation}
a^\dagger_{n_1} \cdots a^\dagger_{n_N}\ket{0},
\end{equation}
belongs to the tensor product of $N$ principal series representations projected onto the subspace that is invariant under the permutation group $S_N$. In general, the representations that appear in the tensor product decomposition include principal and discrete series representations. See Appendix \ref{sec:appendixcg} for more details.

The Wightman function is defined by
\begin{equation}
G_s(t_2,\theta_2 | t_1, \theta_1) := \braket{0|\varphi(t_2,\theta_2) \varphi(t_1,\theta_1)|0},
\label{eq:wightman}
\end{equation}
and it is invariant under the isometries of dS$_2$. An explicit formula is given in Appendix \ref{sec:appendixformulas}.

\section{Observers in dS$_2$}
\label{sec:observers}

\subsection{A model of an observer}

We discuss our model for an observer in dS$_2$. Starting from the action of a free massive particle,
\begin{equation}
S_{\text{particle}} = -m \int d\tau \sqrt{-g_{\tau \tau}},
\label{eq:sparticle}
\end{equation}
the action for an observer, $S_{\text{observer}}$, is obtained by simply promoting the mass parameter $m$ to a dynamical field $Q(\tau)$. To make the mass dynamical, we must introduce a conjugate momentum $P(\tau)$. We have that
\begin{equation}
S_{\text{observer}} = \int d\tau \, \left[ P \partial_\tau Q - Q \sqrt{-g_{\tau \tau}}  \right],
\end{equation}
which is the same observer action that was used in \cite{CLPW,Witten}.\footnote{This is (2.2) in \cite{Witten} with $m = 0$.} Here, $\tau$ is a parameter on the observer's worldline, and $\sqrt{-g_{\tau \tau}}$ is derived from the metric \eqref{eq:metric}.
More explicitly, the observer's action is
\begin{equation}
   S_{\text{observer}} = \int d\tau \, \left[ P \partial_\tau Q - Q \sqrt{(\partial_\tau t)^2 - (\partial_\tau \theta)^2\cosh^2 t}  \right].
\end{equation}
For the purposes of quantization, it helps to write this action in first-order form,
\begin{equation}
       S_{\text{observer}} = \int d\tau \, \left[ P\partial_\tau Q + P_t \partial_\tau t + P_\theta \partial_\tau \theta - \frac{e}{2} \left(-P_t^2 + \frac{P_\theta^2}{\cosh^2 t}  + Q^2\right) \right],
\end{equation}
where the canonical momenta $P_t$ and $P_\theta$ have been introduced, and $e$ is a Lagrange multiplier. We see that wavefunctions are functions of $(Q,t,\theta)$ that obey the Klein-Gordon equation, which constrains the dependence on $t$ and $\theta$. Thus, the Hilbert space of an observer should be the same as the Hilbert space of a free particle, but with an extra quantum number that labels the mass.\footnote{
As an aside, one might be tempted to consider the observer's quantum mechanics where the target space is a static patch, whose metric is
\begin{equation}
    ds^2 = -(1-r^2) d\eta^2 + (1-r^2)^{-1} dr^2,
\end{equation}
such that the action is
\begin{equation}
       S_{\text{observer}} = \int d\tau \, \left[ P\partial_\tau Q + P_\eta \partial_\tau \eta + P_r \partial_\tau r - \frac{e}{2} \left(
    -(1-r^2)^{-1} P_\eta^2 + (1-r^2) P_r^2
     + Q^2  \right) \right].
     \label{eq:staticaction}
\end{equation}
However, geodesics in the static patch generically exit the static patch. This means that classical solutions to \eqref{eq:staticaction} may exit the phase space in a finite amount of time, so we will not attempt to quantize \eqref{eq:staticaction}.
}

Physically, the observer should be thought of as a localized quantum mechanical system that propagates through spacetime. The spectrum of the observer's Hamiltonian can be arbitrary, and we have taken it to be continuous for simplicity. When the spectrum is sufficiently dense, the system can be used to define a clock. Since the observer's Hamiltonian should be bounded from below on physical grounds, \cite{CLPW} projected onto the subspace where $Q > \Lambda$, where $\Lambda$ is a large mass parameter.\footnote{Because \cite{CLPW} worked in the semiclassical limit, their mass parameter $q$ is related to $Q$ by $Q = q + \Lambda$. That is, they restricted to $q > 0$.} We will instead restrict to the subspace $Q > \frac{1}{2}$, such that we only need to consider principal series representations. The Hilbert space of the observer is the Hilbert space of a single particle augmented with an additional quantum number, the mass. Since mass is naturally labeled by the parameter $s$ (see equation \eqref{eq:masss}), we define the Hilbert space of the observer, $\calhobs$, to be spanned by the states
\begin{equation}
\ket{s \, n}, \quad \quad s > 0, \quad n \in \mathbb{Z},
\end{equation}
which are normalized according to
\begin{equation}
\braket{s^\prime \, n^\prime | s \, n} := \delta(s - s^\prime) \delta_{n,n^\prime}. \label{eq:innerproduct}
\end{equation}
The parameter $s$ labels the mass of the observer, while $n$ labels the momentum of a single-particle state whose mass is given by \eqref{eq:masss}.

We now describe a convenient way to represent the inner-product \eqref{eq:innerproduct}. If we define
\begin{equation}
	\label{eq:psidef}
\Psi^s_n (t,\theta) := \psi^s_n(t,\theta) \sqrt{\frac{s \tanh \pi s}{\pi}}, \quad \tilde{\Psi}^s_n (t,\theta) := \psi_{n}^{s}(-t,\theta + \pi) \sqrt{\frac{s \tanh \pi s}{\pi}},
\end{equation}
then it follows that
\begin{equation}
\label{eq:represent}
	\int dt d\theta \sqrt{-g}  \left(\Psi_{n_1}^{s_1}(t,\theta)\right)^* \Psi_{n_2}^{s_2}(t,\theta) 
 = 	\int dt d\theta \sqrt{-g}  \left(\tilde{\Psi}_{n_1}^{s_1}(t,\theta)\right)^* \tilde{\Psi}_{n_2}^{s_2}(t,\theta) =  \delta_{nm} \delta(s_1 - s_2),
\end{equation}
\begin{equation}
\int dt d\theta \sqrt{-g}  \left(\Psi_{n_1}^{s_1}(t,\theta)\right)^* \tilde{\Psi}_{n_2}^{s_2}(t,\theta) = 0.
\end{equation}
Here, the integrals are performed over all of spacetime, whereas the integral in \eqref{eq:KGinner} is performed on a slice of constant $t$.

The function $\Psi^s_n(t,\theta)$ is a positive-frequency wavefunction and has positive norm according to the KG inner product \eqref{eq:KGinner}. The function $\tilde{\Psi}^s_n(t,\theta)$ is the negative-frequency counterpart\footnote{By construction, $\tilde{\Psi}_{n}^{s}(t,\theta)$ and $\Psi_{n}^{s}(t,\theta)$ transform the same way under the isometries of dS$_2$. This is because $(-t,\theta + \pi)$ is the antipode of $(t,\theta)$, and the isometries commute with the $\mathbb{Z}_2$ antipodal map. Hence, $\tilde{\Psi}^s_n(t,\theta)$ is the ``negative-frequency counterpart'' to $\Psi^s_n(t,\theta)$ in a natural sense.}  to $\Psi^s_n(t,\theta)$,  and it has negative norm according to \eqref{eq:KGinner}. Using \eqref{eq:represent}, either $\Psi^s_n(t,\theta)$ or $\tilde{\Psi}^s_n(t,\theta)$ may be used to represent the inner-product \eqref{eq:innerproduct}. We will use the positive-frequency wavefunction $\Psi_n^s(t,\theta)$, because the observer's Hilbert space in a fixed-mass sector should agree with the Hilbert space of a free particle, which is constructed using positive-frequency wavefunctions.\footnote{In Section \ref{sec:candidatetypeii}, we will introduce a generalized notion of an observer that incorporates $\tilde{\Psi}^s_n(t,\theta)$.}

It will be convenient to define the formal braket $\braket{t \, \theta | s \, n}$ as
\begin{equation}
\label{eq:braketdef}
\braket{t \, \theta | s \, n} := \Psi^s_n(t,\theta).
\end{equation}
Because \eqref{eq:braketdef} obeys the KG equation, we stress that the bras $\bra{t \, \theta}$ are not linearly independent. Using this notation, we may write
\begin{equation}
    \braket{s^\prime \, n^\prime | s \, n} = \int dt d\theta \sqrt{-g} \, \braket{s^\prime \, n^\prime | t \, \theta } \braket{t \, \theta | s \, n},
\end{equation}
which shows that the identity operator on $\calhobs$ may be formally represented as
\begin{equation}
    1 = \int dt d\theta \sqrt{-g} \, \ket{ t \, \theta } \bra{t \, \theta }.
\end{equation}
Another way to write the identity operator is\footnote{For a separate but equivalent definition of $\calhobs$, consider $L^2(dS_2)$, which is the space of square-integrable functions on dS$_2$. Given two functions $f_1, f_2 \in L^2(dS_2)$, their inner product is
\begin{equation}
    \braket{f_2|f_1} := \int dt d\theta \sqrt{-g} \, f_2^*(t,\theta) f_1(t,\theta).
\end{equation}
We may view \eqref{eq:braketdef} as the definition of $\ket{s \, n} \in L^2(dS_2)$. We may define $\calhobs$ as the subspace of $L^2(dS_2)$ that is preserved by \eqref{eq:3.10}, which is a projection.
\label{ft:f15}}
\begin{equation}
1 = \int_0^\infty ds \, P_s,
\label{eq:3.10}
\end{equation}
where the projection $P_s$ is defined by
\begin{equation}
	\label{eq:psdef}
P_s := \sum_{n \in \mathbb{Z}} \ket{s \, n}\bra{s \, n}.
\end{equation}

The amplitude for an observer in a fixed-mass state to propagate from $(t_1,\theta_1)$ to $(t_2,\theta_2)$ is given by
\begin{equation}
P_s(t_2,\theta_2|t_1,\theta_1) := \sum_n \Psi^s_n(t_2,\theta_2) \left( \Psi^s_n(t_1,\theta_1) \right)^* = \braket{t_2 \, \theta_2 | P_s | t_1 \, \theta_1},
\label{eq:prop}
\end{equation}
where $s$ labels the mass according to \eqref{eq:masss}, and we have used the formal braket representation in the last equality. This is \eqref{eq:wightman} up to a rescaling. It follows that
\begin{equation}
\int dt d\theta \sqrt{-g} \, P_s(t_2,\theta_2|t,\theta) P_{s^\prime}(t,\theta|t_1,\theta_1) = P_s(t_2,\theta_2|t_1,\theta_1) \delta(s - s^\prime).
\label{eq:integral}
\end{equation}
Heuristically, \eqref{eq:prop} is a sum over worldlines with fixed endpoints, weighted by $e^{- i m \tau}$, where $\tau$ is the proper length of the worldline. Integrating over all of spacetime in \eqref{eq:integral} is equivalent to summing over all worldlines between $(t_1,\theta_1)$ and $(t_2,\theta_2)$ and also summing over a marked point on each worldline. Usually, for a particle with action \eqref{eq:sparticle}, we do not want to sum over this marked point, because the symmetry that reparameterizes each worldline is gauged. However, because our observer can be in a superposition of states with different masses, our observer is effectively a clock, and different locations along each worldline are physically distinct.

From the perspective of quantum gravity, none of the states in $\calhobs$ are physical because they are not invariant under the isometries of dS$_2$. Thus, we view $\calhobs$ as a ``pre-Hilbert space,'' or a building block that is used to construct physical Hilbert spaces. We construct a physical Hilbert space of two observers in the next section.

\subsection{Two observers}

\label{sec:twoentangled}

In Section \ref{sec:CLPW}, we discussed a spacetime in which an observer is moving along a future-directed geodesic at each of the North and South poles. After quantizing the observers as described above, their states are no longer specified by geodesics. In this section, we discuss the gauge-invariant Hilbert space that describes the configurations of two observers. When we need to distinguish the observers, we will refer to them as the left and right observers.

Imposing the gravitational constraints, physical states naively ought to lie in the subspace of $\calhobs \otimes \calhobs$ that is invariant under the spacetime isometry group. The most general expression that is invariant under $SO^+(2,1)$ is
\begin{equation}
\label{eq:psifstate}
\ket{\Psi_f} \propto \int_0^\infty ds \, f(s) \, \sum_{n \in \mathbb{Z}} (-1)^n \ket{s \, n}\ket{s, - n},
\end{equation} 
for some function $f(s)$. This expression is also invariant under parity. It is invariant under time reversal when $f(s)$ is real.

In the classical theory of a relativistic particle, the geodesic that describes the particle's worldline is charged under the spacetime isometry group. The charge for a given Killing vector field $\xi^\mu$ is $V^\mu \xi_\mu$, where $V^\mu$ is a unit vector tangent to the geodesic. There is a single configuration (up to isometries) of two future-directed geodesics with vanishing total charge. In this configuration, the geodesics are located at the North and South poles. Hence, $\ket{\Psi_f}$ should be physically interpreted as a state in which the two observers are at each other's antipodes. The observers are evolving along future-directed geodesics because the wavefunctions $\ket{s \, n}$ are represented using the positive frequency wavefunctions $\Psi^s_n(t,\theta)$.

An immediate concern is that $\ket{\Psi_f}$ is not normalizable, and in particular is not even delta-function normalizable. Thus, $\ket{\Psi_f}$ is not actually a state in $\calhobs \otimes \calhobs$. The tensor product of two principal series representations does not contain any singlets. This does not imply that there are no physical states of two observers, because $\calhobs$ is unphysical by itself from the perspective of quantum gravity, as noted earlier.

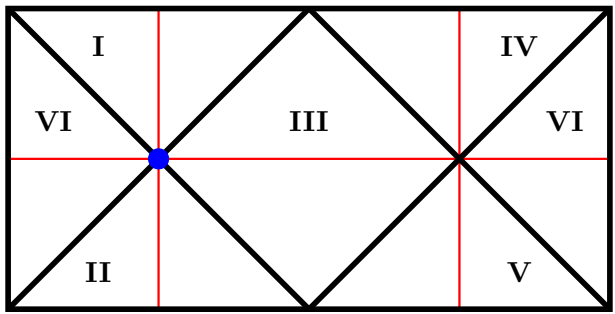
\begin{figure}[ht]
	\centering
	\begin{tikzpicture}[scale=2, line width=0.75mm]
		\draw[red, line width=0.3mm] (1,2) -- (1,0); 
		\draw[red, line width=0.3mm] (3,2) -- (3,0); 
		\draw[red, line width=0.3mm] (0,1) -- (4,1); 
		
		\draw[black] (0,0) rectangle (4,2);
		\draw[black] (0,2) -- (2,0);
		\draw[black] (0,0) -- (2,2);
		\draw[black] (2,2) -- (4,0);
		\draw[black] (2,0) -- (4,2);
		
		\node at (0.6,1.75) {\small \textbf{I}};     
		\node at (0.6,0.25) {\small \textbf{II}};    
		\node[above] at (2,1.1) {\small \textbf{III}}; 
		\node at (3.4,1.75) {\small \textbf{IV}};    
		\node at (3.4,0.25) {\small \textbf{V}};     
		\node at (0.3,1.25) {\small \textbf{VI}};    
		\node at (3.7,1.25) {\small \textbf{VI}};    
		
		\fill[blue] (1,1) circle (2pt); 
		
	\end{tikzpicture}
	\caption{A Penrose diagram of dS$_2$. An $SO^+(2,1)$-invariant function of $(t_L,\theta_L,t_R,\theta_R)$ can be gauge-fixed such that $(t_L,\theta_L)$ is fixed to the blue dot (which we take to be the origin, $(0,0)$), and $(t_R,\theta_R)$ is integrated along the red lines only.}
	\label{fig:gaugefix}
\end{figure}

Because the approach of finding a singlet in $\calhobs \otimes \calhobs$ did not produce any normalizable states, we will instead define the Hilbert space of two observers to be $L^2(\mathbb{R}_+,\rhopl)$ for some measure $\rhopl(s)$ that is to be determined. The inner product of two functions $f_1, f_2 \in L^2(\mathbb{R}_+,\rhopl)$ is
\begin{equation}
\braket{f_1|f_2}_{\rhopl} := \int_0^\infty ds \, \rhopl(s) \, f_1^*(s) f_2(s).
\end{equation}
Our task is to find some context in which the following expression makes sense:
\begin{equation}
\braket{\Psi_{f_1} | \Psi_{f_2}} = \braket{f_1 | f_2}_{\rhopl}, \label{eq:psif1psif2}
\end{equation}
for some appropriate choice of $\rhopl$. To do this, we will work in position space. Define
\begin{equation}
\Psi_f(t_L,\theta_L;t_R,\theta_R) := \int_0^\infty ds \, f(s) \, \sum_{n \in \mathbb{Z}} (-1)^n \braket{t_L \, \theta_L  | s \, n}\braket{t_R,\theta_R | s, - n}.
\label{eq:psifdef}
\end{equation}
The sum is an antipodal Green's function, which converges as long as $(t_L,\theta_L)$ is not null-separated from the antipode of $(t_R,\theta_R)$. In position space, the inner product becomes
\begin{equation}
	\braket{\Psi_{f_1} | \Psi_{f_2}} \propto \int dt_L d\theta_L \sqrt{-g_L} \, dt_R d\theta_R \sqrt{-g_R} \, \left(\Psi_{f_1}(t_L,\theta_L;t_R,\theta_R)\right)^* \Psi_{f_2}(t_L,\theta_L;t_R,\theta_R). 
\end{equation}
This integral diverges because the integrand is invariant under $SO^+(2,1)$, which is noncompact and thus has infinite volume $V$. Using the standard Fadeev-Popov procedure,\footnote{See \cite{Chakraborty:2023los,Chakraborty:2023yed} for another construction of the Hilbert space of de Sitter quantum gravity that uses Fadeev-Popov gauge-fixing.} one can formally rewrite this divergent integral as $V$ times a sum over six convergent integrals. If we gauge-fix $(t_L,\theta_L)$ to be at $(0,0)$, then each of these six integrals refers to the cases when $(t_R,\theta_R)$ is in regions I through VI in Figure \ref{fig:gaugefix}. Due to the residual boost symmetry that fixes $(0,0)$, we can furthermore gauge-fix $(t_R,\theta_R)$ to be on a line within each region. Let $F(t_L,\theta_L|t_R,\theta_R)$ be a function of two points that is invariant under $SO^+(2,1)$. The result of the gauge fixing procedure is that
\begin{equation}
\int_{GF} F(t_L,\theta_L|t_R,\theta_R) := \frac{1}{V}\int dt_L d\theta_L\sqrt{-g_L} \,  dt_R d\theta_R \sqrt{-g_R}	F(t_L,\theta_L|t_R,\theta_R)
\label{eq:GFdef}
\end{equation}
is equivalent to
\begin{align}
&\int_1^\infty d(\cosh t)  \left[F(0,0|t,0) + F(0,0|-t,0)\right] \label{eq:3.19}
\\
+ &\int_1^\infty d(\cosh t)  \left[F(0,0|t,\pi) + F(0,0|-t,\pi)\right] \label{eq:3.20}
\\
+ &\int_{-1}^{1} d(\cos \theta)  \left[F(0,0|0,\theta) + F(0,0|0,-\theta)\right], \label{eq:3.21}
\end{align}
where line \eqref{eq:3.19} includes the integrals over regions I and II, \eqref{eq:3.20} integrates over regions IV and V, and \eqref{eq:3.21} integrates over regions III and VI.

We have shown that the origin of the non-normalizability of $\braket{\Psi_{f_1}|\Psi_{f_2}}$ is the infinite volume of $SO^+(2,1)$. A better definition of $\ket{\Psi_f}$ is then given by
\begin{equation}
	\ket{\Psi_f} := \frac{1}{\sqrt{V}} \int_0^\infty ds \, f(s) \, \sum_{n \in \mathbb{Z}} (-1)^n \ket{s \, n}\ket{s, - n},
	\label{eq:3.23}
\end{equation}
where we have in mind that the inner product should be computed in the position basis and gauge-fixed using the procedure described above. Thus, the gauge-fixed Hilbert space for two observers is the space of functions of $(t_L,\theta_L,t_R,\theta_R)$ that can be written as \eqref{eq:psifdef} for some $f(s)$, and the Fadeev-Popov procedure leads to a well-defined inner product, which must take the form of \eqref{eq:psif1psif2} for some choice of $\rhopl$. Explicitly, the inner product is
\begin{equation}
\braket{\Psi_{f_1}|\Psi_{f_2}} = \int_{GF} \Psi^*_{f_1}(t_L,\theta_L,t_R,\theta_R) \Psi_{f_2}(t_L,\theta_L,t_R,\theta_R).
\label{eq:fixinner}
\end{equation}
In Appendix \ref{sec:rhopl}, we determine $\rhopl$ to be
\begin{equation}
\rhopl(s) = \frac{s \tanh \pi s}{4 \pi^2}.
\label{eq:rhopl}
\end{equation}

When doing computations, it helps to use the following shortcut. If we use \eqref{eq:3.23} and \eqref{eq:innerproduct} to compute $\braket{\Psi_{f_1}|\Psi_{f_2}}$, then we find that
\begin{equation}
\braket{\Psi_{f_1}|\Psi_{f_2}} =  \int_0^\infty ds \, f_1^*(s) f_2(s)  \, \frac{\delta(0) \sum_{n \in \mathbb{Z}} }{V}.
\end{equation}
The $\sum_{n \in \mathbb{Z}}$ should be interpreted as the trace of the identity element in a principal series representation whose Casimir is labeled by $s$.\footnote{See our conventions for principal series representations in Appendix \ref{sec:appendixcg}.} This trace is infinite, but $V$ is also infinite, such that we may make the replacement
\begin{equation}
\frac{\delta(0) \sum_{n \in \mathbb{Z}} }{V} \rightarrow \rhopl(s).
\label{eq:prescription}
\end{equation}
All of the correlators we compute in this paper may be written as integrals over spacetime points that are connected by propagators. These integrals will be $SO^+(2,1)$ invariant and will require Fadeev-Popov gauge-fixing, just as described above. In practice, the simplest way to gauge-fix these integrals is to work in the $\ket{s \, n}$ basis and apply \eqref{eq:prescription} to properly divide by the volume $V$. This technique has been used before in \cite{Suh:2020lco,Jafferis:2019wkd,Blommaert:2018iqz}.

Next, we define the operator $\hat{G}_s$ by the following expression:
\begin{align}
\braket{\Psi_{f_1}| \hat{G}_s | \Psi_{f_2}} &:= 
	\int_{GF} \Psi_{f_1}^*(t_L,\theta_L|t_R,\theta_R) G_s(t_L,\theta_L|t_R,\theta_R) \Psi_{f_2}(t_L,\theta_L|t_R,\theta_R),
	\\
	&= \frac{1}{64 \pi^7} \int_0^\infty ds_1 f^*_1(s_1) \int_0^\infty ds_2 f_2(s_2) \,\, s_2 \sinh \pi s_2 \, s_1 \sinh \pi s_1  \, \Gamma\left( \frac{1}{2} \left(\frac{1}{2} \pm i s \pm i s_1 \pm i s_2\right)\right).
	\label{eq:3.46}
\end{align}
The computation that leads to the second line is explained in Appendix \ref{sec:rhopl}. We define the function $\Gamma\left( \frac{1}{2} \left(\frac{1}{2} \pm i s \pm i s_1 \pm i s_2\right)\right)$ to be the product of the eight gamma functions that correspond to the different choices of signs. As mentioned earlier, $G_s(t_L,\theta_L|t_R,\theta_R)$ is the amplitude for a particle of mass-squared $m^2 = \frac{1}{4} + s^2$ to propagate from the point $(t_R,\theta_R)$ to the point $(t_L,\theta_L)$. To specify the initial and final points in a gauge-invariant way, we must use the observers. The operator $\hat{G}_s$ is the gauge-invariant operator that creates a particle at the location of the right observer, and annihilates the same particle at the location of the left observer.\footnote{This operator is completely analogous to the operator $e^{- \Delta \ell}$ in JT gravity, where $\Delta$ is analogous to $s$.} When the initial and final states of the observers are given by $f_2$ and $f_1$, \eqref{eq:3.46} represents the complete transition amplitude between the initial and final states, where a particle emitted by the right observer is absorbed by the left observer. Note that we could have exchanged $G_s(t_L,\theta_L|t_R,\theta_R)$ with $G_s(t_R,\theta_R|t_L,\theta_L)$ in \eqref{eq:3.46} without changing the result, so the amplitude for left-right propagation is the same as right-left propagation. This indicates that the two observers are spacelike separated.

Our eventual aim is to develop a graphical language for computing correlators and representing states. We graphically represent $\ket{\Psi_f}$ as
\begin{equation*}
	\includegraphics[width=0.1\linewidth]{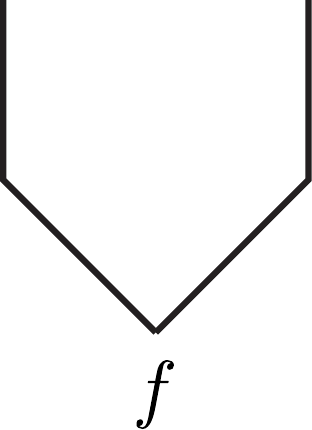}
\end{equation*}
which is meant to pictorally capture the entanglement between the two observers in \eqref{eq:3.23}.

\subsection{Adding matter}
\label{sec:addmatter}

In the previous subsection, we constructed a gauge-invariant Hilbert space that describes two observers in dS$_2$. We now extend this Hilbert space to include a free scalar field with mass-squared $m^2 = \frac{1}{4} + s^2$. The term ``matter'' is used to distinguish the free scalar from the observers. Let $\calh$ denote the physical $SO^+(2,1)$-invariant Hilbert space of two observers plus the free scalar, which is meant to capture all  states that are invariant under the gravitational constraints.\footnote{We leave the discrete isometries ungauged for now.}  $\calh$ contains a sector $\calh_0$ where the free scalar is in the vacuum state $\ket{0}$. $\calh_0$ is equivalent to the Hilbert space of two observers that was constructed in the previous subsection. More generally, we write
\begin{equation}
	\label{eq:calhdef}
	\calh = \bigoplus_{N = 0}^\infty \calh_N,
\end{equation}
where $\calh_N$ denotes the sector of $\calh$ with $N$ matter particles.

To construct $\calh_N$, we consider the tensor product
\begin{equation}
	\calhobs \otimes \calh_{N,m} \otimes \calhobs,
	\label{eq:3.49}
\end{equation}
where $\calh_{N,m}$ is the $N$-particle subsector of the matter theory. An orthonormal basis of $\calh_{1,m}$ is given by $\ket{n}_s$, for $n \in \mathbb{Z}$, and these states transform in a principal series representation of $SO^+(2,1)$.  Thus, a general state in \eqref{eq:3.49} may be written as
\begin{equation}
	\ket{\Psi_\cala} := \int_0^\infty ds_L ds_R \, \sum_{n_L,n_1,\cdots,n_N,n_R \in \mathbb{Z}} \ket{s_L \, n_L} \ket{n_1}_s \cdots \ket{n_N}_s \ket{s_R \, n_R} \cala^{s_L,s,s_R}_{n_L,n_1,\cdots,n_N,n_R},
	\label{eq:3.50}
\end{equation}
where $\cala^{s_L,s,s_R}_{n_L,n_1,\cdots,n_N,n_R}$ is a set of coefficients. Due to Bose symmetry, $\cala^{s_L,s,s_R}_{n_L,n_1,\cdots,n_N,n_R}$ must be symmetric under permutations of $n_1$ through $n_N$. We stress that \eqref{eq:3.49} is a ``pre-Hilbert space'' that is used to construct the physical Hilbert space $\calh_N$.\footnote{Because we start with a pre-Hilbert space, we may view the observers as a Quantum Reference Frame for the QFT. See \cite{AliAhmad:2024wja,Fewster:2024pur,AliAhmad:2024vdw,Susskind:2023rxm} for recent related discussions on the role of Quantum Reference Frames in constructing gauge-invariant Hilbert spaces.}

We are interested in the case when \eqref{eq:3.50} is invariant under $SO^+(2,1)$. As one might anticipate, there are no normalizable singlets. Following the logic of the previous subsection, we define a new inner product on the space of non-normalizable singlets. The singlets that are normalizable with respect to this new inner product comprise $\calh$.\footnote{Note that there exist singlets that are not even normalizable with respect to the new inner product, and hence not in $\calh$. A two-particle example is $\cala^{s_L,s,s_R}_{n_L,n_1,n_2,n_R} = (-1)^{n_L + n_1} \delta_{n_L,-n_R} \delta_{n_1,-n_2}.$ For this example, the sum \eqref{eq:3.51} does not converge to a normalizable state in $\calh_{N,m}$.} Given a singlet $\ket{\Psi_\cala}$, we define
\begin{align}
	&\ket{\Psi_\cala(t_L,\theta_L|t_R,\theta_R)} 
	\\
	&:= \int_0^\infty ds_L ds_R \,\sum_{n_L,n_1,\cdots,n_N,n_R \in \mathbb{Z}} \braket{t_L \, \theta_L | s_L \, n_L} \ket{n_1}_s \cdots \ket{n_N}_s \braket{t_R \, \theta_R | s_R \, n_R} \cala^{s_L,s,s_R}_{n_L,n_1,\cdots,n_N,n_R} \in \calh_{N,m},
	\label{eq:3.51}
\end{align}
which is invariant under $SO^+(2,1)$ transformations that act jointly on $\calh_{N,m}$ and the spacetime points $(t_L,\theta_L)$, $(t_R,\theta_R)$. Because the transformations act unitarily on $\calh_{N,m}$, we have that
\begin{equation}
	\braket{\Psi_{\cala_1}(t_L,\theta_L|t_R,\theta_R) | \Psi_{\cala_2}(t_L,\theta_L|t_R,\theta_R)}
\end{equation}
is an $SO^+(2,1)$-invariant function of its arguments. Hence, we define the inner product on the singlets to be
\begin{equation}
	\braket{\Psi_{\cala_1}|\Psi_{\cala_2}} := \int_{GF} \braket{\Psi_{\cala_1}(t_L,\theta_L|t_R,\theta_R) | \Psi_{\cala_2}(t_L,\theta_L|t_R,\theta_R)}.
	\label{eq:GFinnerproduct}
\end{equation}

Having defined an inner product on the space of singlets that leads to a nontrivial $\calh$, we now describe how states in $\calh$ may be constructed, beginning with $\calh_1$.

States in $\calh_1$ are represented by tensors $\cala^{s_1,s,s_R}_{n_L,n_1,n_R}$ that are invariant under $SO^+(2,1)$. There are only two such tensors, and they correspond to the two independent Clebsch-Gordan (CG) coefficients that couple three principal series representations.\footnote{See Appendix \ref{sec:appendixcg} for a more detailed discussion of CG coefficients.} They both lead to normalizable states following the procedure above. These states are also invariant under parity and time-reversal.\footnote{This may be derived using the fact that \eqref{eq:ISdef} and \eqref{eq:IAdef} are invariant under $\theta \rightarrow - \theta$. Time-reversal invariance also requires that the coefficients $\cala^{s_1,s,s_R}_{n_L,n_1,n_R}$ are real. The CG coefficients can be taken to be real. The CG coefficients in \eqref{eq:cpsppdef} and \eqref{eq:cpsadef} are equal to their own conjugates up to phases that do not depend on the $n$ indices.
}

Next, we consider $\calh_{2}$. Let $\cala^{s_L,s,s_R}_{n_L,n_1,n_2,n_R}$ be a set of coefficients that leads to a normalizable state in $\calh_2$. Let $\left(\calc_s^j\right)_{n_1,n_2}^n$ be a set of CG coefficients that couples two principal series representations with Casimir $\frac{1}{4} + s^2$ to a third representation $j$, which could be from the principal or discrete series. The normalizability of $\cala^{s_L,s,s_R}_{n_L,n_1,n_2,n_R}$ implies that the following sum must converge:
\begin{equation}
 \tilde{\cala}^{s_L,j,s_R}_{n_L,n_1,n_R} := \sum_{n_1,n_2 \in \mathbb{Z}} 	\left(\left(\calc_s^j\right)_{n_1,n_2}^n \right)^* \cala^{s_L,s,s_R}_{n_L,n_1,n_2,n_R}.
\end{equation}
This is because for fixed $n_L$ and $n_R$, $\cala^{s_L,s,s_R}_{n_L,n_1,n_2,n_R}$ is the wavefunction in the $\ket{n_1}_s \ket{n_2}_s$ basis of a normalizable state in $\calh_{2,m}$, and the CG coefficients by definition are wavefunctions of normalizable states in $\calh_{2,m}$ that transform irreducibly under $SO^+(2,1)$. Having established that $\tilde{\cala}^{s_L,j,s_R}_{n_L,n_1,n_R}$ exists, $SO^+(2,1)$ invariance of $\tilde{\cala}^{s_L,j,s_R}_{n_L,n_1,n_R}$ uniquely fixes it in terms of CG coefficients, which lead to normalizable states.

The subspaces $\calh_N$ for $N > 2$ may be constructed recursively. Given an invariant, normalizable $\cala^{s_L,s,s_R}_{n_L,n_1,\cdots,n_N,n_R}$ tensor, one may contract it with a CG coefficient to get a tensor that corresponds to a normalizable state in $\calh_{N-1}$. Iterating, one concludes that $\cala^{s_L,s,s_R}_{n_L,n_1,\cdots,n_N,n_R}$ must be formed by contracting CG coefficients together in a connected manner\footnote{By this, we mean that the tensor diagrams that we will introduce shortly must be connected.} and symmetrizing the indices $n_1$ through $n_N$. 

There is a nice way to graphically depict the contraction of various CG coefficients. We will consider some examples that involve only one kind of CG coefficient, $$\left(\calc_{PP}^{P,S}\right)_{s_1, n_1 | s_2, n_2}^{s_3, n_3}.$$ We explain our notation for CG coefficients in Appendix \ref{sec:appendixcg}. Each diagram computes a tensor that corresponds to a wavefunction, and a CG coefficient is associated to every trivalent vertex. An example of a state in $\calh_1$ is
\begin{equation}
	\ket{\Psi^{(1)}_{f_a,f_b}} := \int_0^\infty ds_a \, f_a(s_a) \, ds_b \, f_b(s_b) \sum_{n_a,n_1,n_b \in \mathbb{Z}} (-1)^{n_a} \ket{s_a \, n_a} \ket{n_1}_s \ket{s_b \, n_b} \left(\calc_{PP}^{P,S}\right)_{s, n_1 | s_b, n_b}^{s_a, -n_a},
\end{equation}
which we graphically represent as follows:
\begin{equation*}
	\includegraphics[width=0.2\linewidth]{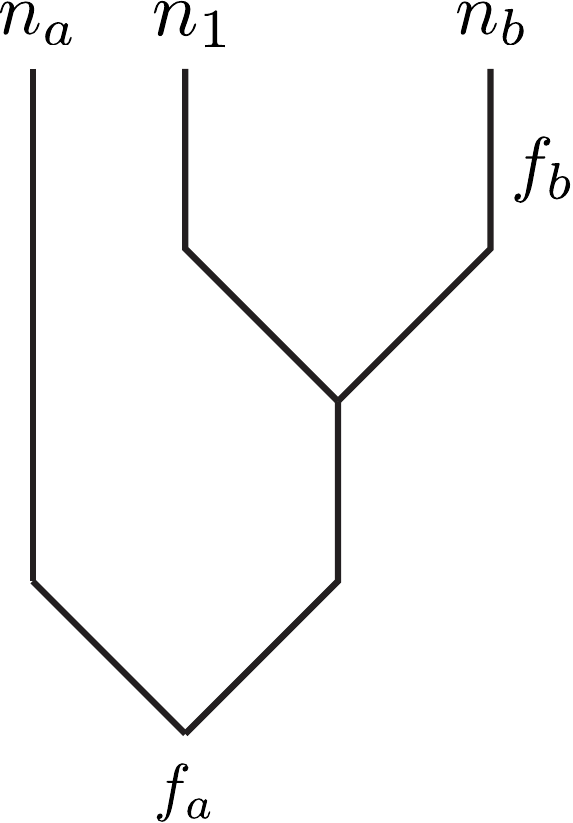}.
\end{equation*}
The inner product of two such states is given by
\begin{equation}
	\label{eq:3.54}
	\braket{\Psi^{(1)}_{\tilde{f}_a,\tilde{f}_b}|\Psi^{(1)}_{f_a,f_b}} = \frac{1}{\pi^3 }\int_0^\infty ds_a \, \tilde{f}_a^*(s_a) f_a(s_a) \, \int_0^\infty ds_b \,   \, \tilde{f}_b^*(s_b) f_b(s_b),
\end{equation}
where we have used \eqref{eq:innerproduct} and \eqref{eq:prescription} as well as orthogonality of the CG coefficients, \eqref{eq:cgnorm}.

An example of a state in $\calh_2$ is
\begin{equation}
	\begin{aligned}
		\ket{\Psi^{(2)}_{f_a,f_b,f_c}} := & P_{2,m} \int_0^\infty ds_a \, f_a(s_a) \, ds_b \, f_b(s_b) \, ds_c \, f_c(s_c) \\
		& \times \sum_{n_a,n_b,n_c,n_1,n_2 \in \mathbb{Z}} 
		\Bigg[
		(-1)^{n_a} 
		\ket{s_a , -n_a} \ket{n_1}_s \ket{n_2}_s \ket{s_c , n_c} \\
		& \qquad \times \left(\calc_{PP}^{P,S}\right)_{s_b, n_b | s, n_1}^{s_a,n_a}
		\left(\calc_{PP}^{P,S}\right)_{s, n_2 | s_c, n_c}^{s_b,n_b}
		\Bigg],
	\end{aligned}
\label{eq:3.55}
\end{equation}
which we graphically represent as follows:
\begin{equation*}
	\includegraphics[width=0.2\linewidth]{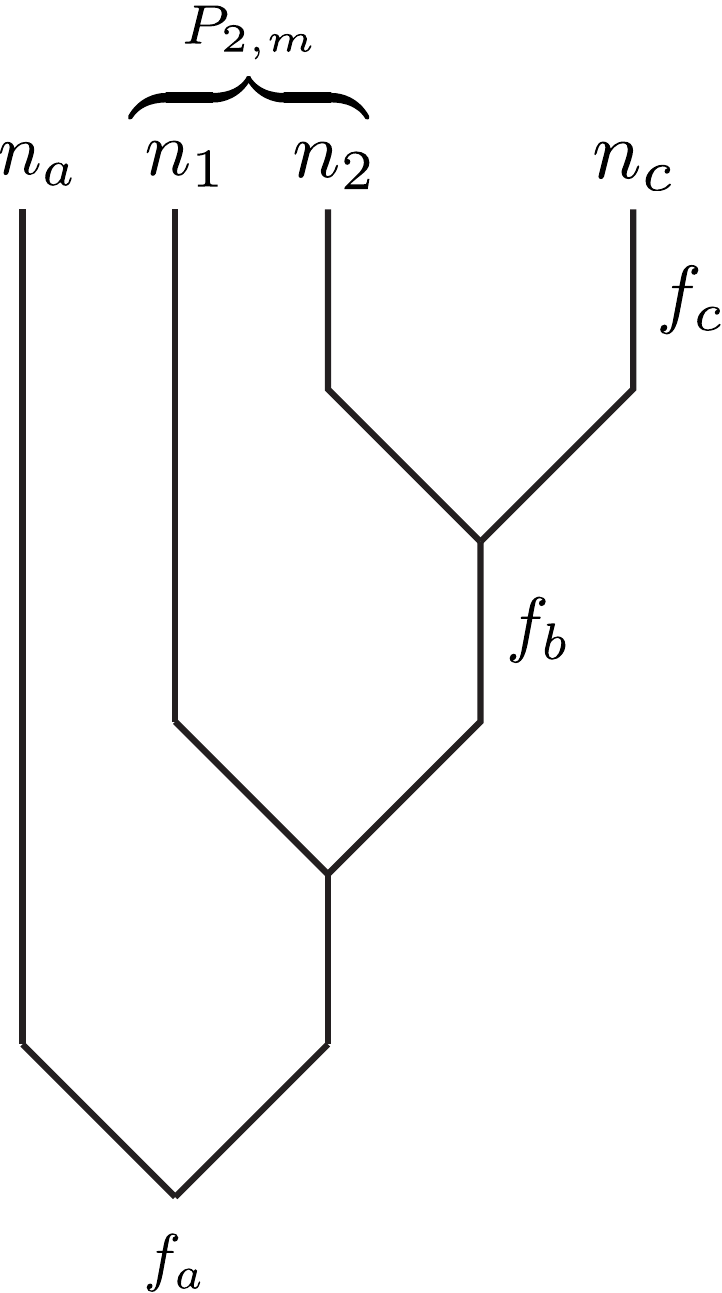}.
\end{equation*}
The operator $P_{2,m}$ is defined to be a projection onto the subspace of $\calh_{2,m}$ that is invariant under swapping the two matter particles. Applying $P_{2,m}$ is equivalent to symmetrizing over $n_1$ and $n_2$.

Here is another way to construct a state in $\calh_2$,
\begin{equation}
	\vcenter{\hbox{\includegraphics[width=0.2\linewidth]{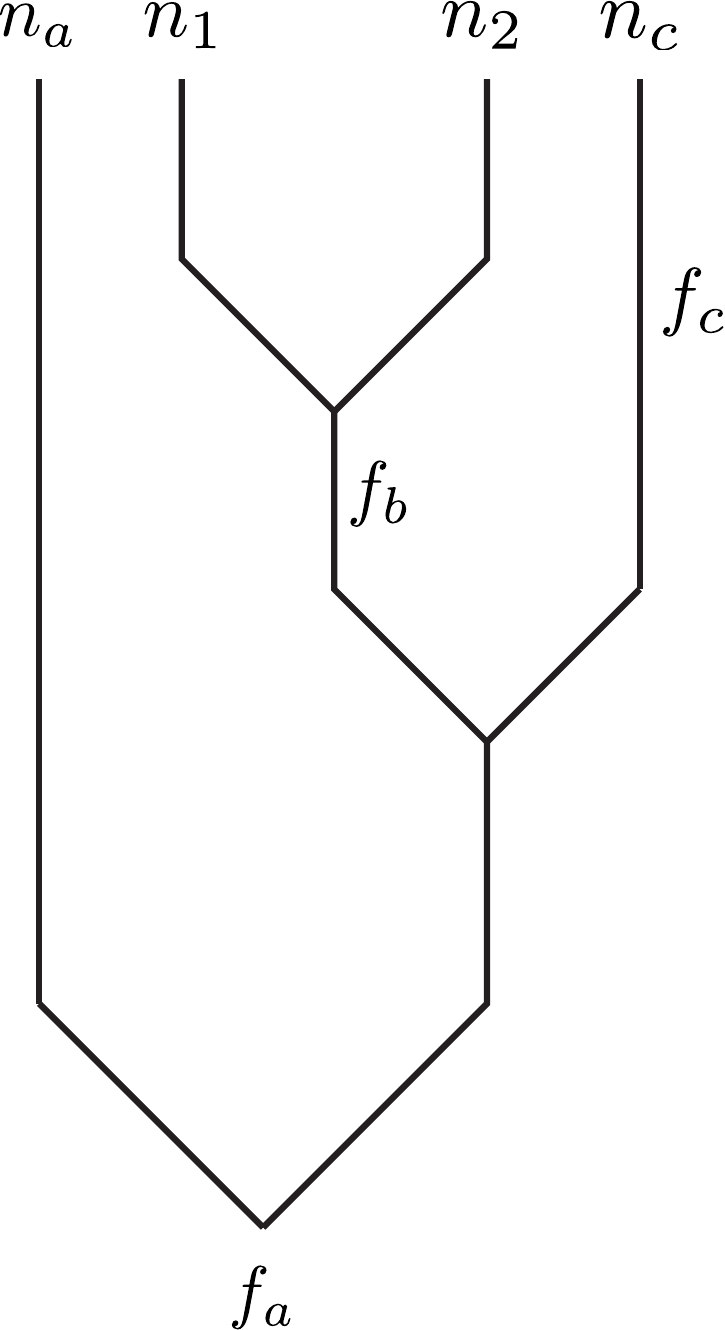}}}
	\label{eq:leftrightvertexdiagram}
\end{equation}
which corresponds to the state
\begin{equation}
\begin{aligned}
	&\int_0^\infty ds_a \, f_a(s_a) \, ds_b \, f_b(s_b) \, ds_c \, f_c(s_c) \\
	&\times \sum_{n_a,n_b,n_c,n_1,n_2 \in \mathbb{Z}} 
	\Bigg[
	(-1)^{n_a} 
	\ket{s_a , -n_a} \ket{n_1}_s \ket{n_2}_s \ket{s_c , n_c} \\
	& \qquad \times \left(\calc_{PP}^{P,S}\right)_{s_b, n_b | s_c, n_c}^{s_a,n_a}
	\left(\calc_{PP}^{P,S}\right)_{s, n_1 | s, n_2}^{s_b,n_b}
	\Bigg].
	\label{eq:3.56}
\end{aligned}
\end{equation}
We did not write $P_{2,m}$ because $\left(\calc_{PP}^{P,S}\right)_{s, n_1 | s, n_2}^{s_b,n_b}$ is already symmetric in $n_1$ and $n_2$. Comparing \eqref{eq:3.56} and \eqref{eq:3.55}, we see that the tensor product of four principal series representations can be decomposed in different ways to get different states in $\calh_2$. More generally, if we choose to construct states in $\calh_N$ by first decomposing $\calh_{N,m}$ into irreps and then using each irrep to form a singlet with $\calhobs \otimes \calhobs$, we will have a diagram like \eqref{eq:leftrightvertexdiagram} where the left- and right-most branches directly meet at a vertex. Let $N \ge 1$. If the decomposition of $\calh_{N,m}$ into irreps is given by
\begin{equation}
	\label{eq:4.58}
	\calh_{N,m}= \bigoplus_j \calh_{N,m,j},
\end{equation}
where $j$ labels an irrep,\footnote{In general, a given irrep will appear with multiplicity. So, specifying the Casimir is not enough to uniquely refer to an irrep. In \eqref{eq:4.58}, $j$ refers to not just the Casimir label, but also the other labels that are needed to select a single irrep from the tensor product. } then $\calh_N$ may be written as\footnote{In \eqref{eq:4.59}, when $j$ is a principal series representation, there should be an additional discrete index in the direct sum that selects one of the two types of CG coefficients that couple three principal series representations. We will absorb this extra index into what is meant by $j$ in \eqref{eq:4.59}.}
\begin{equation}
	\calh_N = \bigoplus_{j} L^2(\mathbb{R}_+) \otimes L^2(\mathbb{R}_+),
	\label{eq:4.59}
\end{equation}
where each $L^2(\mathbb{R}_+)$ refers to functions of either $s_L$ or $s_R$.

An important point is that states in $\calh$ may be constructed using connected diagrams only. If a diagram has, for example, two disconnected components, then its norm using the un-gauge-fixed inner product will be proportional to $V^2$, and the gauge-fixing procedure will only introduce a factor of $\frac{1}{V}$, resulting in a state that is non-normalizable with respect to the gauge-fixed inner product, which thus is not in $\calh$.

In this section, we have given specific examples of invariant states in $\calh$, constructed using a single type of CG coefficient. Of course, more general states can be constructed from CG coefficients other than $\calc^{PS}_{PP}$. In this case, we need to add labels to the trivalent vertices to distinguish between different CG coefficients. Our graphical language will help us discuss the properties of gauge-invariant operators on $\calh$, which we turn to in the next section.

In $\eqref{eq:4.59}$, $\calh_N$ is the physical space of $SO^+(2,1)$-invariant states, normalized using the gauge-fixed inner product. Before closing this section, we discuss the effects of imposing invariance under the discrete isometries, beginning with parity. Parity will mix states in different $j$ subspaces. For example, when $j$ is a  discrete series representation, the subspace at fixed $j$ cannot be invariant under parity, because a discrete series representation contains states with a fixed sign of the momentum quantum number $n$. Thus, after projecting onto parity-invariant states, $\calh_N$ takes the same form as \eqref{eq:4.59}, except the $j$ index runs over parity-invariant singlets.

We now discuss time-reversal symmetry. Let $\calh_P$ be the Hilbert space that is invariant under $SO^+(2,1)$ and parity. Consider a state $\ket{\Psi_\cala}$, defined in \eqref{eq:3.50}, that is in $\calh_P$. Then time-reversal acts by sending
\begin{equation}
	\cala^{s_L,s,s_R}_{n_L,n_1,\cdots,n_N,n_R} \rightarrow \left(\cala^{s_L,s,s_R}_{n_L,n_1,\cdots,n_N,n_R}\right)^*,
\end{equation}
which defines a new state, $\ket{\Psi_{\cala^*}}$, that is also in $\calh_P$.\footnote{A way to see this is to work in the position space representation of an $(N+2)$-particle state, and then use \eqref{eq:3.7}, \eqref{eq:3.8} and \eqref{eq:prop312}.} Thus, once $\calh_P$ is constructed, one can construct a time-reversal-invariant subspace by taking the real part of $\cala$ for each $\ket{\psi_\cala} \in \calh_P$.

To conclude, after imposing invariance under parity and time-reversal, $\calh_N$ may still be expressed as \eqref{eq:4.59}, where each $L^2(\mathbb{R}_+)$ is a real Hilbert space, and the $\bigoplus_j$ is a direct sum of real Hilbert spaces.

\section{Gauge-invariant operators}

\label{sec:gaugeinvariant}

In the previous section we discussed the construction of $\calh$, the physical Hilbert space of two observers and a free scalar field that is invariant under the gravitational constraints. In this section, we explain how we dress local operators in the matter QFT to one of the observers, leading us to a definition of an algebra that describes the local measurements available to the observer.

Although the Hilbert space $\calh$ is not equivalent to $\calhobs \otimes \calh_m \otimes \calhobs$, we have seen in the previous section that it is convenient to write normalizable states in $\calh$ as non-normalizable singlets in $\calhobs \otimes \calh_m \otimes \calhobs$. Likewise, the gauge-invariant operators we are interested in may be written as operators that act on $\calhobs \otimes \calh_m \otimes \calhobs$ and are invariant under the spacetime isometries. The first operator we define, $\hat{s}_R$, acts only on the right $\calhobs$ factor. We have 
\begin{equation}
	\hat{s}_R \ket{s_R \, n_R} := s_R \ket{s_R \, n_R}.
\label{eq:srdef}
\end{equation}
We can think of $\hat{s}_R$ as the Hamiltonian (or mass) of the right observer, although strictly speaking this is only correct in the large $s_R$ limit. Following \eqref{eq:masss}, the precise formula for the Hamlitonian is
\begin{equation}
\label{eq:HRdef}
    H_R := \sqrt{\hat{s}_R^2 + \frac{1}{4}}.
\end{equation}
If $g(s)$ is an arbitrary function, we have for example that
\begin{equation}
	g(\hat{s}_R) \ket{\Psi_{f}} = \ket{\Psi_{g f}}, \quad g(\hat{s}_R) \ket{\Psi^{(1)}_{f_a,f_b}} = \ket{\Psi^{(1)}_{f_a,g f_b}},
\end{equation}
where these states were defined in \eqref{eq:3.23} and \eqref{eq:3.54}. We define $\hat{s}_L$ analogously.

Next, we wish to define a gauge invariant operator given by the scalar field $\varphi$ evaluated at the location of the right observer. It is tempting to try to define the spacetime position operators $\hat{t}_R$ and $\hat{\theta}_R$ on $\calhobsR$\footnote{We define $\calhobsR$ to be the right $\calhobs$ factor in $\calhobs \otimes \calh_m \otimes \calhobs$.} so that the dressed operator may be given by $\varphi(\hat{t}_R,\hat{\theta}_R)$, but this is not possible because $\hat{t}_R$ and $\hat{\theta}_R$ are not well-defined operators on $\calhobsR$. This is because the bras $\bra{t \, \theta}$ are not linearly independent, as noted below \eqref{eq:braketdef}.\footnote{If we view $\calhobs$ as a subspace of $L^2(dS_2)$, following Footnote \ref{ft:f15}, then the position operators $\hat{t}$ and $\hat{\theta}$ are well-defined on $L^2(dS_2)$, but they do not map $\calhobs$ to itself, since $\calhobs$ is a proper subspace of $L^2(dS_2)$ that is only spanned by the positive-frequency wavefunctions $\Psi^s_n(t,\theta)$.} To define the operator $\phi_R$, we write
\begin{equation}
	\braket{s_R^\prime \, n_R^\prime|\phi_R| s_R \, n_R} :=  \int dt_R d\theta_R\sqrt{-g_R} \left(\Psi^{s_R^\prime}_{n_R^\prime}(t_R,\theta_R)\right)^*\varphi(t_R,\theta_R) \Psi^{s_R}_{n_R}(t_R,\theta_R)
	\label{eq:4.4}
\end{equation}
which may be viewed as a matrix element of an operator on $\calhobsR$ that is valued in the space of operators on $\calh_m$. This is sufficient to define the action of $\phi_R$ on a non-normalizable singlet in $\calhobs \otimes \calh_m \otimes \calhobs$, which in turn defines the action of $\phi_R$ on $\calh$. As mentioned in the previous section, a non-normalizable singlet in $\calhobs \otimes \calh_m \otimes \calhobs$ may be represented using a graph, and the non-normalizable singlets that lead to normalizable operators in $\calh$ have connected graphs. One may check that $\phi_R$ maps singlets with connected graphs to singlets with connected graphs. Hence, $\phi_R$ is a well-defined operator on the physical Hilbert space $\calh$.

One may show that $\phi_R$ is invariant under parity and time-reversal. The operator $\phi_L$ is defined analogously. Clearly, $\phi_R$ and $\phi_L$ are Hermitian. We can write $\phi_R$ as
\begin{equation}
    \phi_R = \int_0^\infty ds \, ds^\prime \sum_{n, n^\prime \in \mathbb{Z}} \, \braket{s^\prime \, n^\prime|\phi_R|s \, n} \ket{s^\prime \, n^\prime} \bra{s \, n} = \int_0^\infty ds \, ds^\prime P_{R,s^\prime} \phi_R P_{R,s}.
\end{equation}
We define $\phi_R(\tau)$ by analogy with \eqref{eq:timeevolvewithhamiltonian},
\begin{equation}
    \phi_R(\tau) := e^{- i H_{R} \tau} \phi_R e^{i H_R \tau} = \int_0^\infty ds \, ds^\prime \, e^{-i\left(\sqrt{s^{\prime \, 2} + \frac{1}{4}}  - \sqrt{s^2 + \frac{1}{4}} \right)\tau} P_{R,s^\prime} \phi_R P_{R,s},
\end{equation}
and $\tau$ represents proper time on the observer's worldline.

Having defined $\phi_R$ and $\hat{s}_R$, we now discuss the computation of correlation functions. We will consider expressions such as
\begin{equation}
	\bra{0}\braket{\Psi_{f_2}|P_{R,s_b}  \phi_R P_{R,s_a} \phi_R |\Psi_{f_1}} \ket{0},
	\label{eq:4.5}
\end{equation}
where $\ket{\Psi_f}$ was defined in \eqref{eq:3.23} and $P_{s,R}$ is the projection \eqref{eq:psdef} for the right observer. The matter vacuum bra $\bra{0}$ and ket $\ket{0}$ are present because we are computing a correlation function for states in $\calh_0$. Going forward, we will omit the $\bra{0}$ and $\ket{0}$ symbols in expressions, leaving them implicit. Using the matrix elements \eqref{eq:4.4} as well as \eqref{eq:scalarfield}, we find that \eqref{eq:4.5} is given by
\begin{equation}
	\begin{aligned}
		\frac{1}{V} f_2^*(s_b) f_1(s_b) \sum_{n_a,n_b,n \in \mathbb{Z}} \Bigg[ & \int dt_R d\theta_R\sqrt{-g_R} \left(\Psi^{s_a}_{n_a}(t_R,\theta_R)\right)^* (\psi^s_n)^*(t_R,\theta_R) \Psi^{s_b}_{n_b}(t_R,\theta_R) \\
		& \int dt_R d\theta_R\sqrt{-g_R} \left(\Psi^{s_b}_{n_b}(t_R,\theta_R)\right)^* \psi^s_n(t_R,\theta_R) \Psi^{s_a}_{n_a}(t_R,\theta_R)
		\Bigg].
	\end{aligned}
\end{equation}
If we perform the sum first before performing the integrals, we obtain an integrand that is $SO^+(2,1)$-invariant, leading to a divergence after integrating. The Fadeev-Popov procedure for gauge-fixing this has been described in Section \ref{sec:twoentangled}. To compute this expression, we instead use \eqref{eq:usefulintegral} to perform the integrals first and then compute the sums. We use \eqref{eq:prescription} to cancel the delta function divergences that arise from the CG coefficient orthogonality relations.

We can represent correlation functions with diagrams that are closely related to the diagrams discussed in the previous section. We make the following graphical definitions:\footnote{In our diagrams, vertically oriented straight lines with arrows correspond to an observer's propagator, while other lines with arrows represent the propagator of a particle in the matter QFT.}
\begin{equation}
	\label{eq:4.7}
	\begin{aligned}
		\vcenter{\hbox{\includegraphics[scale=0.4]{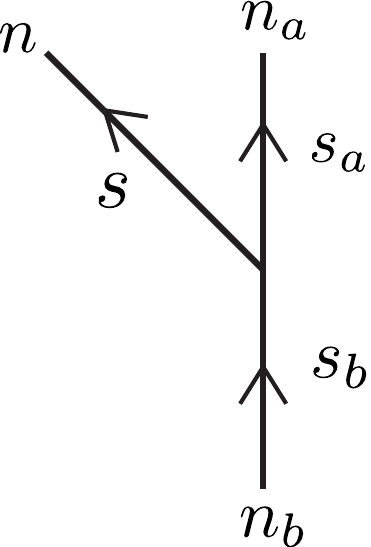}}} & \quad := \int dt\, d\theta\, \sqrt{-g} \left(\Psi^{s_a}_{n_a}(t,\theta)\right)^* (\psi^s_n)^*(t,\theta) \Psi^{s_b}_{n_b}(t,\theta),
	\end{aligned}
\end{equation}

\begin{equation}
	\vcenter{\hbox{\includegraphics[scale=0.4]{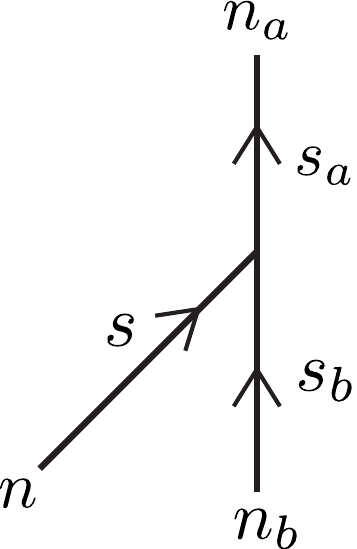}}} \quad := \quad \int dt\, d\theta\, \sqrt{-g} \left(\Psi^{s_a}_{n_a}(t,\theta)\right)^* \psi^s_n(t,\theta) \Psi^{s_b}_{n_b}(t,\theta),
\label{eq:5.8}
\end{equation}

\begin{equation}
	\vcenter{\hbox{\includegraphics[scale=0.4]{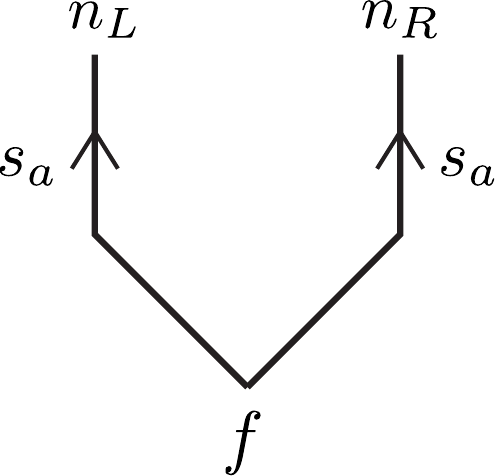}}} \quad := \quad \delta_{n_L,-n_R} (-1)^{n_L} f(s_a),
\end{equation}

\begin{equation}
	\label{eq:4.10}
	\vcenter{\hbox{\includegraphics[scale=0.4]{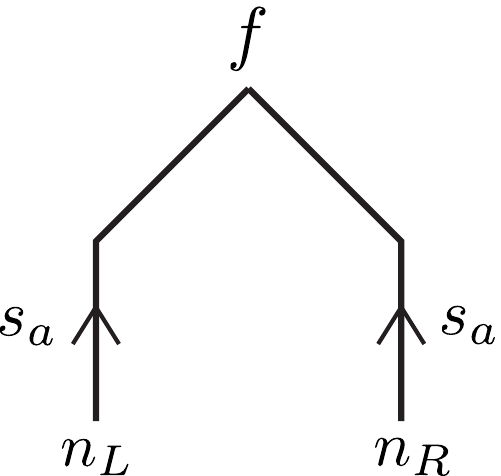}}} \quad := \quad \delta_{n_L,-n_R} (-1)^{n_L} f^*(s_a).
\end{equation}
The two-point function \eqref{eq:4.5} becomes
\begin{equation*}
	\includegraphics[scale=0.4]{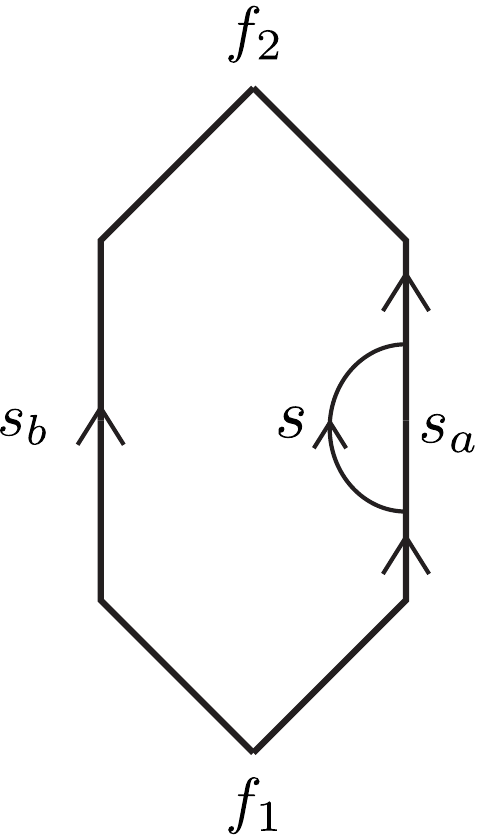}.
\end{equation*}
The four-point correlator,
\begin{equation}
	\braket{\Psi_{f_2}| P_{R,s_d} \phi_R P_{R,s_c} \phi_R P_{R,s_b} \phi_R P_{R,s_a} \phi_R |\Psi_{f_1}} \quad = \quad  \vcenter{\hbox{\includegraphics[scale=0.3]{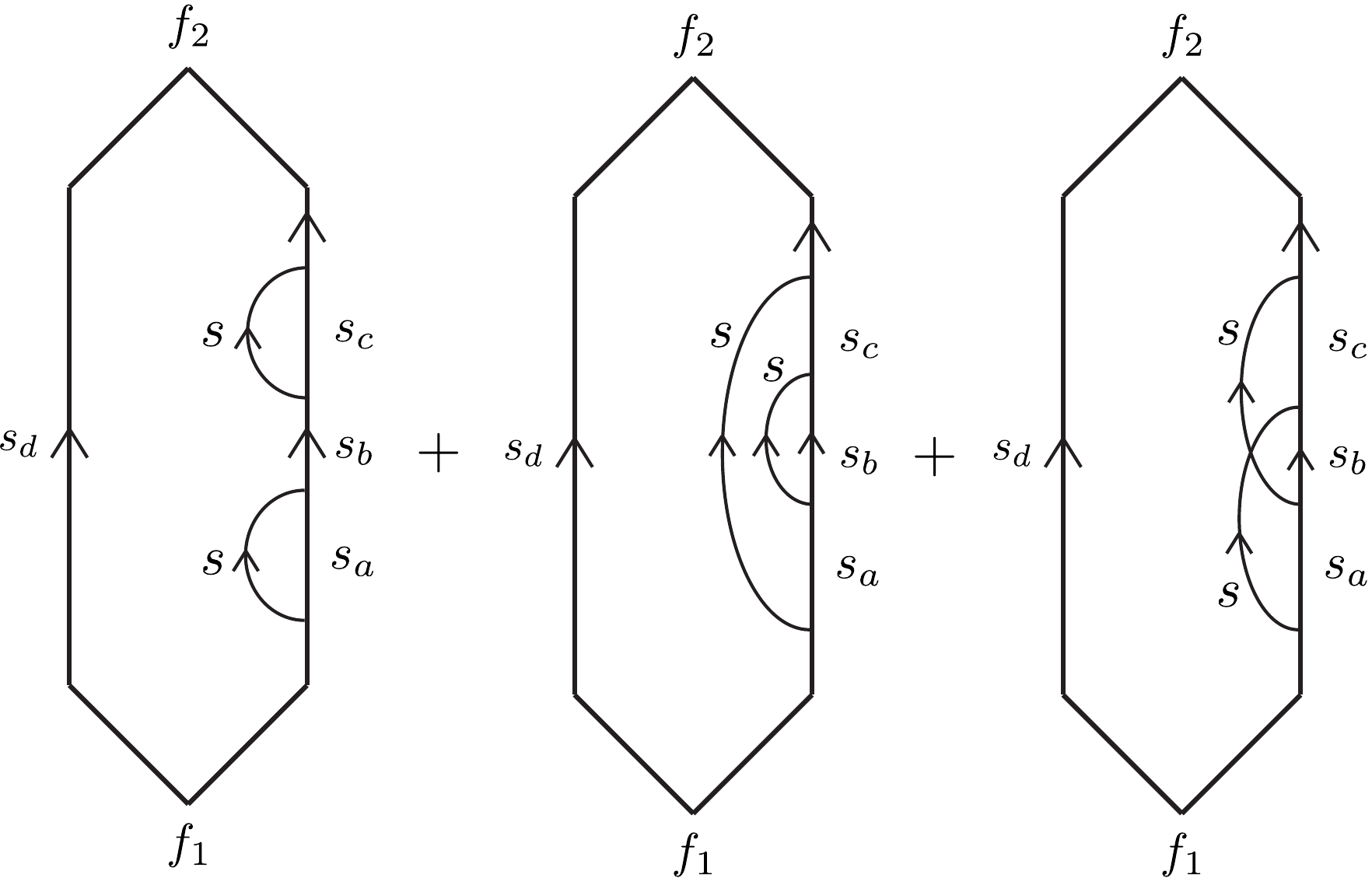}}},
	\label{eq:fourpoint}
\end{equation}
is then represented by the sum of three diagrams, each corresponding to a different Wick contraction of the matter operators. Each diagram is evaluated by making the substitutions according to \eqref{eq:4.7}-\eqref{eq:4.10} and summing over all of the indices. The dependence of each vertex on its indices is given by a CG coefficient (see \eqref{eq:usefulintegral}), such that the sum over the indices is completely fixed by symmetries.

There are a few more useful identities that are conveniently expressed in the diagrammatic language:

\begin{equation}
\label{eq:leftbecomesright}
\quad \quad \quad \quad \quad \quad	\includegraphics[scale=0.4]{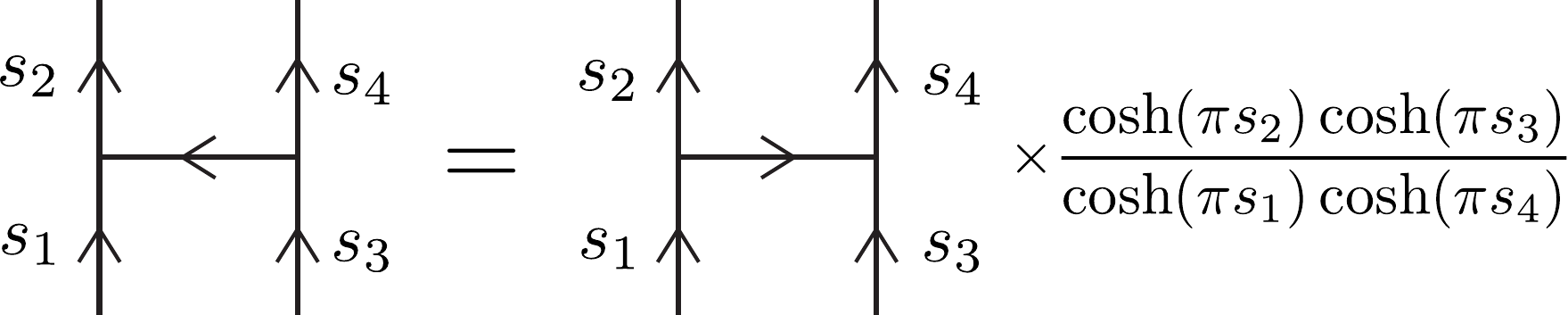},
\end{equation}
\begin{equation}
	\includegraphics[scale=0.4]{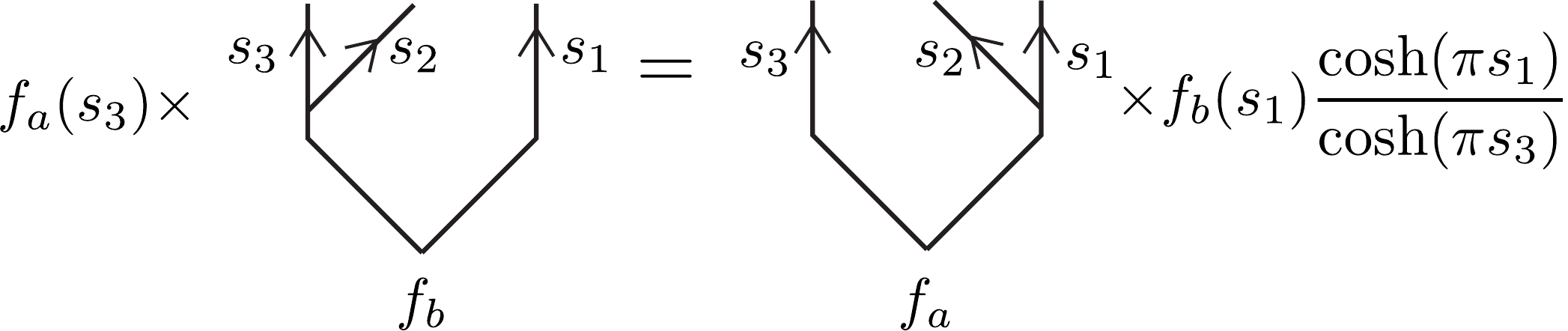},
	\label{eq:philphirequation}
\end{equation}
and
\begin{equation}
	\includegraphics[scale=0.4]{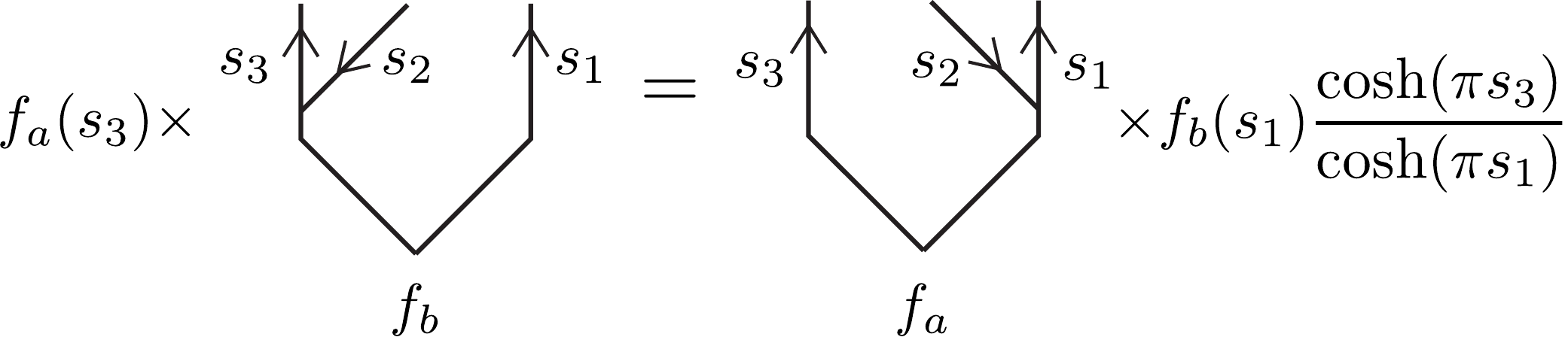}.
\end{equation}
These allow us to rewrite any diagram in terms of a diagram with no vertices on the left branch. For example,
\begin{align}
	&\braket{\Psi_{f_2}|P_{L,s_b} \phi_L \phi_R P_{R,s_a} |\Psi_{f_1}} 
	\\
	\label{eq:5.16line}
	&\vcenter{\hbox{=}} \vcenter{\hbox{\includegraphics[scale=0.4]{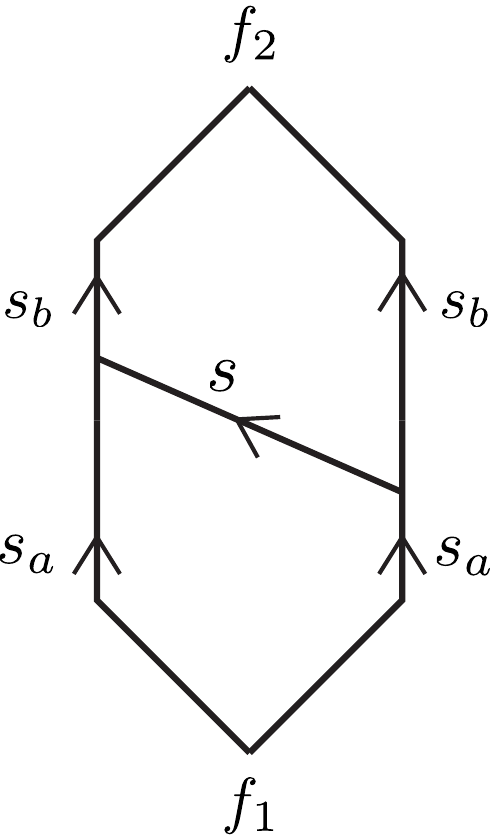}}} \vcenter{\hbox{=}} \vcenter{\hbox{\includegraphics[scale=0.4]{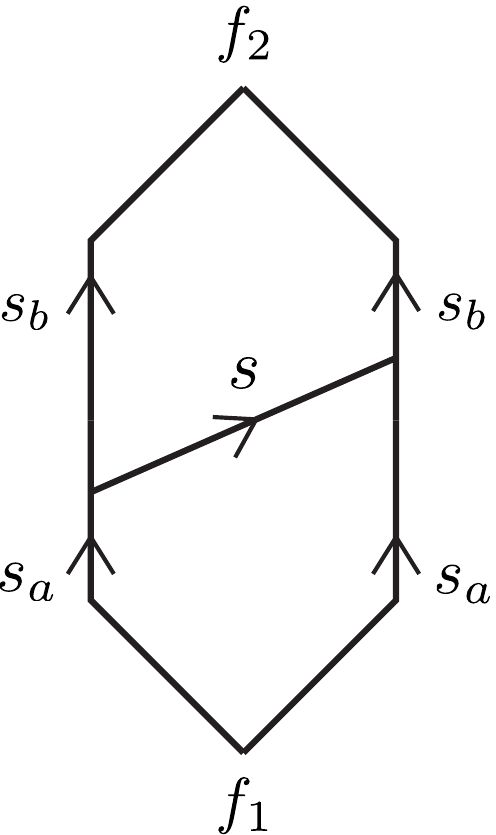}}}  = \braket{\Psi_{f_2}|P_{L,s_b} \phi_R \phi_L P_{R,s_a} |\Psi_{f_1}}
	\\
	&\vcenter{\hbox{=}} \vcenter{\hbox{\includegraphics[scale=0.4]{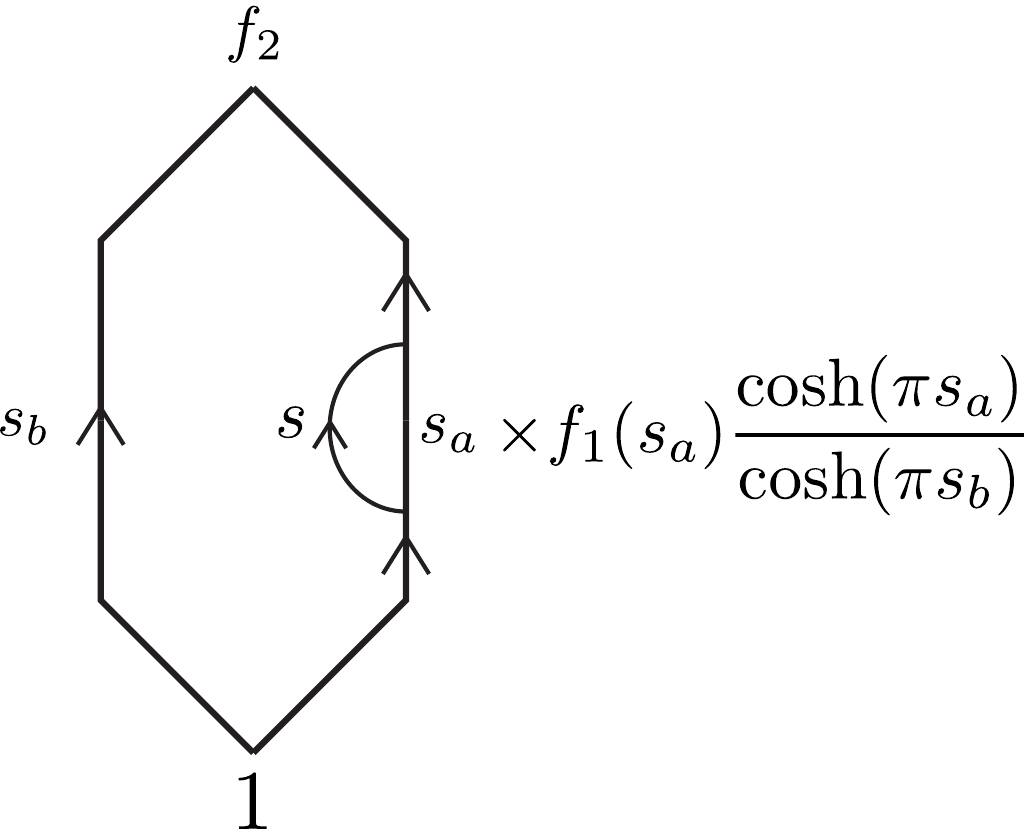}}} \quad  \vcenter{\hbox{=}} \quad \braket{\Psi_{f_2}|P_{R,s_b} \phi_R P_{R,s_a}\phi_R | \Psi_{1}} f_1(s_a) \frac{\cosh(\pi s_a)}{\cosh(\pi s_b)}.
\end{align}
The ordering of $\phi_R$ and $\phi_L$ determines whether the arrows on the lines labeled $s$ in \eqref{eq:5.16line} are pointing to the left or the right. While the order does not matter in this example, we will see soon that $\phi_L$ and $\phi_R$ do not commute. However, $\phi_L$ and $\phi_R$ do commute when acting on state in $\calh_0$. That is,
\begin{equation}
[\phi_L,\phi_R] \ket{\Psi_f} = 0.
\end{equation}
This can be seen from \eqref{eq:leftbecomesright}. When the two lower lines are connected, then $s_1 = s_3$ and $s_2 = s_4$. Hence, the ordering of $\phi_L$ and $\phi_R$ does not matter when acting on $\ket{\Psi_f}$. This suggests that in the subspace $\calh_0$, the two observers are effectively spacelike separated. In states with matter particles, this is no longer the case.

\subsection{The semiclassical limit}

We now discuss how correlators of $\phi_R$ reproduce correlators of the quantum field operator $\varphi_R(t)$ (defined in Section \ref{sec:CLPW}) in the semiclassical limit. The key point is that an observer can move along a classical geodesic to a good approximation when their mass is large. We define the operator\footnote{For our purposes, we also could have used $\Theta_{\Lambda,\infty}$ (which is more similar to what \cite{CLPW} used to get a type II$_1$ algebra) instead of $\Theta_{\Lambda - \frac{\epsilon}{2},\Lambda + \frac{\epsilon}{2}}$. All that matters is that there is a lower bound on the energy.}
\begin{equation}
\label{eq:pidef}
    \Pi_{\Lambda,\epsilon} := \Theta_{\Lambda - \frac{\epsilon}{2},\Lambda + \frac{\epsilon}{2}}(H_R) \Theta_{\Lambda - \frac{\epsilon}{2},\Lambda + \frac{\epsilon}{2}}(H_L),
\end{equation}
where $\Theta_{a,b}(x)$ is a function that equals $1$ for $x \in (a,b)$ and zero otherwise. The usual Heaviside theta function is $\Theta = \Theta_{0,\infty}$. The operator $\Pi_{\Lambda,\epsilon}$ is a projection onto the subspace of the physical Hilbert space where both the left and right observers have energies in a microcanonical window centered at $\Lambda$ with width $\epsilon$. We claim that for any $f(s)$ that is nonzero for all $s > s_0$ for some $s_0 > 0$,
\begin{equation}
 \lim_{\Lambda \rightarrow \infty}   \frac{\braket{\Psi_{f}|\phi_R(\tau_1) \phi_R(\tau_2) \cdots \phi_R(\tau_N) \Pi_{\Lambda,\epsilon}|\Psi_{f}}}{\braket{\Psi_{f}| \Pi_{\Lambda,\epsilon}|\Psi_{f}}} = \braket{0|\varphi_R(\tau_1) \varphi_R(\tau_2) \cdots \varphi_R(\tau_N) |0},
 \label{eq:semiclassicalverify}
\end{equation}
where the correlator on the right hand side is a QFT correlator.

Recall from Section \ref{sec:twoentangled} that $\ket{\Psi_f}$ is a superposition of singlets, where each singlet is labeled by the mass of the two observers (see \eqref{eq:3.23}). If the mass is very large, then the two observers may be treated classically. The classical configuration that corresponds to a two-observer singlet is a pair of antipodal geodesics. Because we only consider operators dressed to the right observer, the relative timeshift mode between the two observers' clocks does not affect the correlator. Hence, the limit should produce a QFT correlator. In Section \ref{sec:hhstate}, we explicitly verify \eqref{eq:semiclassicalverify} in some examples. We may alternatively define the operator
\begin{equation}
    \phi^{\Lambda,\epsilon}_R(\tau) := \Pi_{\Lambda,\epsilon} \phi_R(\tau) \Pi_{\Lambda,\epsilon},
\end{equation}
and it follows that
\begin{equation}
 \lim_{\epsilon \rightarrow \infty} \lim_{\Lambda \rightarrow \infty}   \frac{\braket{\Psi_{f}|\phi^{\Lambda,\epsilon}_R(\tau_1) \phi^{\Lambda,\epsilon}_R(\tau_2) \cdots \phi^{\Lambda,\epsilon}_R(\tau_N) |\Psi_{f}}}{\braket{\Psi_{f}| \Pi_{\Lambda,\epsilon}|\Psi_{f}}} = \braket{0|\varphi_R(\tau_1) \varphi_R(\tau_2) \cdots \varphi_R(\tau_N) |0}.
 \label{eq:semi2}
\end{equation}
The $\Lambda \rightarrow \infty$ limit is taken first, so that $\epsilon \ll \Lambda$. The reason why \eqref{eq:semiclassicalverify} agrees with \eqref{eq:semi2} is because the amplitude for $\phi_R$ to induce a transition between a state with order $\Lambda$ energy and order one energy is exponentially suppressed in $\Lambda$, so inserting a single projection in the numerator in \eqref{eq:semiclassicalverify} is equivalent to conjugating each $\phi_R$ operator with $\Pi_{\Lambda,\epsilon}$.

\subsection{More general operators}

In this section, we have focused our attention on a single dressed matter field operator, $\phi_R$. We now comment on other dressed QFT operators. The normal-ordered QFT operator $:\varphi^n:$ can be dressed to an observer in the same way as $\varphi$. Following \eqref{eq:4.4}, we define the gauge-invariant operator $:\phi^n:_{R}$ as follows,
\begin{equation}
\braket{s^\prime_R n^\prime_R | :\phi^n:_{R} | s_R n_R} :=  \int dt_R d\theta_R\sqrt{-g_R} \left(\Psi^{s_R^\prime}_{n_R^\prime}(t_R,\theta_R)\right)^*:\varphi^n:(t_R,\theta_R) \Psi^{s_R}_{n_R}(t_R,\theta_R).
\label{eq:phin}
\end{equation}
We can also dress spinning operators to an observer. For instance, to dress $\partial_\mu \varphi$, we define $\overrightarrow{\partial} \phi_R$ as follows:
\begin{equation}
	\braket{s^\prime_R n^\prime_R | \overrightarrow{\partial} \phi_R | s_R n_R} :=  \int dt_R d\theta_R\sqrt{-g_R} \left(\Psi^{s_R^\prime}_{n_R^\prime}(t_R,\theta_R)\right)^* g^{\mu \nu}\partial_\mu \varphi (t_R,\theta_R) \, \partial_\nu \Psi^{s_R}_{n_R}(t_R,\theta_R).
	\label{eq:5.19}
\end{equation}
These operators are invariant under the isometries of dS$_2$ (including the discrete ones) because they are manifestly diffeomorphism invariant, and the isometries of dS$_2$ are the diffeomorphisms that do not change the metric. Likewise, we may define
\begin{equation}
	\braket{s^\prime_R n^\prime_R | \overleftarrow{\partial} \phi_R | s_R n_R} :=  \int dt_R d\theta_R\sqrt{-g_R} \, \partial_\mu \left(\Psi^{s_R^\prime}_{n_R^\prime}(t_R,\theta_R)\right)^* g^{\mu \nu}\partial_\nu \varphi \,  \Psi^{s_R}_{n_R}(t_R,\theta_R).
	\label{eq:5.20}
\end{equation}
Integration by parts reveals the following identity:
\begin{equation}
	\label{eq:5.22}
	\overleftarrow{\partial} \phi_R + \overrightarrow{\partial} \phi_R = - m^2 \phi_R.   
\end{equation}
Furthermore,
\begin{equation}
		\label{eq:5.23}
	\left(\overleftarrow{\partial} \phi_R\right)^\dagger = \overrightarrow{\partial} \phi_R.
\end{equation}
Correlators involving these more general operators may be computed using the same techniques discussed above. First, make the substitution \eqref{eq:scalarfield}, then evaluate the correlator of the raising and lowering operators in the Bunch-Davies vacuum, and then integrate over all of the spacetime points $(t_R,\theta_R)$. Finally, use \eqref{eq:prescription} to cancel the divergence that appears in the integration. In practice, one needs to be able to evaluate the integrals in \eqref{eq:5.19} and \eqref{eq:5.20}, where each appearance of $\varphi$ is replaced by either of the modefunctions $\psi^s_n$ or $(\psi^s_n)^*$. We have not evaluated these integrals analytically, but they can be evaluated numerically.

The operators $:\phi^n:_L$, $	\overleftarrow{\partial} \phi_L$, and $\overrightarrow{\partial} \phi_L$ are defined just as above, with the $L$ subscript in place of $R$. Above, we showed analytically that $[\phi_L,\phi_R]$ annihilates any state in $\calh_0$. We have been able to numerically verify that
\begin{equation}
	[\phi_L, \overleftarrow{\partial} \phi_R] \ket{\Psi_f} = [\overleftarrow{\partial} \phi_L,\overleftarrow{\partial} \phi_R] \ket{\Psi_f} = \braket{\Psi_{f_1}|[:\phi^{n}:_L,:\phi^{n}:_R]|\Psi_{f_2}} = 0.
	\label{eq:leftrightcommute}
\end{equation}
It is natural to ask whether in any QFT, a commutator of a local operator dressed to the left observer and a (possibly different) local operator dressed to the right observer annihilates $\calh_0$.  The examples we considered in \eqref{eq:leftrightcommute} are not varied enough to fully probe the general case, so we will leave this question open.

\section{The Hartle-Hawking state}

\label{sec:hhstate}

In the previous section, we defined gauge-invariant matter operators dressed to an observer and discussed the computation of correlation functions. In this section, we introduce the Hartle-Hawking state, a canonical state that may be intuitively prepared using the Euclidean path integral, as described briefly in Section \ref{sec:CLPW}. Understanding the Hartle-Hawking state should be essential for understanding de Sitter holography, just as the thermofield double state is essential for AdS/CFT.

Because the Hartle-Hawking state is intuitively prepared using the Euclidean path integral, it is natural to demand that the Hartle-Hawking state is in $\calh_0$, such that the matter QFT is in the Bunch-Davies vacuum. Thus, the Hartle-Hawking state is $\ket{\psifhh}$ for some function $f_{HH}$. To find $f_{HH}$, we will impose \eqref{eq:lr}, or
\begin{equation}
\phi_L \ket{\psifhh} = \phi_R \ket{\psifhh}.
\label{eq:6.1}
\end{equation}
Equation \eqref{eq:philphirequation} implies that
\begin{equation}
f_a(s_3) P_{L,s_3} P_{R,s_1} \phi_L  \ket{\Psi_{f_b}} = f_b(s_1) \frac{\cosh(\pi s_1)}{\cosh (\pi s_3)} P_{L,s_3} P_{R,s_1} \phi_R \ket{\Psi_{f_a}}.
\end{equation}
Setting $f_a = f_b = f_{HH}$, we obtain
\begin{equation}
f_{HH}(s_3) P_{L,s_3} P_{R,s_1} \phi_L  \ket{\psifhh} = f_{HH}(s_1) \frac{\cosh(\pi s_1)}{\cosh (\pi s_3)} P_{L,s_3} P_{R,s_1} \phi_R \ket{\psifhh}.
\end{equation}
Thus, \eqref{eq:6.1} holds precisely when
\begin{equation}
f_{HH}(s) = \frac{1}{\cosh (\pi s)},
\label{eq:6.4}
\end{equation}
up to a constant of proportionality. Because $s$ is equivalent to the observer's mass at large $s$ (according to \eqref{eq:masss}), a comparison with \eqref{eq:hh} suggests that $f_{HH}(s)$ ought to be proportional to $e^{- \pi s}$ at large $s$, which is the case.

We now consider other dressed QFT operators. By numerically evaluating the appropriate integrals, we have verified that
\begin{align}
	:\phi^n:_L \ket{\psifhh} \, \, &= \, \,  :\phi^n:_R \ket{\psifhh}, \label{eq:propertyphin}
	\\
		\overleftarrow{\partial} \phi_L \ket{\psifhh} \, \, &= \, \,  \overrightarrow{\partial} \phi_R \ket{\psifhh}.
\end{align}
The general lesson appears to be the following. Let $\calo_L$ (resp. $\calo_R$) be a local QFT operator\footnote{To be clear, $\calo_L$ and $\calo_R$ are constructed from the same local QFT operator.} dressed to the left (resp. right) observer. Then the Hartle-Hawking state has the property that
\begin{equation}
	\label{eq:HHproperty}
	\calo_L \ket{\psifhh} =  \Theta \left(\calo_R\right)^\dagger \Theta \ket{\psifhh},
\end{equation}
where $\Theta$ is the antiunitary CPT operator. The operators \eqref{eq:phin}, \eqref{eq:5.19}, and \eqref{eq:5.20} are all invariant under conjugation by $\Theta$. Even though we have been working with a free QFT, we expect that \eqref{eq:HHproperty} holds for general operators in general QFTs. This is because the operators $:\varphi^n:$ for $n \ge 2$ are not free fields, yet \eqref{eq:propertyphin} is still true.

Next, we compute correlation functions in the Hartle-Hawking state.

\subsection{Zero-point function}
\label{sec:zero-point}

We begin by computing \begin{equation}
\braket{\psifhh | g(\hat{s}_R) | \psifhh},
\end{equation}
for an arbitrary function $g(s)$. Using \eqref{eq:prescription} to gauge-fix (or divide by $V$), it is given by
\begin{equation}
\braket{\psifhh | g(\hat{s}_R) | \psifhh} = \int_0^\infty ds \, |f_{HH}(s)|^2 \rhopl(s) \, g(s).
\label{eq:6.6}
\end{equation}
We define
\begin{equation}
\rho(s) := |f_{HH}(s)|^2 \rhopl(s) = \frac{s \tanh \pi s}{4 \pi^2 \cosh^2 (\pi s)}. \label{eq:rhos}
\end{equation}

In the semiclassical setup that was described in Section \ref{sec:CLPW}, the Hartle-Hawking state is a thermal state on the commutative algebra generated by $H_R$ alone. We can recover this from
\begin{equation}
\braket{\psifhh | g(\hat{s}_R) \Pi_{\Lambda,\epsilon}| \psifhh}    = \int_0^\infty ds \, \rho(s) \, g(s) \Theta_{\Lambda - \frac{\epsilon}{2},\Lambda + \frac{\epsilon}{2}}\left(\sqrt{s^2 + \frac{1}{4}}\right),
\label{eq:gwithpi}
\end{equation}
where $\Pi_{\Lambda,\epsilon}$ was defined in \eqref{eq:pidef}. Next, we define $q := s - \Lambda$, $\tilde{g}(q) := g(s)$ and take the large $\Lambda$ limit. Then, \eqref{eq:gwithpi} becomes
\begin{equation}
\frac{\Lambda e^{- 2 \pi \Lambda}}{\pi^2}  \int_{-\epsilon/2}^{\epsilon/2} dq \, e^{- 2 \pi q} \, \tilde{g}(q).
\label{eq:6.11}
\end{equation}
We have arbitrarily restricted $q$ to be in $(-\frac{\epsilon}{2},\frac{\epsilon}{2})$, but we could have instead chosen any other range by modifying the projection $\Pi_{\Lambda,\epsilon}$. For instance, \cite{CLPW} restricted to $q > 0$.

Physically, a detector that is moving along a geodesic in de Sitter and interacting with its environment will experience a thermal state as if it were inside a bath of temperature $\frac{1}{2 \pi}$. Observables may be computed in the canonical ensemble. When the motion of the detector is instead described by a relativistic quantum particle, we expect that the ensemble is modified to $\rho(s)$, such that the canonical ensemble is recovered when $s$ is large.

We may interpret $\rho(s)$ as the density of states in a putative dual holographic theory. We deduced $\rho(s)$ by requiring that $\ket{\psifhh}$ obeys \eqref{eq:HHproperty}. The analogue of $\rho(s)$ in JT gravity is the disk density of states, $\rho_{JT}(s) := s \sinh 2 \pi s$. In JT gravity, the density of states can be determined by requiring that the Hartle-Hawking state furnishes a trace on the algebra associated to an AdS boundary \cite{Kolchmeyer:2023gwa, Penington:2023dql}. The density of states can also be determined by computing the Euclidean disk path integral. We do not know of a Euclidean path integral that computes \eqref{eq:rhos}. However, the large $s$ behavior of $\rho(s)$ can be computed in Euclidean signature as a closed geodesic instanton on the two-sphere. The geodesic may be interpreted as the ``thermal circle'' of the observer. We perform this calculation in Appendix \ref{sec:oneloopappendix}.

\subsection{Two-point function}
\label{sec:two-pointfunction}

Next, we consider the two-point function. In the time domain, it is
\begin{equation}
        \braket{\psifhh| e^{ -i H_R \tau_2}  \phi_R  e^{-i H_R \tau_1} \phi_R |\psifhh},
    \label{eq:timedomain}
\end{equation}
while in the energy domain, it is
\begin{equation}
\braket{\psifhh| P_{R,s_2} \phi_R P_{R,s_1} \phi_R |\psifhh}.
\label{eq:thetwopointfunction}
\end{equation}
We defined $H_R$ in \eqref{eq:HRdef}. The two domains are related via
\begin{equation}
 \braket{\psifhh| e^{-i H_R \tau_2}  \phi_R  e^{-i H_R \tau_1} \phi_R |\psifhh} =  \int_0^\infty ds_1 \, ds_2 \,  e^{-i \tau_2 \sqrt{s_2^2 + \frac{1}{4}}}e^{-i \tau_1 \sqrt{s_1^2 + \frac{1}{4}}} \, \braket{\psifhh| P_{R,s_2} \phi_R P_{R,s_1} \phi_R |\psifhh}.
 \label{eq:twodomains}
\end{equation}

To compute \eqref{eq:thetwopointfunction}, it is helpful to use the following equation, written in the graphical language of the previous section:
\begin{equation}
	\vcenter{\hbox{\includegraphics[width=0.10\linewidth]{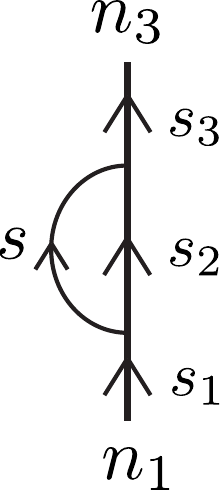}}} =  \frac{1}{64 \pi^7} s_1 \tanh \pi s_1 \,  s_2 \tanh \pi s_2 \, \Gamma\left(\frac{1}{2}\left(\frac{1}{2} \pm i s \pm i s_1 \pm i s_2\right)\right) \frac{\delta(s_1 - s_3)}{\rho(s_1)} \delta_{n_1,n_3}.
\end{equation}
It follows that
\begin{equation}
	\label{eq:6.14}
\braket{\psifhh| P_{R,s_2} \phi_R P_{R,s_1} \phi_R |\psifhh} =  \frac{1}{64 \pi^7} s_1 \tanh \pi s_1 \,  s_2 \tanh \pi s_2 \,  \Gamma\left(\frac{1}{2}\left(\frac{1}{2} \pm i s \pm i s_1 \pm i s_2\right)\right),
\end{equation}
where \eqref{eq:prescription} has been used. Interestingly, we see at the level of the two-point function that the Hartle-Hawking state is tracial, because \eqref{eq:6.14} is invariant under $s_1 \leftrightarrow s_2$. This is only possible when $f_{HH}$ is proportional to \eqref{eq:6.4}. As we will soon see, this tracial property does not hold for higher-point functions due to crossed Wick contractions.
 
The integrals in \eqref{eq:twodomains} converge for all $\tau_1$ and $\tau_2$. This is in accord with the observation in \cite{Witten} that dressing an operator to an observer regulates the UV divergence that ordinarily arises when two operators are close together. At late times, the two-point function decays as $\tau^{-3}$, in agreement with JT gravity \cite{Mertens:2022irh, Bagrets:2017pwq}.

Next, we take the semiclassical limit of the two-point function and compare with \eqref{eq:comparefirst}. We take the semiclassical limit by writing $q_1 = s_1 - \Lambda$, $q_2 = s_2 - \Lambda$. Using
\begin{align}
	&\Gamma\left(\frac{1}{2}\left(\frac{1}{2} \pm i s \pm i s_1 \pm i s_2\right)\right) \frac{1}{4 \pi^3} \rightarrow \frac{e^{- 2\pi \Lambda} e^{-\pi(q_1 + q_2)}}{\Lambda} \frac{1}{\pi}\Gamma\left(\frac{1}{2}\left(\frac{1}{2} \pm i s \pm i (q_1 - q_2)\right)\right)
\end{align}
at large $\Lambda$, we find that
\begin{equation}
\braket{\psifhh| P_{R,s_2} \phi_R P_{R,s_1} \phi_R 
 \Pi_{\Lambda,\epsilon}|\psifhh} \rightarrow  \frac{\Lambda}{16 \pi^5}  e^{- 2\pi \Lambda} e^{-\pi(q_1 + q_2)} \Gamma\left(\frac{1}{2}\left(\frac{1}{2} \pm i s \pm i (q_1 - q_2)\right)\right),
\end{equation}
where it is assumed that $q_1 \in (-\infty,\infty)$ and $q_2 \in \left(- \frac{\epsilon}{2},\frac{\epsilon}{2}\right)$.
To work in the time domain, we multiply this by $\int_{-\infty}^\infty dq_1 \int_{-\epsilon/2}^{\epsilon/2} dq_2 \, e^{- i q_1 \tau_1} e^{-i q_2 \tau_2}$. That is, to compare with Section \ref{sec:CLPW}, time evolution should be generated by $H_R - \Lambda$. We make a change of variables to $\lambda = q_1 - q_2$ and $q = q_2$. The result is
\begin{equation}
\frac{\Lambda}{16 \pi^5}  e^{- 2\pi \Lambda} \int_{-\frac{\epsilon}{2}}^{\frac{\epsilon}{2}}  dq \, e^{- 2 \pi q} e^{- i q (\tau_1 + \tau_2)}  \int_{-\infty}^\infty d\lambda \, e^{- i \lambda \tau_1} e^{-\pi \lambda} \Gamma\left(\frac{1}{2}\left(\frac{1}{2} \pm i s \pm i \lambda \right)\right).
\label{eq:6.17}
\end{equation}
Meanwhile, the Wightman function in the QFT is given by
\begin{equation}
\frac{1}{16 \pi^3} \int_{-\infty}^\infty d\lambda \, e^{- i \lambda t} e^{-\pi \lambda} \Gamma\left(\frac{1}{2}\left(\frac{1}{2} \pm i s \pm i \lambda \right)\right) = \braket{0|\varphi_R(0) \varphi_R(t)|0},
\label{eq:integralrepwightmann}
\end{equation}
where $\varphi_R(t)$ was defined in \eqref{eq:varphiR}, when $\varphi$ is taken to be the scalar field in Section \ref{sec:freescalar}.

By comparing \eqref{eq:6.17} and \eqref{eq:comparefirst}, we see that the semiclassical limit of the correlator agrees with the correlator of the analogous operators in the crossed-product algebra, with a prefactor of \begin{equation}
\frac{\Lambda e^{-2 \pi \Lambda}}{\pi^2}.
\label{eq:prefactor}
\end{equation}
In particular, the $q$ and $\lambda$ integrals in \eqref{eq:6.17} should be viewed as integrals over observer energies in microcanonical windows centered on $\Lambda$.

The prefactor \eqref{eq:prefactor} also appears when comparing correlators with no matter insertions (see \eqref{eq:6.11}).
In section \ref{sec:CLPW}, we pointed out that the $e^{- \pi q}$ factor in $\hhket$ can be understood from a semiclassical path integral perspective. The $e^{- 2\pi \Lambda}$ term above can likewise be interpreted as a closed geodesic instanton. The $\Lambda$ prefactor is the one-loop correction, which we verify in Appendix \ref{sec:oneloopappendix}.

\subsection{Four-point function}

The four-point function is graphically depicted in \eqref{eq:fourpoint}. There are two uncrossed Wick contractions, and one crossed Wick contraction. We discuss them separately.

\subsubsection{Uncrossed Wick contractions}
\label{sec:uncrossed}

As long as a given diagram has no crossed Wick contractions, we can represent the diagram as a standard chord diagram with no arrows. The uncrossed Wick contractions become

\begin{equation}
	\label{eq:6.20}
	\vcenter{\hbox{\includegraphics[width=0.20\textwidth]{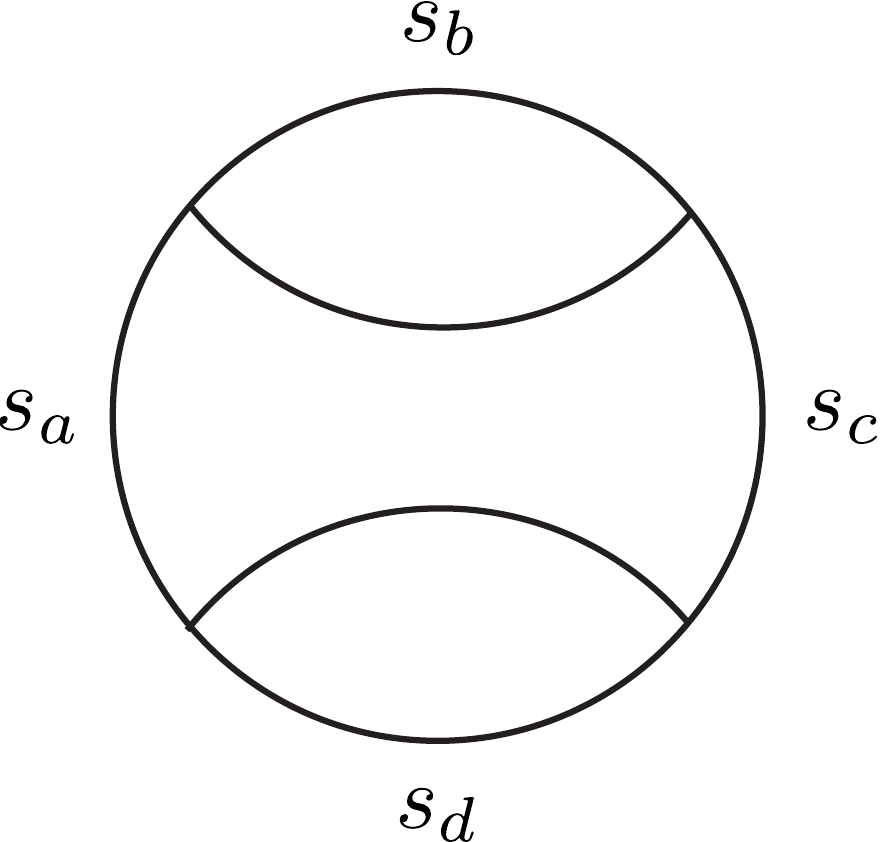}}} \, = \,
	\begin{aligned}[t]
		&\Gamma^s_{s_a s_b} \Gamma^s_{s_a s_d} \frac{1}{\rho(s_a)}\delta(s_c - s_a),
	\end{aligned}
\end{equation}

\begin{equation}
	\label{eq:6.21}
	\vcenter{\hbox{\includegraphics[width=0.20\textwidth]{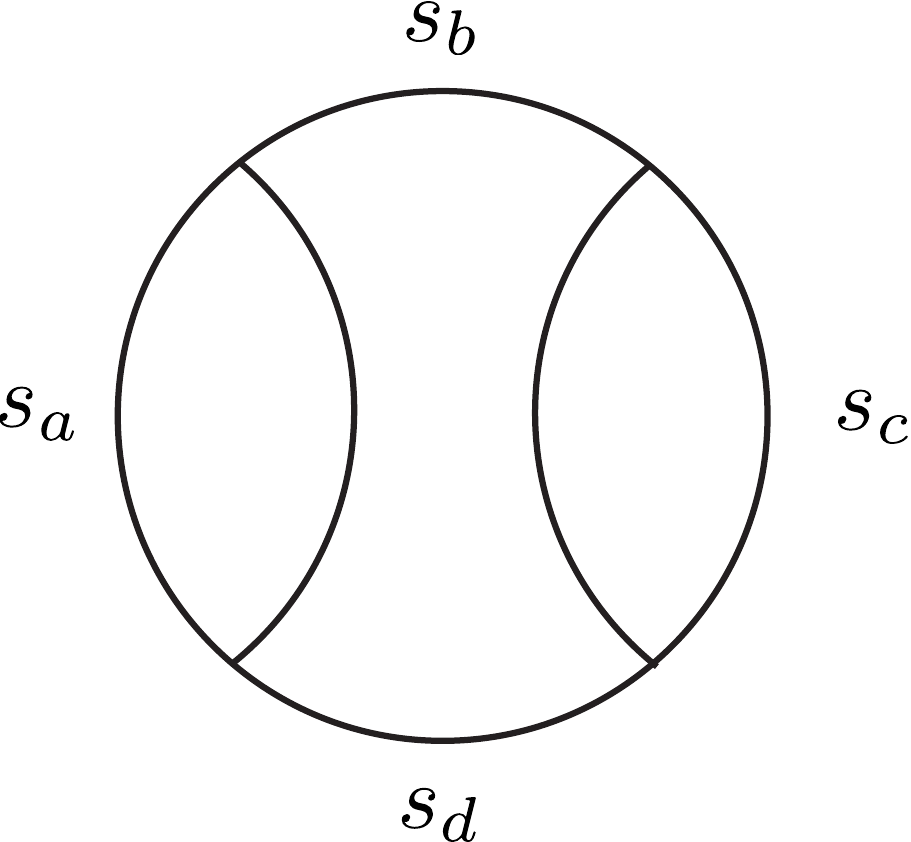}}} \, = \,
	\begin{aligned}[t]
	\Gamma^s_{s_b s_c} \Gamma^s_{s_b s_a} \frac{1}{\rho(s_b)}\delta(s_d - s_b),
	\end{aligned}
\end{equation}
where
\begin{equation}
	\Gamma^s_{s_1 s_2} := \frac{1}{64 \pi^7} s_1 \tanh \pi s_1 \,  s_2 \tanh \pi s_2 \Gamma\left(\frac{1}{2}\left(\frac{1}{2} \pm i s \pm i s_1 \pm i s_2\right)\right). \label{eq:gs1s2}
\end{equation}

For uncrossed Wick contractions, one may evaluate the diagrams using a simpler set of rules. All of the $s$ variables that border the same region in the diagram are to be set equal using delta functions, and for each delta function there is a factor of $\frac{1}{\rho(s)}$ (defined in \eqref{eq:rhos}). A chord that separates two regions contributes a factor of \eqref{eq:gs1s2}.

\subsubsection{Crossed Wick contraction}
\label{sec:crossed}

We now turn to the crossed Wick contraction. Using the result of Appendix \ref{sec:appendixcrossed}, we find that the crossed diagram in \eqref{eq:fourpoint} (with $f_1 = f_2 = f_{HH}$) is given by
\begin{equation}
	\text{cross}(s_a,s_b,s_c,s_d) := s_a \sinh \pi s_a \, s_b \tanh \pi s_b \, s_c \sinh \pi s_c \, s_d \tanh \pi s_d  \, \cals^s(s_a,s_b,s_c,s_d),
	\label{eq:cross}
\end{equation}
where $\cals^s(s_a,s_b,s_c,s_d)$ is a real function that contains a 6j symbol\footnote{Two of the representation labels are set to be the same, $s$, which is why there are only five arguments.} and is symmetric under cyclic permutations of $(s_a, s_b,s_c, s_d)$ as well as $(s_a, s_b,s_c, s_d) \rightarrow (s_d, s_c,s_b, s_a)$. An explicit expression is provided in \eqref{eq:seq}. Due to the prefactors, the crossed diagram is not a cyclic function of $(s_a,s_b,s_c,s_d)$. Instead, we have that
\begin{equation}
	\text{cross}(s_a,s_b,s_c,s_d) = \text{cross}(s_d,s_c,s_b,s_a) \frac{\cosh (\pi s_a) \cosh (\pi s_c)}{\cosh (\pi s_d) \cosh (\pi s_b)}.
	\label{eq:crossrelation}
\end{equation}
We now consider the semiclassical limit of the crossed diagram. For large values of $s_a$ through $s_d$, the factor that spoils the cyclicity of the crossed diagram becomes
\begin{equation}
	\frac{\cosh (\pi s_a) \cosh (\pi s_c)}{\cosh (\pi s_d) \cosh (\pi s_b)} \rightarrow e^{\pi (s_a + s_c - s_b - s_d)}.
	\label{eq:6.25}
\end{equation}
Because this does not equal $1$, it at first appears that the crossed Wick contraction is not cyclic even in the semiclassical limit, which would contradict the result of \cite{CLPW} that the Hartle-Hawking state furnishes a trace on the algebra. It turns out that in this limit, the 6j symbol is proportional to a delta function $\delta(s_a + s_c - s_b - s_d)$, such that there is no contradiction. In JT gravity plus matter, a delta function also appears when taking the semiclassical limit of the 6j symbol. See Appendix C of \cite{Jafferis:2022wez}. We show in Appendix \ref{sec:appotocorderone} that the four-point function in the semiclassical limit agrees with the four-point function of $\phi_R$ in \eqref{eq:2.3} up to the prefactor \eqref{eq:prefactor}, as expected.

\subsubsection{OTOC}

\label{sec:otoc}

Having studied the crossed Wick contraction of the four-point function, it is now natural to consider the out-of-time-ordered correlator (OTOC). As explained in \cite{Maldacena:2015waa}, the OTOC is a useful probe of quantum chaos. The OTOC decays on the order of the scrambling time, which is logarithmic in the entropy of the system. The rate at which the system scrambles information is characterized by the Lyapunov exponent. The authors of \cite{Maldacena:2015waa} showed that for a unitary large-$N$ quantum-mechanical system at inverse temperature $\beta$, the Lyapunov exponent may not exceed $\frac{2 \pi}{\beta}$. Our goal in this section is to compute the OTOC and find the Lyapunov exponent. 

The correlator we are interested in is
\begin{equation}
\frac{\braket{\psifhh| \phi^B_R(\frac{\tau}{2}) \phi^A_R(-\frac{\tau}{2}) \phi^B_R(\frac{\tau}{2}+u_2) \phi^A_R(-\frac{\tau}{2}+u_1) \Pi_{\Lambda,\epsilon} |\psifhh}}{\braket{\psifhh|\Pi_{\Lambda,\epsilon}|\psifhh}}
	\label{eq:OTOC1}
\end{equation}
where the $A$ and $B$ superscripts denote two flavors of the massive scalar field (so that there are no uncrossed contractions), with masses given by $s_A$ and $s_B$ (using \eqref{eq:masss}). The projection $\Pi_{\Lambda,\epsilon}$ projects onto the subspace of the Hilbert space where the right observer's mass is in a microcanonical window of width $\epsilon$ centered on $\Lambda$ (see \eqref{eq:pidef}).\footnote{It does not matter if $\Pi_{\Lambda,\epsilon}$ restricts either the right or left observer's energy, because we are acting with $\Pi_{\Lambda,\epsilon}$ on a state where the two energies are equal.} We divide by the normalization factor $\braket{\psifhh|\Pi_{\Lambda,\epsilon}|\psifhh}$, computed in Section \ref{sec:zero-point}.

When $\Lambda$ is fixed and $\tau$ is large, the OTOC \eqref{eq:OTOC1} decays to zero. To see this, simply note that \eqref{eq:cross} is a smooth function, so once the time evolution phases are put in, the integrals over the $s$ parameters will go to zero at large times due to the oscillations of the integrand. When $\tau$ is fixed and $\Lambda$ is large, the OTOC becomes a four-point function in the QFT,
\begin{equation}
    \braket{0| \varphi_R^B\left(\frac{\tau}{2}\right)
                \varphi_R^A\left(-\frac{\tau}{2}\right)
                \varphi_R^B\left(\frac{\tau}{2} + u_2\right)
                \varphi_R^A\left(-\frac{\tau}{2} + u_1\right)
   |0 },
\end{equation}
where $\varphi_R$ was defined in Section \ref{sec:CLPW}. We show this in Appendix \ref{sec:appotocorderone}. This is the same semiclassical limit that was considered in Section \ref{sec:two-pointfunction}, when we considered the two-point function.

Now, we consider a different limit, where $\tau$ and $\Lambda$ are taken large simultaneously. We will define $T$ via
\begin{equation}
    e^T := \frac{e^{\tau}}{2 \Lambda},
    	\label{eq:tscr}
\end{equation}
and consider a limit where $\tau$ and $\Lambda$ both go to infinity, keeping $T$ fixed. We define the scrambling time to be
\begin{equation}
	\tau_{scr} := \log 2 \Lambda,
\end{equation}
which corresponds to $T = 0$. After taking this limit, we will obtain a function that is a product of two-point functions when $T \rightarrow -\infty$, and is zero when $T \rightarrow \infty$. We have performed this calculation using two separate methods. In the first method, we begin with the exact answer for the crossed Wick contraction and then simplify it in the above limit. The details are given in Appendix \ref{sec:appotocscrambling}. We will devote the rest of this section to explaining the second method, which is inspired by the computation of the OTOC in JT gravity in \cite{Maldacena:2016upp}. The first method is more straightfoward but highly technical. The second method is meant to provide a physical interpretation that nontrivially agrees with the first method.

We begin by considering the four-point function on the observer's worldline in the regime where the backreaction on the observer is completely negligible. The observer is sitting on the North Pole. This is a four-point function in the QFT. It is
\begin{align}
&\braket{0| \overset{4}{\overbrace{\varphi_R^B\left(\frac{\tau}{2}\right)}}
		 \overset{3}{\overbrace{\varphi_R^A\left(-\frac{\tau}{2}\right)}}
		   \overset{2}{\overbrace{\varphi_R^B\left(\frac{\tau}{2} + u_2\right)}}
	   \overset{1}{\overbrace{\varphi_R^A\left(-\frac{\tau}{2} + u_1\right)}}
   |0 } 
\label{eq:6.30}
\\
&= \braket{0|\varphi_R^B(\frac{\tau}{2}) \varphi_R^B(\frac{\tau}{2} + u_2)|0} \braket{ 0| \varphi_R^A(-\frac{\tau}{2})  \varphi_R^A(-\frac{\tau}{2}+u_1) |0},
\nonumber
\end{align} 
where we have used factorization. We follow the conventions of Section \ref{sec:CLPW}, where $\varphi_R$ refers to a local QFT operator on the observer's worldline that becomes $\phi_R$ after dressing. It will be convenient to refer to each operator by the number that appears above the operator on the left hand side of \eqref{eq:6.30}.

When the observer's mass is not strictly infinite, acting with a dressed operator causes them to recoil. After operator 1 acts, the observer's late-time location on future infinity can in principle be anywhere. Thus, we will integrate over this position using a variable $x_+$. Likewise, after operator 2 acts, the observer's trajectory is also deflected. Their location on past infinity is denoted by a parameter $x_-$. The variables $x_+$ and $x_-$ represent two off-shell modes in the observer's worldline path integral that capture the recoil effects. All other modes are on-shell.  

\begin{figure}[ht!]
	\centering
	\includegraphics[width=0.6\textwidth]{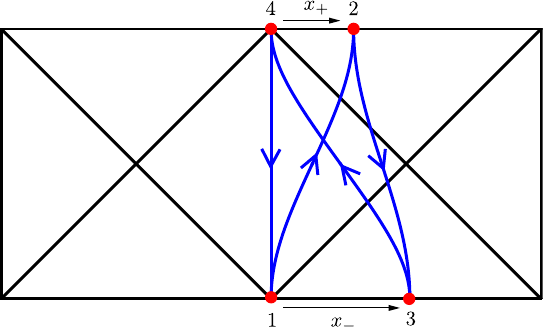}
	\caption{The center vertical line connecting points 1 and 4 is the North Pole. Each red dot represents one of the four operators in the OTOC. As we evolve along the observer's worldline to compute the OTOC, the observer travels from point 1 back to point 1 following the blue arrows. The red dots are at future and past infinity because we are working in the limit $\tau \rightarrow \infty$. The distances labeled by $x_+$ and $x_-$ are greatly exaggerated because \eqref{eq:6.31} will be dominated by order $\frac{1}{\sqrt{\Lambda}}$ values of $x_+$ and $x_-$.}
	\label{fig:foursegment}
\end{figure}

\begin{figure}[ht!]
	\centering
	\includegraphics[width=0.6\textwidth]{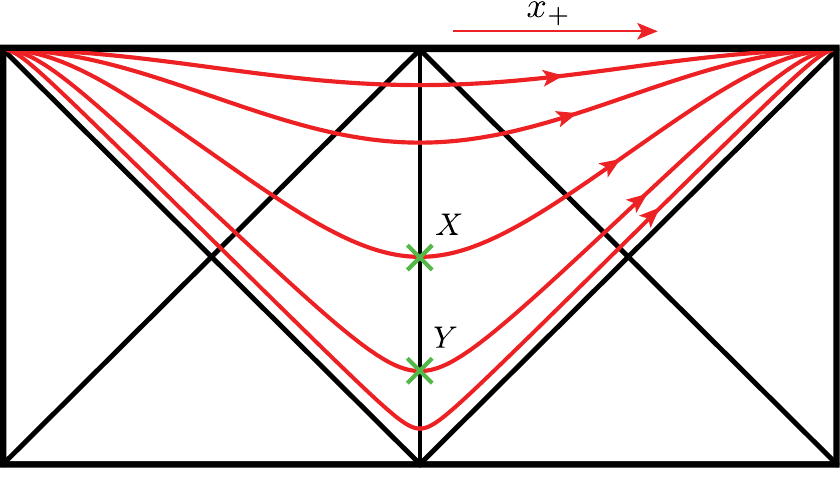}
	\caption{The Hermitian operator $\hat{P}^+$ is defined to be the symmetry generator whose flow lines are shown above. This isometry is used to define the $x_+$ coordinate along future infinity. A constant shift of $x_+$ moves an operator along a flow line. Some examples of flow lines are given in red. We label two arbitrary points $X,Y$ on the center line (the North Pole), which is the observer's worldline before backreaction is taken into account. Note that the flow lines are transverse to the observer's worldline. When a transformation labeled by $x_+$ is applied to point $X$ only, the change in the geodesic distance between $X$ and  $Y$ is of order $x_+^2$ for small $x_+$, due to symmetry. Next, note that when $x_+$ is large enough, point $X$ will be null-separated from point $Y$. Because $x_+$ is proportional to $e^\tau$, this indicates that for a certain choice of kinematics, the OTOC will initially grow and then decay.}
	\label{fig:flow}
\end{figure}

We now discuss our precise method for computing the OTOC using the observer's worldline path integral. In the semiclassical limit, it suffices to assume that the observer's trajectory is given by four geodesic segments, which we depict in Figure \ref{fig:foursegment}. The first and fourth operators are on the North Pole at the same positions as operators 1 and 4 in \eqref{eq:6.30} because they are adjacent to the Hartle-Hawking state in \eqref{eq:OTOC1}. The positions of the second and third operators are integrated along future and past infinity. We label the position of operator 2 along future infinity by the coordinate $x_+$. A constant shift of $x_+$ transforms operator 2 by the isometry that is generated by the null momentum $\hat{P}^+$. We define $\hat{P}^+$ to be a Hermitian operator that generates a flow along the lines shown in Figure \ref{fig:flow}. The point $x_+ = 0$ represents the point on the North Pole where operator 2 is inserted in \eqref{eq:6.30}. That is, we may construct a coordinate system in the Poincare patch by specifying a point on the North Pole and specifying a value of $x_+$, which instructs us to move this point along a flow line in Figure \ref{fig:flow} by a certain amount. This coordinate system is used to locate operator 2. When $x_+$ is nonzero, we write operator 2 as $\varphi_R^B\left(\frac{\tau}{2} + u_2 ; x_+\right)$ to indicate that it is located away from the North Pole. The coordinate $x_-$ is defined analogously using the isometry generated by $\hat{P}^-$, which is the time-reflection of $\hat{P}^+$ (that is, the flow lines that correspond to the Hermitian generator $\hat{P}^-$ are given by the ones in Figure \ref{fig:flow} reflected about a horizontal line that passes through the two bifurcation points). The position of operator 3 is thus labeled by $x_-$, which is integrated over (as well as the point on the North Pole where operator 3 is located in \eqref{eq:6.30}, which is fixed). When $x_-$ is nonzero, we write operator 3 as $\varphi_R^A\left(-\frac{\tau}{2};x_-\right)$ to indicate its position. We emphasize that this prescription for computing the observer's path integral in this limit is an ansatz which is only fully justified by the fact that it agrees with the more direct method used in Appendix \ref{sec:appotocscrambling}, which simplifies the exact formula for \eqref{eq:OTOC1}.

Using the above prescription, the semiclassical path integral is proportional to
\begin{align}
&\int_{-\infty}^\infty dx_+ dx_- e^{i S(x_+,x_-)}
\braket{0|\varphi_R^B\left(\frac{\tau}{2}\right) \varphi_R^A\left(-\frac{\tau}{2} ; x_-\right) \varphi_R^B \left(\frac{\tau}{2} + u_2;x_+\right)   \varphi_R^A\left(-\frac{\tau}{2}+u_1 \right) |0},
 \\
= &\int_{-\infty}^\infty dx_+ dx_- e^{i S(x_+,x_-)} \nonumber 
\\
&\times \braket{0|\varphi_R^B\left(\frac{\tau}{2}\right) e^{i x_+ \hat{P}^+ } \varphi_R^B \left(\frac{\tau}{2} + u_2\right)|0}
\braket{ 0| \varphi_R^A\left(-\frac{\tau}{2}\right) e^{i x_- \hat{P}^- } \varphi_R^A\left(-\frac{\tau}{2}+u_1\right) |0},
 \\
 =
 &\int_{-\infty}^\infty dx_+ dx_- e^{i S(x_+,x_-)}
	\nonumber\\ 
 &\times\int_{-\infty}^\infty \, dP^+ \,	e^{i x_+ P^+ }
	\zeta_{s_B,\frac{\tau}{2}}(P^+)
	\zeta^*_{s_B,\frac{\tau}{2} + u_2}(P^+)
	\int_{-\infty}^\infty \, dP^- \,	e^{i x_- P^- }
	\zeta^*_{s_A,\frac{\tau}{2}}(P^-)
 \zeta_{s_A,\frac{\tau}{2} - u_1}(P^-)		\label{eq:6.31},
\end{align}
where $S(x_+,x_-)$ is the observer's worldline action evaluated for the four geodesic segments shown in Figure \ref{fig:foursegment}, and $\zeta_{s,t}(P^+)$ is a wavefunction in the $P^+$ basis for the state $\varphi_R(t)\ket{0}$. As mentioned above, $\hat{P}^+$ is a Hermitian operator that generates the isometry shown in Figure \ref{fig:flow}. \\

\noindent Explicitly, we have 
\begin{equation}
\zeta_{s,t}(P^+) := \frac{\sqrt{\pi}  e^{-t/2} H_{-i s}^{(1)}(e^{-t} |P^+|)}{2 \sqrt{1 + e^{-2 \pi s}} \, \Gamma\left(\frac{1}{2} + i s\right)}.
\label{eq:zetadef}
\end{equation}
The analogous wavefunction in the $P^-$ basis is given by $\zeta_{s,-t}(P^-)^*$.
Thus, the insertions of $e^{i x_+ P^+}$ and $e^{i x_- P^-}$ in \eqref{eq:6.31} have the effect of displacing operators 2 and 3 from the North Pole to their positions shown in Figure \ref{fig:foursegment}.

The action is given by
\begin{equation}
	e^{iS(x_+,x_-)} = \left(\frac{1}{(1 + x_+ x_-)^2}\right)^{- i \Lambda} \approx e^{2 i \Lambda x_+ x_-},
\end{equation}
where we have used the approximation that the semiclassical path integral is dominated by small values of $x_+ x_-$ at large $\Lambda$. After evaluating the $x_+$, $x_-$ integrals in \eqref{eq:6.31}, we obtain
\begin{equation}
	\int_{-\infty}^\infty \, dP^+ dP^- \, e^{ - i \frac{P^+ P^-}{2  \Lambda}  }	
	\zeta_{s_B,\frac{\tau}{2}}(P^+)
	\zeta^*_{s_B,\frac{\tau}{2} + u_2}(P^+)	
	\zeta_{s_A,\frac{\tau}{2} - u_1}(P^-)
	\zeta^*_{s_A,\frac{\tau}{2}}(P^-).
	\label{eq:6.34}
\end{equation}
This expression indicates that the OTOC is a scattering amplitude for highly boosted particles near the observer's cosmological horizon. Particle $A$ with momentum $P^-$ along the past cosmological horizon scatters against particle $B$ with momentum $P^+$ along the future cosmological horizon. The S-matrix is just a phase, as we would expect for scattering in the eikonal limit. Although the two particles do not directly interact, the observer mediates an interaction between them. This result is analogous to the case of AdS gravity, which has been explained in \cite{Shenker:2014cwa, Maldacena:2016upp, Mertens:2022irh}.

Note that after performing the variable redefinition $(P^+,P^-) \rightarrow (e^{\tau_0} P^+,e^{-\tau_0} P^-)$, \eqref{eq:6.34} becomes
\begin{equation}
	\int_{-\infty}^\infty \, dP^+ dP^- \, e^{ - i \frac{P^+ P^-}{2  \Lambda}  }	
	\zeta_{s_B,\frac{\tau}{2}-\tau_0}(P^+)
	\zeta^*_{s_B,\frac{\tau}{2} + u_2-\tau_0}(P^+)	
	\zeta_{s_A,\frac{\tau}{2} - u_1 + \tau_0}(P^-)
	\zeta^*_{s_A,\frac{\tau}{2} + \tau_0}(P^-),
\end{equation}
which corresponds to invariance under time translation by $\tau_0$ along the observer's worldline. Likewise, if we perform $(P^+,P^-) \rightarrow (e^{\tau_0} P^+,e^{\tau_0} P^-)$, we obtain
\begin{equation}
	\int_{-\infty}^\infty \, dP^+ dP^- \, e^{ - i \frac{P^+ P^- e^{2 \tau_0}}{2  \Lambda}  }	
	\zeta_{s_B,\frac{\tau}{2}-\tau_0}(P^+)
	\zeta^*_{s_B,\frac{\tau}{2} + u_2-\tau_0}(P^+)	
	\zeta_{s_A,\frac{\tau}{2} - u_1 - \tau_0}(P^-)
	\zeta^*_{s_A,\frac{\tau}{2} - \tau_0}(P^-).
\end{equation}
If we set $\tau_0 = \frac{\tau}{2}$, then the integrand only depends on $\tau$ in the exponent,
\begin{equation}
	\int_{-\infty}^\infty \, dP^+ dP^- \, e^{ - i \frac{P^+ P^- e^{\tau}}{2  \Lambda}  }	
	\zeta_{s_B,0}(P^+)
	\zeta^*_{s_B,u_2}(P^+)	
	\zeta_{s_A,- u_1}(P^-)
	\zeta^*_{s_A,0}(P^-).
 \label{eq:6.42}
\end{equation}
Also, each $\zeta$ function above is invariant under $P^+ \rightarrow - P^+$ or $P^- \rightarrow - P^-$ (see \eqref{eq:zetadef}). We may compute corrections to \eqref{eq:6.30} by expanding the exponent. To zeroth order, we recover \eqref{eq:6.30}. The first order term in the expansion, which is proportional to $e^{\tau}$, vanishes due to the odd powers of $P^+$ and $P^-$. Only the terms with even powers of $P^+$ and $P^-$ are nonzero after integration. Thus, the corrections are given by a power series in $e^{2 \tau}$. The leading correction to the OTOC is proportional to $e^{2\tau}$, or $e^{\frac{4 \pi}{\beta_{\text{dS}}} \tau}$, where $\beta_{\text{dS}} = 2\pi$ is the inverse de Sitter temperature. The Lyapunov exponent is $\frac{4 \pi}{\beta_{\text{dS}}}$.

This Lyapunov exponent has an intuitive explanation. We rewrite  \eqref{eq:6.42} as follows:
\begin{equation}
	\int_{-\infty}^\infty \, dP^+ dP^- \, 	\braket{0|\varphi^B_R(0)
        e^{ - i \frac{\hat{P}^+ P^- e^{\tau}}{2  \Lambda}  }    
        \varphi^B_R(u_2)|0} 
 \zeta_{s_A,- u_1}(P^-)
	\zeta^*_{s_A,0}(P^-).
 \label{eq:rewriteotoc}
\end{equation}
For fixed $P^-$, we act with an isometry on one of the $\varphi^B_R$ operators that moves it away from the North Pole along one of the flow lines in Figure \ref{fig:flow} by an amount $x_+ = \frac{P^- e^{\tau}}{2  \Lambda} $. Let the two green crosses in Figure \ref{fig:flow} refer to the insertions of $\varphi^B_R(0)$ and $\varphi^B_R(u_2)$ on the North Pole. As explained in Figure \ref{fig:flow}, the change in the two-point function is proportional to $x_+^2$ for small $x_+$.\footnote{In AdS JT gravity, the change in the two-point function is linear in $x_+$. See (6.54) and (6.55) in \cite{Maldacena:2016hyu} for the transformed two-point functions. This is why the leading correction to the OTOC in JT gravity is proportional to $e^t$.} Thus, the leading correction to the OTOC is order $e^{2\tau}$. We expect that this intuition generalizes to higher dimensions.\footnote{Consider three dimensions. Although we have not explicitly repeated our calculations in three dimensions, there is a natural way to generalize \eqref{eq:6.34}. In addition to the $P^+$ and $P^-$ integrals, we also integrate over $\phi^+$ and $\phi^-$, which label angles in the transverse circle. In the exponent, we replace $P^+ P^-$ with $P^+ P^- \cos\left(\phi^+ - \phi^-\right)$. We only integrate $P^+$ and $P^-$ in the range $(0,\infty)$. The $\zeta$ functions only depend on the momenta and not the $\phi^{\pm}$ coordinates. This generalized integral may be rewritten in a form analogous to \eqref{eq:rewriteotoc}, and it follows that the leading correction to the OTOC is the change in a two-point function after applying an isometry that is transverse to the observer's worldline to one of the operators, as discussed in Figure \ref{fig:flow}. To recover the two-dimensional integral studied in the main text, one should restrict $\phi^{\pm}$ to be either $0$ or $\pi$, and sum over all four combinations. For readers who prefer to think of the OTOC as a scattering amplitude, note that a shockwave along the horizon is a radial null geodesic, labeled by a momentum and an angle $\phi$ along the transverse circle. There is an isometry that moves points along the horizon such that the eikonal phase is proportional to $\cos(\phi^+ - \phi^-)$, where $\phi^+$ and $\phi^-$ are the transverse positions of the two scattering shockwaves. In analogy to our two-dimensional example, the magnitude of the phase is maximized when the shockwaves are located at either the same or opposite points on the transverse circle, with only a sign difference between the two cases. This isometry acts transverse to the observer's worldline.}

This same exponent was observed in \cite{Aalsma:2020aib} in three dimensions. See equation (3.29). There, the OTOC is also an eikonal scattering amplitude. However, the scattering is mediated by graviton exchange instead of a dynamical observer. There is an important physical difference between these two effects. For graviton exchange, \cite{Aalsma:2020aib} found that the eikonal phase has a definite sign (see (3.21)), which reflects the fact that a geodesic crossing a positive-energy shockwave always experiences a time advance. In our OTOC, the eikonal phase can have either sign (note the integration ranges in \eqref{eq:6.34}). When the right observer emits a particle, they will recoil in the opposite direction. Suppose that at early times, the left observer fires a test null geodesic in one of the two directions around the spatial circle. Depending on the direction, the test geodesic may or may not reach the right observer. If the test geodesic reaches the right observer, we may say that it experiences a time advance. Otherwise, it experiences a time delay. Due to the different eikonal phases, our intuition for why our OTOCs have Lyapunov exponent $\frac{4 \pi}{\beta_{\text{dS}}}$ does not immediately apply to the case where the eikonal scattering is mediated by graviton exchange. Because this exponent was nonetheless observed in \cite{Aalsma:2020aib}, it would be interesting to find more physical connections between their calculation and ours.\footnote{Another difference between the two calculations is that their scrambling time is proportional to $\log \frac{1}{G_N}$, while our scrambling time is proportional to $\log \Lambda$.}

\begin{figure}[ht]
	\centering
	\begin{subfigure}[b]{0.45\textwidth}
		\centering
		\includegraphics[width=\textwidth]{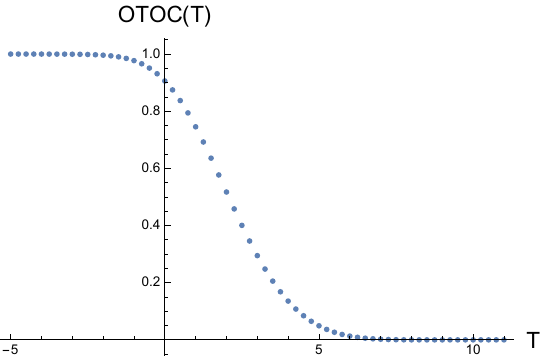}
		\label{fig:image1}
	\end{subfigure}
	\hfill
	\begin{subfigure}[b]{0.45\textwidth}
		\centering
		\includegraphics[width=\textwidth]{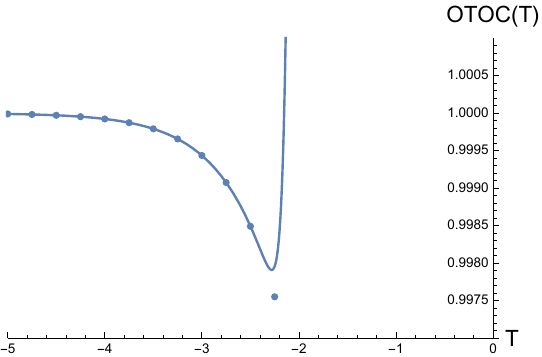}
		\label{fig:image2}
	\end{subfigure}
	\caption{On the left, we plot the numerical integration of \eqref{eq:otocforplot}. On the right, we focus on early times and also plot the sum of the first eleven terms of the asymptotic series \eqref{eq:6.38} as a solid line, finding good agreement for sufficiently early times. This demonstrates the validity of the asymptotic series, which is a series in powers of $e^{2 T}$. We have picked order one values for $s_A$, $s_B$, and $\epsilon$.}
	\label{fig:otocplots}
\end{figure}

Next, we demonstrate that \eqref{eq:6.34} is equal to the result of the calculation performed in Appendix \ref{sec:appotocscrambling}, which instead starts with the explicit expression for the 6j symbol in the crossed Wick contraction. We will derive an asymptotic series expansion for \eqref{eq:6.34}, which will be helpful for analyzing the OTOC in more detail. To simplify \eqref{eq:6.34}, we begin by removing the explicit dependence on $\Lambda$. We make the redefinition $(P^+,P^-) \rightarrow (e^{\tau_{scr}/2}P^+,e^{\tau_{scr}/2} P^-)$ and find that \eqref{eq:6.34} is proportional to
\begin{equation}
	\int_{-\infty}^\infty \, dP^+ dP^- \, e^{ - i P^+ P^-  }	
	\zeta_{s_B,\frac{T}{2}}(P^+)
	\zeta^*_{s_B,\frac{T}{2} + u_2}(P^+)	
	\zeta_{s_A,\frac{T}{2} - u_1}(P^-)
	\zeta^*_{s_A,\frac{T}{2}}(P^-),
\end{equation}
where $T$ was defined in \eqref{eq:tscr}. Then, we perform a Fourier transform on $u_2$, using 6.561.14 of \cite{gradshteyn2007table}. We then perform the $P^+$ integral using 6.621.1 of \cite{gradshteyn2007table}. We find that
\begin{align}
	&\int_{-\infty}^\infty \, dP^+ \,  e^{ - i P^+ P^-  }	
	\zeta_{s_B,\frac{T}{2}}(P^+)
	\zeta^*_{s_B,\frac{T}{2} + u_2}(P^+)	
		 \nonumber
	\\
	&= \int_{-\infty}^\infty d\lambda_2  \, \frac{e^{-(\pi + i u_2) \lambda_2}}{16 \pi^3}
	 \Gamma\left(\frac{1}{2}\left(\frac{1}{2} \pm i s_B \pm i \lambda_2\right)\right) 
	 \label{eq:6.36}
	 \,\\
	 &\times {}_2F_1\left(\frac{1}{4} \left(1 - 2 i s_B + 2 i \lambda_2\right), \frac{1}{4} \left(1 + 2 i s_B + 2 i \lambda_2\right), \frac{1}{2}, e^{T} \left(P^-\right)^2 \right).
	 	\nonumber
\end{align}
Because we are interested in the OTOC for $T$ large and negative, we will expand this hypergeometric function in powers of $e^T$.\footnote{Strictly speaking, this expansion is not allowed due to the infinite range of $P^-$. The price we pay for doing this is that the series we obtain is an asymptotic series. This asymptotic series faithfully approximates the original function, as demonstrated in Figure \ref{fig:otocplots}. The exact answer may be written as a sum over the first $n$ terms of the series plus an error term that decays faster than the $n$th term as $T \rightarrow -\infty$, as we explain in Appendix \ref{sec:appotocscrambling}. Our method for producing this asymptotic series has been validated on JT gravity, where the OTOC has been computed in \cite{Maldacena:2016upp}.} The $n$th order term contains a factor of $(P^-)^{2 n}$. Using 6.561.14 of \cite{gradshteyn2007table}, we find that\footnote{To make these integrals converge, $u_1$ should be given a small positive imaginary part.}
\begin{align}
	\label{eq:6.37}
	&\int_{-\infty}^\infty dP^- \, (P^-)^{2 n}	
	\zeta_{s_A,\frac{T}{2} - u_1}(P^-)
	\zeta^*_{s_A,\frac{T}{2}}(P^-) 
	\\
	&= e^{T n} \int_{-\infty}^\infty d\lambda_1 \, \frac{e^{-i u_1 \lambda_1 - \pi \lambda_1}}{16 \pi^3} \Gamma\left(\frac{1}{2}\left(\frac{1}{2} \pm  i s_A \pm i \lambda_1\right)\right) (-1)^n 4^{n}  \, \left(\frac{1}{4} \left(1 \pm 2 i s_A - 2 i \lambda_1\right)\right)_n .
	\nonumber
\end{align}
Putting \eqref{eq:6.36} and \eqref{eq:6.37} together, we find that the OTOC may be represented by the following asymptotic series:
\begin{align}
& \int_{-\infty}^\infty d\lambda_1  d\lambda_2  \, \frac{e^{- i u_2 \lambda_2 -\pi \lambda_2  }}{16 \pi^3} \frac{e^{-i u_1 \lambda_1 - \pi \lambda_1}}{16 \pi^3} \Gamma\left(\frac{1}{2}\left(\frac{1}{2} \pm  i s_A \pm i \lambda_1\right)\right) 
	 \Gamma\left(\frac{1}{2}\left(\frac{1}{2} \pm i s_B \pm i \lambda_2\right)\right) 
	 \label{eq:6.38}
	 \,\\
	 &
	 \sum_{n = 0}^\infty \left(\frac{1}{4} \left(1 \pm 2 i s_B + 2 i \lambda_2\right)\right)_n  \left(\frac{1}{4} \left(1 \pm 2 i s_A - 2 i \lambda_1\right)\right)_n  \frac{1}{(2 n)!} e^{2 n T}
	 \, (-16)^n  . \nonumber
\end{align}
In Appendix \ref{sec:appotocscrambling}, we compute the OTOC by simplifying the explicit result for the crossed Wick contraction that was discussed in Section \ref{sec:crossed}. The result is presented as a contour integral along the real axis. An asymptotic series in powers of $e^{2 T}$ may be obtained by closing the contour in the upper-half-plane. This asymptotic series is a good representation of the OTOC when $T$ is large and negative, and it agrees with \eqref{eq:6.38}, which confirms the physical interpretation of the OTOC in Figure \ref{fig:foursegment}. The contour could also be closed in the lower-half-plane to obtain an asymptotic series that is applicable for large positive $T$. From the contour integral, one may check that the OTOC decays to zero for large $T$.

To plot the OTOC, it is convenient to evolve the operators in imaginary time. For example, consider
\begin{equation}
\frac{\braket{\psifhh | \phi_R^B\left(\frac{\tau}{2} - i \epsilon\right) \phi_R^A\left(-\frac{\tau}{2} - i \epsilon \right) \phi_R^B\left(\frac{\tau}{2}  + i \epsilon \right) \phi_R^A\left(-\frac{\tau}{2}  + i \epsilon\right) \Pi_{\Lambda,\tilde{\epsilon}} | \psifhh}}{\braket{\psifhh|\Pi_{\Lambda,\tilde{\epsilon}}|\psifhh}},
\end{equation}
where $\epsilon > 0$ and $\tilde{\epsilon} > 0$. After taking the semiclassical limit and normalizing to one at early times, this OTOC becomes
\begin{equation}
\text{OTOC}(T) := 	\frac{\int_{-\infty}^\infty dP^+ dP^- \, e^{-i P^+ P^-}	\left|\zeta_{s_B,\frac{T}{2} - i \epsilon}(P^+) \right|^2  \left|\zeta_{s_A,\frac{T}{2}  - i \epsilon}(P^-) \right|^2}{\int_{-\infty}^\infty dP^+  \, \left|\zeta_{s_B,\frac{T}{2} - i \epsilon}(P^+) \right|^2  \int_{-\infty}^\infty dP^- \, \left|\zeta_{s_A,\frac{T}{2}  - i \epsilon}(P^-) \right|^2}.
\label{eq:otocforplot}
\end{equation}
This OTOC is manifestly real and bounded from above by $1$. We plot this in Figure \ref{fig:otocplots} together with the first eleven terms of the asymptotic series \eqref{eq:6.38}, finding good agreement for sufficiently early times. 

This OTOC violates the chaos bound (for the de Sitter temperature) by a factor of two. It was shown in \cite{Maldacena:2015waa} that an OTOC of simple operators in a many-body chaotic quantum system with a large hierarchy between dissipation and scrambling times cannot have a Lyapunov exponent greater than $\frac{2 \pi}{\beta}$. Because the de Sitter inverse temperature is $\beta_{\text{dS}} = 2 \pi$, the maximum allowed exponent (naively) is 1, but we have found an exponent of 2. The putative holographic dual must violate one of the assumptions that entered into the proof of the chaos bound. We offer more comments in the Discussion.

\begin{figure}[h!]
	\centering
	\includegraphics[width=0.5\textwidth]{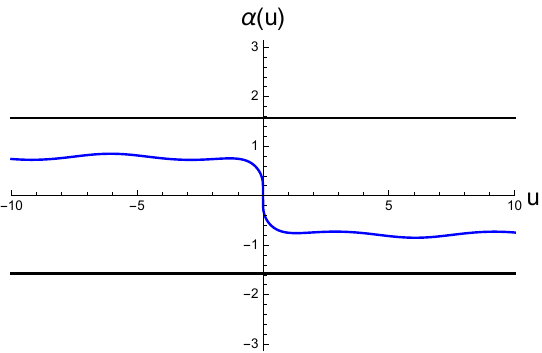}
	\caption{The blue curve is a plot of \(\alpha(u)\) as a function of \(u\). We also plotted $\frac{\pi}{2}$ and $-\frac{\pi}{2}$ in black. As long as $\alpha(u)$ remains in $\left(-\frac{\pi}{2},\frac{\pi}{2}\right)$, the OTOC \eqref{eq:6.41} grows initially before decaying. We set $s_A = s_B = \epsilon = \frac{1}{2}$.}
	\label{fig:alpha_u_plot}
\end{figure}

For other kinematics, the OTOC can grow first before decaying. When two operators are timelike separated along the observer's worldline, the $\hat{P}^+$ isometry applied to one operator will cause the distance between the operators to initially decrease. This corresponds to an OTOC that grows first before decaying to zero. See Figure \ref{fig:flow}. To see this effect quantitatively, consider the OTOC
\begin{equation}
	\frac{\braket{\psifhh | \phi_R^B\left(\frac{\tau}{2} \right) \phi_R^A\left(-\frac{\tau}{2} - i \epsilon \right) \phi_R^B\left(\frac{\tau}{2}  + u \right) \phi_R^A\left(-\frac{\tau}{2}  + i \epsilon\right) \Pi_{\Lambda,\tilde{\epsilon}}| \psifhh}}{\braket{\psifhh|\Pi_{\Lambda,\tilde{\epsilon}}|\psifhh}},
	\label{eq:6.41}
\end{equation}
which becomes in the semiclassical limit
\begin{equation}
	\int_{-\infty}^\infty dP^+ dP^- \, e^{-i P^+ P^-}	\,
	\zeta_{s_B,\frac{T}{2}}(P^+) \zeta^*_{s_B,\frac{T}{2} + u}(P^+)
	  \left|\zeta_{s_A,\frac{T}{2}  - i \epsilon}(P^-) \right|^2.
\end{equation}
Using \eqref{eq:6.38}, we may express this as a power series in $e^{2 T}$. If we normalize the series such that the leading term is one, then we may write it as
\begin{equation}
	1 + r(u) e^{i \alpha(u)} e^{2 T} + \cdots,
\end{equation}
and we are only interested in the coefficient of the first correction, which we have decomposed into a positive real part $r(u)$ and phase $e^{i \alpha(u)}$. When $\alpha(u) \in \left(-\frac{\pi}{2},\frac{\pi}{2}\right)$, the OTOC grows initially.  See Figure \ref{fig:alpha_u_plot} for a plot of $\alpha(u)$. We find that for both positive and negative $u$, the OTOC grows initially before decaying, which confirms our expectations.

Consider the same OTOC \eqref{eq:6.41} in JT gravity, where the expectation value is taken in a thermal state of inverse temperature $2 \pi$, and all operators have scaling dimension $\Delta$. The normalized OTOC becomes \cite{Maldacena:2016upp}
\begin{equation}
	1 -   \frac{\Delta^2}{2 C} \frac{ e^{ u /2}}{\epsilon \sinh \frac{u}{2}   } e^{\tau} + \cdots,
\end{equation}
where $C$ is a large coupling constant in the semiclassical limit, and $\frac{1}{C} e^{\tau}$ is fixed and small. When $u > 0$, the OTOC initially decays, whereas when $u< 0$, the OTOC initially grows.

In de Sitter space, the OTOC \eqref{eq:6.41} initially grows regardless of the sign of $u$, whereas in AdS, the OTOC can initially grow or decay, depending on the sign of $u$. This effect can be used to probe whether an emergent spacetime is AdS or dS.

We reiterate that we have computed an expression that schematically takes the form
\begin{equation}
\braket{ \text{HH} | V_R(\tau) W_R(0) V_R(\tau) W_R(0) | \text{HH} },
\end{equation}
and {\it not} an expression of the form
\begin{equation}
\text{Tr } e^{- 2 \pi H} V(\tau) W(0) V(\tau) W(0).
\end{equation}
That is, the Hartle-Hawking state does not furnish a trace on the observer's algebra. From this perspective, it is not surprising that the OTOC might violate the chaos bound.

In Section \ref{sec:algebra}, we discuss the theory of a massless chiral scalar. One may also compute the OTOC for this choice of matter field. In this example, the chaos bound is saturated. The answer is similar to JT gravity with matter because the crossed Wick contraction is governed by the same 6j symbol as in JT gravity, as we explain in more detail in Section \ref{sec:algebra}. As mentioned earlier, we expect that higher-dimensional examples with dynamical observers will have a Lyapunov exponent of $\frac{4 \pi}{\beta_{\text{dS}}}$. It would also be interesting to compute the OTOC for the case when the scalar transforms in the complementary series representation (with mass $\frac{1}{2} > m > 0$), since the masses considered in \cite{Narovlansky:2023lfz} are in the complementary series.

\section{An algebra for an observer}

\label{sec:algebra}

In previous sections, we constructed the gauge-invariant Hilbert space $\calh$ of two observers and a matter QFT, and we showed how local QFT operators can be dressed to the worldline of an observer. We also defined the Hartle-Hawking state and showed that it shares some properties with the Hartle-Hawking state of \cite{CLPW} (such as \eqref{eq:HHproperty}), but not others (in particular, it does not define a trace). In this section, we define the algebra associated to an observer and study its properties.

In \cite{CLPW,Witten}, the observer's algebra is generated from the local QFT operators in the vicinity of their geodesic worldline, as well as their Hamiltonian. Note that due to the timelike tube theorem \cite{Strohmaier:2023opz}, the algebra includes QFT operators in the entire static patch. This construction leads to a crossed-product algebra.

In Section \ref{sec:gaugeinvariant}, we gave examples of local QFT operators dressed to an observer (namely $\phi_R$, $:\phi^n:_R$, and $\overrightarrow{\partial} \phi_R$). We define the right observer's algebra $\cala_R$ to be the von Neumann algebra acting on $\calh$ that is generated by all such dressed operators together with $\hat{s}_R$, which was defined in \eqref{eq:srdef} and represents the right observer's Hamiltonian.\footnote{When we say that a vN algebra $\cala$ is generated by a set of operators $S$, we mean that $\cala$ is the weak closure (or the double-commutant) of the $\star$-algebra given by finite sums of finite products of operators in $S$, operators in $S^*$, and the identity. We refer to this $\star$-algebra as the ``generating algebra'' of $\cala$. $S^*$ is the set of adjoints of the operators in $S$. One should use bounded functions of any unbounded operators in $S$. \label{ft:genfoot}} The algebra $\cala_L$ for the left observer is defined analogously.

Having defined these algebras, there are various questions one can ask, such as
\begin{itemize}
	\item Is the Hartle-Hawking state cyclic and/or separating for $\cala_R$?
	\item What is the commutant of $\cala_R$?
	\item Is $\cala_R$ a factor? If so, what is its type?
\end{itemize}
We cannot rigorously answer any of these questions for the algebras as they are defined above, so we will now specialize to a simplified model where we can give more compelling arguments and build our intuition. The conclusion we will reach is that the Hartle-Hawking state is cyclic but not separating, and that the commutant of $\cala_R$ is generated by $\hat{s}_L$, the left observer's Hamiltonian. Thus, $\cala_R$ is a direct integral of type I$_\infty$ factors, and the center is generated by $\hat{s}_L$. This is in stark contrast to \cite{CLPW}, which found that $\cala_L$ and $\cala_R$ are each other's commutants, and are type II$_1$ factors.

In our simplified model, we use the same setup described earlier in this paper, but we substitute the massive scalar QFT with a massless chiral scalar. The benefit of considering a massless chiral scalar is that the matter QFT Hilbert space consists of only one type of representation: the positive discrete series representation.\footnote{The positive discrete series corresponds to right-movers, while the negative discrete series corresponds to left-movers} In contrast, the Hilbert space of the massive scalar described in Section \ref{sec:freescalar} contains principal series representations as well as positive and negative discrete series representations. The model with a massless chiral scalar is structurally very similar to JT gravity with matter. In JT gravity with matter, the matter Hilbert space is comprised of positive discrete series representations \cite{Kolchmeyer:2023gwa}, while the ``boundary particle'' \cite{Kitaev:2018wpr,Yang:2018gdb} that encapsulates the gravitational dynamics is in the principal series. Likewise, the observers in this paper are modeled as free particles that transform in the principal series. Thus, without having to do any calculations, we can already anticipate that correlators with the massless chiral scalar will look similar to correlators in JT gravity with matter. In particular, crossed Wick contractions will be governed by the same 6j symbol.

Because the chiral scalar only contains right-moving excitations, it is not possible to construct states that are separately invariant under parity and time-reversal, because these isometries map right-movers to left-movers. Thus, our model with a chiral scalar should be viewed as a toy model where the Hilbert space $\calh$ is invariant only under $SO^+(2,1)$. We may choose to additionally restrict to CPT-invariant states to get a real Hilbert space. As mentioned in Section \ref{sec:gaugediscrete}, in this paper we are mostly interested in the case where only $SO^+(2,1)$ is gauged. In this case, the commutant of $\cala_R$ will include the discrete spacetime symmetry generators in addition to the left observer's Hamiltonian. This is because the generators of $\cala_R$ are fully diff-invariant. Then, the direct integral of type I$_\infty$ factors becomes a direct integral and direct sum that includes additional discrete superselection sectors. Throughout this paper, we will simply say that $\cala_R$ is a direct integral of type I$_\infty$ factors. When the discrete isometries are gauged, the discrete superselection sectors go away.

In the remainder of this section, we will discuss the chiral scalar in more detail and explain how it simplifies the analysis of $\cala_R$.

\subsection{Simplified model: Chiral scalar field}

\label{sec:chiralscalar}

Working in global coordinates as in \eqref{eq:metric}, we expand the chiral scalar field as follows:
\begin{equation}
	\varphi(t,\theta) = \sum_{n \ge 1 } \chi_n(t,\theta) a_n + (\chi_n(t,\theta))^* a_n^\dagger,
\label{eq:chiral}
\end{equation}
where the modefunctions are given by
\begin{equation}
	\chi_n(t,\theta) = e^{i n \theta} \frac{1}{\sqrt{ n}} \left(\frac{1 -  i \tanh \frac{t}{2}}{1 + i \tanh \frac{t}{2}}\right)^n,
\end{equation}
and we impose the canonical commutation relations
\begin{equation}
	[a_n, a_m^\dagger] = \delta_{nm}.
\end{equation}
We deliberately omitted the zero modes in \eqref{eq:chiral}. As with any 2D CFT, there are symmetry generators acting on the Hilbert space that correspond to the Killing vectors and the conformal Killing vectors of \eqref{eq:metric}. Together, these six operators are denoted by $L_n$ and $\bar{L}_n$ for $n \in \{-1,0,1\}$. Given a field $\calo_{h,\bar{h}}$ labeled by conformal weights $(h,\bar{h})$, they act as follows:
\begin{align}
	\label{eq:7.4}
	[L_n, \calo_{h,\bar{h}}] &= V_n^\mu \partial_\mu \calo_{h,\bar{h}} + h\left(\partial_\mu V_n^\mu\right) \calo_{h,\bar{h}},
	\\
	[\bar{L}_n, \calo_{h,\bar{h}}] &= \bar{V}_n^\mu \partial_\mu \calo_{h,\bar{h}} + \bar{h} \left(\partial_\mu \bar{V}_n^\mu\right) \calo_{h,\bar{h}},	
	\label{eq:7.5} 
\end{align}
where the (conformal) Killing vectors $V_n$ and $\bar{V}_n$ are given in Appendix \ref{sec:ckvs}. The commutation relations are
\begin{equation}
	[L_n,L_m] = (n-m)L_{n+m}, \quad \quad [\bar{L}_n,\bar{L}_m] = (n-m) \bar{L}_{n+m}.
\end{equation}
We are interested in the operator $\partial_\mu \varphi$, which has weight $(h,\bar{h}) = (1,0)$. To see this, we explicitly compute that
\begin{align}
	\partial_\mu\left(V_{-1}^\nu \partial_\nu \varphi\right) &= \sum_{n \ge 2 } \sqrt{n(n-1)} \partial_\mu \chi_n \, a_{n-1} - \sum_{n \ge 1} \sqrt{n(n+1)} \partial_\mu\chi_n^* \, a_{n+1}^\dagger,
	\\
	\partial_\mu\left(V_{1}^\nu \partial_\nu \varphi\right) &= \sum_{n \ge 1 } \sqrt{n(n+1)} \partial_\mu \chi_n \, a_{n+1} - \sum_{n \ge 2} \sqrt{n(n-1)} \partial_\mu \chi_n^* \, a_{n-1}^\dagger,
	\\
	\partial_\mu\left(V_{0}^\nu \partial_\nu \varphi\right) &= \sum_{n \ge 1 } -n \partial_\mu \chi_n \, a_n + n \partial_\mu \chi_n^* \, a_n^\dagger,
\end{align}
and then apply \eqref{eq:7.4} to conclude that
\begin{align}
	[L_{-1},a_n] &= \sqrt{n(n-1)} a_{n-1}, \quad n \ge 2,
	\\
	[L_{-1},a_1] &= 0,
	\\
	[L_{-1},a_n^\dagger] &= - \sqrt{n(n+1)} a_{n+1}^\dagger,
	\\    
	[L_1,a_n] &= \sqrt{n(n+1)}  a_{n+1},
	\\
	[L_1,a_n^\dagger] &= -\sqrt{n(n-1)}  a_{n-1}^\dagger, \quad n \ge 2,
	\\
	[L_1,a_1^\dagger] &= 0,
	\\
	[L_0,a_n] &= -n a_n,
	\\
	[L_0,a_n^\dagger] &= n a_n^\dagger.
\end{align}
Comparing with \eqref{eq:c6} through \eqref{eq:c8}, we conclude that the conformal generators act unitarily on the Hilbert space, and single-particle states transform in a positive discrete series representation with vanishing Casimir.\footnote{In particular, $\Delta = 1$. See Appendix \ref{sec:appendixcg} for our conventions.} Thus, the $N$-particle sector of the matter QFT $\calh_{N,m}$ is the symmetrized tensor product of $N$ positive discrete series representations. A nice aspect of the positive (or negative) discrete series representations is that their tensor product does not contain any new types of representations, so the Hilbert space is easier to study.

For our purposes, it is necessary to study only one operator, $i \overrightarrow{\partial} \phi_R$, which has been defined in \eqref{eq:5.19}.\footnote{The $i$ is included to make this operator Hermitian. See \eqref{eq:5.22} and \eqref{eq:5.23}.} To compute correlators, we are interested in two integrals. As before, we represent them graphically:

\begin{align}
	\raisebox{-0.5\height}{\includegraphics[scale=0.5]{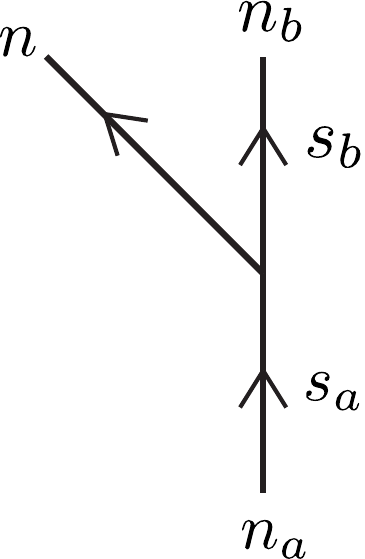}} 
	&= i\int d\theta dt \sqrt{-g} 
	\left(\Psi^{s_b}_{n_b}(t,\theta)\right)^* g^{\mu \nu} \partial_\mu \chi_n^* 
	\partial_\nu \Psi^{s_a}_{n_a}(t,\theta),
	\nonumber
	\\
	&= i\pi \frac{\Gamma\left(1 + i s_a - i s_b\right)}
	{\Gamma\left(\frac{1}{2} + i s_a\right) \Gamma\left(\frac{1}{2} - i s_b\right)}
	\sqrt{ \frac{\cosh \pi s_a}{\cosh \pi s_b} 
		\Gamma\left(1 \pm i (s_a + s_b)\right)}
	\nonumber
	\\
	&\quad \times \sqrt{\frac{s_a}{\pi} \tanh \pi s_a} 
	\sqrt{\frac{s_b}{\pi} \tanh \pi s_b} 
	(-1)^n \left(\calctilde_{PP}^{D,-}\right)^{1,n}_{-s_a,-n_a,s_b,n_b}
	\nonumber
	\\
	&\quad \times e^{i \phi_{n_a,s_a}/2} e^{- i \phi_{n_b,s_b}/2}.
	\label{eq:7.18}
\end{align}

\begin{align}
	\raisebox{-0.5\height}{\includegraphics[scale=0.5]{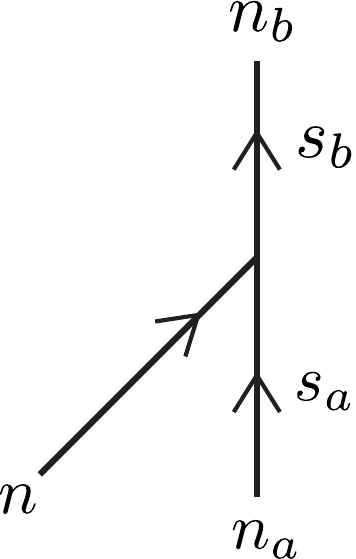}} 
	&= i\int d\theta dt \sqrt{-g} 
	\left(\Psi^{s_b}_{n_b}(t,\theta)\right)^* g^{\mu \nu} \partial_\mu \chi_n 
	\partial_\nu\Psi^{s_a}_{n_a}(t,\theta),
	\nonumber
	\\
	&= -i\pi \frac{\Gamma\left(1 - i s_b + i s_a\right)}
	{\Gamma\left(\frac{1}{2} - i s_b\right) \Gamma\left(\frac{1}{2} + i s_a\right)}
	\sqrt{ \frac{\cosh \pi s_b}{\cosh \pi s_a} 
		\Gamma\left(1 \pm i (s_a + s_b)\right)}
	\nonumber
	\\
	&\quad \times \sqrt{\frac{s_a}{\pi} \tanh \pi s_a} 
	\sqrt{\frac{s_b}{\pi} \tanh \pi s_b} 
	(-1)^n \left(\left(\calctilde_{PP}^{D,-}\right)^{1,n}_{-s_b,-n_b,s_a,n_a}\right)^*
	\nonumber
	\\
	&\quad \times e^{-i \phi_{n_b,s_b}/2} e^{i \phi_{n_a,s_a}/2}.
	\label{eq:7.19}
\end{align}
The Clebsch-Gordan coefficients are defined in Appendix \ref{sec:ppdcg}. The phase $e^{i \phi_{n,s}}$ is defined in \eqref{eq:eiphi}. Using these vertices, we may verify that
\begin{equation}
	i \overrightarrow{\partial} \phi_L \ket{\psifhh} = -i \overrightarrow{\partial} \phi_R \ket{\psifhh}, \quad \quad [i \overrightarrow{\partial} \phi_L,i \overrightarrow{\partial} \phi_R] \ket{\psifhh} =0,
\end{equation}
which agrees with our expectations in \eqref{eq:leftrightcommute} and \eqref{eq:HHproperty}.

Next, we discuss correlators of $i \overrightarrow{\partial} \phi_R$ in the Hartle-Hawking state. We assemble the above vertices into diagrams just as in Section \ref{sec:gaugeinvariant}. A diagram with no crossed Wick contractions can be represented as a standard chord diagram as in \eqref{eq:6.20}, \eqref{eq:6.21} and evaluated using a simple set of diagrammatic rules. As before, all of the $s$ variables that border the same region are to be set equal using delta functions, and for each delta function there is a factor of $\frac{1}{\rho(s)}$, defined in \eqref{eq:rhos}. A chord that separates two regions contributes a factor of
\begin{equation}
	\label{eq:7.21}
	 \frac{1}{8 \pi^4} s_1 \tanh \pi s_1 \, s_2 \tanh \pi s_2 \, \Gamma_{s_1 s_2},
\end{equation}
where we define
\begin{equation}
	\Gamma_{s_1 s_2} := \Gamma\left(1 \pm i s_1 \pm i s_2\right)
\end{equation}

Unlike the case of the massive scalar QFT considered earlier, there is a fairly simple prescription for computing diagrams with crossed Wick contractions, which follows from the following identities:

\begin{equation}
	\label{eq:7.23}
	\raisebox{-0.5\height}{\scalebox{0.3}{\includegraphics{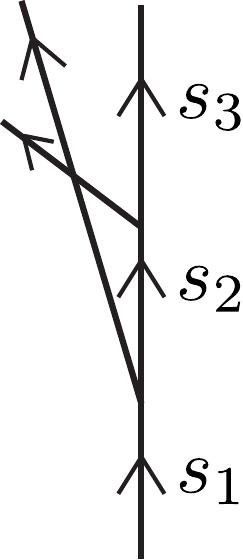}}} 
	\quad = \int_0^\infty ds_4 \, 
	\raisebox{-0.5\height}{\scalebox{0.3}{\includegraphics{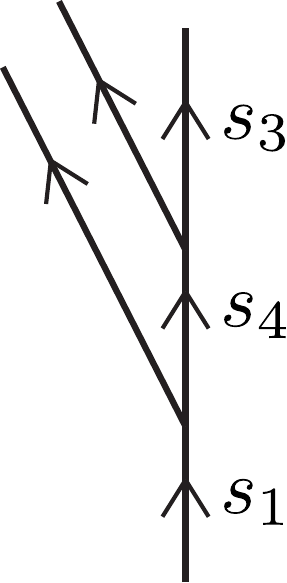}}} 
	\quad \sqrt{\frac{\Gamma_{s_1 s_2} \Gamma_{s_2 s_3}}{\Gamma_{s_1 s_4} \Gamma_{s_4 s_3}}}  
	\left\{
	\begin{array}{ccc}
		1 & s_2 & s_3 \\
		1 & s_4 & s_1
	\end{array}
	\right\} 
	\frac{s_2 \sinh 2 \pi s_2}{\pi^2}.
\end{equation}

\begin{equation}
	\label{eq:7.24}
	\raisebox{-0.5\height}{\scalebox{0.3}{\includegraphics{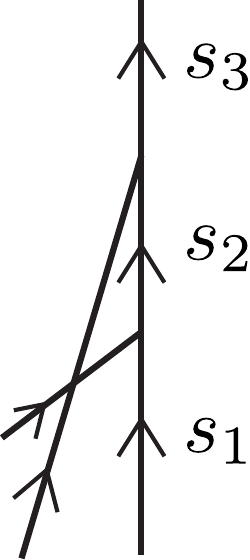}}} 
	\quad = \int_0^\infty ds_4 \, 
	\raisebox{-0.5\height}{\scalebox{0.3}{\includegraphics{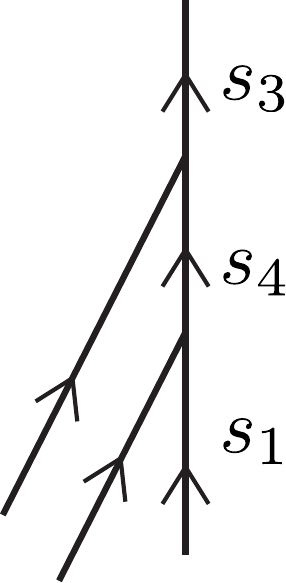}}} 
	\quad \sqrt{\frac{\Gamma_{s_1 s_2} \Gamma_{s_2 s_3}}{\Gamma_{s_1 s_4} \Gamma_{s_4 s_3}}}  
	\left\{
	\begin{array}{ccc}
		1 & s_2 & s_3 \\
		1 & s_4 & s_1
	\end{array}
	\right\} 
	\frac{s_2 \sinh 2 \pi s_2}{\pi^2}.
\end{equation}
The expression involving the braces is a special function called the 6j symbol, which was defined in (A.18) of \cite{Jafferis:2022wez}. It is a real function that is invariant under cyclic permutations of $(s_1,s_2,s_3,s_4)$ as well as $(s_1,s_2,s_3,s_4) \rightarrow (s_4,s_3,s_2,s_1)$. In each graphic above, note that the two slanted lines, which represent the propagator for the massless chiral scalar, are either both going into or out of the vertical line, which represents the propagator of the observer. For the case where one slanted line is going in and the other slanted line is going out, there is no simple identity that ``uncrosses'' the two slanted lines. In JT gravity with matter, there exists a similar identity that is insensitive to the arrows \cite{Mertens:2017mtv}.

Using these results, the crossed diagram in the four-point function \eqref{eq:fourpoint} becomes\footnote{
	Because we are now applying the rules in \eqref{eq:7.18} and \eqref{eq:7.19} to the diagram, the ``$s$'' label on the internal lines should be disregarded. We are working with a massless scalar. We set $f_1 = f_2 = f_{HH}$.}
\begin{equation}
	\frac{1}{32 \pi^8} s_a \sinh 2 \pi s_a \, s_b \tanh \pi s_b \, s_c \sinh 2 \pi s_c \, s_d \tanh \pi s_d \, \sqrt{\Gamma_{s_a s_b}\Gamma_{s_b s_c} \Gamma_{s_c s_d} \Gamma_{s_d s_a}}  	\left\{
	\begin{array}{ccc}
		1 & s_a & s_b \\
		1 & s_c & s_d
	\end{array}
	\right\},
	\label{eq:7.25}
\end{equation}
which, as expected from the results of Section \ref{sec:crossed}, is not invariant under cyclic permutations of $(s_a,s_b,s_c,s_d)$ due to the hyperbolic trig functions. A correlator with an arbitrary number of $i \overrightarrow{\partial} \phi_R$ insertions may be computed by using \eqref{eq:7.23} and/or \eqref{eq:7.24} to rewrite any diagram with crossed Wick contractions in terms of a diagram with no crossed contractions, which may then be evaluated using the rules described just above \eqref{eq:7.21}.

The reader familiar with JT gravity with matter will note that \eqref{eq:7.25} looks very similar to the crossed contraction of four operators with scaling dimensions $\Delta = 1$. In JT gravity with matter, each subregion of a chord diagram is assigned the function $s \sinh 2 \pi s$. In \eqref{eq:7.25}, two of the $s \sinh 2 \pi s$ functions have been replaced with $s \tanh \pi s$. A general diagram that represents a contraction of multiple $i \overrightarrow{\partial} \phi_R$ operators is equivalent to the corresponding chord diagram in JT gravity, except that some of the $s \sinh 2 \pi s$ functions belonging to regions that touch the boundary are replaced by other hyperbolic trig functions.

\subsection{The Hartle-Hawking state is cyclic}

In the previous subsection, we introduced the diagrammatic rules for computing correlators of $i \overrightarrow{\partial} \phi_R$ in the Hartle-Hawking state. We now argue that the Hartle-Hawking state is cyclic with respect to the right observer's algebra. In fact, the argument given below extends to any state $\ket{\Psi_f} \in \calh_0$ (defined in \eqref{eq:3.23}) where $f(s)$ is supported on the entirety of $\mathbb{R}^+$. We will focus on the Hartle-Hawking state for concreteness.

We first comment on the structure of the gauge-invariant Hilbert space $\calh$. As discussed in Section \ref{sec:addmatter}, wavefunctions for states in $\calh$ may be constructed by contracting together CG coefficients. For the massive scalar, multiple types of CG coefficients must be used to construct a complete basis. In contrast, for the case of a massless chiral scalar, only the CG coefficient that couples two principal series irreps and a positive discrete series irrep is needed. To see why, consider the task of constructing singlets\footnote{To be clear, we mean singlets that are normalizable using the inner product in \eqref{eq:GFinnerproduct}. This means that we may only consider connected diagrams.} from the tensor product of two principal series representations (that correspond to the observers) and two discrete series representations (that correspond to matter particles). The two principal series irreps will be denoted by $s_a$ and $s_c$, while each discrete series irrep\footnote{The $1$ indicates that $\Delta = 1$, where the Casimir is given by $\Delta(1-\Delta)$. The plus superscript indicates a positive discrete series representation.} is denoted by $1^+$. Following our diagrammatic conventions in Section \ref{sec:addmatter}, one way to construct a singlet is given by
\begin{equation}
	\raisebox{-0.5\height}{\scalebox{0.25}{\includegraphics{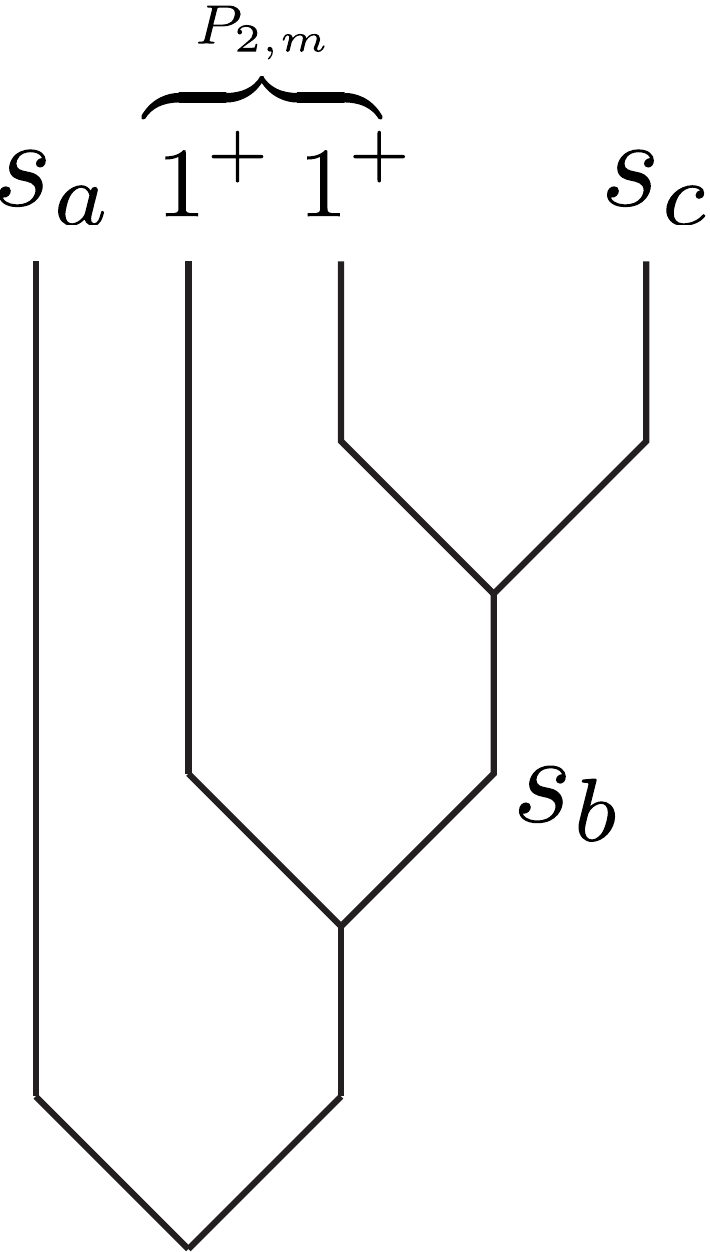}}}.
	\label{eq:7.26}
\end{equation}
As in Section \ref{sec:addmatter}, the $P_{2,m}$ indicates a projection onto states that are invariant under permutations of the discrete irreps. The state represented by this diagram is constructed by first fusing the $s_c$ irrep with a $1^+$ irrep, and projecting onto the principal series irrep given by $s_b$, which is subsequently fused with the next $1^+$ irrep. The result must be projected into a principal series irrep with Casimir specified by $s_a$ in order to form a singlet with the leftmost principal series irrep. Note that the tensor product of a principal and positive discrete series irrep also produces positive discrete series irreps, so the reader might expect that in order to construct a complete basis of $\calh_2$, one must also consider diagrams where the $s_b$ irrep is replaced by a positive discrete series irrep. However, the tensor product of two positive discrete series irreps only produces positive discrete series irreps, which is incompatible with forming a singlet with the $s_a$ irrep. Thus, a complete basis\footnote{In fact, this basis is over-complete due to the projection $P_{2,m}$.} of $\calh_2$ may be constructed by contracting a single type of CG coefficient together as shown in the above diagram. This result generalizes straightforwardly to $\calh_N$, where the representations in the diagrams are fused together starting from the right and proceeding to the left, as in \eqref{eq:7.26}.

Note that the CG coefficient that appears in the vertex \eqref{eq:7.18} is exactly the same coefficient that is used to construct singlets in $\calh_N$. Thus, by acting on the Hartle-Hawking state with $P_{R,s_a} i \overrightarrow{\partial} \phi_R P_{R,s_b}$ and integrating over $s_a$ and $s_b$ with a suitable wavefunction, one can construct an arbitrary state in $\calh_1$. To construct an arbitrary state in $\calh_2$, one must take appropriate linear combinations of states of the form
\begin{equation}
	P_{R,s_a} i \overrightarrow{\partial} \phi_R P_{R,s_b} i \overrightarrow{\partial} \phi_R P_{R,s_c} \ket{\psifhh} \quad \quad \text{and} \quad \quad P_{R,s_a}\ket{\psifhh}.
\end{equation}
A general state in $\calh_N$ may be constructed by acting on the Hartle-Hawking state with a linear combination of operators that each involve at most $N$ products of $i \overrightarrow{\partial} \phi_R$. This demonstrates that the Hartle-Hawking state is cyclic with respect to the algebra generated by $i \overrightarrow{\partial} \phi_R$ and $\hat{s}_R$. To obtain this result, we did not have to consider other matter operators dressed to the observer's worldline. In contrast, for the case of the massive scalar QFT of Section \ref{sec:freescalar}, multiple types of dressed operators must be used to demonstrate that the Hartle-Hawking state is cyclic.

\subsection{The Hartle-Hawking state is not separating}

Having established that the Hartle-Hawking state is cyclic with respect to $\cala_R$, we now show that it is not separating. We will find a family of nontrivial operators that annihilate the Hartle-Hawking state. As in the previous subsection, the logic here can be extended to any state in $\calh_0$.

We begin by constructing a family of operators that annihilate $\ket{\psifhh}$. First, consider the state
\begin{equation}
	P_{R,s_a} \iphir P_{R,s_b} \iphir P_{R,s_c} \ket{\psifhh}.
	\label{eq:7.28}
\end{equation}
In a general correlation function, the two $\iphir$ operators above can Wick contract into each other or into other operators. We will subtract a state in $\calh_0$ from \eqref{eq:7.28} that has the effect of canceling the contraction of the two $\iphir$ operators in \eqref{eq:7.28}. Define
\begin{equation}
	\ket{\Psi^{s_a,s_c}_{s_b}} := P_{R,s_a} \iphir P_{R,s_b} \iphir P_{R,s_c} \ket{\psifhh} - P_{R,s_a}\ket{\psifhh} \delta(s_a-s_c) \frac{s_b \tanh \pi s_b}{2 \pi^2} \cosh^2 (\pi s_a) \Gamma_{s_a s_b}. 
\end{equation}
Now, consider a general correlator of the form
\begin{equation}
	\braket{\psifhh|\cdots|\Psi^{s_a,s_c}_{s_b}},
\end{equation}
where the $\cdots$ represents arbitrary insertions of $P_{R,s}$ and $\iphir$. We may graphically depict the Wick contractions as follows:
\begin{equation}
	\braket{\psifhh|\cdots|\Psi^{s_a,s_c}_{s_b}} = 
	\vcenter{\hbox{\scalebox{0.3}{\includegraphics{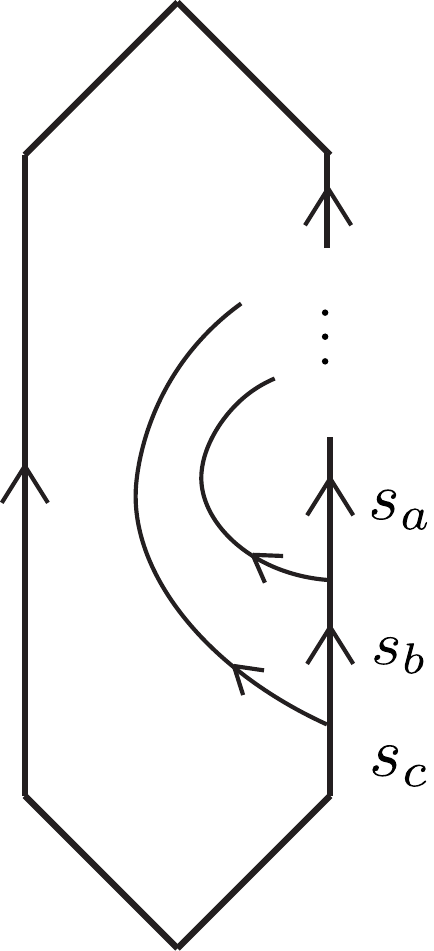}}}} + 
	\vcenter{\hbox{\scalebox{0.3}{\includegraphics{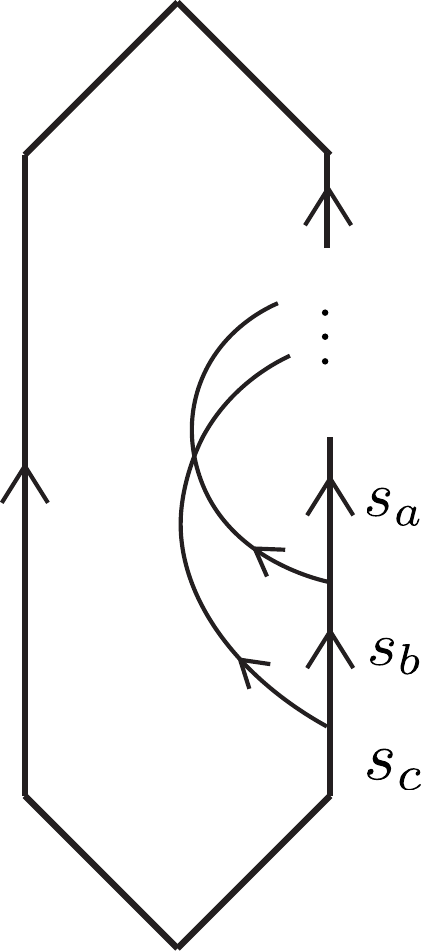}}}}.
\end{equation}
We only need to keep track of whether the two matter propagators shown are crossed or uncrossed. Using \eqref{eq:7.23}, we may rewrite this as follows:
\begin{equation}
	\braket{\psifhh|\cdots|\Psi^{s_a,s_c}_{s_b}} = \int_0^\infty ds_d \, 
	\vcenter{\hbox{\scalebox{0.3}{\includegraphics{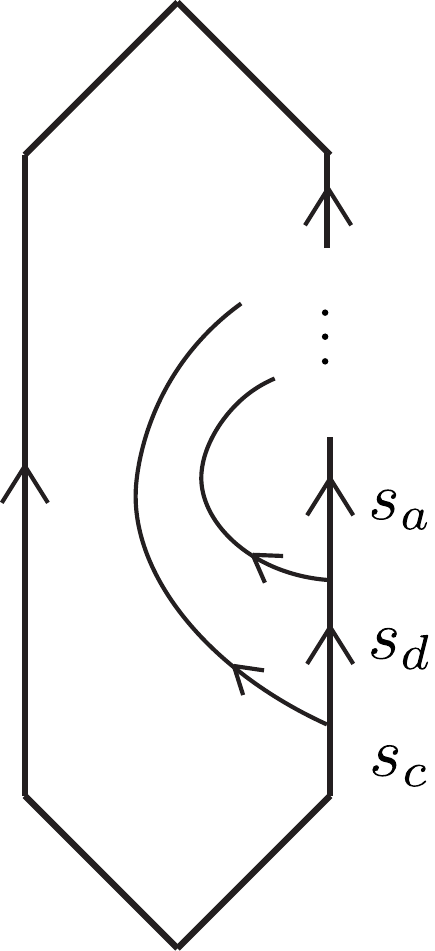}}}} 
	\left[ \delta(s_d - s_b) + \sqrt{\frac{\Gamma_{s_a s_b} \Gamma_{s_b s_c}}{\Gamma_{s_a s_d} \Gamma_{s_d s_c}}} 
	\left\{\begin{array}{ccc}
		1 & s_a & s_b \\
		1 & s_c & s_d
	\end{array} \right\} 
	\frac{s_b \sinh 2 \pi s_b}{\pi^2} \right].
\end{equation}
The 6j symbols obey the following orthogonality relation \cite{Mertens:2017mtv},
\begin{equation}
	\int_0^\infty ds_d \, \frac{s_d \sinh 2 \pi s_d}{\pi^2} \left\{\begin{array}{ccc}
	1 & s_a & s_b \\
	1 & s_c & s_d
\end{array} \right\} 	\left\{\begin{array}{ccc}
1 & s_a & s_b^\prime \\
1 & s_c & s_d
\end{array} \right\} = \frac{\pi^2}{s_b \sinh 2 \pi s_b} \delta(s_b - s_b^\prime).
\end{equation}
It follows that
\begin{equation}
	\int_0^\infty ds_d \, \frac{\braket{\psifhh|\cdots|\Psi^{s_a,s_c}_{s_d}}}{\sqrt{\Gamma_{s_a s_d} \Gamma_{s_d s_c}}}  \left[ \delta(s_d - s_b) - \frac{s_b \sinh 2 \pi s_b}{\pi^2} \left\{\begin{array}{ccc}
		1 & s_a & s_b \\
		1 & s_c & s_d
	\end{array} \right\} \right] = 0.
\end{equation}
To rewrite this result in a more compact form, define
\begin{equation}
	\cale_{s_a,s_c}(s_b) := \int_0^\infty ds_d \,  \calk_{s_a,s_c}(s_b,s_d) \calm_{s_a,s_c}(s_d)
\label{eq:7.35}
\end{equation}
where
\begin{equation}
	\calm_{s_a,s_c}(s_d) := \frac{P_{R,s_a} \iphir P_{R,s_d} \iphir P_{R,s_c}}{\sqrt{\Gamma_{s_a s_d} \Gamma_{s_d s_c}}}  
		- P_{R,s_c} \delta(s_a - s_c) \frac{s_d \tanh \pi s_d}{2 \pi^2} \cosh^2 (\pi s_a)
		\label{eq:7.36}
\end{equation}
and
\begin{equation}
	\calk_{s_a,s_c}(s_b,s_d) := \frac{s_b \sinh 2 \pi s_b}{\pi^2} \left\{\begin{array}{ccc} 1 & s_a & s_b \\ 1 & s_c & s_d \end{array} \right\} - \delta(s_d - s_b) 
\end{equation}
We have thus shown that
\begin{equation}
	\cale_{s_a,s_c}(s_b) \ket{\psifhh} = 0.
	\label{eq:operatoreq}
\end{equation}
In JT gravity with matter, there is an analogue of $\cale_{s_a,s_c}(s_b)$ that vanishes identically, due to the uncrossing identities \cite{Jafferis:2022wez}. In the present setup, $\cale_{s_a,s_c}(s_b)$ does not annihilate the entire Hilbert space. In fact, we will argue that for $N \ge 1$, there is no state in $\calh_N$ that is annihilated by $\cale_{s_a,s_c}(s_b)$ for all choices of $(s_a,s_b,s_c)$.

Consider a correlator
\begin{equation}
	\braket{\psifhh|\cdots \calm_{s_a,s_c}(s_d) \cdots |\psifhh},
\label{eq:7.38}
\end{equation}
where each $\cdots$ represents arbitrary insertions of $P_{s,R}$ and $\iphir$ with at least one $\iphir$. We depict the Wick contractions as follows,
\begin{equation}
	\label{eq:7.39}
	\begin{aligned}
		&\braket{\psifhh|\cdots \calm_{s_a,s_c}(s_d) \cdots |\psifhh} \\
		&=\raisebox{-0.5\height}{\includegraphics[scale=0.375]{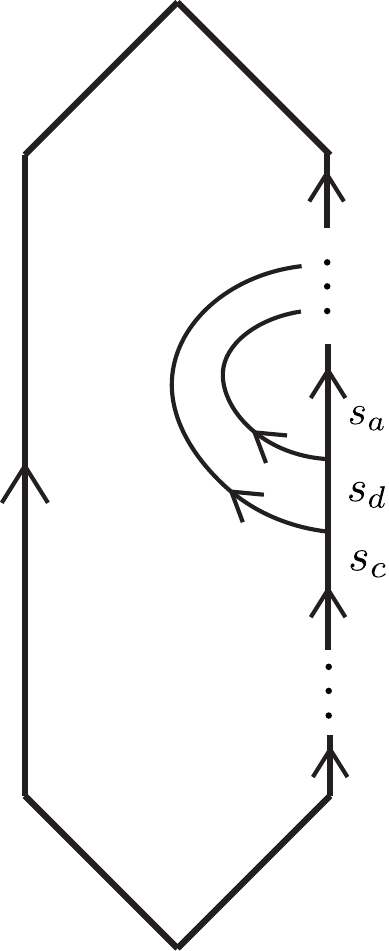}} + 
		\raisebox{-0.5\height}{\includegraphics[scale=0.375]{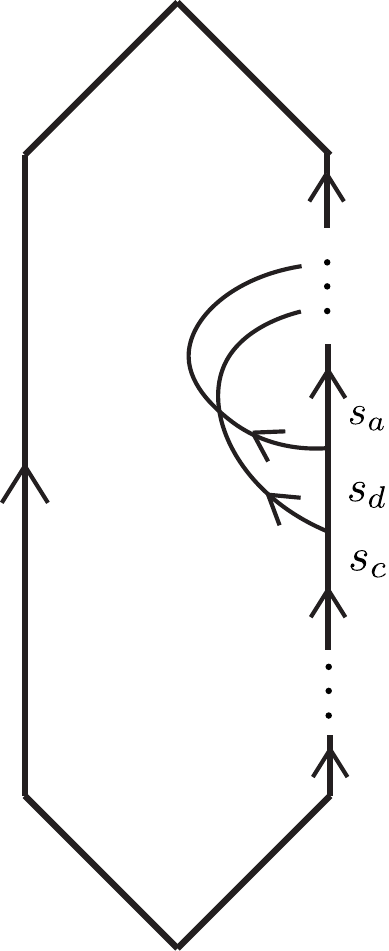}} + 
		\raisebox{-0.5\height}{\includegraphics[scale=0.375]{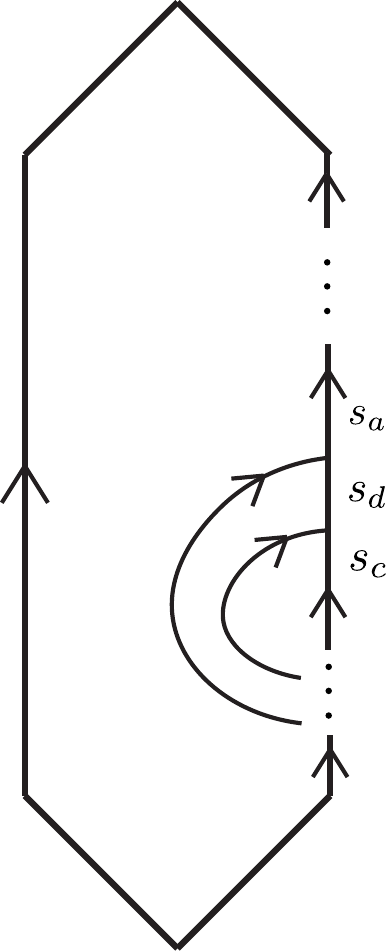}} + 
		\raisebox{-0.5\height}{\includegraphics[scale=0.375]{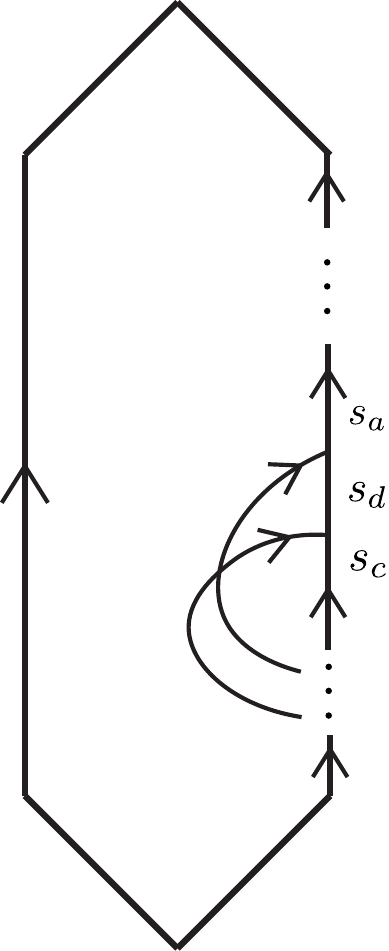}} \\
		&+ \raisebox{-0.5\height}{\includegraphics[scale=0.375]{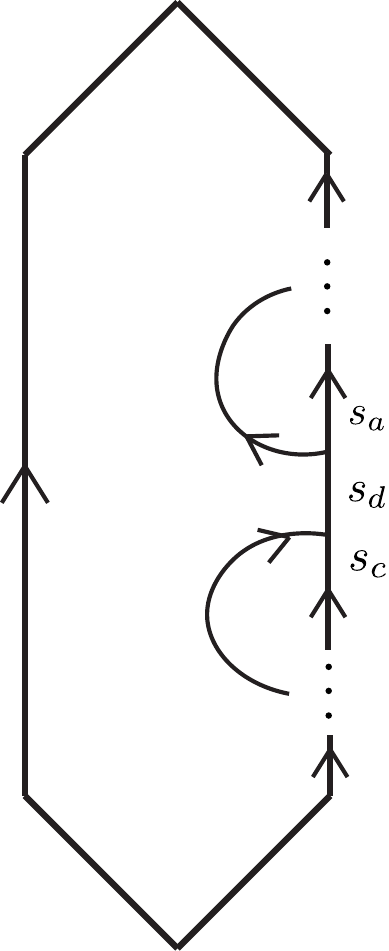}} + 
		\raisebox{-0.5\height}{\includegraphics[scale=0.375]{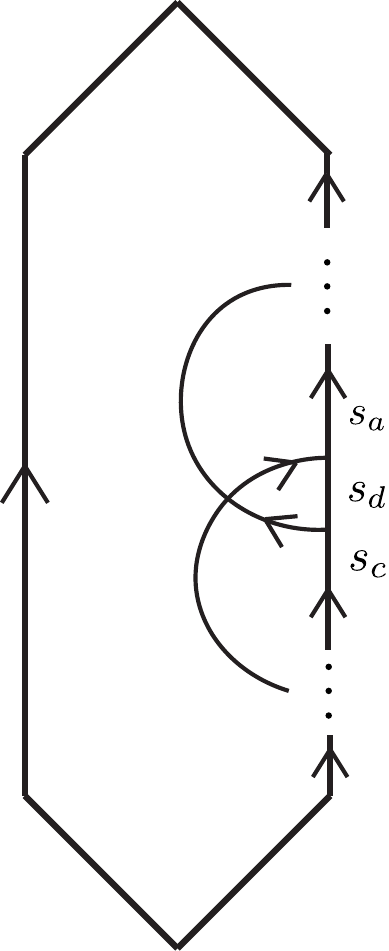}}
	\end{aligned}
\end{equation}
We have only explicitly shown the matter propagators that neighbor the $s_d$ boundary segment, and we have organized the diagrams based on whether these two lines contract into operators in the upper $\cdots$ or the lower $\cdots$, and whether these two lines are crossed or uncrossed.\footnote{In the case where one of the $\cdots$ contains only one $\iphir$ insertion, the diagrams in \eqref{eq:7.39} that have two lines contracting into that $\cdots$ should be disregarded.} To compute
\begin{equation}
	\braket{\psifhh|\cdots \cale_{s_a,s_c}(s_b) \cdots |\psifhh},
	\label{eq:caleeq}
\end{equation}
we begin with \eqref{eq:7.38} and integrate $s_d$ against $\calk_{s_a,s_c}(s_b,s_d)$, as shown in \eqref{eq:7.35}. Due to the uncrossing identities \eqref{eq:7.23} and \eqref{eq:7.24}, the Wick contractions on the first row of \eqref{eq:7.39} do not contribute to \eqref{eq:caleeq}. They return zero after performing the $s_d$ integral. On the other hand, the uncrossing identity does not apply to the last two diagrams in \eqref{eq:7.39}, because the arrows are going in different directions.\footnote{For the case of four operators, this can be seen explicitly from \eqref{eq:7.25}.} Thus, for a given choice of $(s_a, s_b, s_c)$, \eqref{eq:caleeq} can only be zero if the operators in the $\cdots$ are fine-tuned. If \eqref{eq:caleeq} vanishes for a fixed choice of operators in the $\cdots$, then \eqref{eq:caleeq} will no longer vanish as $(s_a,s_b,s_c)$ are varied. Because $\ket{\psifhh}$ is cyclic, we conclude that the only states that are annihilated by $\cale_{s_a,s_c}(s_b)$ for all choices of $(s_a,s_b,s_c)$ are in $\calh_0$. Thus, all of the states in $\calh_0$, including the Hartle-Hawking state, are not separating.

\subsection{The commutant of $\cala_R$ is generated by the left observer's Hamiltonian}

Next, we find the commutant of $\cala_R$, which we denote by $\cala_R^\prime$. Up to now, we have showed that a dense set of states in $\calh_0$ (including $\ket{\psifhh}$) is cyclic with respect to $\cala_R$ but not separating. We have also showed that for every state $\ket{\Psi} \in \calh$ that is not in $\calh_0$, there exists a choice of $(s_a,s_b,s_c)$ such that $\cale_{s_a,s_c}(s_b) \ket{\Psi} \neq 0$. Note that $\cale_{s_a,s_c}(s_b) \in \cala_R$ and $\cale_{s_a,s_c}(s_b) \ket{\Psi_f} = 0 \quad \forall \ket{\Psi_f} \in \calh_0$.

First, we show that any operator in $\cala_R^\prime$ maps $\calh_0$ into $\calh_0$. Let $\ket{\Psi_f} \in \calh_0$, and let $\calo \in \cala_R^\prime$. It follows that
\begin{equation}
	0 = \calo \cale_{s_a,s_c}(s_b) \ket{\Psi_f} = 	 \cale_{s_a,s_c}(s_b) \calo \ket{\Psi_f}.
\end{equation}
If $\calo \ket{\psifhh} \notin \calh_0$, then there is a choice of $(s_a,s_b,s_c)$ such that the above equation cannot hold, since $\cale_{s_a,s_c}(s_b) \calo \ket{\psifhh} \neq 0$. Thus, every operator in $\cala_R^\prime$ maps $\calh_0$ to $\calh_0$.

In fact, an operator $\calo \in \cala_R^\prime$ is completely characterized by its restriction to $\calh_0$. Consider how $\calo$ acts on a dense set of states which take the form $a_R \ket{\psifhh}$ for $a_R \in \cala_R$,
\begin{equation}
	\calo a_R \ket{\psifhh} = a_R \calo \ket{\psifhh}. 
\end{equation}
Thus, $\calo$ is completely determined by its restriction to $\calh_0$.

We now argue that $\calo$ must be a function of $\hat{s}_L$. Suppose that $\calo$ acts on $\calh_0$ as follows,
\begin{equation}
	\calo P_{R,s_a} \ket{\psifhh} = \int_0^\infty ds_b \,  F(s_b,s_a) P_{R,s_b} \ket{\psifhh},
		\label{eq:7.44}
\end{equation}
where $F(s_b,s_a)$ is some arbitrary function. Next, we use \eqref{eq:7.35} and \eqref{eq:operatoreq} to write 
\begin{equation}
	\int_0^\infty ds_d \,  \calk_{s_a,s_c}(s_b,s_d) \calm_{s_a,s_c}(s_d) \ket{\psifhh} = 0,
\end{equation}
and it follows that
\begin{align}
	0 &= \int_0^\infty ds_d \,  \calk_{s_a,s_c}(s_b,s_d) \calo \calm_{s_a,s_c}(s_d) \ket{\psifhh}
	\\
	&= \int_0^\infty ds_d \,  \calk_{s_a,s_c}(s_b,s_d) \calo \left[\frac{P_{R,s_a} \iphir P_{R,s_d} \iphir P_{R,s_c}}{\sqrt{\Gamma_{s_a s_d} \Gamma_{s_d s_c}}}
	\right. \notag
	\\
	& \quad \left. - P_{R,s_c} \delta(s_a - s_c) \frac{s_d \tanh \pi s_d}{2 \pi^2} \cosh^2 (\pi s_a)\right] \ket{\psifhh}
	\\
	&= \int_0^\infty ds_d \,  \calk_{s_a,s_c}(s_b,s_d)  \left[\frac{P_{R,s_a} \iphir P_{R,s_d} \iphir \calo P_{R,s_c}}{\sqrt{\Gamma_{s_a s_d} \Gamma_{s_d s_c}}}
	\right. \notag
	\\
	& \quad \left. - \calo P_{R,s_c} \delta(s_a - s_c) \frac{s_d \tanh \pi s_d}{2 \pi^2} \cosh^2 (\pi s_a)\right] \ket{\psifhh}
	\\
	&= \int_0^\infty ds_d \, ds_e \, F(s_e,s_c) \,  \calk_{s_a,s_c}(s_b,s_d)  \left[\frac{P_{R,s_a} \iphir P_{R,s_d} \iphir  P_{R,s_e}}{\sqrt{\Gamma_{s_a s_d} \Gamma_{s_d s_c}}}
	\right. \notag
	\\
	& \quad \left. -  P_{R,s_e} \delta(s_a - s_c) \frac{s_d \tanh \pi s_d}{2 \pi^2} \cosh^2 (\pi s_a)\right] \ket{\psifhh}
	\\
	&= \int_0^\infty ds_d \, ds_e \, F(s_e,s_c) \,  \calk_{s_a,s_c}(s_b,s_d)  \bigg[
	\calm_{s_a,s_e}(s_d) \frac{\sqrt{\Gamma_{s_a s_d} \Gamma_{s_d s_e}}}{\sqrt{\Gamma_{s_a s_d} \Gamma_{s_d s_c}}}
	 \notag
	\\
	& \quad + P_{R,s_e} \frac{s_d \tanh \pi s_d}{2 \pi^2} \cosh^2(\pi s_a)
	\left(\delta(s_a - s_e)  \frac{\sqrt{ \Gamma_{s_a s_d}}}{\sqrt{ \Gamma_{s_d s_c}}} -   \delta(s_a - s_c) \right) \bigg] \ket{\psifhh}.
	\label{eq:7.50}
\end{align}
Equation \eqref{eq:7.50} only vanishes for all $(s_a,s_b,s_c)$ when $F(s_e,s_c)$ contains a delta function $\delta(s_e - s_c)$. Hence, $\cala_R^\prime$ is generated by $\hat{s}_L$, or the left observer's Hamiltonian.

\subsection{$\cala_R$ is a direct integral of type I$_\infty$ factors}

Having found the commutant of $\cala_R$, we can now explain how $\cala_R$ fits into the type classification of von Neumann algebras. A von Neumann algebra is equal to its double-commutant. Thus, $\cala_R$ is equivalent to the commutant of the commutative algebra generated by $\hat{s}_L$. Recall from Section \ref{sec:addmatter} that if the matter QFT Hilbert space decomposes into $SO^+(2,1)$ irreps as follows,
\begin{equation}
	\calh_m = \bigoplus_j \calh_{m,j},
\end{equation}
then $\calh$ admits the following expansion:
\begin{equation}
\label{eq:structmassless}
	\calh = L^2(\mathbb{R}_+) \oplus \bigoplus_j L^2(\mathbb{R}_+) \otimes L^2(\mathbb{R}_+).
\end{equation}
For the case of a massless chiral scalar, each $j$ refers to a positive discrete series representation. The left observer's Hamiltonian $\hat{s}_L$ acts on the first $L^2(\mathbb{R}_+)$ factor (which represents $\calh_0$) as well as the first $L^2(\mathbb{R}_+)$ factor within each term of the direct sum $\bigoplus_j$. Factoring these $L^2(\mathbb{R}_+)$ factors out, we may write
\begin{equation}
	\calh = L^2(\mathbb{R}_+) \otimes \left[\mathbb{C} \oplus \bigoplus_j  L^2(\mathbb{R}_+)\right] := L^2(\mathbb{R}_+) \otimes \calh^\prime,
\end{equation}
where $\hat{s}_L$ acts only on the first $L^2(\mathbb{R}_+)$ factor, and we have defined $\calh^\prime$ to be the bracketed expression. We may decompose $\calh$ into sectors labeled by the eigenvalue of $\hat{s}_L$. Each sector is isomorphic to $\calh^\prime$. Because $\hat{s}_L \in \cala_R^\prime$, these are superselection sectors for $\cala_R$. Within each sector, $\cala_R$ acts as the algebra of all bounded operators on $\calh^\prime$, because $\cala_R$ is the commutant of the algebra generated by $\hat{s}_L$.  It follows that $\cala_R$ is a direct integral of type I$_\infty$ factors, and the center is generated by $\hat{s}_L$.

\subsection{Emergence of the Cosmological Horizon}

\label{sec:tracecomments}

A key result in this section is that the commutant of $\cala_R$ is generated by the left observer's Hamiltonian. Intuitively, this is related to the fact that in a QFT on global de Sitter, the algebra of observables associated to a Cauchy slice is type I and has trivial commutant. When the observer is moving along a fixed geodesic (as in \cite{CLPW}), their algebra is localized within their cosmological horizon. When the observer is fully dynamical, operators in their algebra can effectively change their trajectory, and thus there is no proper subregion in which their algebra is confined. Thus, local operators dressed to an observer are effectively smeared over an entire Cauchy slice in a state-dependent way. From this perspective, it is not surprising that the commutant of $\cala_R$ is generated by the left observer's Hamiltonian. We showed this above for the case that the matter theory is a chiral scalar, and we expect it to hold more generally.

As explained in the previous subsection, this result together with the structure of the Hilbert space \eqref{eq:structmassless} (or \eqref{eq:calhdef} and \eqref{eq:4.59} for the massive QFT) implies that $\cala_R$ is a direct integral of type I$_\infty$ factors. From the theory of von Neumann algebras, this implies that $\cala_R$ admits a non-unique trace. The different traces simply count the states in the physical Hilbert space $\calh$ weighted by some measure that depends on the left observer's energy. We may define a trace as
\begin{equation}
 \text{tr } a :=   \int dE_L \, p(E_L) \, \text{Tr}_{E_L} a , \quad a \in \cala_R,
 \label{eq:trace}
\end{equation}
where $\text{Tr}_{E_L}$ is the ordinary trace on the subspace of $\calh$ where the left observer has energy $E_L$, and $p(E_L)$ is an arbitrary probability measure. The different choices of $p(E_L)$ correspond to different traces on $\cala_R$. This trace is analogous to the trace that is associated to the type I factor in a QFT that describes all of the quantum fields on a Cauchy slice. Both traces count states in the Hilbert space.

Because the trace on $\cala_R$ is type I, its corresponding entropy should not be interpreted as the generalized entropy of a horizon. This is in contrast to JT gravity with matter \cite{Penington:2023dql,Kolchmeyer:2023gwa}, where the algebra associated to an AdS boundary is type II$_\infty$. The type II entropy found in \cite{Penington:2023dql,Kolchmeyer:2023gwa} may be interpreted as a generalized entropy even though it does not have a state-counting interpretation within JT gravity.  The boundary algebra provides a generalized notion of entanglement wedge.

In \cite{CLPW}, the type II entropy measures algebraic entanglement between the two static patches. In our de Sitter setup, the notion of the cosmological horizon is ill-defined away from the semiclassical limit, so this interpretation of the entropy breaks down. The entropy has a state-counting interpretation, but these states do not provide a statistical explanation of the Gibbons-Hawking entropy of a cosmological horizon. They are simply the gauge-invariant states in $\calh$.

Recall from Section \ref{sec:CLPW} that in the semiclassical limit, the Hilbert space takes the form $L^2(\mathbb{R}) \otimes \calh_{QFT}$, where the $L^2(\mathbb{R})$ factor refers to square-integrable functions of the left observer's Hamiltonian. Thus, \eqref{eq:trace} has a clear analogue in the semiclassical limit, which is given by the same expression where $\text{Tr}_{E_L}$ is the trace over $\calh_{QFT}.$ However, this trace is not a valid trace on $\cala_R$ in the semiclassical limit.\footnote{This is necessary to avoid a contradiction because in the semiclassical limit, $\cala_R$ is a type II factor, which has a unique trace.} The trace-class operators on which \eqref{eq:trace} is well-defined are not in $\cala_R$ in the semiclassical limit. If $\cala_R$ did include such a trace-class operator (which we can take to be self-adjoint w.l.o.g.), then $\cala_R$ would include all of its spectral projections, which are finite rank.\footnote{A finite rank projection is associated to a finite-dimensional subspace.} However, a type II factor cannot have nonzero finite rank projections. A factor with at least one nonzero finite rank projection has a minimal projection, and by definition a factor with a nonzero minimal projection is type I \cite{Jones2009, Sorce:2023fdx}.

We now discuss how the type II algebra of \cite{CLPW} emerges in the semiclassical limit from our type I algebra. We will first discuss the emergence of operators, and then we will discuss the emergence of the trace. The key point is that away from the semiclassical limit, the observer may access the entire spacetime, while in the semiclassical limit, the observer develops a cosmological horizon, and all of the operators in their algebra are restricted to their static patch.

When the observer's mass is not strictly infinite, they have access to the entire spacetime. When an operator localized on the worldline acts, the observer's trajectory is deflected. By acting with local operators and evolving along the worldline (in both the forward and backward directions), the observer can explore all of spacetime. When the observer's mass is large, the evolved time needs to be of order the scrambling time in order for their trajectory to change by an appreciable amount. For instance, suppose that an observer with mass $\Lambda$ sitting at the North Pole at $\theta = 0$ (using the coordinates in \eqref{eq:metric}) emits a particle of mass $m \ll \Lambda$ at coordinate time $-t$. The particle has order one rapidity in the observer's rest frame. After recoiling, the observer's new trajectory extends to a point on future infinity that has an angular coordinate of $\theta \propto \frac{m}{\Lambda} e^{t}$, to leading order in $\frac{m}{\Lambda} e^{t}$. Hence, when $\Lambda$ is large, as long as $t$ is large enough such that $\frac{m}{\Lambda} e^{t}$ is fixed, the change in the observer's trajectory will not be parametrically suppressed. The observer's new trajectory defines a new static patch. On this trajectory, the observer may emit another particle at coordinate time $t$ to deflect their trajectory to a new trajectory which extends to a point on past infinity with angular coordinate $\theta \propto \frac{m}{\Lambda}e^t$ (to leading order in $\theta \propto \frac{m}{\Lambda}e^t$). By iterating this process, the observer can move their static patch further.\footnote{This has implications for  bulk reconstruction in observer-centric de Sitter holography, which is based on the premise that the holographic screen is an observer's worldline. Operators anywhere in the spacetime that are dressed to an observer can be reconstructed using operators localized on their worldline. The operators inside the horizon have order one complexity, while those outside are logarithmically complex in the observer's mass.}

In contrast, if the duration of the time evolution does not scale with the mass of the observer, then in the semiclassical limit, the observer will be confined to the North Pole. By definition, the generating algebra\footnote{See footnote \ref{ft:genfoot} for our definition of generating algebra.} does not include operators that evolve the observer's clock by an amount that scales with the mass. The timelike tube theorem \cite{Strohmaier:2023opz} then implies that the double-commutant of the generating algebra is localized within the static patch associated to the North Pole. Hence, the only operators that are generated are operators in the static patch that are dressed to the observer.

The type II algebra is emergent because the semiclassical limit does not commute with taking the double-commutant of the generating algebra of $\cala_R$. The order of limits determines whether the algebra on $L^2(\mathbb{R}) \otimes \calh_{QFT}$ is a type II factor or a direct integral of type I$_\infty$ factors. The generating algebra consists of operators localized on the observer's worldline. If we take the semiclassical limit after the double-commutant, we obtain the algebra of all operators on $L^2(\mathbb{R}) \otimes \calh_{QFT}$ that commute with $q$, the left observer's Hamiltonian.\footnote{A subalgebra of this is the algebra of all operators on $\calh_{QFT}$. Physically, these operators are all dressed to the right observer, even though they are not all in the right observer's static patch. This is because in the semiclassical limit, the right observer's geodesic (which is physically parameterized by their clock) can be used to define a coordinate system for the global spacetime that may be used to situate local QFT operators (this gauge choice is depicted in Figure \ref{fig:penrose}). Any operator in $\calh_{QFT}$ is thus dressed to the right observer, regardless of whether this operator is in the right observer's static patch.} In contrast, if the semiclassical limit is taken first, then the final algebra is type II, due to the observer's cosmological horizon.

As shown in \cite{CLPW}, the trace on the type II algebra is the expectation value in the Hartle-Hawking state. Away from the semiclassical limit, we showed in Section \ref{sec:hhstate} that the Hartle-Hawking state is not a trace on the observer's algebra. For example, we showed in Section \ref{sec:crossed} that the crossed Wick contraction of four scalar operators is not cyclic, although it becomes cyclic in the semiclassical limit. Thus, the trace on the type II algebra is emergent.

\section{A candidate type II factor}

\label{sec:candidatetypeii}

In this paper, we have taken a ``bottom-up'' approach to de Sitter holography by studying local operators dressed to observers in a simple setup. Unlike analogous work in JT gravity \cite{Penington:2023dql,Kolchmeyer:2023gwa}, we are not aware of a UV completion or nonperturbative dual description. In this section, we will briefly discuss the properties of algebras that might emerge from a top-down construction.

In AdS holography, large $N$ factorization plays an important role in the emergence of bulk spacetime \cite{Gesteau:2024rpt}. In this section, we assume that a physically interesting algebra in de Sitter spacetime can also emerge from large $N$ factorization. That is, we will assume the existence of a single-trace operator $\Phi(x)$ with Gaussian correlation functions in the Hartle-Hawking state.\footnote{For now, we will not specify the space that $x$ lives in. For example, $x$ could be a coordinate on an observer's worldline.}

A principle of static patch holography is that an empty static patch is dual to a maximum entropy state, or a state with unit density matrix. In order to apply this notion, we will assume that the algebra generated by $\Phi(x)$ admits a trace given by the expectation value in the Hartle-Hawking state. This condition places strong constraints on the algebra, and it does not hold for the algebras constructed in Section \ref{sec:algebra}. Together with large $N$ factorization, it implies that the algebra generated by $\Phi(x)$ is real and commutative. It is commutative because
\begin{equation}
	\braket{\Phi(x) \Phi(y)} = \text{tr } \Phi(x) \Phi(y) = \text{tr } \Phi(y) \Phi(x) = \braket{ \Phi(y) \Phi(x)}, 
\end{equation}
and any higher-point function may be expressed in terms of the two-point function using Wick contractions. It is real because
\begin{equation}
	\braket{\Phi(x) \Phi(y)}^* = \braket{\Phi(y) \Phi(x) } = \braket{\Phi(x) \Phi(y)}.  
\end{equation}
In Section \ref{sec:gaugediscrete}, we pointed out that a spacetime algebra that is invariant under time-reversal must be real. Thus, the principles of large $N$ factorization and a maximum entropy state fit nicely with the fact that all spacetime symmetries, including the discrete ones, must be gauged. See \cite{Susskind:2023rxm} for a related discussion.

We now provide an example of a commutative real algebra in the theory of a free scalar on global dS$_2$. Using the conventions of Section \ref{sec:CLPW}, let $\varphi_R(t)$ be the scalar field at a point $t$ on the North pole. Let $\varphi_R^a(t)$ be the same operator evaluated at the antipode. We define
\begin{equation}
	\Phi(t) := \varphi_R(t) + \varphi_R^a(t).
	\label{eq:8.3}
\end{equation}
Because this is symmetric under the operation that exchanges pode and antipode, $\Phi(t)$ naturally lives in a non-time-orientable spacetime, namely Elliptic de Sitter \cite{Parikh:2002py, SANCHEZ19871111, GIBBONS1986497}, which is the de Sitter analogue of the Lorentzian M\"obius strip considered in \cite{Harlow:2023hjb}. The algebra generated by $\Phi(t)$ for all $t \in \mathbb{R}$ is commutative because
\begin{equation}
	[\Phi(t_1),\Phi(t_2)] =  [\varphi_R(t_1),\varphi_R(t_2)] + [\varphi_R^a(t_1),\varphi_R^a(t_2)] = 0.
	\label{eq:8.4}
\end{equation}
The last equality follows from the fact that each of the two commutators is equal to the identity times a purely imaginary function of the time separation, and the two commutators have opposite time-orderings, since $\varphi_R^a(t_1)$ is to the future of $\varphi_R^a(t_2)$ when $\varphi_R(t_1)$ is to the past of $\varphi_R(t_2)$. We can define other operators, such as the normal ordered product  $:\Phi^2:$, as follows:
\begin{equation}
 :\Phi^2:(t) \, := \, \lim_{t^\prime \rightarrow t} \Phi(t) \Phi(t^\prime) - \frac{1}{\pi} \log |t - t^\prime|.
\end{equation}
With respect to the algebra generated by $\Phi$, the Bunch-Davies vacuum is KMS with infinite temperature.

Note that \eqref{eq:8.4} can only hold when $\varphi_R$ is a free field. If $\varphi_R$ is interacting, then the two commutators in \eqref{eq:8.4} are nontrivial\footnote{We mean that the commutator is not proportional to the identity.} elements of the left and right static patch algebras. Because the static patch algebras are factors, the two commutators cannot cancel each other.

The reason why we consider this algebra is because we can generalize it to a more interesting type II algebra with a Hamiltonian. The algebra is similar in some ways to the algebras considered earlier in this paper, which we constructed by dressing operators to a fully quantized observer. Recall that the observer's Hilbert space was constructed using the positive-frequency wavefunctions $\Psi^s_n(t,\theta)$, defined in \eqref{eq:psidef}. The negative-frequency wavefunctions are $\tilde{\Psi}^s_n(t,\theta)$. Note that $\tilde{\Psi}^s_n(t,\theta)$ transforms the same way as $\Psi^s_n(t,\theta)$ under the spacetime isometries. We can generalize the notion of observer by re-defining what is meant by ``positive-frequency.'' We define
\begin{equation}
	\Xi^s_n(t,\theta) := \frac{1}{\sqrt{2}}\left(\Psi^{s}_{n}(t,\theta)  	+  \tilde{\Psi}^{s}_{n}(t,\theta)\right).
\end{equation}
Because $\Xi^s_n$ transforms the same way as $\Psi^s_n$, the propagator
\begin{equation}
	\widetilde{P_{s}}(t_2,\theta_2 | t_1, \theta_1) := \sum_{n \in \mathbb{Z}} \Xi^s_n(t_2,\theta_2) \left(\Xi^s_n\right)^*(t_1,\theta_1)
	\label{eq:newpropagator}
\end{equation}
is invariant under the spacetime isometries. To construct our new algebra, we take \eqref{eq:newpropagator} to be the observer's propagator. We can construct a Hilbert space following the same discussion in Section \ref{sec:observers}, but with $\Xi^s_n$ in place of $\Psi^s_n$. For example, we may define a new zero-particle Hilbert space $\widetilde{\calh_0}$ as the set of wavefunctions
\begin{equation}
	\widetilde{\Psi}_f(t_L,\theta_L;t_R,\theta_R) := \int_0^\infty ds \, f(s) \, \sum_{n \in \mathbb{Z}} (-1)^n \, \Xi_{-n}^s(t_L,\theta_L) \, \Xi_n^s(t_R,\theta_R).
\end{equation}
We define a new dressed operator $\widetilde{\phi_R}$, which is conveniently expressed in the new positive-frequency basis,
\begin{equation}
	\braket{s_R^\prime \, n_R^\prime|\widetilde{\phi_R}| s_R \, n_R} :=  \int dt_R d\theta_R\sqrt{-g_R} \left(\Xi^{s_R^\prime}_{n_R^\prime}(t_R,\theta_R)\right)^*\varphi(t_R,\theta_R) \, \Xi^{s_R}_{n_R}(t_R,\theta_R).
	\label{eq:newphirdef}
\end{equation}
We define $\widetilde{\phi_L}$ analogously. To compute correlators of $\widetilde{\phi_R}$ and $\widetilde{\phi_L}$, we define diagrammatic rules following \eqref{eq:4.7} and \eqref{eq:5.8},
\begin{equation}
	\begin{aligned}
		\vcenter{\hbox{\includegraphics[scale=0.4]{figures/PDF/dfig1.pdf}}} & \quad := \int dt\, d\theta\, \sqrt{-g} \left(\Xi^{s_a}_{n_a}(t,\theta)\right)^* (\psi^s_n)^*(t,\theta) \Xi^{s_b}_{n_b}(t,\theta),
	\end{aligned}
\label{eq:newrule1}
\end{equation}
\begin{equation}
	\vcenter{\hbox{\includegraphics[scale=0.4]{figures/PDF/dfig2.pdf}}} \quad := \quad \int dt\, d\theta\, \sqrt{-g} \left(\Xi^{s_a}_{n_a}(t,\theta)\right)^* \psi^s_n(t,\theta) \Xi^{s_b}_{n_b}(t,\theta).
	\label{eq:newrule2}
\end{equation}
With these new rules, the following identities hold:
\begin{equation}
		\includegraphics[scale=0.4]{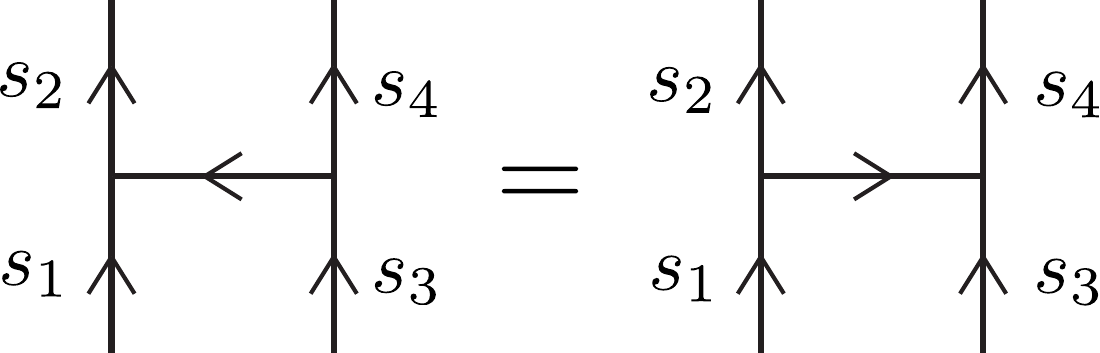},
		\label{eq:newid1}
\end{equation}
\begin{equation}
	\includegraphics[scale=0.4]{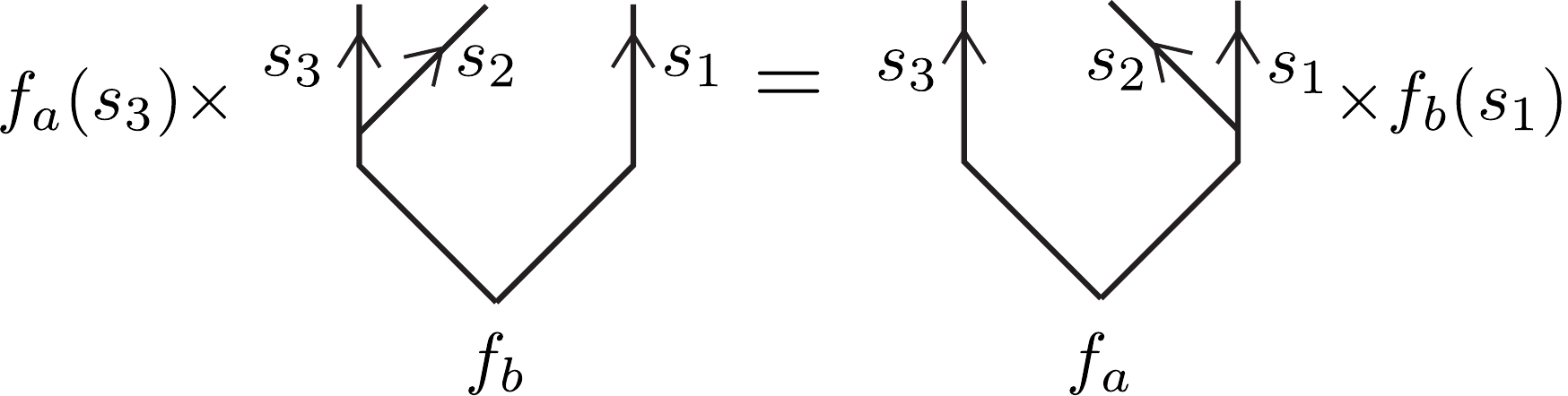}.
	\label{eq:newid2}
\end{equation}
These should be compared to 	\eqref{eq:leftbecomesright} and 	\eqref{eq:philphirequation}. The first identity \eqref{eq:newid1} implies that the arrows on the matter propagator lines can be disregarded, and the second identity \eqref{eq:newid2} implies that if we define a new Hartle-Hawking state to be
\begin{equation}
	\widetilde{\Psi_{f_{HH}}}(t_L,\theta_L;t_R,\theta_R) := \int_0^\infty ds  \, \sum_{n \in \mathbb{Z}} (-1)^n \, \Xi_{-n}^s(t_L,\theta_L) \, \Xi_n^s(t_R,\theta_R),
	\label{eq:newhh}	
\end{equation}
Then the property \eqref{eq:6.1} holds for  $\widetilde{\phi_L}$ and $\widetilde{\phi_R}$. Moreover, \eqref{eq:newid1} implies that $\widetilde{\phi_L}$ and $\widetilde{\phi_R}$ commute. Correlation functions may be computed using the diagrammatic rules explained in Section \ref{sec:gaugeinvariant}, but all of the diagrams may be represented as ordinary chord diagrams. For instance, the diagrams in \eqref{eq:fourpoint} may be rewritten as
\begin{equation}
	\includegraphics[scale=0.3]{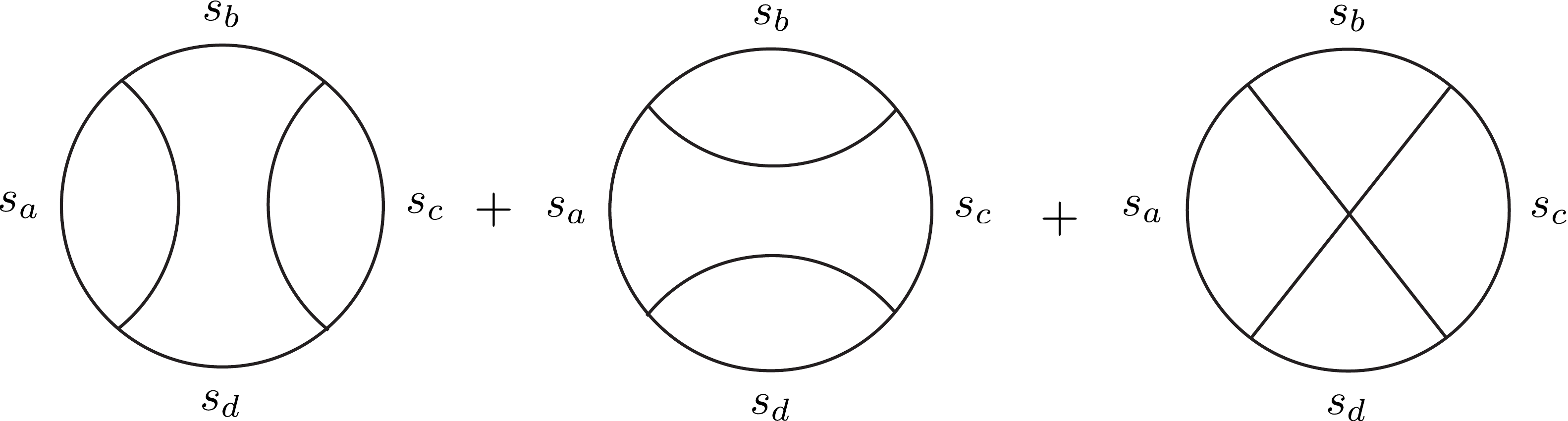},
	\label{eq:threecontractions}
\end{equation}
times a factor of $f_1(s_d) f_2^*(s_d)$. To restore the arrows such that the rules \eqref{eq:newrule1} and \eqref{eq:newrule2} can be directly applied, one should pick an orientation of the boundary and draw the arrows accordingly. The arrows on the chords may be chosen arbitrarily. The result does not depend on the different choices.

The two-point function of $\widetilde{\phi_R}$ is
\begin{equation}
	\begin{aligned}
		\braket{\widetilde{\psifhh}| \widetilde{P_{R,s_2}} \widetilde{\phi_R} \widetilde{P_{R,s_1}} \widetilde{\phi_R} |\widetilde{\psifhh}}
		&=  \left(\cosh \pi s + \cosh \pi s_1 + \cosh \pi s_2\right)^2 \\
		&\quad \times \frac{\left(\prod_{i \in \{s_1, s_2\}} s_i \tanh \pi s_i\right)}{128 \pi^4} \\
		&\quad \times \Gamma\left(\frac{1}{2}\left(\frac{1}{2} \pm i s \pm i s_1 \pm i s_2\right)\right),
	\end{aligned}
\end{equation}
which should be compared to \eqref{eq:6.14}. Here, $\widetilde{P_{R,s}}$ is the new propagator \eqref{eq:newpropagator} of the right observer. After taking the semiclassical limit, one recovers the two-point function of \eqref{eq:8.3} up to normalization.

The conclusion is that correlation functions of $\widetilde{\phi_R}$ and $\widetilde{\phi_L}$ in the Hartle-Hawking state \eqref{eq:newhh} can be computed by applying rules to chord diagrams. We have checked that this conclusion extends to the dressed operators $\widetilde{\overrightarrow{\partial} \phi_R}$ and $\widetilde{\overrightarrow{\partial} \phi_L}$, which are defined analogously to \eqref{eq:5.19} with $\Psi^s_n$ replaced by $\Xi^s_n$. We expect that this conclusion can be extended to all dressed operators that are linear in the free field. In contrast, we have checked that $\widetilde{:\phi^n:_R}$ and $\widetilde{:\phi^n:_L}$ do not commute.

We define the new algebra $\widetilde{\cala_R}$ to be the algebra generated by $\hat{s}_R$, $\widetilde{\phi_R}$, and all other operators dressed to the right observer that are linear in the free scalar field. The left algebra $\widetilde{\cala_L}$ is defined analogously. These two algebras commute with each other, and they have a trace given by the expectation value in the new Hartle-Hawking state \eqref{eq:newhh}. The Hilbert space can be constructed using the GNS construction, where the Hartle-Hawking state is used to define the inner product.\footnote{See \cite{Lin:2022rbf, Xu:2024hoc} for another example of how the GNS construction can be applied to an algebra whose correlators are computed using chord diagrams.} Due to the definition of the Hartle-Hawking state \eqref{eq:newhh}, the trace of the identity comes from the Plancherel measure only,
\begin{equation}
	\braket{\widetilde{\psifhh}|\widetilde{\psifhh}} = \int_0^\infty ds \, \rhopl(s) = \infty,
	\label{eq:traceid}
\end{equation}
where $\rhopl(s)$ was defined in \eqref{eq:rhopl}. We have not checked explicitly whether $\widetilde{\cala_R}$ and $\widetilde{\cala_L}$ are commutants of each other or factors, but we expect that they are because correlators may be computed by applying diagrammatic rules to chord diagrams. However, a more careful analysis is required, which we leave for future work. Examples of type II factors that are constructed from chord diagrams may be found in JT gravity coupled to a free scalar field \cite{Penington:2023dql,Kolchmeyer:2023gwa} and the double-scaled SYK model \cite{Lin:2022rbf, Xu:2024hoc}. Because \eqref{eq:traceid} is $\infty$, we conjecture that $\widetilde{\cala_L}$ and $\widetilde{\cala_R}$ are type II$_\infty$ factors.

In summary, by introducing a generalized notion of an observer, we have constructed a candidate type II factor that reduces to a real commutative algebra in the semiclassical limit. In Sections \ref{sec:freescalar} and \ref{sec:observers}, the matter QFT and observers were constructed by quantizing positive-frequency modes. In this section, the observer's Hilbert space is defined using a different definition of positive-frequency. The commutative real algebra is generated by operators that are supported in both static patches, and it could plausibly emerge from a holographic dual that exhibits large $N$ factorization. In this section, it was important that we worked with a free field.

\section{Discussion}

In this work, we studied a global dS$_2$ background with a scalar field QFT and two observers. The observers are fully-quantized relativistic point-particles with clock degrees of freedom. We showed how local QFT operators can be dressed to an observer to produce gauge-invariant operators. We also proposed an algebraic criteron for finding the Hartle-Hawking state that generalizes the Hartle-Hawking state in \cite{CLPW} away from the semiclassical limit. The correlation functions of dressed operators in the Hartle-Hawking state reduce to the usual semiclassical correlators in the limit that the observers have large masses and are localized along geodesic trajectories at the podes.  When the observer masses are order one in cosmic units, we find that their respective algebras fail to factorize. Our results indicate that their algebras should be viewed as emerging from a single quantum system rather than two decoupled quantum systems.

Our original goal was to investigate possible generalizations of the type II$_1$ entropy discovered in \cite{CLPW}. This entropy is defined using the tracial Hartle-Hawking state. As explained in Section \ref{sec:tracecomments}, the observer's algebra away from the semiclassical limit is a direct integral of type I$_\infty$ factors, and none of the traces on this algebra yield satisfactory generalizations of the observer's generalized entropy. 

The Hartle-Hawking state is not a trace away from the semiclassical limit. Technically, this is due to the crossed Wick contractions discussed in Section \ref{sec:crossed}. As discussed in Section \ref{sec:uncrossed}, the diagrams with no crossed Wick contractions can be computed using a simple set of diagrammatic rules. These rules have direct analogies in JT gravity with matter and a two-matrix model, which are related using the ETH matrix model ansatz developed in \cite{Jafferis:2022wez, Jafferis:2022uhu}. In \cite{Jafferis:2022wez, Jafferis:2022uhu}, crossed Wick contractions were a technical nuisance that forced the matrix model to be non-Gaussian. In our work, the crossed Wick contractions change the type of the algebra, such that the ETH ansatz cannot in principle be applied at order one times. However, when adjacent operators in a correlator are well-separated in time one may neglect the crossed Wick contractions in both our work and in JT gravity with matter, and a Gaussian ETH ansatz is applicable.\footnote{For example, we showed in Section \ref{sec:otoc} that for an OTOC, the crossed Wick contraction decays to zero at late times.} This timescale can be viewed as the Thouless time, and it is well past the scrambling time. We conclude that the putative holographic dual from which the observer's algebra emerges should have a notion of a Thouless time beyond which the Hartle-Hawking state is approximately a trace and the ETH is applicable.

We now compare our gauge-fixing methods to other works. To obtain a gauge-invariant Hilbert space, we began with a space of unnormalizable singlets in a tensor product pre-Hilbert space and then used the Fadeev-Popov procedure to gauge-fix the divergent inner product. Our method of gauge-fixing is closely related to group averaging, which has been applied in a similar context in the recent work \cite{Kaplan:2024xyk}, which builds upon earlier work \cite{Marolf:2008it,Marolf:2008hg,Marolf:2000iq,Higuchi_19911,Higuchi_19912}. As reviewed in \cite{Kaplan:2024xyk}, group averaging is a rigorous technique for dividing the original inner-product on invariant states by the infinite volume of the isometry group. We believe that a careful application of the group averaging formalism to our setup would lead to the same results. In particular, \cite{Penington:2023dql} noted that in JT gravity with matter, group averaging is equivalent to working with unnormalizable states and formally dividing by the infinite group volume. The calculations in Section \ref{sec:hhstate} closely resemble the JT calculations in \cite{Suh:2020lco}.

Our calculations shed light on the observation of \cite{Kaplan:2024xyk} that two-particle states have a divergent group averaging norm. As noted in \cite{Kaplan:2024xyk}, this is related to the fact that classical gauge-invariant two-particle states have an unbroken boost isometry. In Section \ref{sec:twoentangled}, we constructed a Hilbert space of two observers where all states have finite norms. The difference between our two-observer Hilbert space $\calh_0$ and a two-particle Hilbert space is that in the former, the mass of the particles is promoted to a continuous quantum number, $s$. The singlet condition requires that the two observers have the same mass. As noted in Section \ref{sec:twoentangled}, a wavefunction in $\calh_0$ is a function of the mass, $f(s)$. To make contact with the two-particle Hilbert space of \cite{Kaplan:2024xyk}, we must choose a particular mass $s_0$. The wavefunction of two particles with definite mass in our Hilbert space is $f(s) = \delta(s - s_0)$, which has infinite norm. This points to the origin of the divergence noted by \cite{Kaplan:2024xyk}. Promoting the particles to observers is equivalent to introducing clocks that break the residual boost isometry and allow for finite-norm states.

Our computations involve Clebsch-Gordan coefficients and 6j symbols of $\mathfrak{so}(2,1)$, following similar calculations in \cite{Suh:2020lco}. One of our technical results is a direct link from a 6j symbol of $\mathfrak{so}(2,1)$ to an integral over two effective modes, $x_+$ and $x_-$, which gives a physical picture of the OTOC that resembles the work of \cite{Stanford:2021bhl, Maldacena:2016upp}. The OTOC physics we observe in Section \ref{sec:otoc} is related to (but strictly different from) the physics that was observed in three-dimensional de Sitter gravity in \cite{Aalsma:2020aib}. The scrambling time in \cite{Aalsma:2020aib} is proportional to $\log \frac{1}{G_N}$, whereas the scrambling time in our OTOCs is proportional to the log of the observer's mass. The eikonal phase has a definite sign in \cite{Aalsma:2020aib}, whereas the phase can have either sign in our work. In future work, it would be interesting to study OTOCs with both dynamical observers and metrics. For instance, one could consider the setup of \cite{Verlinde:2024znh}, where the observer sources a conical deficit.

We observed that the scrambling time increases logarithmically with the observer's mass. This may be compared with AdS/CFT, where the scrambling time is logarithmic in the size of the system. This naively suggests that the dimension of the code subspace in which an observer with mass $m$ is present increases exponentially with $m$. On the other hand, the generalized entropy of the observer's cosmological horizon decreases with $m$ \cite{Witten}. A successful proposal for a holographic dual must reconcile these seemingly incompatible observations. 

In Section \ref{sec:otoc}, we observed that the chaos bound is violated by a factor of 2. That is, in the semiclassical limit, the putative holographic dual must exhibit both thermal correlations at the de Sitter temperature and a Lyapunov exponent of $\frac{4 \pi}{\beta_\text{dS}}$. This could pose a challenge for candidate models of de Sitter holography that are based on the chord diagram structure of the double-scaled SYK model \cite{Almheiri:2024xtw, Rahman:2024iiu, Narovlansky:2023lfz}, which has instead been linked to sub-maximal chaos \cite{Maldacena:2016hyu,Choi:2019bmd,Streicher:2019wek,Lin:2023trc}. At the same time, we emphasize that we have not found a strict paradox. It has been proposed that the dual of an empty static patch is at infinite temperature, and the finite de Sitter temperature is emergent \cite{Lin:2022nss}. It would be very interesting to investigate whether systems with finite ``tomperature'' \cite{Lin:2022nss} obey bounds analogous to the chaos bound. We have given a simple explanation for the exponent $\frac{4 \pi}{\beta_{\text{dS}}}$ in our OTOCs, which suggests that the putative holographic dual might saturate a new bound that is analogous to the chaos bound of \cite{Maldacena:2015waa} but has different assumptions.

To speculate further on how the chaos bound might be violated, we consider the work of \cite{Narovlansky:2023lfz}, which proposed a dual description of the $s$-wave sector of de Sitter gravity in three dimensions. The model of \cite{Narovlansky:2023lfz} consists of two copies of the double-scaled SYK model projected onto the sector where the two Hamiltonians are equal. The proposal is that an operator along the observer's worldline is dual to $\calo_{phys} := \Pi \calo_\Delta \calo_{1-\Delta} \Pi$, where $\calo_\Delta$ is a simple operator of scaling dimension $\Delta$ in one SYK model, $\calo_{1-\Delta}$ is a simple operator of dimension $1-\Delta$ in the other SYK model, and $\Pi$ is the projection operator onto the equal energy sector.\footnote{It was found in \cite{Narovlansky:2023lfz} that the two-point function of $\calo_{phys}$ is the {\it antipodal} Green's function of a massive scalar. Thus it is a priori not obvious that the holographic screen is given by a single observer's worldline at the  North pole. The authors of \cite{Narovlansky:2023lfz} identified Euclidean time-evolved versions of $\calo_{phys}$ with complex operators evaluated on the North pole in order to maintain the interptation that the holographic screen is at the North pole.} The operator $\calo_{phys}$ is not simple due to the projection, and thus it is not obvious whether an OTOC involving $\calo_{phys}$ will obey the chaos bound. It is important to investigate higher-point correlators in \cite{Narovlansky:2023lfz} (see \cite{Verlinde:2024znh} for a concrete proposal).

As mentioned in the Introduction, it is important to study as many backgrounds and parameter ranges as possible. Another interesting context would be de Sitter JT gravity with observers. In this case, we can make both the observers and the metric fully dynamical. We expect that the results in \cite{Alonso-Monsalve:2024oii,Nanda:2023wne,Levine:2022wos,Maldacena:2019cbz,Held:2024rmg} will be useful.

\paragraph{Acknowledgements}

We thank Adam Levine and Ying Zhao for stimulating and helpful discussions. We also thank Edward Witten for comments on the draft. We thank the NYU Abu Dhabi Institute and the organizers of the workshop ``A Quantum Al-Khawarizm for Spacetime: A Workshop on von Neumann Algebras in Quantum Field Theory \& Gravity'' for hospitality and for the stimulating environment in which this work was initiated. This material is based upon work supported by the U.S. Department of Energy, Office of Science, Office of High Energy Physics of U.S. Department of Energy under grant Contract Number DE-SC0012567 and DE-SC0020360 (MIT contract \# 578218), and by the Packard Foundation award for Quantum Black Holes from Quantum Computation and Holography.

	\appendix

\section{Explicit Formulas}
\label{sec:appendixformulas}

In this Appendix, we provide explicit formulas for some quantities referenced in the main text.

\subsection{Wightman function}

When the two points are spacelike separated, the Wightman function \eqref{eq:wightman} is given by
\begin{equation}
\braket{0|\varphi(t_2,\theta_2) \varphi(t_1,\theta_1)|0} = \frac{1}{4 \pi} \Gamma\left(\frac{1}{2} - i s\right) \Gamma\left(\frac{1}{2} + i s\right) \, {}_2F_1\left(\frac{1}{2} + i s, \frac{1}{2} - i s; 1; \frac{1 + c}{2}\right).
\end{equation}
 Here, $c$ is defined by
\begin{align}
	c &:= \cos(\theta_1 - \theta_2) \cosh(t_1) \cosh(t_2) - \sinh(t_1) \sinh(t_2),
	\\
	&= 	\cosh(t_1 - t_2) \cos(\theta_1 - \theta_2) - 2 \sin^2\left(\frac{\theta_1 - \theta_2}{2}\right) \sinh(t_1) \sinh(t_2). 
\end{align}
When one point is spacelike separated from the antipode of the other point, we have $c \in (-1,1).$ When one point is to the future or past of the antipode of the other point, we have $c \in (-\infty,-1)$.

When $(t_2,\theta_2)$ is to the future of $(t_1,\theta_1)$, the Wightman function is given by\footnote{See also the integral representation given in \eqref{eq:integralrepwightmann}.}
\begin{equation}
	\braket{0|\varphi(t_2,\theta_2) \varphi(t_1,\theta_1)|0} = \frac{1}{4 \pi} \Gamma\left(\frac{1}{2} - i s\right) \Gamma\left(\frac{1}{2} + i s\right) \, {}_2F_1\left(\frac{1}{2} + i s, \frac{1}{2} - i s; 1; \frac{1 + c}{2} - i \epsilon\right),
\end{equation}
and $c \in (1,\infty)$. The $i \epsilon$ indicates which side of the hypergeometric branch cut we want to consider. When $(t_2,\theta_2)$ is to the past of $(t_1,\theta_1)$, we replace $i \epsilon \rightarrow - i \epsilon$.

\subsection{Modefunctions}

The modefunctions $\psi^s_n(t,\theta)$ that were introduced in Section \ref{sec:freescalar} are defined as follows:
\begin{align}
	\psi^s_n(t,\theta) &:= e^{i \phi_{n,s}/2} e^{i n \theta} \frac{(1 - i)  (2i)^n  \cosh(t)^{-n} \Gamma\left(\frac{1}{2} - i s\right) }{2\sqrt{e^{- \pi s} + e^{ \pi s}} \, \Gamma\left(\frac{1}{2} + n - i s\right)} \label{eq:modefunction}
	\\
	&\times {}_2\tilde{F}_1\left(\frac{1}{2} - n - i s, \frac{1}{2} - n + i s; 1 - n; \frac{1}{2} (1 - i \sinh(t))\right)
	\nonumber
\end{align}
where ${}_2\tilde{F}_1$ refers to the regulated hypergeometric function. We define
\begin{equation}
	e^{i \phi_{n,s}} := \frac{i \cosh (\pi  s) \Gamma \left(i s+\frac{1}{2}\right) \Gamma \left(\frac{1}{2} - i s \pm n \right)}{\pi 
		\Gamma \left(\frac{1}{2}-i s\right)} ,
	\label{eq:eiphi}
\end{equation}
which is a pure phase. Note that $e^{i \phi_{n,s}} = e^{i \phi_{-n,s}}$ and $e^{i \phi_{n,-s}} = -e^{-i \phi_{n,s}}.$

\subsection{Vertex function}

The integral \eqref{eq:usefulintegral} is used to determine the diagrammatic rule associated to the vertices \eqref{eq:4.7} and \eqref{eq:5.8}. The $\calv^{s_1}_{s_2,s_3}$ function that appears there is defined to be
\begin{align}
	\label{eq:vertexintegraldef}
\calv^{s_1}_{s_2,s_3} &:=	-\frac{
		e^{\frac{1}{4} \pi (-5i + 2s_1 + 2s_2 + 2s_3)}
		\cosh(\pi s_1)
		\Gamma\left( -\frac{i}{4} (i + 2s_1 - 2s_2 - 2s_3) \right)
	}{
		2 \sqrt{2} \sqrt{1 + e^{2 \pi s_1}} \sqrt{1 + e^{2 \pi s_2}} \sqrt{1 + e^{2 \pi s_3}} \pi
		\Gamma\left( \frac{1}{2} - i s_1 \right) \Gamma\left( \frac{1}{2} + i s_2 \right)
	}  \nonumber \\
	&\times
		\Gamma\left( -\frac{i}{4} (i + 2s_1 + 2s_2 - 2s_3) \right)
		\Gamma\left( -\frac{i}{4} (i + 2s_1 - 2s_2 + 2s_3) \right)
		\Gamma\left( \frac{i}{4} (-i + 2s_1 + 2s_2 + 2s_3) \right)
 \nonumber \\
	&\times
		\Gamma\left( \frac{1}{2} + i s_3 \right)
		\sqrt{s_1 \tanh(\pi s_1)} \sqrt{s_2 \tanh(\pi s_2)} \sqrt{s_3 \tanh(\pi s_3)}
\end{align}

\subsection{Conformal Killing vectors}

\label{sec:ckvs}

The conformal Killing vectors that appear in \eqref{eq:7.4} and \eqref{eq:7.5} are

\begin{align}
	V_{1} &= \left( -\frac{1}{2} e^{-i \theta} \left( -i + \sinh(t) \right) \right) \partial_t + \left( \frac{1}{2} e^{-i \theta} \text{sech}(t) \left( -i + \sinh(t) \right) \right) \partial_\theta,
	\\
	V_{0} &= \left( -\frac{i}{2} \cosh(t) \right) \partial_t + \left( \frac{i}{2} \right) \partial_\theta
	\\
	V_{-1} &= \left( \frac{1}{2} e^{i \theta} \left( i + \sinh(t) \right) \right) \partial_t + \left( -\frac{1}{2} e^{i \theta} \text{sech}(t) \left( i + \sinh(t) \right) \right) \partial_\theta
	\\
	\bar{V}_{1} &= \left( \frac{1}{2} e^{-i \theta} \left( i + \sinh(t) \right) \right) \partial_t + \left( \frac{1}{2} e^{-i \theta} \text{sech}(t) \left( i + \sinh(t) \right) \right) \partial_\theta
	\\
	\bar{V}_{0} &= \left( \frac{1}{2} i \cosh(t) \right) \partial_t + \left( \frac{i}{2} \right) \partial_\theta
	\\
	\bar{V}_{-1} &= \left( -\frac{1}{2} e^{i \theta} \left( -i + \sinh(t) \right) \right) \partial_t + \left( -\frac{1}{2} e^{i \theta} \text{sech}(t) \left( -i + \sinh(t) \right) \right) \partial_\theta
\end{align}

\subsection{Crossed Wick contraction}

	The expression for the crossed Wick contraction in Section \ref{sec:crossed} makes use of the following function,
\begin{equation}	
	\begin{aligned}
		\label{eq:seq}
		\cals^s(s_1,s_2,s_3,s_4) = &\, \frac{1}{4096 \pi^{11}} S(s_1,s_2,s_3,s_4,s,s) \\
		&\times \Gamma\left(\frac{1}{2} \left(\frac{1}{2} - i s \pm (i s_1 + i s_2)\right)\right) \\
		&\times \Gamma\left(\frac{1}{2} \left(\frac{1}{2} \pm i s + i s_1 - i s_2\right)\right) \\
		&\times \Gamma\left(\frac{1}{2} \left(\frac{1}{2} - i s \pm (i s_2 + i s_3)\right)\right) \\
		&\times \Gamma\left(\frac{1}{2} \left(\frac{1}{2} \pm i s + i s_2 - i s_3\right)\right) \\
		&\times \Gamma\left(\frac{1}{2} \left(\frac{1}{2} + i s \pm (i s_3 + i s_4)\right)\right) \\
		&\times \Gamma\left(\frac{1}{2} \left(\frac{1}{2} \pm i s + i s_3 - i s_4\right)\right) \\
		&\times \Gamma\left(\frac{1}{2} \left(\frac{1}{2} + i s \pm (i s_4 + i s_1)\right)\right) \\
		&\times \Gamma\left(\frac{1}{2} \left(\frac{1}{2} \pm i s + i s_4 - i s_1\right)\right)
	\end{aligned}
\end{equation}
where $S(s_1,s_2,s_3,s_4,s,s)$ is defined in \eqref{eq:sdef}.

\section{Crossed Wick contraction}

\label{sec:appendixcrossed}

In this appendix, we discuss the evaluation of the following diagram:
\begin{equation}
	\label{eq:crosseddiagramappendix}
\includegraphics[width=0.1\linewidth]{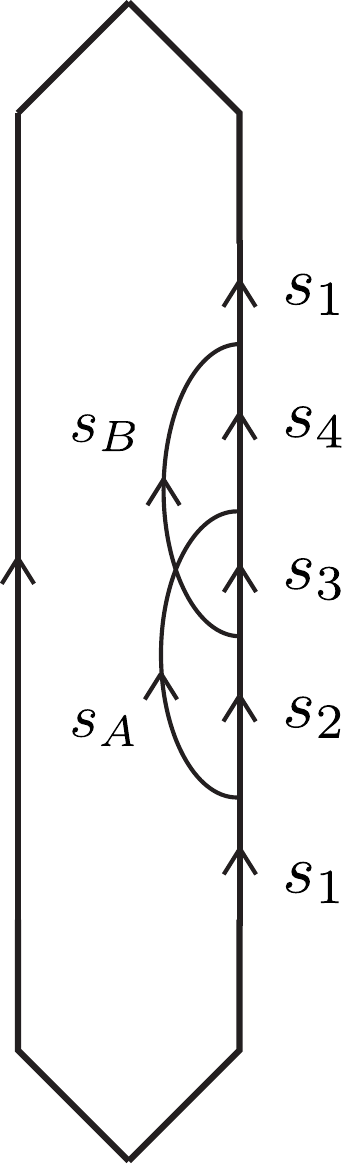}.
\end{equation}
Using \eqref{eq:4.7}, \eqref{eq:5.8}, \eqref{eq:usefulintegral}, and \eqref{eq:cpspp}, we find that this becomes

\begin{equation}
	\begin{aligned}
		\label{eq:firsteqB}
		\sum_{n_1,n_A,n_2} &(C_S)^{1-\Delta_{s_A},1-\Delta_{s_2},\Delta_{s_1}}_{-n_A,-n_2,n_1} \left(\calv^{s_1}_{s_A,s_2}\right)^* \left(\calv^{s_2}_{s_B,s_3}\right)^* \calv^{s_4}_{s_A,s_3} \calv^{s_1}_{s_B,s_4}\\
		&\times \left[\sum_{n_B,n_3,n_4}  (C_S)^{1-\Delta_{s_B},1-\Delta_{s_3},\Delta_{s_2}}_{-n_B,-n_3,n_2}  
		\times (C_S)^{\Delta_{s_A},\Delta_{s_3},1-\Delta_{s_4}}_{n_A,n_3,-n_4}  
		\times (C_S)^{\Delta_{s_B},\Delta_{s_4},1-\Delta_{s_1}}_{n_B,n_4,-n_1}  \right]
	\end{aligned}
\end{equation}
Using \eqref{eq:ISdef}, the part of this expression in square brackets is equal to
\begin{align}
	&\int_{0}^{2 \pi} \frac{d\theta_B}{2 \pi}  \frac{d\theta_3}{2 \pi} \frac{d\theta_4}{2 \pi}
	\\
	&I_S^{1-\Delta_B,1-\Delta_3,\Delta_2}(\theta_B,\theta_3,\theta_2)
	I_S^{\Delta_A,\Delta_3,1-\Delta_4}(\theta_A,\theta_3,\theta_4)
	I_S^{\Delta_B,\Delta_4,1-\Delta_1}(\theta_B,\theta_4,\theta_1)
\end{align}
which becomes
\begin{align}
	&	|1+t_1^2|^{1-\Delta_1} |1+t_2^2|^{\Delta_2} |1+t_A^2|^{\Delta_A}
	\int_0^\infty \frac{dt_B}{ \pi}  \frac{dt_3}{ \pi} \frac{dt_4}{ \pi}
	\label{eq:thisintegral}
	\\
	&I^{1-\Delta_B,1-\Delta_3,\Delta_2}(t_B,t_3,t_2)
	I^{\Delta_A,\Delta_3,1-\Delta_4}(t_A,t_3,t_4)
	I^{\Delta_B,\Delta_4,1-\Delta_1}(t_B,t_4,t_1)
\end{align}
after the change of coordinates $t = \tan \frac{\theta}{2}$ for each of the $\theta$ variables. We have defined
\begin{equation}
	I^{\Delta_1,\Delta_2,\Delta_3}(t_1,t_2,t_3) = 	\left|t_{31}\right|^{-\Delta_3 - \Delta_1 + \Delta_2} \left|t_{32}\right|^{-\Delta_3 - \Delta_2 + \Delta_1} \left|t_{21}\right|^{-\Delta_2 - \Delta_1 + \Delta_3}.
\end{equation}
Due to symmetry, \eqref{eq:thisintegral} should be proportional to
\begin{equation}
	I^{1-\Delta_1,\Delta_2,\Delta_A}(t_1,t_2,t_A).
\end{equation}
Thus, let us set $t_1 = 0$, $t_2 = 1$, $t_A = \infty$. After making the following redefinitions,
\begin{equation}
	t_4 \rightarrow \frac{t_4 - 1}{t_4(1-t_B)} \quad t_3 \rightarrow \frac{1-t_3 t_B}{1-t_B} \quad t_B \rightarrow \frac{1}{1-t_B}
\end{equation}
we find that \eqref{eq:thisintegral} becomes
\begin{align}
	S(s_1,s_2,s_3,s_4,s_A,s_B) := &\int_{-\infty}^\infty \frac{dt_B}{ \pi}  \frac{dt_3}{ \pi} \frac{dt_4}{ \pi}
	\\
	&|1-t_3|^{-\Delta_2 + \Delta_3 - \Delta_B} |t_3|^{-2 + \Delta_2 + \Delta_3 + \Delta_B}
	\\
	&|1-t_4|^{-1 + \Delta_1 - \Delta_4 + \Delta_B} |t_4|^{-1 + \Delta_3 + \Delta_4 - \Delta_A}
	\\
	&|1-t_B|^{ - \Delta_1 + \Delta_2 - \Delta_A} |t_B|^{-1 - \Delta_2 + \Delta_3 + \Delta_B}
	\\
	&|1 - t_3 t_4 t_B|^{-1 - \Delta_3 + \Delta_4 + \Delta_A}
\end{align}
This integral has been evaluated in (B.7) of \cite{Liu:2018jhs}.\footnote{See also \cite{Derkachev2023}.} We thus have that
\begin{align}
	&\int \frac{d\theta_B}{2 \pi}  \frac{d\theta_3}{2 \pi} \frac{d\theta_4}{2 \pi}
	\\
	&I_S^{1-\Delta_B,1-\Delta_3,\Delta_2}(\theta_B,\theta_3,\theta_2)
	I_S^{\Delta_A,\Delta_3,1-\Delta_4}(\theta_A,\theta_3,\theta_4)
	I_S^{\Delta_B,\Delta_4,1-\Delta_1}(\theta_B,\theta_4,\theta_1)
	\\
	&= I_S^{1-\Delta_1,\Delta_2,\Delta_A}(\theta_1,\theta_2,\theta_A) \times S(s_1,s_2,s_3,s_4,s_A,s_B)
\end{align}
which implies that
\begin{align}
	&\int \frac{d\theta_A}{2 \pi} \frac{d\theta_B}{2 \pi} \frac{d\theta_1}{2 \pi} \frac{d\theta_2}{2 \pi} \frac{d\theta_3}{2 \pi} \frac{d\theta_4}{2 \pi}
	\\
	&I_S^{1-\Delta_A,1-\Delta_2,\Delta_1}(\theta_A,\theta_2,\theta_1)
	I_S^{1-\Delta_B,1-\Delta_3,\Delta_2}(\theta_B,\theta_3,\theta_2)
	I_S^{\Delta_A,\Delta_3,1-\Delta_4}(\theta_A,\theta_3,\theta_4)
	I_S^{\Delta_B,\Delta_4,1-\Delta_1}(\theta_B,\theta_4,\theta_1)
\end{align}
equals
\begin{equation}
	\delta(s_1 - s_1) \sum_{n_1} \frac{8}{2 \pi s_1 \tanh \pi s_1} \times S(s_1,s_2,s_3,s_4,s_A,s_B).
	\label{eq:B20}
\end{equation}
This is equal to the part of 	\eqref{eq:firsteqB} aside from the $\calv$ and $\left(\calv\right)^*$ factors. Using the results of \cite{Liu:2018jhs}, we have that

\begin{equation}
	\begin{aligned}
		\label{eq:sdef}
		\pi^3 S(s_1,s_2,s_3,s_4,s_A,s_B) = &\, G\left(\frac{1}{2} \left(-i s_2 + i s_4 - i s_A - i s_B + 1\right), \frac{1}{2} \left(i s_1 + i s_3 + i s_A - i s_B + 1\right), \right. \\
		&\phantom{G\big(} \frac{1}{2} \left(-i s_2 + i s_4 + i s_A + i s_B + 1\right), \frac{1}{2} \left(-i s_1 - i s_3 + i s_A - i s_B + 1\right), \\
		&\phantom{G\big(} \frac{1}{2} \left(i s_1 + i s_2 - i s_3 + i s_4 + 1\right), \frac{1}{2} \left(-i s_1 + i s_2 + i s_3 + i s_4 + 1\right) \biggr) \\
		&+ G\left(\frac{1}{2} \left(-i s_2 + i s_4 + i s_A + i s_B + 1\right), \frac{1}{2} \left(-i s_1 - i s_3 + i s_A - i s_B + 1\right), \right. \\
		&\phantom{G\big(} \frac{1}{2} \left(-i s_2 + i s_4 - i s_A - i s_B + 1\right), \frac{1}{2} \left(i s_1 + i s_3 + i s_A - i s_B + 1\right), \\
		&\phantom{G\big(} 1 + \frac{1}{2} \left(-i s_1 - i s_2 + i s_3 - i s_4 - 1\right), 1 + \frac{1}{2} \left(i s_1 - i s_2 - i s_3 - i s_4 - 1\right) \biggr)
	\end{aligned}
\end{equation}

\begin{equation}
	\begin{aligned}
		G(h_1,h_2,h_3,h_4,h,h') = &\, B(-h-h_3-h_4+2, 2h_4) B(h-h_1-h_2+1, h_1-h_4-h'+1) \\
		&\times B(h_2+h_3-h', h-h_3-h_4+1) F(h_1, h_2, h_3, 1-h_4, h, h') \\
		&+ B(h_1+h_4+h'-1, -h_1+h_4-h'+1) B(h-h_1-h_2+1, h_1+h_4-h') \\
		&\times B(h_2+h_3-h', h-h_3+h_4) F(h_1, h_2, h_3, h_4, h, h')
	\end{aligned}
\end{equation}

\begin{equation}
	B(a_2,a_1) = \frac{\sqrt{\pi} \, \Gamma \left(\frac{1-a_1}{2}\right) \Gamma \left(\frac{1-a_2}{2}\right) \Gamma \left(\frac{a_1 + a_2 - 1}{2}\right)}
	{\Gamma \left(\frac{a_1}{2}\right) \Gamma \left(\frac{a_2}{2}\right) \Gamma \left(\frac{2 - a_1 - a_2}{2}\right)}
\end{equation}
\begin{equation}
	\begin{aligned}
		F(h_1, h_2, h_3, h_4, h, h') = 
		\, {}_4F_3\bigg(&\left\{
		1 + h' - h_1 - h_4, \, h' + h_1 - h_4, \, 2 - h_3 - h_4 - h, \, 1 + h_3 - h_4 - h
		\right\}; \, \\
		&\left\{
		2 - 2 h_4, \, 2 + h' - h_2 - h_4 - h, \, 1 + h' + h_2 - h_4 - h
		\right\}; \, 1\bigg)
	\end{aligned}
\end{equation}

Beginning from \eqref{eq:B20}, the contribution from the crossed Wick contraction to the four-point function in the Hartle-Hawking state may be evaluated by including the $\calv$ and $\left(\calv\right)^*$ factors that appear in \eqref{eq:firsteqB}, dividing by $V$ using \eqref{eq:prescription}, and multiplying by $\frac{1}{\cosh^2 \pi s_1}$, the wavefunction of the Hartle-Hawking state \eqref{eq:6.4}. The result of this is
\begin{equation}
\label{eq:B25}
	A_1 A_2 S(s_1,s_2,s_3,s_4,s_A,s_B),
\end{equation}
where $A_1$ is defined to be
\begin{equation}
	A_1 := \frac{s_1 s_2 s_3 s_4 \sinh(\pi s_2) \sinh(\pi s_4) \tanh(\pi s_1) \tanh(\pi s_3)}{4096 \pi^{11}}
\end{equation}
and $A_2$ is defined to be
\begin{align}
	A_2 := & \Gamma\left(\frac{1}{2} \left(\frac{1}{2} - i s_2 - i s_3 - i s_B\right)\right)
	\Gamma\left(\frac{1}{2} \left(\frac{1}{2} + i s_2 + i s_3 - i s_B\right)\right)
	\Gamma\left(\frac{1}{2} \left(\frac{1}{2} + i s_2 - i s_3 - i s_B\right)\right) \nonumber \\
	& \times \Gamma\left(\frac{1}{2} \left(\frac{1}{2} + i s_2 - i s_3 + i s_B\right)\right)
	\Gamma\left(\frac{1}{2} \left(\frac{1}{2} - i s_1 + i s_4 - i s_B\right)\right)
	\Gamma\left(\frac{1}{2} \left(\frac{1}{2} - i s_1 + i s_4 + i s_B\right)\right) \nonumber \\
	& \times \Gamma\left(\frac{1}{2} \left(\frac{1}{2} + i s_1 + i s_4 + i s_B\right)\right)
	\Gamma\left(\frac{1}{2} \left(\frac{1}{2} - i s_1 - i s_4 + i s_B\right)\right)
	\Gamma\left(\frac{1}{2} \left(\frac{1}{2} + i s_3 - i s_4 - i s_A\right)\right) \nonumber \\
	& \times \Gamma\left(\frac{1}{2} \left(\frac{1}{2} + i s_3 - i s_4 + i s_A\right)\right)
	\Gamma\left(\frac{1}{2} \left(\frac{1}{2} + i s_3 + i s_4 + i s_A\right)\right)
	\Gamma\left(\frac{1}{2} \left(\frac{1}{2} - i s_3 - i s_4 + i s_A\right)\right) \nonumber \\
	& \times \Gamma\left(\frac{1}{2} \left(\frac{1}{2} - i s_1 - i s_2 - i s_A\right)\right)
	\Gamma\left(\frac{1}{2} \left(\frac{1}{2} + i s_1 + i s_2 - i s_A\right)\right)
	\Gamma\left(\frac{1}{2} \left(\frac{1}{2} + i s_1 - i s_2 + i s_A\right)\right) \nonumber \\
	& \times \Gamma\left(\frac{1}{2} \left(\frac{1}{2} + i s_1 - i s_2 - i s_A\right)\right).
\end{align}
Meanwhile, $S(s_1,s_2,s_3,s_4,s_A,s_B)$ is a sum of four terms, each involving a hypergeometric function. We write
\begin{equation}
	S(s_1,s_2,s_3,s_4,s_A,s_B) = B_1 + B_2 + B_3 + B_4
\end{equation}
where
\begin{align}
	B_1 := & -\frac{1}{\pi^3} 
	8 i \cos\left(\frac{1}{4} \pi (1 - 2 i s_3 + 2 i s_4 + 2 i s_A)\right) \nonumber \\
	& \times \cosh\left(\frac{1}{4} \pi (i + 2 s_1 - 2 s_2 + 2 s_A)\right) 
	\cosh\left(\frac{1}{4} \pi (i + 2 s_3 - 2 s_4 + 2 s_A)\right) \nonumber \\
	& \times \cosh\left(\frac{1}{4} \pi (i + 2 s_3 + 2 s_4 + 2 s_A)\right)
	\cosh\left(\frac{1}{4} \pi (i + 2 s_2 + 2 s_3 - 2 s_B)\right) \nonumber \\
	& \times \cosh\left(\frac{1}{4} \pi (-i + 2 s_1 - 2 s_4 + 2 s_B)\right)
	\csch(\pi s_3) 
	\csch\left(\frac{\pi}{2} (s_1 + s_3 + s_A - s_B)\right) \nonumber \\
	& \times \csch\left(\frac{\pi}{2} (s_1 - s_3 - s_A + s_B)\right)
	\Gamma\left(\frac{1}{2} - i s_1 + i s_2 - i s_A\right) 
	\Gamma\left(\frac{1}{2} - i s_3 + i s_4 - i s_A\right) \nonumber \\
	& \times \Gamma\left(\frac{1}{2} - i s_3 + i s_4 + i s_A\right)
	\Gamma\left(-\frac{1}{2} i (i + 2 s_3 + 2 s_4 + 2 s_A)\right) 
	\Gamma\left(\frac{1}{2} - i s_2 - i s_3 + i s_B\right) \nonumber \\
	& \times \Gamma\left(\frac{1}{2} + i s_1 - i s_4 + i s_B\right)
	{}_4\tilde{F}_3\left(\left\{-\frac{1}{2} i (i + 2 s_3 + 2 s_4 + 2 s_A), 
	-\frac{1}{2} i (i + 2 s_3 - 2 s_4 + 2 s_A), \right. \right. \nonumber \\
	& \left. \quad -\frac{1}{2} i (i + 2 s_2 + 2 s_3 - 2 s_B), \frac{1}{2} + i s_2 - is_3 + i s_B\right\}, \nonumber \\
	& \left. \quad \{1 - 2 i s_3, -i (i + s_1 + s_3 + s_A - s_B), i (-i + s_1 - s_3 - s_A + s_B)\}, 1\right)
\end{align}
\begin{align}
	B_2 := & \frac{1}{\pi^3} 8 i 
	\cos\left(\frac{1}{4} \pi (1 - 2 i s_2 + 2 i s_3 + 2 i s_B)\right) \nonumber \\
	& \times \cosh\left(\frac{1}{4} \pi (-i + 2 s_3 + 2 s_4 - 2 s_A)\right) 
	\cosh\left(\frac{1}{4} \pi (i + 2 s_1 - 2 s_2 + 2 s_A)\right) \nonumber \\
	& \times \cosh\left(\frac{1}{4} \pi (i + 2 s_2 - 2 s_3 + 2 s_B)\right)
	\cosh\left(\frac{1}{4} \pi (-i + 2 s_2 + 2 s_3 + 2 s_B)\right) \nonumber \\
	& \times \cosh\left(\frac{1}{4} \pi (-i + 2 s_1 - 2 s_4 + 2 s_B)\right)
	\csch(\pi s_3)
	\csch\left(\frac{\pi}{2} (s_1 - s_3 + s_A - s_B)\right) \nonumber \\
	& \times \csch\left(\frac{\pi}{2} (s_1 + s_3 - s_A + s_B)\right)
	\Gamma\left(\frac{1}{2} - i s_1 + i s_2 - i s_A\right)
	\Gamma\left(\frac{1}{2} + i s_3 + i s_4 - i s_A\right) \nonumber \\
	& \times \Gamma\left(\frac{1}{2} - i s_2 + i s_3 - i s_B\right)
	\Gamma\left(\frac{1}{2} - i s_2 + i s_3 + i s_B\right)
	\Gamma\left(\frac{1}{2} + i s_2 + i s_3 + i s_B\right) \nonumber \\
	& \times \Gamma\left(\frac{1}{2} + i s_1 - i s_4 + i s_B\right)
	{}_4\tilde{F}_3\left(\left\{\frac{1}{2} + i s_3 - i s_4 - i s_A, 
	\frac{1}{2} + i s_3 + i s_4 - i s_A, \right. \right. \nonumber \\
	& \left. \quad \frac{1}{2} - i s_2 + i s_3 + i s_B, 
	\frac{1}{2} + i s_2 + i s_3 + i s_B\right\}, \nonumber \\
	& \left. \quad \{1 + 2 i s_3, -i (i + s_1 - s_3 + s_A - s_B), 
	i (-i + s_1 + s_3 - s_A + s_B)\}, 1\right)
\end{align}

\begin{align}
	B_3 := & \frac{1}{\pi^3} 8 i 
	\cos\left(\frac{1}{4} \pi (1 - 2 i s_1 + 2 i s_2 + 2 i s_A)\right) 
	\cos\left(\frac{1}{4} \pi (1 - 2 i s_3 + 2 i s_4 + 2 i s_A)\right) \nonumber \\
	& \times \cosh\left(\frac{1}{4} \pi (i + 2 s_1 + 2 s_2 - 2 s_A)\right) 
	\cosh\left(\frac{1}{4} \pi (i + 2 s_1 - 2 s_2 + 2 s_A)\right) 
	\cosh\left(\frac{1}{4} \pi (i + 2 s_2 - 2 s_3 + 2 s_B)\right) \nonumber \\
	& \times \cosh\left(\frac{1}{4} \pi (i + 2 s_1 + 2 s_4 + 2 s_B)\right) 
	\csch(\pi s_1) 
	\csch\left(\frac{\pi}{2} (s_1 - s_3 - s_A + s_B)\right) \nonumber \\
	& \times \csch\left(\frac{\pi}{2} (s_1 + s_3 - s_A + s_B)\right) 
	\Gamma\left(\frac{1}{2} - i s_1 + i s_2 - i s_A\right) 
	\Gamma\left(\frac{1}{2} - i s_1 - i s_2 + i s_A\right) \nonumber \\
	& \times \Gamma\left(\frac{1}{2} - i s_1 + i s_2 + i s_A\right) 
	\Gamma\left(\frac{1}{2} - i s_3 + i s_4 + i s_A\right) 
	\Gamma\left(\frac{1}{2} - i s_2 + i s_3 - i s_B\right) \nonumber \\
	& \times \Gamma\left(-\frac{1}{2} i (i + 2 s_1 + 2 s_4 + 2 s_B)\right) 
	{}_4\tilde{F}_3\left(\left\{-\frac{1}{2} i (i + 2 s_1 + 2 s_2 - 2 s_A), 
	\frac{1}{2} - i s_1 + i s_2 + i s_A, \right. \right. \nonumber \\
	& \left. \quad -\frac{1}{2} i (i + 2 s_1 + 2 s_4 + 2 s_B), 
	-\frac{1}{2} i (i + 2 s_1 - 2 s_4 + 2 s_B)\right\}, \nonumber \\
	& \left. \quad \{1 - 2 i s_1, -i (i + s_1 + s_3 - s_A + s_B), 
	-i (i + s_1 - s_3 - s_A + s_B)\}, 1\right)
\end{align}

\begin{align}
	B_4 := & -\frac{1}{\pi^3} 
	8 i \cos\left(\frac{1}{4} \pi (1 - 2 i s_3 + 2 i s_4 + 2 i s_A)\right) \nonumber \\
	& \times \cosh\left(\frac{1}{4} \pi (-i + 2 s_1 + 2 s_2 + 2 s_A)\right) \cosh\left(\frac{1}{4} \pi (-i + 2 s_1 - 2 s_4 - 2 s_B)\right)  \nonumber \\
	& \times 
	\cosh\left(\frac{1}{4} \pi (-i + 2 s_1 + 2 s_4 - 2 s_B)\right) 
	\cosh\left(\frac{1}{4} \pi (i + 2 s_2 - 2 s_3 + 2 s_B)\right) \nonumber \\
	& \times \cosh\left(\frac{1}{4} \pi (-i + 2 s_1 - 2 s_4 + 2 s_B)\right) 
	\csch(\pi s_1) 
	\csch\left(\frac{\pi}{2} (s_1 - s_3 + s_A - s_B)\right) \nonumber \\
	& \times \csch\left(\frac{\pi}{2} (s_1 + s_3 + s_A - s_B)\right) 
	\Gamma\left(\frac{1}{2} + i s_1 + i s_2 + i s_A\right) 
	\Gamma\left(\frac{1}{2} - i s_3 + i s_4 + i s_A\right) \nonumber \\
	& \times \Gamma\left(\frac{1}{2} - i s_2 + i s_3 - i s_B\right) 
	\Gamma\left(\frac{1}{2} + i s_1 - i s_4 - i s_B\right) 
	\Gamma\left(\frac{1}{2} + i s_1 + i s_4 - i s_B\right) \nonumber \\
	& \times \Gamma\left(\frac{1}{2} + i s_1 - i s_4 + i s_B\right) 
	{}_4\tilde{F}_3\left(\left\{\frac{1}{2} + i s_1 - i s_2 + i s_A, 
	\frac{1}{2} + i s_1 + i s_2 + i s_A, \right. \right. \nonumber \\
	& \left. \quad \frac{1}{2} + i s_1 - i s_4 - i s_B, 
	\frac{1}{2} + i s_1 + i s_4 - i s_B\right\}, \nonumber \\
	& \left. \quad \{1 + 2 i s_1, i (-i + s_1 - s_3 + s_A - s_B), 
	i (-i + s_1 + s_3 + s_A - s_B)\}, 1\right)
\end{align}
and ${}_4\tilde{F}_3$ is the regulated Hypergeometric function.

\subsection{Semiclassical Limit}

We now set $s_i = \Lambda + q_i$ for $i \in \{1,2,3,4\}$. We do not take any limits on $s_A$ or $s_B$. First, we consider how $B_1$ behaves when $\Lambda$ is large. The most nontrivial part of $B_1$ is the Hypergeometric function. We simplify it in the same way as (C.30) of \cite{Jafferis:2022wez}. When $\Lambda$ is large, it becomes\footnote{Here we are referring to the Hypergeometric function, not the regularized Hypergeometric function}
\begin{equation}
	\begin{aligned}
		{}_4F_3 & \left( 
		\left\{ 
		-\frac{1}{2} i (i + 2 s_3 + 2 s_4 + 2 s_A), 
		-\frac{1}{2} i (i + 2 s_3 - 2 s_4 + 2 s_A), 
		-\frac{1}{2} i (i + 2 s_2 + 2 s_3 - 2 s_B), 
		\frac{1}{2} + i s_2 - i s_3 + i s_B 
		\right\}, \right. \\
		& \left. \{ 
		1 - 2 i s_3, 
		-i (i + s_1 + s_3 + s_A - s_B), 
		i (-i + s_1 - s_3 - s_A + s_B) 
		\}, 1 \right) \\
		& \rightarrow {}_2F_1 \left(  
		\frac{1}{2} - i s_3 + i s_4 - i s_A
		, 
		\frac{1}{2} + i s_2 - i s_3 + i s_B 
		; 
		 1 + i s_1 - i s_3 - i s_A + i s_B
		 , 1 \right)
		 \\
		 &= \frac{\Gamma(i s_1 - i s_2 + i s_3 - i s_4) \Gamma(1 + i s_1 - i s_3 - i s_A + i s_B)}
		 {\Gamma\left(\frac{1}{2} + i s_1 - i s_2 - i s_A\right) \Gamma\left(\frac{1}{2} + i s_1 - i s_4 + i s_B\right)}
	\end{aligned}
\end{equation}
Simplifying the rest of $B_1$ is straightforward. In the large $\Lambda$ limit, $B_1$ becomes
\begin{align*}
	B_1 &\rightarrow \frac{1}{\Lambda} \left(\frac{2 \Lambda}{e}\right)^{i(s_1 - s_2 + s_3 - s_4)} 
	\\
	&\times \frac{
		\Gamma(i (s_1 - s_2 + s_3 - s_4)) 
		\Gamma\left(\frac{1}{2} - i s_1 + i s_2 - i s_A\right) 
		\Gamma\left(\frac{1}{2} - i s_3 + i s_4 - i s_A\right) 
		\Gamma\left(\frac{1}{2} - i s_3 + i s_4 + i s_A\right)}
	{\pi^3 \Gamma\left(\frac{1}{2} + i s_1 - i s_2 - i s_A\right)}
	\\
	&\times 4 i \cos\left(\frac{1}{4} \pi (1 - 2 i s_3 + 2 i s_4 + 2 i s_A)\right) 
	\cosh\left(\frac{1}{4} \pi (i + 2 s_1 - 2 s_2 + 2 s_A)\right)
	\cosh\left(\frac{1}{4} \pi (i + 2 s_3 - 2 s_4 + 2 s_A)\right)
	\\
	&\times \cosh\left(\frac{1}{4} \pi (-i + 2 s_1 - 2 s_4 + 2 s_B)\right)
	\csch\left(\frac{1}{2} \pi (s_1 - s_3 - s_A + s_B)\right).
\end{align*}
We make the following redefinitions:
\begin{equation}
    s_{ij} := s_i - s_j = q_i - q_j, \quad b_{\pm} := s_{23} \pm s_B,  \quad a_\pm := s_{12} \pm s_A, \quad x := s_{12} + s_{34},
\end{equation}
\begin{align}
C_1 &:= \frac{4i}{\pi^2 \Lambda}  \Gamma(i x) 
		\Gamma\left(\frac{1}{2} - i a_+ \right) \Gamma\left(\frac{1}{2} + i (a_+ -x) \right) \\ \nonumber
		&\times \quad \cos \left(\frac{\pi}{4} -  {i \frac{\pi}{2} } a_+\right) 
		\cos\left(\frac{\pi}{4}  +  {i \frac{\pi}{2}} (a_+ - x) \right) \csch\left(\frac{1}{2} \pi (s_{13} - s_A + s_B)\right), \\
		C_2 &:= \frac{4i }{\pi^2 \Lambda}  \Gamma(-i x) \Gamma\left(\frac{1}{2} - i b_+  \right) \Gamma\left(\frac{1}{2} + i (b_+ +x) \right) \\
&\times\quad \nonumber
	 \cos \left(\frac{\pi}{4}  - {i \frac{\pi}{2}} b_+ \right)
	\cos\left(\frac{ \pi}{4} + {i \frac{\pi}{2}} (b_+ +x) \right) \csch\left(\frac{1}{2} \pi (s_{13} + s_A - s_B)\right).
\end{align} 
We may simplify $B_2$, $B_3$ and $B_4$ in the same way as $B_1$. Using the above redefinitions, the results are

\begin{align}
B_1 &\rightarrow \left(\frac{2 \Lambda}{e}\right)^{ix} C_1  \frac{\cos\left(\frac{\pi}{4}  +  {i \frac{\pi}{2}} (b_+ +x) \right) 
	\cos\left(\frac{\pi}{4}  +  {i \frac{\pi}{2}} (a_- - x) \right) 
		}
	{ \Gamma\left(\frac{1}{2} + i a_- \right) \Gamma\left(\frac{1}{2} - i (a_- -x) \right)  \cosh \pi (a_- -x)} ,
	\\
B_3 &\rightarrow - \left(\frac{2 \Lambda}{e}\right)^{ix}  C_1 
\frac{		\cos\left(\frac{\pi}{4}  -  {i \frac{\pi}{2}} b_+ \right) \cos \left(\frac{\pi}{4}  -  {i \frac{\pi}{2}} a_-\right)}
			{ \Gamma\left(\frac{1}{2} - i (a_- -x) \right) \Gamma\left(\frac{1}{2} + i a_- \right) \cosh \pi a_-},  \\
B_2 &\rightarrow - \left(\frac{2 \Lambda}{e}\right)^{-i x} C_2 \frac{
		 \cos \left(\frac{\pi}{4}  - {i \frac{\pi}{2}} b_-  \right)
	\cos\left(\frac{ \pi}{4} - {i \frac{\pi}{2}} a_+ \right)}
	{\Gamma\left(\frac{1}{2} - i (b_-  +x) \right) \Gamma\left(\frac{1}{2} + i b_-  \right)  \cosh \pi b_- }  ,
	 \\
B_4 &\rightarrow \left(\frac{2 \Lambda}{e}\right)^{-i x} C_2 \frac{	
		 \cos \left(\frac{\pi}{4}  + {i \frac{\pi}{2}} (a_+ - x) \right)
	\cos\left(\frac{ \pi}{4} + {i \frac{\pi}{2}} (b_-+x) \right)}
	{\Gamma\left(\frac{1}{2} + i b_-  \right) \Gamma\left(\frac{1}{2} - i (b_- +x) \right) \cosh \pi (b_-+x)}.
\end{align}

Meanwhile, $A_1$ and $A_2$ simplify to
\begin{equation}
	A_1 \rightarrow \frac{e^{\pi (s_2 + s_4)} \Lambda^4}{16384 \pi^{11}}
\end{equation}
\begin{align}
	A_2 &\rightarrow \frac{16  \pi^4}{\Lambda^2} e^{-\pi (s_1 + s_2 + s_3 + s_4)}
	\Gamma\left(\frac{1}{4}  + {\frac{i}{2}} b_-  \right)
	\Gamma\left(\frac{1}{4}  + {\frac{i}{2}} b_+  \right)
	 \Gamma\left(\frac{1}{4}  - {\frac{i}{2}} (b_+ +x) \right)
	\Gamma\left(\frac{1}{4}  - {\frac{i}{2}} (b_-+x)  \right) \\ \nonumber
	&\times \Gamma\left(\frac{1}{4}  - {\frac{i}{2}} (a_+ -x) \right)
	\Gamma\left(\frac{1}{4}  - {\frac{i}{2}} (a_--x)  \right)	
		  \Gamma\left(\frac{1}{4}  + {\frac{i}{2}} a_+  \right)
	\Gamma\left(\frac{1}{4}  + {\frac{i}{2}} a_-  \right)	.
\end{align} 

At this point, there are two different ways we may proceed. First, we will consider the case where all the operators are time-evolved by order one times. Second, we will consider the case where there is a large time separation on the order of $\log \Lambda$.

The correlator we are interested in is
\begin{equation}
\frac{\braket{\psifhh| \phi^B_R(\tau) \phi^A_R(0) \phi^B_R(\tau+u_2) \phi^A_R(u_1) \Pi_{\Lambda,\epsilon} |\psifhh}}{\braket{\psifhh|\Pi_{\Lambda,\epsilon}|\psifhh}} 
\label{eq:OTOC}
\end{equation}
where the $A$ and $B$ superscripts denote two flavors of the massive scalar field, with masses given by $s_A$ and $s_B$ (using \eqref{eq:masss}). The projection $\Pi_{\Lambda,\epsilon}$ was defined below \eqref{eq:pidef}. We divide by the normalization factor $\braket{\psifhh|\Pi_{\Lambda,\epsilon}|\psifhh}$, computed in Section \ref{sec:zero-point}.  Note that in the semiclassical limit, this normalization factor becomes
\begin{equation}
	\braket{\psifhh|\Pi_{\Lambda,\epsilon}|\psifhh} \rightarrow \frac{\Lambda e^{-2 \pi \Lambda}}{\pi^2} \int_{-\epsilon/2}^{\epsilon/2} dq \, e^{- 2 \pi q}.
\end{equation}
The two cases we consider are the case where all times are order one, and the case where $u_1$ and $u_2$ are order one but $\tau$ is of order the scrambling time,\footnote{To obtain a precise match between the OTOC in this appendix and the OTOC in the main text, which was computed as an eikonal scattering amplitude, we need to use definitions of $\tau_{scr}$ that differ by an additive constant, which is physically meaningless.}
\begin{equation}
\tau_{scr} := \log \frac{2 \Lambda}{e}.
\end{equation}

\subsubsection{OTOC at order one times}

\label{sec:appotocorderone}

Given that all of the operators are evolved for order one times, we further simplify the sum of $B_1$ through $B_4$ under the assumption that $x \approx 0$, due to the oscillations from large $\Lambda$. The result is
\begin{equation}
	\frac{2}{\pi^2 \Lambda}   \frac{1}{x} \times\left[-i \Lambda^{i x}+i \Lambda^{-i x}\right] \rightarrow \frac{4}{\pi \Lambda} \delta(x)
\end{equation}

This is the delta function that was promised below \eqref{eq:6.25}. After putting the time evolution factors in, the answer becomes $2 \cali_1$, where $\cali_1$ is defined below in \eqref{eq:I1}. It is a product of two-point functions, as expected.

\subsubsection{OTOC at the scrambling time}

\label{sec:appotocscrambling}

Before simplifying \eqref{eq:B25} further, we multiply by $\left(\frac{2 \Lambda}{e}\right)^{-i x} e^{-i T x}$, the phase factors associated with the time evolution on the observer's worldline which depend on $x$. The time variable $T$ is defined such that $T = 0$ is the scrambling time. That is,
\begin{equation}
\tau = \tau_{scr} + T.
\end{equation}

We may still simplify $B_2 + B_4$ by assuming that $x \approx 0$. The result is
\begin{equation}
\left(\frac{2 \Lambda}{e}\right)^{-i x}\left(	B_2 + B_4\right) \rightarrow \frac{2 i}{\pi^2 \Lambda x} \Lambda^{-2 i x}.
\end{equation}
Using that $\cos(a x)/x \rightarrow 0$ as $a \rightarrow \infty$ in the distributional sense, this becomes
\begin{equation}
\left(\frac{2 \Lambda}{e}\right)^{-i x} \left(	B_2 + B_4  \right)\rightarrow \frac{2 }{\pi \Lambda }  \delta(x).
\end{equation}
In contrast, multiplying by $\left(\frac{2 \Lambda}{e}\right)^{-i x}$ cancels out the large $\Lambda$ oscillations in $B_1 + B_3$. We have that
\begin{equation}
\begin{aligned}
   \left(\frac{2 \Lambda}{e}\right)^{-i x} (B_1 + B_3) \rightarrow &\frac{1}{\Lambda} \biggl[\frac{2 e^{-2 \pi s_A} \left(1 + i e^{\pi a_+}\right) \left(1 - i e^{\pi (a_+ - x)}\right) \Gamma\left(\frac{1}{2} - i a_+\right) \Gamma\left(\frac{1}{2} + i a_+ - i x\right)}
{\pi^2 \Gamma\left(\frac{1}{2} + i a_-\right) \left(1 - i e^{\pi a_-}\right) \left(1 + i e^{\pi (a_- - x)}\right) \Gamma\left(\frac{1}{2} - i a_- + i x\right)}
\\
&\times\cosh\left(\frac{\pi x}{2}\right) \Gamma(i x)
 \biggr].
\end{aligned}
\end{equation}
Thus, the OTOC \eqref{eq:OTOC} for fixed $T$, $u_1$, $u_2$ and large $\Lambda$ becomes

\begin{align}
&\frac{1}{\int_{-\epsilon/2}^{\epsilon/2} dq \, e^{-2 \pi q}}\int_{-\epsilon/2}^{\epsilon/2} dq_1 \, \int_{-\infty}^\infty  dq_2 \, dq_3 \, dq_4 e^{i u_1 (q_1 - q_2)} e^{iu_2(q_2 - q_3)} e^{-i T x} \, \frac{1}{1024 \pi^{5}}  e^{-\pi (q_1  + q_3 )} \notag \\
&\times \Gamma\left(\frac{1}{2}\left(\frac{1}{2} + i b_{\pm} \right)\right) \Gamma\left(\frac{1}{2}\left(\frac{1}{2} - i x  -i b_{\pm} \right)\right) \Gamma\left(\frac{1}{2}\left(\frac{1}{2} + i x - i a_{\pm} \right)\right)\Gamma\left(\frac{1}{2}\left(\frac{1}{2} + i a_{\pm} \right)\right) \notag \\
\times&\Bigg( \frac{2 }{\pi } \delta(x) + \biggl[\frac{2 e^{-2 \pi s_A} \left(1 + i e^{\pi a_+}\right) \left(1 - i e^{\pi (a_+ - x)}\right) \Gamma\left(\frac{1}{2} - i a_+\right) \Gamma\left(\frac{1}{2} + i a_+ - i x\right)}
{\pi^2 \Gamma\left(\frac{1}{2} + i a_-\right) \left(1 - i e^{\pi a_-}\right) \left(1 + i e^{\pi (a_- - x)}\right) \Gamma\left(\frac{1}{2} - i a_- + i x\right)}
\cosh\left(\frac{\pi x}{2}\right) \Gamma(i x) \biggr] \Bigg)
    \label{eq:B48}
\end{align}
Next, we change integration variables from $(q_1,q_2,q_3,q_4)$ to $(q,x,\lambda_1,\lambda_2)$, defined by
\begin{equation}
\label{eq:B49}
	q := q_1, \quad x := q_1 - q_2 + q_3 - q_4, \quad \lambda_1 := q_2 - q_1, \quad \lambda_2 := q_3 - q_2 
\end{equation}
and write \eqref{eq:B48} as $\cali_1 + \cali_2$, where

\begin{align}
\label{eq:I1}
\cali_1 &:=  \int_{-\infty}^\infty d\lambda_1 \, d\lambda_2   \, e^{-i u_1 \lambda_1} e^{-iu_2\lambda_2}  \, \frac{1}{512 \pi^{6}}  e^{-\pi (  \lambda_2  +\lambda_1 )} \notag \\
&\times \Gamma\left(\frac{1}{2}\left(\frac{1}{2} \pm i \lambda_2 \pm i s_B\right)\right)  \notag \\
&\times \Gamma\left(\frac{1}{2}\left(\frac{1}{2} \pm i \lambda_1 \pm i s_A\right)\right), \notag \\
\end{align}
and
\begin{align}
\cali_2 &:= \int_{-\infty}^\infty  \, d\lambda_1 \, d\lambda_2 \, \frac{1}{2}\left(\int_{-\infty - i \epsilon}^{\infty - i \epsilon} dx + \int_{-\infty + i \epsilon}^{\infty + i \epsilon} dx\right) \, e^{-i u_1 \lambda_1} e^{-iu_2 \lambda_2} e^{- i T x} \, \frac{1}{1024 \pi^{5}}  e^{-\pi (   \lambda_1 + \lambda_2 )} \notag \\
&\times \Gamma\left(\frac{1}{2}\left(\frac{1}{2} - i \lambda_2 \pm i s_B\right)\right) \Gamma\left(\frac{1}{2}\left(\frac{1}{2} + i \lambda_2 - i x \pm i s_B\right)\right) \notag \\
&\times \Gamma\left(\frac{1}{2}\left(\frac{1}{2} + i \lambda_1 + i x \pm i s_A\right)\right)\Gamma\left(\frac{1}{2}\left(\frac{1}{2} - i \lambda_1 \pm i s_A\right)\right) \notag 
 \\
\times&\Bigg( 
\frac{2 \left(e^{\pi s_A}-i e^{\pi \lambda_1}\right)  \left(e^{\pi s_A}+i e^{\pi (\lambda_1+x)}\right)}{\pi^2 \left(e^{\pi (\lambda_1 + s_A)} - i\right) \left(e^{\pi (\lambda_1 + s_A + x)} + i\right)}\frac{ \Gamma \left(-i s_A + i \lambda_1 + \frac{1}{2}\right) \Gamma \left(i s_A - i x - i \lambda_1 + \frac{1}{2}\right)}{\Gamma \left(-i s_A - i \lambda_1 + \frac{1}{2}\right) \Gamma \left(i s_A + i x + i \lambda_1 + \frac{1}{2}\right)}
\\
&\times \cosh\left( \frac{\pi x}{2} \right) \Gamma(i x)
\Bigg).
\label{eq:I2}
\end{align}
$\cali_1$ is the contribution from the $\delta(x)$ term, and $\cali_2$ represents the rest of \eqref{eq:B48}. There is a pole singularity at $x = 0$ from $\Gamma(ix)$. The $x$ integral should be performed using a principal value prescription, as indicated by the two contours for the $x$ integral above. For some readers, a more familiar notation for the principal value prescription is
\begin{equation}
    \mathcal{P} \int_{-\infty}^\infty dx := \frac{1}{2}\left(\int_{-\infty - i \epsilon}^{\infty - i \epsilon} dx + \int_{-\infty + i \epsilon}^{\infty + i \epsilon} dx\right).
\end{equation}
That is, the principal value prescription involves averaging over contours that pass over and under the pole at $x = 0$.

Note that
\begin{equation}
	\lim_{T \rightarrow -\infty} \cali_2 = \cali_1, \quad \quad \lim_{T \rightarrow \infty} \cali_2 = - \cali_1.
 \label{eq:Tlimits}
\end{equation}
To evaluate both of these limits, we can assume that $x \approx 0$ in the integrand. Equation \eqref{eq:Tlimits} implies that the OTOC, $\cali_1 + \cali_2$, decays to zero at large times, as expected. We are interested in corrections to the leading order result $2 \cali_1$ in the regime where $T$ is large and negative. We can derive an asymptotic expansion in powers of $e^{2T}$ by closing the contour for the $x$ integral in the upper-half-plane, and evaluating the contributions from the poles. This expansion is asymptotic and not convergent, meaning that strictly speaking, one may not close the contour in the upper-half-plane. Instead, one should deform the $x$ contour to a contour that runs parallel to the real axis and has a positive imaginary part, $y > 0$. The $x$ integral converges along this deformed contour. If $y$ is between $n-1$ and $n$ for some $n \in \mathbb{Z}_{\ge 1}$, then the contour deformation picks up contributions from $n$ poles, which represent the first $n$ terms of the asymptotic series. Thus, the exact result of the $x$ integral can be written as a sum over the first $n$ terms of the asymptotic series plus an error term that decays faster than the $n$th term as $T \rightarrow -\infty$.

If we were instead interested in the behavior for large positive $T$, we would close the contour in the lower-half-plane. Note that all of the poles in the upper-half-plane come from the $\Gamma(ix)$ function, whereas the locations of the poles in the lower-half plane are not universal (they depend on $s_A$ and $s_B$). The would-be pole at $x = i$ is canceled by the $\cosh\left(\frac{\pi x }{2}\right)$ function, so the first pole is at $x = 2 i$, which means that the leading correction to the OTOC is proportional to $e^{2 T}$, which indicates that the Lyapunov exponent is 2.\footnote{As always, we work in units where the de Sitter temperature is $\frac{1}{2 \pi}$.} When we compute the OTOC, $\cali_1 + \cali_2$, by closing the $x$ contour in the upper half plane, we obtain an asymptotic series that reproduces \eqref{eq:6.38}.

An analogous calculation may be performed in JT gravity, and the results agree with \cite{Maldacena:2016upp,Mertens:2017mtv,Mertens:2022irh,Lam:2018pvp}. In particular, one may derive an analogous integral representation for the OTOC that becomes an asymptotic series after closing the contour in the upper-half plane. This asymptotic series is Borel-resummable.

\section{One loop correction to density of states}

\label{sec:oneloopappendix}

In this appendix, we compute the one-loop correction to the geodesic instanton that wraps a two-sphere once. The metric is given by
\begin{equation}
	ds^2 = dr^2 + \cos^2 r \, d\theta^2.
\end{equation}
We write a general path as
\begin{equation}
	r(s) = r_{cl}(s) + \delta r(s), \quad \theta(s) = \theta_{cl}(s) + \delta \theta(s),
\end{equation}
where the classical solution is given by
\begin{equation}
	r_{cl}(s) = 0, \quad \theta_{cl}(s) = s, \quad s \in (0,2 \pi).
\end{equation}
We can gauge-fix $\delta\theta(s) = 0$ due to worldline reparameterization invariance. The worldline action is
\begin{equation}
	I = m\int_{0}^{2 \pi} ds \sqrt{\dot{r}^2 + \cos^2 r \, \dot{\theta}^2} \approx  2 \pi m + \frac{m}{2}\int_{0}^{2 \pi} ds \,\left[ (\delta\dot{r})^2 - (\delta r)^2 \right],
\end{equation}
where we have expanded to quadratic order. We view $\delta r$ as a field that lives on a circle (the worldline). We expand in modes:
\begin{equation}
	\delta r(s) = \sum_{n \in \mathbb{Z}} a_n e^{i n s},
\end{equation}
with the condition that $a_n = a_{-n}^*$. There are two zero-modes at $n  = 1$. These correspond to symmetries of the target space. We should mod out by these symmetries, because they correspond to equivalent worldline configurations. Thus, we do not integrate over these modes. The third symmetry corresponds to the rotation that preserves the classical solution. We mod out by this symmetry by dividing the final answer by $\frac{1}{2 \pi}$.

There is a negative (or wrong-sign) mode at $n = 0$. This mode exists because the geodesic is contractible. We Wick-rotate this mode to make it positive.

The remaining modes, $n \ge 2$, are positive.

The semiclassical path integral with the one-loop correction becomes
\begin{equation}
	\frac{1}{2 \pi} e^{- 2\pi m} \frac{1}{\sqrt{m}}\prod_{n \ge 2} \frac{1}{m(n^2 - 1)}
	\label{eq:oneloopcorrection}
\end{equation}
To get a finite answer, we write
\begin{equation}
	\log \prod_{n \ge 2} \frac{1}{m(n^2 - 1)} =  -\sum_{n \ge 2} e^{-\epsilon n} \log \left[m(n^2 - 1)\right] = -\frac{\log m}{\epsilon} + \frac{3}{2} \log m  + \text{constants} + O(\epsilon),
\end{equation}
where $e^{-\epsilon n}$ serves as a regulator. We only keep the $\frac{3}{2} \log m$ term in the above. Then, 	\eqref{eq:oneloopcorrection} becomes
\begin{equation}
	\frac{m}{2 \pi} e^{- 2\pi m},  
\end{equation}
which should be compared to \eqref{eq:prefactor}.

\section{Plancherel Measure}

\label{sec:rhopl}

In this Appendix, we determine $\rhopl$. Our strategy is to evaluate
\begin{equation}
	\int_{GF} \Psi_{f_1}^*(t_L,\theta_L|t_R,\theta_R) G_s(t_L,\theta_L|t_R,\theta_R) \Psi_{f_2}(t_L,\theta_L|t_R,\theta_R) \label{eq:3.24}
\end{equation}
using two different methods.\footnote{The definition of $G_s$ was given in \eqref{eq:wightman}.} The first method is to directly evaluate the integrals in \eqref{eq:3.19} to \eqref{eq:3.21}.\footnote{The reader might wonder why we did not directly evaluate \eqref{eq:3.24} without the $G_s$, which seems simpler. The answer is that including the $G_s$ allows us to avoid working with the delta function that sets the arguments of $f_1$ and $f_2$ to be equal. In practice, the second method was performed first, and it allowed us to correctly guess the result of \eqref{eq:3.24}. We verified our guess numerically. The numerics would not have converged without the $G_s$, due to the delta function.} The second method is to assume that
\begin{equation}
G_s(t_L,\theta_L|t_R,\theta_R) \Psi_{f_2}(t_L,\theta_L|t_R,\theta_R)
\label{eq:3.25}
\end{equation}
may be replaced in \eqref{eq:3.24} by
\begin{equation}
\Psi_{\tilde{f}}(t_L,\theta_L|t_R,\theta_R),
\end{equation}
for some function $\tilde{f}$ that we will find, such that  \eqref{eq:3.24} is unchanged. We will then apply \eqref{eq:fixinner} and \eqref{eq:psif1psif2} to find the choice of $\rhopl$ that is required for the two methods to agree.

To deduce $\tilde{f}$, we compute the overlap of \eqref{eq:3.25} with $\bra{s_L n_L} \bra{s_R n_R}$, and obtain
\begin{align}
	&\int dt_L d\theta_L \sqrt{-g_L} dt_R d\theta_R \sqrt{-g_R} \,  \braket{s_L \, n_L | t_L \, \theta_L} \braket{s_R \, n_R | t_R \, \theta_R} G_s(t_L,\theta_L|t_R,\theta_R) \Psi_{f_2}(t_L,\theta_L|t_R,\theta_R)
	\\
	&= \int_0^\infty ds_2 f_2(s_2) \sum_{n_2,n \in \mathbb{Z}} (-1)^{n_2}
	\\
	&\times \int dt_L d\theta_L \sqrt{-g_L} \, \left(\Psi^{s_L}_{n_L}\right)^*(t_L,\theta_L) \psi^s_n(t_L,\theta_L) \Psi^{s_2}_{n_2}(t_L,\theta_L) \label{eq:3.27}
	\\
	&\times \int dt_R d\theta_R \sqrt{-g_R} \,  \left(\Psi^{s_R}_{n_R}\right)^*(t_R,\theta_R) (\psi^s_n)^*(t_R,\theta_R) \Psi^{s_2}_{-n_2}(t_R,\theta_R) \label{eq:3.28}.
\end{align}
Lines \eqref{eq:3.27} and \eqref{eq:3.28} may be explicitly evaluated. The following integral will be used multiple times in this paper:
\begin{align}
	&\int dt d\theta \, \sqrt{-g} \, \left(\Psi^{s_1}_{n_1}\right)^*(t,\theta)
	\, 
	\Psi^{s_2}_{n_2}(t,\theta)
	\,
	\Psi^{s_3}_{n_3}(t,\theta) = \calv^{s_1}_{s_2,s_3} 	\left(\left(\calc_{PP}^{P,S}\right)_{s_3, n_3 | s_2, n_2}^{s_1, n_1}\right)^*,
	\label{eq:usefulintegral}
\end{align}
where $\calv^{s_1}_{s_2,s_3}$ is defined in \eqref{eq:vertexintegraldef} and $\calc_{PP}^{P,S}$ is a Clebsch-Gordan coefficient that is defined in \eqref{eq:cpsppdef}. Applying this to \eqref{eq:3.27} and \eqref{eq:3.28}, we find that
\begin{align}
	&\int dt_L d\theta_L \sqrt{-g_L} dt_R d\theta_R \sqrt{-g_R} \,  \braket{s_L \, n_L | t_L \, \theta_L} \braket{s_R \, n_R | t_R \, \theta_R} G_s(t_L,\theta_L|t_R,\theta_R) \Psi_{f_2}(t_L,\theta_L|t_R,\theta_R)
	\\
	&= \frac{\pi}{s \tanh \pi s} \int_0^\infty ds_2 f_2(s_2) \sum_{n_2,n \in \mathbb{Z}} (-1)^{n_2} \,  \calv^{s_L}_{s,s_2} 
	\left(\left(\calc_{PP}^{P,S}\right)_{s, n | s_2, n_2}^{s_L, n_L}\right)^* \, \left(\calv^{s_2}_{s,s_R}\right)^* 	
		\left(\calc_{PP}^{P,S}\right)_{s, n | s_R, n_R}^{s_2, -n_2} .
\end{align}
Using \eqref{eq:movelegidentity} and the orthogonality of the Clebsch-Gordan coefficients \eqref{eq:cgnorm}, this becomes
\begin{align}
	&\int dt_L d\theta_L \sqrt{-g_L} dt_R d\theta_R \sqrt{-g_R} \,  \braket{s_L \, n_L | t_L \, \theta_L} \braket{s_R \, n_R | t_R \, \theta_R} G_s(t_L,\theta_L|t_R,\theta_R) \Psi_{f_2}(t_L,\theta_L|t_R,\theta_R)	\\
	&= \delta(s_L - s_R) \delta_{n_L,n_R} (-1)^{n_L}  \frac{1}{16 \pi^5} \int_0^\infty ds_2 f_2(s_2) \, \, s_2 \sinh \pi s_2 \,  \cosh \pi s_L \Gamma\left( \frac{1}{2} \left(\frac{1}{2} \pm i s \pm i s_L \pm i s_2\right)\right),
\end{align}
where $\Gamma\left( \frac{1}{2} \left(\frac{1}{2} \pm i s \pm i s_L \pm i s_2\right)\right)$ stands for a product of eight gamma functions with all possible combinations of the signs. Formally, we can use this to project \eqref{eq:3.25} into $\calhobs \otimes \calhobs$ to obtain the following invariant expression,
\begin{equation}
\ket{\Psi_{\tilde{f}}} = \int_0^\infty d\tilde{s}   \sum_{\tilde{n} \in \mathbb{Z}} (-1)^{\tilde{n}} \ket{\tilde{s} \, \tilde{n}} \ket{\tilde{s}, - \tilde{n}} \tilde{f}(\tilde{s}),
\label{eq:3.39}
\end{equation}
where
\begin{equation}
\tilde{f}(\tilde{s}) := \frac{1}{16 \pi^5} \int_0^\infty ds_2 f_2(s_2) \,\, s_2 \sinh \pi s_2 \,  \cosh \pi \tilde{s} \, \Gamma\left( \frac{1}{2} \left(\frac{1}{2} \pm i s \pm i \tilde{s} \pm i s_2\right)\right).
\end{equation}
Although in the last step we pretended that \eqref{eq:3.39} is in $\calhobs \otimes \calhobs$, which is not strictly true, our result for $\tilde{f}$ is correct. By the definition of $\rhopl$, \eqref{eq:3.24} evaluates to
\begin{equation}
\int_0^\infty ds_1 \, \rhopl(s_1) \, f_1^*(s_1) \tilde{f}(s_1).
\end{equation}
Comparing this with the result of the first method, we conclude that
\begin{equation}
\rhopl(s) = \frac{s \tanh \pi s}{4 \pi^2}.
\end{equation}

To recapitulate, the position-space wavefunction \eqref{eq:psifdef} is, strictly speaking, not in $\calhobs \otimes \calhobs$, because it is not normalizable using the inner product \eqref{eq:innerproduct}. However, by instead using the inner product \eqref{eq:fixinner}, we may define a new gauge-fixed Hilbert space, which we have shown is equivalent to $L^2(\mathbb{R}_+,\rhopl)$ with $\rhopl$ given by \eqref{eq:rhopl}.

\section{Clebsch-Gordan coefficients for $\mathfrak{so}(2,1)$}

\label{sec:appendixcg}

In this appendix, we collect results on the Clebsch-Gordan (CG) coefficients of unitary irreducible representations of $\mathfrak{so}(2,1)$, the symmetry algebra of dS$_2$. We restrict our attention to the discrete and principal series representations. The three generators of the group are denoted by $L_{-1}$, $L_0$, and $L_1$, and they obey the commutation relation
\begin{equation}
[ L_n, L_m ] = (n - m) L_{n + m}.
\end{equation}
This algebra has a Casimir given by
\begin{equation}
-L_0^2 + \frac{1}{2}\left\{ L_{-1}, L_1\right\},
\end{equation}
and it is convenient to write the value of the Casimir for a given representation as $\Delta(1-\Delta)$. We are only interested in unitary representations, for which $L_0^\dagger = L_0$ and $L_{\pm 1}^\dagger = L_{\mp 1}$.

The principal series representations are labeled by a parameter $s > 0$ such that $\Delta = \Delta_s := \frac{1}{2} + i s$. We denote a state in a principal series representation by $\ket{n}_s$, where $n \in \mathbb{Z}$ labels the eigenvalue of $L_0$. In our conventions, these states transform under the action of the generators as follows:
\begin{align}
	 L_0 \ket{n}_s &= n \ket{n}_s,
	\\
	 L_1\ket{n}_s &= - i (n  - 1+ \Delta_s)   \ket{n-1}_s,
	\\
	 L_{-1}\ket{n}_s &= i (n  +1- \Delta_s)    \ket{ n + 1}_s.
\end{align}
The principal series representations are self-conjugate.

There are two types of discrete series representations, which we will refer to as the positive and negative discrete series. For the positive (resp. negative) discrete series, the $L_0$ generator is strictly positive (resp. negative). The positive and negative discrete series are exchanged under complex conjugation.

A state in a positive (or lowest-weight) discrete series representation is denoted by $\ket{n}_{\Delta}$, where $\Delta \in \mathbb{Z}_{\ge 1}$, and $n \in \mathbb{Z}_{\ge \Delta}$. These states transform as follows:
\begin{align}
	\label{eq:c6}
	L_0 \ket{n}_\Delta &=  n \ket{n}_\Delta,
	\\
	\label{eq:c7}	
	L_1 \ket{n}_\Delta &= -\sqrt{(n - \Delta)( \Delta + n-1)} \ket{n-1}_\Delta,
	\\
	\label{eq:c8}	
	L_{-1}\ket{n}_\Delta &= -\sqrt{(\Delta + n) (n - \Delta + 1)} \ket{n+1}_\Delta.
\end{align}
Note that $L_1 \ket{\Delta}_\Delta = 0$.

A state in a negative (or highest-weight) discrete series representation is denoted by $\ket{-n}_\Delta$, where $\Delta \in \mathbb{Z}_{\ge 1}$, and $n \in \mathbb{Z}_{\ge \Delta}$. These states transform as follows:
\begin{align}
	L_0 \ket{-n}_\Delta &= -n \ket{-n}_\Delta,
	\\
	L_1 \ket{-n}_\Delta &= -\sqrt{(n - \Delta +1)(\Delta + n)} \ket{-n-1}_\Delta,
	\\
	L_{-1} \ket{-n}_\Delta &= -\sqrt{(\Delta + n - 1 ) (n - \Delta )} \ket{-n+1}_\Delta.
\end{align}
Note that $L_{-1} \ket{- \Delta}_\Delta = 0$.

\subsection{Principal-principal}

Consider the tensor product of two principal series representations, labeled by $s_1, s_2 > 0$. After decomposing into irreps, one obtains two copies of the principal series  as well as both the positive and negative discrete series. We explicitly list the relevant CG coefficients as well as the identities they obey.

\subsubsection{Principal-principal $\rightarrow$ principal}

The first set of CG coefficients we consider is denoted by $ \left(\calctilde_{PP}^{P,S}\right)_{s_1, n_1 | s_2, n_2}^{s, n} $. Their main property is that the state $\ket{n}_s^S$, defined by
\begin{equation}
	\label{eq:cbs}
\ket{n}^S_s := \sum_{ \substack{n_1,n_2 \in \mathbb{Z} \\  n_1 + n_2 = n} } \left(\calctilde_{PP}^{P,S}\right)_{s_1, n_1 | s_2, n_2}^{s, n} \ket{n_1}_{s_1} \ket{n_2}_{s_2},
\end{equation}
transforms in a principal series representation with Casimir $\Delta_s(1 - \Delta)_s$. The $S$ superscript stands for ``symmetric'' and is meant to distinguish this state from another state, $\ket{n}_s^A$, which is defined by
\begin{equation}
		\label{eq:cba}
\ket{n}_s^A := \sum_{ \substack{n_1,n_2 \in \mathbb{Z} \\  n_1 + n_2 = n} } \left(\calctilde_{PP}^{P,A}\right)_{s_1, n_1 | s_2, n_2}^{s, n} \ket{n_1}_{s_1} \ket{n_2}_{s_2},
\end{equation}
where we have introduced a second set of CG coefficients. The $S$ and $A$ superscripts refer to how the coefficients behave under a swap operation:
\begin{equation}
\left(\calctilde_{PP}^{P,S}\right)_{s_1, n_1 | s_2, n_2}^{s, n} = \left(\calctilde_{PP}^{P,S}\right)_{s_2, n_2 | s_1, n_1}^{s, n}, \quad \quad \left(\calctilde_{PP}^{P,A}\right)_{s_1, n_1 | s_2, n_2}^{s, n} = -\left(\calctilde_{PP}^{P,A}\right)_{s_2, n_2 | s_1, n_1}^{s, n}.
\end{equation}
To derive an explicit expression for these coefficients, start by defining:
\begin{equation}
	\label{eq:ISdef}
I_S^{\Delta_1,\Delta_2,\Delta_3}(\theta_1,\theta_2,\theta_3) := 	\left|\sin \frac{\theta_{31}}{2}\right|^{-\Delta_3 - \Delta_1 + \Delta_2} \left|\sin \frac{\theta_{32}}{2}\right|^{-\Delta_3 - \Delta_2 + \Delta_1} \left|\sin \frac{\theta_{21}}{2}\right|^{-\Delta_2 - \Delta_1 + \Delta_3}
\end{equation}
and
\begin{equation}
I_A^{\Delta_1,\Delta_2,\Delta_3}(\theta_1,\theta_2,\theta_3) := 	I_S^{\Delta_1,\Delta_2,\Delta_3}(\theta_1,\theta_2,\theta_3) \cdot \text{sign} \left(\sin \frac{\theta_{31}}{2} \cdot \sin \frac{\theta_{32}}{2}  \cdot \sin \frac{\theta_{21}}{2}  \right),
\label{eq:IAdef}
\end{equation}
where $\theta_{ij} = \theta_i - \theta_j$. These are conformally invariant expressions for three-point functions in 1D CFT on a circle. Next, define $\left(C_S\right)^{\Delta_1,\Delta_2,\Delta_3}_{n_1,n_2,n_3}$ and $\left(C_A\right)^{\Delta_1,\Delta_2,\Delta_3}_{n_1,n_2,n_3}$ to be the discrete Fourier transforms, such that
\begin{align}
	\sum_{n_1,n_2,n_3 \in \mathbb{Z}}	\left(C_S\right)^{\Delta_1,\Delta_2,\Delta_3}_{n_1,n_2,n_3} e^{-in_1 \theta_1} e^{-in_2 \theta_2} e^{-in_3 \theta_3}  &:= I_S^{\Delta_1,\Delta_2,\Delta_3}(\theta_1,\theta_2,\theta_3),
	\\
	\sum_{n_1,n_2,n_3 \in \mathbb{Z}}	\left(C_A\right)^{\Delta_1,\Delta_2,\Delta_3}_{n_1,n_2,n_3} e^{-in_1 \theta_1} e^{-in_2 \theta_2} e^{-in_3 \theta_3}  &:= I_A^{\Delta_1,\Delta_2,\Delta_3}(\theta_1,\theta_2,\theta_3).
\end{align}

The CG coefficients are given by
\begin{align}
	\label{eq:cpspp}
	 \left(\calctilde_{PP}^{P,S}\right)_{s_1, n_1 | s_2, n_2}^{s, n}
	&:= \left(C_S\right)^{1-\Delta_{s_1},1-\Delta_{s_2},\Delta_s}_{-n_1,-n_2,n},
	\\
	 \left(\calctilde_{PP}^{P,A}\right)_{s_1, n_1 | s_2, n_2}^{s, n} 
	&:= \left(C_A\right)^{1-\Delta_{s_1},1-\Delta_{s_2},\Delta_s}_{-n_1,-n_2,n} 
	\label{eq:cpspa}
\end{align}

They are normalized such that
\begin{align}
	\sum_{ \substack{n_1,n_2 \in \mathbb{Z} \\ n_1 + n_2 = n}} \left( \left(\calctilde_{PP}^{P,S}\right)_{s_1, n_1 | s_2, n_2}^{s^\prime, n}\right)^*  \left(\calctilde_{PP}^{P,S}\right)_{s_1, n_1 | s_2, n_2}^{s, n} &= \sum_{ \substack{n_1,n_2 \in \mathbb{Z} \\ n_1 + n_2 = n}} \left( \left(\calctilde_{PP}^{P,A}\right)_{s_1, n_1 | s_2, n_2}^{s^\prime, n}\right)^*  \left(\calctilde_{PP}^{P,A}\right)_{s_1, n_1 | s_2, n_2}^{s, n}
	\\
	&= \frac{4}{  \pi s \tanh \pi s} \delta(s - s^\prime).
	\label{eq:cgnorm}
\end{align}

The explicit formulas for the CG coefficients are
\begin{align*}
\left(C_S\right)^{\Delta_1,\Delta_2,\Delta_3}_{n_1,n_2,n_3}&= \delta_{n_1 + n_2 + n_3}  
\frac{
2^{\Delta_1 + \Delta_2 + \Delta_3 - 1}
\Gamma(\Delta_1 - \Delta_2 - \Delta_3 + 1)
\Gamma(-\Delta_1 + \Delta_2 - \Delta_3 + 1) \Gamma(-\Delta_1 - \Delta_2 + \Delta_3 + 1)
}
{\pi^2} \\
&\quad \times 
\Bigg[
\frac{
(-1)^{n_2} (\cos(\pi \Delta_3) - \cos(\pi (\Delta_2 - \Delta_1)))
\Gamma\left(\frac{1}{2} (-\Delta_1 + \Delta_2 + \Delta_3 + 2)\right)
}
{
\Gamma\left(\frac{1}{2} (-2n_2 - \Delta_1 - \Delta_2 + \Delta_3)\right)
} \\
&\quad \times 
\Gamma\left(\frac{1}{2} (-2n_3 + \Delta_1 - \Delta_2 + \Delta_3 + 2)\right) \\
&\quad \times 
{}_4\tilde{F}_3\left(
1, 
\frac{1}{2} (2n_2 + \Delta_1 + \Delta_2 - \Delta_3 + 2), \right. \\
&\quad \left. 
\frac{1}{2} (-2n_3 + \Delta_1 - \Delta_2 + \Delta_3 + 2), 
\frac{1}{2} (-\Delta_1 + \Delta_2 + \Delta_3 + 2);
\frac{1}{2} (\Delta_1 - \Delta_2 - \Delta_3 + 4), \right. \\
&\quad \left.
\frac{1}{2} (-2n_3 - \Delta_1 + \Delta_2 - \Delta_3 + 4),
\frac{1}{2} (2n_2 - \Delta_1 - \Delta_2 + \Delta_3 + 4); 1
\right) \\
&\quad - 
\frac{
(-1)^{n_3} (\cos(\pi \Delta_2) - \cos(\pi (\Delta_3 - \Delta_1)))
\Gamma\left(\frac{1}{2} (-\Delta_1 + \Delta_2 + \Delta_3)\right)
}
{
\Gamma\left(\frac{1}{2} (-2n_3 - \Delta_1 + \Delta_2 - \Delta_3 + 2)\right)
} \\
&\quad \times 
\Gamma\left(\frac{1}{2} (-2n_2 + \Delta_1 + \Delta_2 - \Delta_3)\right) \\
&\quad \times 
{}_4\tilde{F}_3\left(
1, 
\frac{1}{2} (-2n_2 + \Delta_1 + \Delta_2 - \Delta_3), \right. \\
&\quad \left. 
\frac{1}{2} (2n_3 + \Delta_1 - \Delta_2 + \Delta_3), 
\frac{1}{2} (-\Delta_1 + \Delta_2 + \Delta_3);
\frac{1}{2} (\Delta_1 - \Delta_2 - \Delta_3 + 2), \right. \\
&\quad \left.
\frac{1}{2} (2n_3 - \Delta_1 + \Delta_2 - \Delta_3 + 2),
\frac{1}{2} (-2n_2 - \Delta_1 - \Delta_2 + \Delta_3 + 2); 1
\right)
\Bigg]
\end{align*}

where $\, _4\tilde{F}_3$ refers to the regularized Hypergeometric function, and

\begin{align*}
\left(C_A\right)^{\Delta_1,\Delta_2,\Delta_3}_{n_1,n_2,n_3} &= \delta_{n_1 + n_2 + n_3}  \frac{e^{ i \frac{\pi}{2} \left(-\Delta_1 - \Delta_2 - \Delta_3 + 1\right)} 
2^{\Delta_1 + \Delta_2 + \Delta_3 - 3} 
\left(-1 + e^{i \pi (\Delta_1 + \Delta_2 - \Delta_3 - 1)} \right)}{\Gamma(\Delta_1 + \Delta_2 - \Delta_3) 
\Gamma\left(\frac{1}{2} (-\Delta_1 - \Delta_2 + \Delta_3 + 1)\right)} \\
&\quad \times \frac{\left(-1 + e^{i \pi (\Delta_1 - \Delta_2 + \Delta_3 - 1)} \right) 
\left(-1 + e^{i \pi (-\Delta_1 + \Delta_2 + \Delta_3 - 1)} \right)}
{\Gamma(\Delta_1 - \Delta_2 + \Delta_3) 
\Gamma(-\Delta_1 + \Delta_2 + \Delta_3)} \\
&\quad \times \frac{\csc(\pi (\Delta_1 + \Delta_2 - \Delta_3)) 
\csc(\pi (\Delta_1 - \Delta_2 + \Delta_3)) 
\csc(\pi (-\Delta_1 + \Delta_2 + \Delta_3))}
{\Gamma\left(\frac{1}{2} (-2 n_1 - \Delta_1 + \Delta_2 - \Delta_3 + 3)\right) 
\Gamma\left(\frac{1}{2} (-2 n_2 + \Delta_1 - \Delta_2 - \Delta_3 + 1)\right)} \\
&\quad \times \Gamma\left(\frac{1}{2} (\Delta_1 + \Delta_2 - \Delta_3 - 1)\right) 
\\
&\quad \times \Gamma\left(\frac{1}{2} (-2 n_1 + \Delta_1 - \Delta_2 + \Delta_3 + 1)\right) 
\Gamma\left(\frac{1}{2} (-2 n_2 - \Delta_1 + \Delta_2 + \Delta_3 - 1)\right) \\
&\quad \times \Bigg[ 
{}_4F_3\left(1, \frac{\Delta_1}{2} + \frac{\Delta_2}{2} - \frac{\Delta_3}{2} - \frac{1}{2}, 
n_1 + \frac{\Delta_1}{2} - \frac{\Delta_2}{2} + \frac{\Delta_3}{2} - \frac{1}{2}, \right. \\
&\quad \left. -n_2 - \frac{\Delta_1}{2} + \frac{\Delta_2}{2} + \frac{\Delta_3}{2} - \frac{1}{2}; 
-n_2 + \frac{\Delta_1}{2} - \frac{\Delta_2}{2} - \frac{\Delta_3}{2} + \frac{1}{2}, \right. \\
&\quad \left. n_1 - \frac{\Delta_1}{2} + \frac{\Delta_2}{2} - \frac{\Delta_3}{2} + \frac{1}{2}, 
-\frac{\Delta_1}{2} - \frac{\Delta_2}{2} + \frac{\Delta_3}{2} + \frac{1}{2}; 1 \right) \\
&\quad + {}_4F_3\left(1, \frac{\Delta_1}{2} + \frac{\Delta_2}{2} - \frac{\Delta_3}{2} + \frac{1}{2}, 
-n_1 + \frac{\Delta_1}{2} - \frac{\Delta_2}{2} + \frac{\Delta_3}{2} + \frac{1}{2}, \right. \\
&\quad \left. n_2 - \frac{\Delta_1}{2} + \frac{\Delta_2}{2} + \frac{\Delta_3}{2} + \frac{1}{2}; 
n_2 + \frac{\Delta_1}{2} - \frac{\Delta_2}{2} - \frac{\Delta_3}{2} + \frac{3}{2}, \right. \\
&\quad \left. -n_1 - \frac{\Delta_1}{2} + \frac{\Delta_2}{2} - \frac{\Delta_3}{2} + \frac{3}{2}, 
-\frac{\Delta_1}{2} - \frac{\Delta_2}{2} + \frac{\Delta_3}{2} + \frac{3}{2}; 1 \right)  - 1 \Bigg]
\end{align*}

We have also shown that
\begin{align}
	\label{eq:movelegidentity}
	(C_S)^{\Delta_1,\Delta_2,1-\Delta_3}_{n_1,n_2,n_3} &= -i 
	\frac{
		\left( \cos\left( \pi (\Delta_1 - \Delta_2) \right) + \cos(\pi \Delta_3) \right)
	}{
		2^{-1 + 2\Delta_3} \pi \Gamma(\Delta_3)
	} \nonumber \\
	&\times
		\Gamma(1 - \Delta_3) \Gamma(\Delta_1 - \Delta_2 + \Delta_3) \Gamma(-\Delta_1 + \Delta_2 + \Delta_3)
	\nonumber \\
	&\times (C_S)^{\Delta_1,\Delta_2,\Delta_3}_{n_1,n_2,n_3} (-1)^{n_3} e^{i \phi_{n_3,s_3}} 
\end{align}
where $e^{i \phi_{n,s}}$ is defined in \eqref{eq:eiphi}.

Our conventions for Clebsch-Gordan coefficients in Section \ref{sec:observers} differ slightly from our conventions in this Appendix. We define
\begin{equation}
	\left(\calc_{PP}^{P,S}\right)_{s_3, n_3 | s_2, n_2}^{s_1, n_1} := \left(\calctilde_{PP}^{P,S}\right)_{s_3, n_3 | s_2, n_2}^{s_1, n_1} e^{i \phi_{n_1,s_1}/2} e^{-i \phi_{n_2,s_2}/2} e^{-i \phi_{n_3,s_3}/2}
	\label{eq:cpsppdef}
\end{equation}
\begin{equation}
	\left(\calc_{PP}^{P,A}\right)_{s_3, n_3 | s_2, n_2}^{s_1, n_1} := \left(\calctilde_{PP}^{P,A}\right)_{s_3, n_3 | s_2, n_2}^{s_1, n_1} e^{i \phi_{n_1,s_1}/2} e^{-i \phi_{n_2,s_2}/2} e^{-i \phi_{n_3,s_3}/2}
	\label{eq:cpsadef}
\end{equation}
The untilded coefficients appear in Section \ref{sec:observers}.

\subsubsection{Principal-principal $\rightarrow$ discrete}

\label{sec:ppdcg}

To extract the discrete series representations from the tensor product of two principal series representations, we begin with expressions analogous to \eqref{eq:cbs} and \eqref{eq:cba},
\begin{equation}
	\ket{n}_\Delta := \sum_{ \substack{n_1,n_2 \in \mathbb{Z} \\  n_1 + n_2 = n} } \left(\calctilde_{PP}^{D,+}\right)_{s_1, n_1 | s_2, n_2}^{\Delta, n} \ket{n_1}_{s_1} \ket{n_2}_{s_2},
\end{equation}
\begin{equation}
	\ket{-n}_\Delta := \sum_{ \substack{n_1,n_2 \in \mathbb{Z} \\  n_1 + n_2 = -n} } \left(\calctilde_{PP}^{D,-}\right)_{s_1, n_1 | s_2, n_2}^{\Delta, n} \ket{n_1}_{s_1} \ket{n_2}_{s_2}.
\end{equation}
The CG coefficients are normalized such that
\begin{align}
	&\sum_{ \substack{n_1,n_2 \in \mathbb{Z} \\  n_1 + n_2 = n} } \left(\left(\calctilde_{PP}^{D,+}\right)_{s_1, n_1 | s_2, n_2}^{\Delta^\prime, n}\right)^* \left(\calctilde_{PP}^{D,+}\right)_{s_1, n_1 | s_2, n_2}^{\Delta, n} \\
	&= \sum_{ \substack{n_1,n_2 \in \mathbb{Z} \\  n_1 + n_2 = -n} } \left(\left(\calctilde_{PP}^{D,-}\right)_{s_1, n_1 | s_2, n_2}^{\Delta^\prime, n}\right)^* \left(\calctilde_{PP}^{D,-}\right)_{s_1, n_1 | s_2, n_2}^{\Delta, n} = \delta_{\Delta,\Delta^\prime}.
\end{align}
Their sign under the swap operation depends on the parity of $\Delta$,
\begin{equation}
	\left(\calctilde_{PP}^{D,\pm}\right)_{s_2, n_2 | s_1, n_1}^{\Delta, n} = (-1)^\Delta \left(\calctilde_{PP}^{D,\pm}\right)_{s_1, n_1 | s_2, n_2}^{\Delta, n}
\end{equation}
Their explicit expressions are given by
\begin{align*}
	\left(\calctilde_{PP}^{D,+}\right)_{s_1, n_1 | s_2, n_2}^{\Delta, n} =& \frac{(-i)^{\Delta +n} \cosh (\pi s_1) \Gamma \left(-n_1-i s_1+\frac{1}{2}\right) 
		\text{csch}(\pi (s_1+s_2)) \left( i s_1-i s_2+\Delta \right)_{n-\Delta } }
	{\Gamma \left(-n_1+i s_2+\Delta +\frac{1}{2}\right) \Gamma (-i (s_1+s_2-i (\Delta -1)))} \\
	& \times \sqrt{\frac{(2 \Delta -1) \Gamma (i s_1-i s_2+\Delta ) \Gamma (-i s_1+i s_2+\Delta )}{\Gamma (n-\Delta +1) 
			\Gamma (n+\Delta )}} \\
	& \times \, _3F_2\left(-n_1-i s_1+\frac{1}{2}, \Delta -n,-i s_1+i s_2+\Delta ; 
	-n-i s_1+i s_2+1, -n_1+i s_2+\Delta +\frac{1}{2};1\right)
\end{align*}

\begin{align*}
\left(\calctilde_{PP}^{D,-}\right)_{s_1, n_1 | s_2, n_2}^{\Delta, n}	&=	\frac{
		i^{\Delta +n} \cosh(\pi s_2) \text{csch}(\pi (s_1 + s_2)) 
		\Gamma\left(-n - n_1 - i s_2 + \frac{1}{2}\right)
		(-n + i s_1 - i s_2 + 1)_{n-\Delta}
	}
	{
		\Gamma(-i s_1 - i s_2 - \Delta + 1)
		\Gamma\left(-n - n_1 + i s_1 + \Delta + \frac{1}{2}\right)
	}
\\
	& \times
	\sqrt{\frac{(2 \Delta -1) \Gamma(i s_1 - i s_2 + \Delta) \Gamma(-i s_1 + i s_2 + \Delta)}
		{\Gamma(n - \Delta + 1) \Gamma(n + \Delta)}} \\
	& \times
	_3F_2\left(
	-n - n_1 - i s_2 + \frac{1}{2}, \Delta - n, i s_1 - i s_2 + \Delta; \right. \\
	& \qquad \left. 
	-n + i s_1 - i s_2 + 1, -n - n_1 + i s_1 + \Delta + \frac{1}{2}; 1
	\right)
\end{align*}

\subsubsection{Completeness relation}

To summarize, the CG coefficients for the tensor product of two principal series representations are given by
\begin{equation}
\left(\calctilde_{PP}^{P,S}\right)_{s_1, n_1 | s_2, n_2}^{s, n}, \quad 	 \left(\calctilde_{PP}^{P,A}\right)_{s_1, n_1 | s_2, n_2}^{s, n}, \quad  \left(\calctilde_{PP}^{D,+}\right)_{s_1, n_1 | s_2, n_2}^{\Delta, n}, \quad \left(\calctilde_{PP}^{D,-}\right)_{s_1, n_1 | s_2, n_2}^{\Delta, n},
\end{equation}
and have been defined above. The completeness relations ensure that we haven't missed any irreps in the decomposition. Let $m \in \mathbb{Z}$. For $n_1 + n_2 > 0$, we have that
\begin{align*}
	&\sum_{x \in \{S,A\}} \int_0^\infty ds 
	\frac{  \pi s \tanh \pi s}{4}\, \left(\left(\calctilde_{PP}^{P,x}\right)_{s_1, n_1 + m  | s_2, n_2 - m}^{s, n_1 + n_2}\right)^* \left(\calctilde_{PP}^{P,x}\right)_{s_1, n_1 | s_2, n_2}^{s, n_1 + n_2} 
	\\
	&+ \sum_{\Delta = 1}^{n_1 + n_2} \left(\left(\calctilde_{PP}^{D,+}\right)_{s_1, n_1 + m | s_2, n_2 - m}^{\Delta, n_1 + n_2}\right)^* \left(\calctilde_{PP}^{D,+}\right)_{s_1, n_1 | s_2, n_2}^{\Delta, n_1 + n_2} = \delta_{m,0}.
\end{align*}
while for $n_1 + n_2 < 0$ we have that
\begin{align*}
	&\sum_{x \in \{S,A\}} \int_0^\infty ds 
	\frac{  \pi s \tanh \pi s}{4}\, \left(\left(\calctilde_{PP}^{P,x}\right)_{s_1, n_1 + m  | s_2, n_2 - m}^{s, n_1 + n_2}\right)^* \left(\calctilde_{PP}^{P,x}\right)_{s_1, n_1 | s_2, n_2}^{s, n_1 + n_2} 
	\\
	&+ \sum_{ \Delta =  1}^{-(n_1 + n_2)} \left(\left(\calctilde_{PP}^{D,-}\right)_{s_1, n_1 + m | s_2, n_2 - m}^{\Delta, -(n_1 + n_2)}\right)^* \left(\calctilde_{PP}^{D,-}\right)_{s_1, n_1 | s_2, n_2}^{\Delta, -(n_1 + n_2)} = \delta_{m,0}.
\end{align*}
and for $n_1 + n_2 = 0$, we have that
\begin{align*}
	\sum_{x \in \{S,A\}} \int_0^\infty ds 
	\frac{  \pi s \tanh \pi s}{4}\, \left(\left(\calctilde_{PP}^{P,x}\right)_{s_1, n_1 + m  | s_2, -n_1 - m}^{s,0}\right)^* \left(\calctilde_{PP}^{P,x}\right)_{s_1, n_1 | s_2, -n_1}^{s,0} 
 = \delta_{m,0}.
\end{align*}

\subsection{Principal-discrete}

Next, we consider the tensor product of a principal with a positive discrete series representation.\footnote{The case with a negative discrete series representation follows from complex conjugation and will not be presented explicitly.} In the decomposition into irreps, only the principal and positive discrete series appear.

\subsubsection{Principal-discrete $\rightarrow$ principal}

A principal series state is constructed from the tensor product as follows:
\begin{equation}
	\ket{n}_s = \sum_{ \substack{n_1,n_2 \in \mathbb{Z} \\  n_1 + n_2 = n} } \left(\calctilde_{P,+}^{P}\right)_{s_1, n_1 | \Delta_2, n_2}^{s, n} \ket{n_1}_{s_1} \ket{n_2}_{\Delta_2}.
\end{equation}
The CG coefficient is expressible in terms of CG coefficients that appeared earlier. We have
\begin{equation}
\left(\calctilde_{P,+}^{P}\right)_{s_1, n_1 | \Delta_2, n_2}^{s, n} = (-1)^{n_2} \left(\calctilde_{PP}^{D,-}\right)^{\Delta_2,n_2}_{s_1,n_1 | - s, - n}. 	
\end{equation}

\subsubsection{Principal-discrete $\rightarrow$ discrete}

A discrete series state is constructed from the tensor product as follows:
\begin{equation}
	\ket{n}_\Delta = \sum_{ \substack{n_1,n_2 \in \mathbb{Z} \\  n_1 + n_2 = n} } \left(\calctilde_{P,+}^{+}\right)_{s_1, n_1 | \Delta_2, n_2}^{\Delta, n} \ket{n_1}_{s_1} \ket{n_2}_{\Delta_2}.
\end{equation}
The explicit expression for the CG coefficient is given by

\begin{equation}
	\begin{aligned}
		&\left(\calctilde_{P,+}^{+}\right)_{s_1, n_1 | \Delta_2, n_2}^{\Delta, n} :=
		\\
		&(-i)^{n_1} \Gamma(1 - n_1 - \Delta_1) \Gamma(1 - n_1 - n_2 - \Delta_1 + \Delta_2) \, 
			\left(\Delta + \Delta_1 - \Delta_2\right)_{n_1 + n_2 - \Delta}
		 \\
		&\times \sqrt{\frac{\Gamma(2 \Delta_2) \Gamma(\Delta - \Delta_1 + \Delta_2) 
				\Gamma(\Delta + \Delta_1 + \Delta_2 - 1)}
			{\Gamma(2 \Delta - 1) \Gamma(1 - \Delta - \Delta_1 + \Delta_2) 
				\Gamma(-\Delta + \Delta_1 + \Delta_2)}} \\
		&\times \sqrt{\frac{\Gamma(2 \Delta) \left(2 \Delta_2\right)_{n_2 - \Delta_2} }
			{\left(n_1 + n_2 - \Delta\right)!\left(n_2 - \Delta_2\right)! \Gamma(n_1 + n_2 + \Delta)}} \\
		&\times \, _3\tilde{F}_2\left( -n_1 - n_2 + \Delta, 1 - n_1 - \Delta_1, \Delta - \Delta_1 + \Delta_2; 
		-n_1 + \Delta + \Delta_2, 1 - n_1 - n_2 - \Delta_1 + \Delta_2; 1 \right)
	\end{aligned}
\end{equation}

where $\, _3\tilde{F}_2$ refers to the regulated Hypergeometric function.

\subsubsection{Completeness relation}

To summarize, there are two CG coefficients, given by
\begin{equation}
	\left(\calctilde_{P,+}^{P}\right)_{s_1, n_1 | \Delta_2, n_2}^{s, n} \quad \text{and} \quad \left(\calctilde_{P,+}^{+}\right)_{s_1, n_1 | \Delta_2, n_2}^{\Delta, n}.
\end{equation}
The completeness relations are as follows. For $n \leq 0$, we have that
\begin{equation}
\int_0^\infty ds \, \frac{s \tanh \pi s}{\Delta_2 - \frac{1}{2}} \,  \left(\left(\calctilde_{P,+}^{P}\right)_{s_1, n - n_2^\prime | \Delta_2, n_2^\prime}^{s, n}\right)^* \left(\calctilde_{P,+}^{P}\right)_{s_1, n - n_2 | \Delta_2, n_2}^{s, n} = \delta_{n_2,n_2^\prime}
\end{equation}
while for $n \ge 1$, we have that
\begin{align}
&\int_0^\infty ds \, \frac{s \tanh \pi s}{\Delta_2 - \frac{1}{2}} \, \left(\left(\calctilde_{P,+}^{P}\right)_{s_1, n - n_2^\prime | \Delta_2, n_2^\prime}^{s, n}\right)^* \left(\calctilde_{P,+}^{P}\right)_{s_1, n - n_2 | \Delta_2, n_2}^{s, n} 
\\
&+ \sum_{\Delta = 1}^n \left(\left(\calctilde_{P,+}^{+}\right)_{s_1, n - n_2^\prime | \Delta_2, n_2^\prime }^{\Delta, n}\right)^* \left(\calctilde_{P,+}^{+}\right)_{s_1, n - n_2 | \Delta_2, n_2}^{\Delta, n}
= \delta_{n_2,n_2^\prime}
\end{align}

	\bibliographystyle{JHEP}
	\nocite{}
	\bibliography{thebibliography}

\providecommand{\href}[2]{#2}\begingroup\raggedright\begin{thebibliography}{100}

\bibitem{CLPW}
V.~Chandrasekaran, R.~Longo, G.~Penington and E.~Witten, \emph{{An algebra of observables for de Sitter space}}, \href{http://dx.doi.org/10.1007/JHEP02(2023)082}{\emph{JHEP} {\bfseries 02} (2023) 082}, [\href{https://arxiv.org/abs/2206.10780}{{\ttfamily 2206.10780}}].

\bibitem{Leutheusser:2021frk}
S.~A.~W. Leutheusser and H.~Liu, \emph{{Emergent Times in Holographic Duality}}, \href{http://dx.doi.org/10.1103/PhysRevD.108.086020}{\emph{Phys. Rev. D} {\bfseries 108} (2023) 086020}, [\href{https://arxiv.org/abs/2112.12156}{{\ttfamily 2112.12156}}].

\bibitem{Leutheusser:2021qhd}
S.~Leutheusser and H.~Liu, \emph{{Causal connectability between quantum systems and the black hole interior in holographic duality}}, \href{http://dx.doi.org/10.1103/PhysRevD.108.086019}{\emph{Phys. Rev. D} {\bfseries 108} (2023) 086019}, [\href{https://arxiv.org/abs/2110.05497}{{\ttfamily 2110.05497}}].

\bibitem{Chandrasekaran:2022eqq}
V.~Chandrasekaran, G.~Penington and E.~Witten, \emph{{Large N algebras and generalized entropy}}, \href{http://dx.doi.org/10.1007/JHEP04(2023)009}{\emph{JHEP} {\bfseries 04} (2023) 009}, [\href{https://arxiv.org/abs/2209.10454}{{\ttfamily 2209.10454}}].

\bibitem{Witten:2021unn}
E.~Witten, \emph{{Gravity and the crossed product}}, \href{http://dx.doi.org/10.1007/JHEP10(2022)008}{\emph{JHEP} {\bfseries 10} (2022) 008}, [\href{https://arxiv.org/abs/2112.12828}{{\ttfamily 2112.12828}}].

\bibitem{Faulkner:2024gst}
T.~Faulkner and A.~J. Speranza, \emph{{Gravitational algebras and the generalized second law}},  \href{https://arxiv.org/abs/2405.00847}{{\ttfamily 2405.00847}}.

\bibitem{Jensen:2023yxy}
K.~Jensen, J.~Sorce and A.~J. Speranza, \emph{{Generalized entropy for general subregions in quantum gravity}}, \href{http://dx.doi.org/10.1007/JHEP12(2023)020}{\emph{JHEP} {\bfseries 12} (2023) 020}, [\href{https://arxiv.org/abs/2306.01837}{{\ttfamily 2306.01837}}].

\bibitem{Kudler-Flam:2023qfl}
J.~Kudler-Flam, S.~Leutheusser and G.~Satishchandran, \emph{{Generalized Black Hole Entropy is von Neumann Entropy}},  \href{https://arxiv.org/abs/2309.15897}{{\ttfamily 2309.15897}}.

\bibitem{Akers:2024bel}
C.~Akers and J.~Sorce, \emph{{Relative state-counting for semiclassical black holes}},  \href{https://arxiv.org/abs/2404.16098}{{\ttfamily 2404.16098}}.

\bibitem{Ouseph:2023juq}
S.~Ouseph, K.~Furuya, N.~Lashkari, K.~L. Leung and M.~Moosa, \emph{{Local Poincar\'e algebra from quantum chaos}}, \href{http://dx.doi.org/10.1007/JHEP01(2024)112}{\emph{JHEP} {\bfseries 01} (2024) 112}, [\href{https://arxiv.org/abs/2310.13736}{{\ttfamily 2310.13736}}].

\bibitem{Furuya:2023fei}
K.~Furuya, N.~Lashkari, M.~Moosa and S.~Ouseph, \emph{{Information loss, mixing and emergent type III$_{1}$ factors}}, \href{http://dx.doi.org/10.1007/JHEP08(2023)111}{\emph{JHEP} {\bfseries 08} (2023) 111}, [\href{https://arxiv.org/abs/2305.16028}{{\ttfamily 2305.16028}}].

\bibitem{Gesteau:2023rrx}
E.~Gesteau, \emph{{Emergent spacetime and the ergodic hierarchy}},  \href{https://arxiv.org/abs/2310.13733}{{\ttfamily 2310.13733}}.

\bibitem{Engelhardt:2023xer}
N.~Engelhardt and H.~Liu, \emph{{Algebraic ER=EPR and complexity transfer}}, \href{http://dx.doi.org/10.1007/JHEP07(2024)013}{\emph{JHEP} {\bfseries 07} (2024) 013}, [\href{https://arxiv.org/abs/2311.04281}{{\ttfamily 2311.04281}}].

\bibitem{Gesteau:2024rpt}
E.~Gesteau and H.~Liu, \emph{{Toward stringy horizons}},  \href{https://arxiv.org/abs/2408.12642}{{\ttfamily 2408.12642}}.

\bibitem{Kang:2018xqy}
M.~J. Kang and D.~K. Kolchmeyer, \emph{{Holographic Relative Entropy in Infinite-Dimensional Hilbert Spaces}}, \href{http://dx.doi.org/10.1007/s00220-022-04627-z}{\emph{Commun. Math. Phys.} {\bfseries 400} (2023) 1665--1695}, [\href{https://arxiv.org/abs/1811.05482}{{\ttfamily 1811.05482}}].

\bibitem{Faulkner:2020hzi}
T.~Faulkner, \emph{{The holographic map as a conditional expectation}},  \href{https://arxiv.org/abs/2008.04810}{{\ttfamily 2008.04810}}.

\bibitem{Gesteau:2021jzp}
E.~Gesteau and M.~J. Kang, \emph{{Nonperturbative gravity corrections to bulk reconstruction}}, \href{http://dx.doi.org/10.1088/1751-8121/acef7d}{\emph{J. Phys. A} {\bfseries 56} (2023) 385401}, [\href{https://arxiv.org/abs/2112.12789}{{\ttfamily 2112.12789}}].

\bibitem{Faulkner:2022ada}
T.~Faulkner and M.~Li, \emph{{Asymptotically isometric codes for holography}},  \href{https://arxiv.org/abs/2211.12439}{{\ttfamily 2211.12439}}.

\bibitem{Bousso:2012sj}
R.~Bousso, S.~Leichenauer and V.~Rosenhaus, \emph{{Light-sheets and AdS/CFT}}, \href{http://dx.doi.org/10.1103/PhysRevD.86.046009}{\emph{Phys. Rev. D} {\bfseries 86} (2012) 046009}, [\href{https://arxiv.org/abs/1203.6619}{{\ttfamily 1203.6619}}].

\bibitem{Czech:2012bh}
B.~Czech, J.~L. Karczmarek, F.~Nogueira and M.~Van~Raamsdonk, \emph{{The Gravity Dual of a Density Matrix}}, \href{http://dx.doi.org/10.1088/0264-9381/29/15/155009}{\emph{Class. Quant. Grav.} {\bfseries 29} (2012) 155009}, [\href{https://arxiv.org/abs/1204.1330}{{\ttfamily 1204.1330}}].

\bibitem{Bousso:2012mh}
R.~Bousso, B.~Freivogel, S.~Leichenauer, V.~Rosenhaus and C.~Zukowski, \emph{{Null Geodesics, Local CFT Operators and AdS/CFT for Subregions}}, \href{http://dx.doi.org/10.1103/PhysRevD.88.064057}{\emph{Phys. Rev. D} {\bfseries 88} (2013) 064057}, [\href{https://arxiv.org/abs/1209.4641}{{\ttfamily 1209.4641}}].

\bibitem{Hubeny:2012wa}
V.~E. Hubeny and M.~Rangamani, \emph{{Causal Holographic Information}}, \href{http://dx.doi.org/10.1007/JHEP06(2012)114}{\emph{JHEP} {\bfseries 06} (2012) 114}, [\href{https://arxiv.org/abs/1204.1698}{{\ttfamily 1204.1698}}].

\bibitem{Almheiri:2014lwa}
A.~Almheiri, X.~Dong and D.~Harlow, \emph{{Bulk Locality and Quantum Error Correction in AdS/CFT}}, \href{http://dx.doi.org/10.1007/JHEP04(2015)163}{\emph{JHEP} {\bfseries 04} (2015) 163}, [\href{https://arxiv.org/abs/1411.7041}{{\ttfamily 1411.7041}}].

\bibitem{Headrick:2014cta}
M.~Headrick, V.~E. Hubeny, A.~Lawrence and M.~Rangamani, \emph{{Causality \& holographic entanglement entropy}}, \href{http://dx.doi.org/10.1007/JHEP12(2014)162}{\emph{JHEP} {\bfseries 12} (2014) 162}, [\href{https://arxiv.org/abs/1408.6300}{{\ttfamily 1408.6300}}].

\bibitem{Soni:2024oim}
R.~M. Soni, \emph{{Extremality as a Consistency Condition on Subregion Duality}},  \href{https://arxiv.org/abs/2403.19562}{{\ttfamily 2403.19562}}.

\bibitem{Leutheusser:2022bgi}
S.~Leutheusser and H.~Liu, \emph{{Subalgebra-subregion duality: emergence of space and time in holography}},  \href{https://arxiv.org/abs/2212.13266}{{\ttfamily 2212.13266}}.

\bibitem{Harlow:2016vwg}
D.~Harlow, \emph{{The Ryu\textendash{}Takayanagi Formula from Quantum Error Correction}}, \href{http://dx.doi.org/10.1007/s00220-017-2904-z}{\emph{Commun. Math. Phys.} {\bfseries 354} (2017) 865--912}, [\href{https://arxiv.org/abs/1607.03901}{{\ttfamily 1607.03901}}].

\bibitem{Bahiru:2022oas}
E.~Bahiru, A.~Belin, K.~Papadodimas, G.~Sarosi and N.~Vardian, \emph{{State-dressed local operators in the AdS/CFT correspondence}}, \href{http://dx.doi.org/10.1103/PhysRevD.108.086035}{\emph{Phys. Rev. D} {\bfseries 108} (2023) 086035}, [\href{https://arxiv.org/abs/2209.06845}{{\ttfamily 2209.06845}}].

\bibitem{Bahiru:2023zlc}
E.~Bahiru, A.~Belin, K.~Papadodimas, G.~Sarosi and N.~Vardian, \emph{{Holography and localization of information in quantum gravity}}, \href{http://dx.doi.org/10.1007/JHEP05(2024)261}{\emph{JHEP} {\bfseries 05} (2024) 261}, [\href{https://arxiv.org/abs/2301.08753}{{\ttfamily 2301.08753}}].

\bibitem{Witten}
E.~Witten, \emph{{A background-independent algebra in quantum gravity}}, \href{http://dx.doi.org/10.1007/JHEP03(2024)077}{\emph{JHEP} {\bfseries 03} (2024) 077}, [\href{https://arxiv.org/abs/2308.03663}{{\ttfamily 2308.03663}}].

\bibitem{Chen:2024rpx}
C.-H. Chen and G.~Penington, \emph{{A clock is just a way to tell the time: gravitational algebras in cosmological spacetimes}},  \href{https://arxiv.org/abs/2406.02116}{{\ttfamily 2406.02116}}.

\bibitem{Kudler-Flam:2024psh}
J.~Kudler-Flam, S.~Leutheusser and G.~Satishchandran, \emph{{Algebraic Observational Cosmology}},  \href{https://arxiv.org/abs/2406.01669}{{\ttfamily 2406.01669}}.

\bibitem{Kaplan:2024xyk}
M.~Kaplan, D.~Marolf, X.~Yu and Y.~Zhao, \emph{{De Sitter quantum gravity and the emergence of local algebras}},  \href{https://arxiv.org/abs/2410.00111}{{\ttfamily 2410.00111}}.

\bibitem{Fewster:2024pur}
C.~J. Fewster, D.~W. Janssen, L.~D. Loveridge, K.~Rejzner and J.~Waldron, \emph{{Quantum reference frames, measurement schemes and the type of local algebras in quantum field theory}},  \href{https://arxiv.org/abs/2403.11973}{{\ttfamily 2403.11973}}.

\bibitem{Gomez:2023jbg}
C.~Gomez, \emph{{On the algebraic meaning of quantum gravity for closed Universes}},  \href{https://arxiv.org/abs/2311.01952}{{\ttfamily 2311.01952}}.

\bibitem{Banks:2023uit}
T.~Banks, \emph{{My Personal History With the Quantum Theory of de Sitter Space}},  \href{https://arxiv.org/abs/2312.10729}{{\ttfamily 2312.10729}}.

\bibitem{Bousso:2000md}
R.~Bousso, \emph{{Bekenstein bounds in de Sitter and flat space}}, \href{http://dx.doi.org/10.1088/1126-6708/2001/04/035}{\emph{JHEP} {\bfseries 04} (2001) 035}, [\href{https://arxiv.org/abs/hep-th/0012052}{{\ttfamily hep-th/0012052}}].

\bibitem{Bousso:2000nf}
R.~Bousso, \emph{{Positive vacuum energy and the N bound}}, \href{http://dx.doi.org/10.1088/1126-6708/2000/11/038}{\emph{JHEP} {\bfseries 11} (2000) 038}, [\href{https://arxiv.org/abs/hep-th/0010252}{{\ttfamily hep-th/0010252}}].

\bibitem{Banks:2000fe}
T.~Banks, \emph{{Cosmological breaking of supersymmetry?}}, \href{http://dx.doi.org/10.1142/S0217751X01003998}{\emph{Int. J. Mod. Phys. A} {\bfseries 16} (2001) 910--921}, [\href{https://arxiv.org/abs/hep-th/0007146}{{\ttfamily hep-th/0007146}}].

\bibitem{Banks:2001yp}
T.~Banks and W.~Fischler, \emph{{M theory observables for cosmological space-times}},  \href{https://arxiv.org/abs/hep-th/0102077}{{\ttfamily hep-th/0102077}}.

\bibitem{Parikh:2004wh}
M.~K. Parikh and E.~P. Verlinde, \emph{{De Sitter holography with a finite number of states}}, \href{http://dx.doi.org/10.1088/1126-6708/2005/01/054}{\emph{JHEP} {\bfseries 01} (2005) 054}, [\href{https://arxiv.org/abs/hep-th/0410227}{{\ttfamily hep-th/0410227}}].

\bibitem{Banks:2005bm}
T.~Banks, \emph{{Some thoughts on the quantum theory of stable de Sitter space}},  \href{https://arxiv.org/abs/hep-th/0503066}{{\ttfamily hep-th/0503066}}.

\bibitem{Banks:2006rx}
T.~Banks, B.~Fiol and A.~Morisse, \emph{{Towards a quantum theory of de Sitter space}}, \href{http://dx.doi.org/10.1088/1126-6708/2006/12/004}{\emph{JHEP} {\bfseries 12} (2006) 004}, [\href{https://arxiv.org/abs/hep-th/0609062}{{\ttfamily hep-th/0609062}}].

\bibitem{Banks:2018ypk}
T.~Banks and W.~Fischler, \emph{{The holographic spacetime model of cosmology}}, \href{http://dx.doi.org/10.1142/S0218271818460057}{\emph{Int. J. Mod. Phys. D} {\bfseries 27} (2018) 1846005}, [\href{https://arxiv.org/abs/1806.01749}{{\ttfamily 1806.01749}}].

\bibitem{Susskind:2021omt}
L.~Susskind, \emph{{De Sitter Holography: Fluctuations, Anomalous Symmetry, and Wormholes}}, \href{http://dx.doi.org/10.3390/universe7120464}{\emph{Universe} {\bfseries 7} (2021) 464}, [\href{https://arxiv.org/abs/2106.03964}{{\ttfamily 2106.03964}}].

\bibitem{Susskind:2021dfc}
L.~Susskind, \emph{{Black Holes Hint towards De Sitter Matrix Theory}}, \href{http://dx.doi.org/10.3390/universe9080368}{\emph{Universe} {\bfseries 9} (2023) 368}, [\href{https://arxiv.org/abs/2109.01322}{{\ttfamily 2109.01322}}].

\bibitem{Dong:2018cuv}
X.~Dong, E.~Silverstein and G.~Torroba, \emph{{De Sitter Holography and Entanglement Entropy}}, \href{http://dx.doi.org/10.1007/JHEP07(2018)050}{\emph{JHEP} {\bfseries 07} (2018) 050}, [\href{https://arxiv.org/abs/1804.08623}{{\ttfamily 1804.08623}}].

\bibitem{Lin:2022nss}
H.~Lin and L.~Susskind, \emph{{Infinite Temperature's Not So Hot}},  \href{https://arxiv.org/abs/2206.01083}{{\ttfamily 2206.01083}}.

\bibitem{Maldacena:2001kr}
J.~M. Maldacena, \emph{{Eternal black holes in anti-de Sitter}}, \href{http://dx.doi.org/10.1088/1126-6708/2003/04/021}{\emph{JHEP} {\bfseries 04} (2003) 021}, [\href{https://arxiv.org/abs/hep-th/0106112}{{\ttfamily hep-th/0106112}}].

\bibitem{Gao:2016bin}
P.~Gao, D.~L. Jafferis and A.~C. Wall, \emph{{Traversable Wormholes via a Double Trace Deformation}}, \href{http://dx.doi.org/10.1007/JHEP12(2017)151}{\emph{JHEP} {\bfseries 12} (2017) 151}, [\href{https://arxiv.org/abs/1608.05687}{{\ttfamily 1608.05687}}].

\bibitem{Penington:2023dql}
G.~Penington and E.~Witten, \emph{{Algebras and States in JT Gravity}},  \href{https://arxiv.org/abs/2301.07257}{{\ttfamily 2301.07257}}.

\bibitem{Kolchmeyer:2023gwa}
D.~K. Kolchmeyer, \emph{{von Neumann algebras in JT gravity}}, \href{http://dx.doi.org/10.1007/JHEP06(2023)067}{\emph{JHEP} {\bfseries 06} (2023) 067}, [\href{https://arxiv.org/abs/2303.04701}{{\ttfamily 2303.04701}}].

\bibitem{Aalsma:2020aib}
L.~Aalsma and G.~Shiu, \emph{{Chaos and complementarity in de Sitter space}}, \href{http://dx.doi.org/10.1007/JHEP05(2020)152}{\emph{JHEP} {\bfseries 05} (2020) 152}, [\href{https://arxiv.org/abs/2002.01326}{{\ttfamily 2002.01326}}].

\bibitem{Anninos:2011af}
D.~Anninos, S.~A. Hartnoll and D.~M. Hofman, \emph{{Static Patch Solipsism: Conformal Symmetry of the de Sitter Worldline}}, \href{http://dx.doi.org/10.1088/0264-9381/29/7/075002}{\emph{Class. Quant. Grav.} {\bfseries 29} (2012) 075002}, [\href{https://arxiv.org/abs/1109.4942}{{\ttfamily 1109.4942}}].

\bibitem{Narovlansky:2023lfz}
V.~Narovlansky and H.~Verlinde, \emph{{Double-scaled SYK and de Sitter Holography}},  \href{https://arxiv.org/abs/2310.16994}{{\ttfamily 2310.16994}}.

\bibitem{Verlinde:2024znh}
H.~Verlinde, \emph{{Double-scaled SYK, Chords and de Sitter Gravity}},  \href{https://arxiv.org/abs/2402.00635}{{\ttfamily 2402.00635}}.

\bibitem{Levine:2020upy}
A.~Levine, A.~Shahbazi-Moghaddam and R.~M. Soni, \emph{{Seeing the entanglement wedge}}, \href{http://dx.doi.org/10.1007/JHEP06(2021)134}{\emph{JHEP} {\bfseries 06} (2021) 134}, [\href{https://arxiv.org/abs/2009.11305}{{\ttfamily 2009.11305}}].

\bibitem{Engelhardt:2021mue}
N.~Engelhardt, G.~Penington and A.~Shahbazi-Moghaddam, \emph{{A world without pythons would be so simple}}, \href{http://dx.doi.org/10.1088/1361-6382/ac2de5}{\emph{Class. Quant. Grav.} {\bfseries 38} (2021) 234001}, [\href{https://arxiv.org/abs/2102.07774}{{\ttfamily 2102.07774}}].

\bibitem{Maldacena:2015waa}
J.~Maldacena, S.~H. Shenker and D.~Stanford, \emph{{A bound on chaos}}, \href{http://dx.doi.org/10.1007/JHEP08(2016)106}{\emph{JHEP} {\bfseries 08} (2016) 106}, [\href{https://arxiv.org/abs/1503.01409}{{\ttfamily 1503.01409}}].

\bibitem{Rahman:2024vyg}
A.~A. Rahman and L.~Susskind, \emph{{Infinite Temperature is Not So Infinite: The Many Temperatures of de Sitter Space}},  \href{https://arxiv.org/abs/2401.08555}{{\ttfamily 2401.08555}}.

\bibitem{Milekhin:2024vbb}
A.~Milekhin and J.~Xu, \emph{{On scrambling, tomperature and superdiffusion in de Sitter space}},  \href{https://arxiv.org/abs/2403.13915}{{\ttfamily 2403.13915}}.

\bibitem{Susskind:2021esx}
L.~Susskind, \emph{{Entanglement and Chaos in De Sitter Space Holography: An SYK Example}}, \href{http://dx.doi.org/10.22128/jhap.2021.455.1005}{\emph{JHAP} {\bfseries 1} (2021) 1--22}, [\href{https://arxiv.org/abs/2109.14104}{{\ttfamily 2109.14104}}].

\bibitem{Shenker:2014cwa}
S.~H. Shenker and D.~Stanford, \emph{{Stringy effects in scrambling}}, \href{http://dx.doi.org/10.1007/JHEP05(2015)132}{\emph{JHEP} {\bfseries 05} (2015) 132}, [\href{https://arxiv.org/abs/1412.6087}{{\ttfamily 1412.6087}}].

\bibitem{Strohmaier:2023opz}
A.~Strohmaier and E.~Witten, \emph{{The Timelike Tube Theorem in Curved Spacetime}}, \href{http://dx.doi.org/10.1007/s00220-024-05009-3}{\emph{Commun. Math. Phys.} {\bfseries 405} (2024) 153}, [\href{https://arxiv.org/abs/2303.16380}{{\ttfamily 2303.16380}}].

\bibitem{Witten:2018zxz}
E.~Witten, \emph{{APS Medal for Exceptional Achievement in Research: Invited article on entanglement properties of quantum field theory}}, \href{http://dx.doi.org/10.1103/RevModPhys.90.045003}{\emph{Rev. Mod. Phys.} {\bfseries 90} (2018) 045003}, [\href{https://arxiv.org/abs/1803.04993}{{\ttfamily 1803.04993}}].

\bibitem{Harlow:2023hjb}
D.~Harlow and T.~Numasawa, \emph{{Gauging spacetime inversions in quantum gravity}},  \href{https://arxiv.org/abs/2311.09978}{{\ttfamily 2311.09978}}.

\bibitem{Li2003}
B.~Li, \emph{Real Operator Algebras}.
\newblock World Scientific, Singapore, 2003.

\bibitem{Li1995}
B.~Li, \emph{Real operator algebras},  in \emph{RIMS Kôkyûroku}, vol.~936, pp.~58--69, Research Institute for Mathematical Sciences, Kyoto University, 1996, \href{https://www.kurims.kyoto-u.ac.jp/~kyodo/kokyuroku/contents/pdf/0936-8.pdf}{https://www.kurims.kyoto-u.ac.jp/~kyodo/kokyuroku/contents/pdf/0936-8.pdf}.

\bibitem{Spradlin:2001pw}
M.~Spradlin, A.~Strominger and A.~Volovich, \emph{{Les Houches lectures on de Sitter space}},  in \emph{{Les Houches Summer School: Session 76: Euro Summer School on Unity of Fundamental Physics: Gravity, Gauge Theory and Strings}}, pp.~423--453, 10, 2001, \href{https://arxiv.org/abs/hep-th/0110007}{{\ttfamily hep-th/0110007}}.

\bibitem{Mottola:1984ar}
E.~Mottola, \emph{{Particle Creation in de Sitter Space}}, \href{http://dx.doi.org/10.1103/PhysRevD.31.754}{\emph{Phys. Rev. D} {\bfseries 31} (1985) 754}.

\bibitem{Allen:1985ux}
B.~Allen, \emph{{Vacuum States in de Sitter Space}}, \href{http://dx.doi.org/10.1103/PhysRevD.32.3136}{\emph{Phys. Rev. D} {\bfseries 32} (1985) 3136}.

\bibitem{Chakraborty:2023los}
T.~Chakraborty, J.~Chakravarty, V.~Godet, P.~Paul and S.~Raju, \emph{{Holography of information in de Sitter space}}, \href{http://dx.doi.org/10.1007/JHEP12(2023)120}{\emph{JHEP} {\bfseries 12} (2023) 120}, [\href{https://arxiv.org/abs/2303.16316}{{\ttfamily 2303.16316}}].

\bibitem{Chakraborty:2023yed}
T.~Chakraborty, J.~Chakravarty, V.~Godet, P.~Paul and S.~Raju, \emph{{The Hilbert space of de Sitter quantum gravity}}, \href{http://dx.doi.org/10.1007/JHEP01(2024)132}{\emph{JHEP} {\bfseries 01} (2024) 132}, [\href{https://arxiv.org/abs/2303.16315}{{\ttfamily 2303.16315}}].

\bibitem{Suh:2020lco}
S.~J. Suh, \emph{{Dynamics of black holes in Jackiw-Teitelboim gravity}}, \href{http://dx.doi.org/10.1007/JHEP03(2020)093}{\emph{JHEP} {\bfseries 03} (2020) 093}, [\href{https://arxiv.org/abs/1912.00861}{{\ttfamily 1912.00861}}].

\bibitem{Jafferis:2019wkd}
D.~L. Jafferis and D.~K. Kolchmeyer, \emph{{Entanglement Entropy in Jackiw-Teitelboim Gravity}},  \href{https://arxiv.org/abs/1911.10663}{{\ttfamily 1911.10663}}.

\bibitem{Blommaert:2018iqz}
A.~Blommaert, T.~G. Mertens and H.~Verschelde, \emph{{Fine Structure of Jackiw-Teitelboim Quantum Gravity}}, \href{http://dx.doi.org/10.1007/JHEP09(2019)066}{\emph{JHEP} {\bfseries 09} (2019) 066}, [\href{https://arxiv.org/abs/1812.00918}{{\ttfamily 1812.00918}}].

\bibitem{AliAhmad:2024wja}
S.~Ali~Ahmad, W.~Chemissany, M.~S. Klinger and R.~G. Leigh, \emph{{Quantum reference frames from top-down crossed products}}, \href{http://dx.doi.org/10.1103/PhysRevD.110.065003}{\emph{Phys. Rev. D} {\bfseries 110} (2024) 065003}, [\href{https://arxiv.org/abs/2405.13884}{{\ttfamily 2405.13884}}].

\bibitem{AliAhmad:2024vdw}
S.~Ali~Ahmad, W.~Chemissany, M.~S. Klinger and R.~G. Leigh, \emph{{Relational Quantum Geometry}},  \href{https://arxiv.org/abs/2410.11029}{{\ttfamily 2410.11029}}.

\bibitem{Susskind:2023rxm}
L.~Susskind, \emph{{A Paradox and its Resolution Illustrate Principles of de Sitter Holography}},  \href{https://arxiv.org/abs/2304.00589}{{\ttfamily 2304.00589}}.

\bibitem{Mertens:2022irh}
T.~G. Mertens and G.~J. Turiaci, \emph{{Solvable models of quantum black holes: a review on Jackiw\textendash{}Teitelboim gravity}}, \href{http://dx.doi.org/10.1007/s41114-023-00046-1}{\emph{Living Rev. Rel.} {\bfseries 26} (2023) 4}, [\href{https://arxiv.org/abs/2210.10846}{{\ttfamily 2210.10846}}].

\bibitem{Bagrets:2017pwq}
D.~Bagrets, A.~Altland and A.~Kamenev, \emph{{Power-law out of time order correlation functions in the SYK model}}, \href{http://dx.doi.org/10.1016/j.nuclphysb.2017.06.012}{\emph{Nucl. Phys. B} {\bfseries 921} (2017) 727--752}, [\href{https://arxiv.org/abs/1702.08902}{{\ttfamily 1702.08902}}].

\bibitem{Jafferis:2022wez}
D.~L. Jafferis, D.~K. Kolchmeyer, B.~Mukhametzhanov and J.~Sonner, \emph{{Jackiw-Teitelboim gravity with matter, generalized eigenstate thermalization hypothesis, and random matrices}}, \href{http://dx.doi.org/10.1103/PhysRevD.108.066015}{\emph{Phys. Rev. D} {\bfseries 108} (2023) 066015}, [\href{https://arxiv.org/abs/2209.02131}{{\ttfamily 2209.02131}}].

\bibitem{Maldacena:2016upp}
J.~Maldacena, D.~Stanford and Z.~Yang, \emph{{Conformal symmetry and its breaking in two dimensional Nearly Anti-de-Sitter space}}, \href{http://dx.doi.org/10.1093/ptep/ptw124}{\emph{PTEP} {\bfseries 2016} (2016) 12C104}, [\href{https://arxiv.org/abs/1606.01857}{{\ttfamily 1606.01857}}].

\bibitem{Maldacena:2016hyu}
J.~Maldacena and D.~Stanford, \emph{{Remarks on the Sachdev-Ye-Kitaev model}}, \href{http://dx.doi.org/10.1103/PhysRevD.94.106002}{\emph{Phys. Rev. D} {\bfseries 94} (2016) 106002}, [\href{https://arxiv.org/abs/1604.07818}{{\ttfamily 1604.07818}}].

\bibitem{gradshteyn2007table}
I.~S. Gradshteyn and I.~M. Ryzhik, \emph{Table of Integrals, Series, and Products}.
\newblock Academic Press, 7th~ed., 2007.

\bibitem{Kitaev:2018wpr}
A.~Kitaev and S.~J. Suh, \emph{{Statistical mechanics of a two-dimensional black hole}}, \href{http://dx.doi.org/10.1007/JHEP05(2019)198}{\emph{JHEP} {\bfseries 05} (2019) 198}, [\href{https://arxiv.org/abs/1808.07032}{{\ttfamily 1808.07032}}].

\bibitem{Yang:2018gdb}
Z.~Yang, \emph{{The Quantum Gravity Dynamics of Near Extremal Black Holes}}, \href{http://dx.doi.org/10.1007/JHEP05(2019)205}{\emph{JHEP} {\bfseries 05} (2019) 205}, [\href{https://arxiv.org/abs/1809.08647}{{\ttfamily 1809.08647}}].

\bibitem{Mertens:2017mtv}
T.~G. Mertens, G.~J. Turiaci and H.~L. Verlinde, \emph{{Solving the Schwarzian via the Conformal Bootstrap}}, \href{http://dx.doi.org/10.1007/JHEP08(2017)136}{\emph{JHEP} {\bfseries 08} (2017) 136}, [\href{https://arxiv.org/abs/1705.08408}{{\ttfamily 1705.08408}}].

\bibitem{Jones2009}
V.~F.~R. Jones, \emph{Von neumann algebras},  2009.

\bibitem{Sorce:2023fdx}
J.~Sorce, \emph{{Notes on the type classification of von Neumann algebras}}, \href{http://dx.doi.org/10.1142/S0129055X24300024}{\emph{Rev. Math. Phys.} {\bfseries 36} (2024) 2430002}, [\href{https://arxiv.org/abs/2302.01958}{{\ttfamily 2302.01958}}].

\bibitem{Parikh:2002py}
M.~K. Parikh, I.~Savonije and E.~P. Verlinde, \emph{{Elliptic de Sitter space: dS/Z(2)}}, \href{http://dx.doi.org/10.1103/PhysRevD.67.064005}{\emph{Phys. Rev. D} {\bfseries 67} (2003) 064005}, [\href{https://arxiv.org/abs/hep-th/0209120}{{\ttfamily hep-th/0209120}}].

\bibitem{SANCHEZ19871111}
N.~Sánchez, \emph{Quantum field theory and the “elliptic interpretation” of de sitter spacetime}, \href{http://dx.doi.org/https://doi.org/10.1016/0550-3213(87)90625-0}{\emph{Nuclear Physics B} {\bfseries 294} (1987) 1111--1137}.

\bibitem{GIBBONS1986497}
G.~Gibbons, \emph{The elliptic interpretation of black holes and quantum mechanics}, \href{http://dx.doi.org/https://doi.org/10.1016/S0550-3213(86)80022-0}{\emph{Nuclear Physics B} {\bfseries 271} (1986) 497--508}.

\bibitem{Lin:2022rbf}
H.~W. Lin, \emph{{The bulk Hilbert space of double scaled SYK}}, \href{http://dx.doi.org/10.1007/JHEP11(2022)060}{\emph{JHEP} {\bfseries 11} (2022) 060}, [\href{https://arxiv.org/abs/2208.07032}{{\ttfamily 2208.07032}}].

\bibitem{Xu:2024hoc}
J.~Xu, \emph{{Von Neumann Algebras in Double-Scaled SYK}},  \href{https://arxiv.org/abs/2403.09021}{{\ttfamily 2403.09021}}.

\bibitem{Jafferis:2022uhu}
D.~L. Jafferis, D.~K. Kolchmeyer, B.~Mukhametzhanov and J.~Sonner, \emph{{Matrix Models for Eigenstate Thermalization}}, \href{http://dx.doi.org/10.1103/PhysRevX.13.031033}{\emph{Phys. Rev. X} {\bfseries 13} (2023) 031033}, [\href{https://arxiv.org/abs/2209.02130}{{\ttfamily 2209.02130}}].

\bibitem{Marolf:2008it}
D.~Marolf and I.~Morrison, \emph{{Group Averaging of massless scalar fields in 1+1 de Sitter}}, \href{http://dx.doi.org/10.1088/0264-9381/26/3/035001}{\emph{Class. Quant. Grav.} {\bfseries 26} (2009) 035001}, [\href{https://arxiv.org/abs/0808.2174}{{\ttfamily 0808.2174}}].

\bibitem{Marolf:2008hg}
D.~Marolf and I.~A. Morrison, \emph{{Group Averaging for de Sitter free fields}}, \href{http://dx.doi.org/10.1088/0264-9381/26/23/235003}{\emph{Class. Quant. Grav.} {\bfseries 26} (2009) 235003}, [\href{https://arxiv.org/abs/0810.5163}{{\ttfamily 0810.5163}}].

\bibitem{Marolf:2000iq}
D.~Marolf, \emph{{Group averaging and refined algebraic quantization: Where are we now?}},  in \emph{{9th Marcel Grossmann Meeting on Recent Developments in Theoretical and Experimental General Relativity, Gravitation and Relativistic Field Theories (MG 9)}}, 7, 2000, \href{https://arxiv.org/abs/gr-qc/0011112}{{\ttfamily gr-qc/0011112}}.

\bibitem{Higuchi_19911}
A.~Higuchi, \emph{Quantum linearization instabilities of de sitter spacetime. i}, \href{http://dx.doi.org/10.1088/0264-9381/8/11/009}{\emph{Classical and Quantum Gravity} {\bfseries 8} (nov, 1991) 1961}.

\bibitem{Higuchi_19912}
A.~Higuchi, \emph{Quantum linearization instabilities of de sitter spacetime. ii}, \href{http://dx.doi.org/10.1088/0264-9381/8/11/010}{\emph{Classical and Quantum Gravity} {\bfseries 8} (nov, 1991) 1983}.

\bibitem{Stanford:2021bhl}
D.~Stanford, Z.~Yang and S.~Yao, \emph{{Subleading Weingartens}}, \href{http://dx.doi.org/10.1007/JHEP02(2022)200}{\emph{JHEP} {\bfseries 02} (2022) 200}, [\href{https://arxiv.org/abs/2107.10252}{{\ttfamily 2107.10252}}].

\bibitem{Almheiri:2024xtw}
A.~Almheiri, A.~Goel and X.-Y. Hu, \emph{{Quantum gravity of the Heisenberg algebra}}, \href{http://dx.doi.org/10.1007/JHEP08(2024)098}{\emph{JHEP} {\bfseries 08} (2024) 098}, [\href{https://arxiv.org/abs/2403.18333}{{\ttfamily 2403.18333}}].

\bibitem{Rahman:2024iiu}
A.~A. Rahman and L.~Susskind, \emph{{$p$-Chords, Wee-Chords, and de Sitter Space}},  \href{https://arxiv.org/abs/2407.12988}{{\ttfamily 2407.12988}}.

\bibitem{Choi:2019bmd}
C.~Choi, M.~Mezei and G.~S\'arosi, \emph{{Exact four point function for large $q$ SYK from Regge theory}}, \href{http://dx.doi.org/10.1007/JHEP05(2021)166}{\emph{JHEP} {\bfseries 05} (2021) 166}, [\href{https://arxiv.org/abs/1912.00004}{{\ttfamily 1912.00004}}].

\bibitem{Streicher:2019wek}
A.~Streicher, \emph{{SYK Correlators for All Energies}}, \href{http://dx.doi.org/10.1007/JHEP02(2020)048}{\emph{JHEP} {\bfseries 02} (2020) 048}, [\href{https://arxiv.org/abs/1911.10171}{{\ttfamily 1911.10171}}].

\bibitem{Lin:2023trc}
H.~W. Lin and D.~Stanford, \emph{{A symmetry algebra in double-scaled SYK}}, \href{http://dx.doi.org/10.21468/SciPostPhys.15.6.234}{\emph{SciPost Phys.} {\bfseries 15} (2023) 234}, [\href{https://arxiv.org/abs/2307.15725}{{\ttfamily 2307.15725}}].

\bibitem{Alonso-Monsalve:2024oii}
E.~Alonso-Monsalve, D.~Harlow and P.~Jefferson, \emph{{Phase space of Jackiw-Teitelboim gravity with positive cosmological constant}},  \href{https://arxiv.org/abs/2409.12943}{{\ttfamily 2409.12943}}.

\bibitem{Nanda:2023wne}
K.~K. Nanda, S.~K. Sake and S.~P. Trivedi, \emph{{JT gravity in de Sitter space and the problem of time}}, \href{http://dx.doi.org/10.1007/JHEP02(2024)145}{\emph{JHEP} {\bfseries 02} (2024) 145}, [\href{https://arxiv.org/abs/2307.15900}{{\ttfamily 2307.15900}}].

\bibitem{Levine:2022wos}
A.~Levine and E.~Shaghoulian, \emph{{Encoding beyond cosmological horizons in de Sitter JT gravity}}, \href{http://dx.doi.org/10.1007/JHEP02(2023)179}{\emph{JHEP} {\bfseries 02} (2023) 179}, [\href{https://arxiv.org/abs/2204.08503}{{\ttfamily 2204.08503}}].

\bibitem{Maldacena:2019cbz}
J.~Maldacena, G.~J. Turiaci and Z.~Yang, \emph{{Two dimensional Nearly de Sitter gravity}}, \href{http://dx.doi.org/10.1007/JHEP01(2021)139}{\emph{JHEP} {\bfseries 01} (2021) 139}, [\href{https://arxiv.org/abs/1904.01911}{{\ttfamily 1904.01911}}].

\bibitem{Held:2024rmg}
J.~Held and H.~Maxfield, \emph{{The Hilbert space of de Sitter JT: a case study for canonical methods in quantum gravity}},  \href{https://arxiv.org/abs/2410.14824}{{\ttfamily 2410.14824}}.

\bibitem{Liu:2018jhs}
J.~Liu, E.~Perlmutter, V.~Rosenhaus and D.~Simmons-Duffin, \emph{{$d$-dimensional SYK, AdS Loops, and $6j$ Symbols}}, \href{http://dx.doi.org/10.1007/JHEP03(2019)052}{\emph{JHEP} {\bfseries 03} (2019) 052}, [\href{https://arxiv.org/abs/1808.00612}{{\ttfamily 1808.00612}}].

\bibitem{Derkachev2023}
S.~Derkachev and A.~Ivanov, \emph{Racah coefficients for the group sl(2,r)}, \href{http://dx.doi.org/10.1007/s10958-023-06681-x}{\emph{Journal of Mathematical Sciences} {\bfseries 275} (2023) 289--298}.

\bibitem{Lam:2018pvp}
H.~T. Lam, T.~G. Mertens, G.~J. Turiaci and H.~Verlinde, \emph{{Shockwave S-matrix from Schwarzian Quantum Mechanics}}, \href{http://dx.doi.org/10.1007/JHEP11(2018)182}{\emph{JHEP} {\bfseries 11} (2018) 182}, [\href{https://arxiv.org/abs/1804.09834}{{\ttfamily 1804.09834}}].

\end{thebibliography}\endgroup

\end{document}